\documentclass[a4paper, 11pt, oneside]{book}
\usepackage[DIV=12,BCOR=0.0cm,headinclude=true]{typearea}

\usepackage[titletoc]{appendix}
\usepackage[english]{babel}
\usepackage[utf8]{inputenc}
\usepackage{prettyref}
\usepackage[backend=bibtex,style=phys, biblabel=brackets]{biblatex}
\usepackage{amsmath,amssymb}
\usepackage{color}
\usepackage{float}
\usepackage{graphicx}
\usepackage{mathtools}
\usepackage[dvipsnames]{xcolor}
\usepackage[font=small,labelfont=bf,labelsep=space]{caption}
\usepackage{wrapfig}
\usepackage{subcaption}
\usepackage{array}
\usepackage{dsfont}
\usepackage{makecell}
\usepackage[babel]{csquotes}
\usepackage[hidelinks,colorlinks,allcolors=blue]{hyperref} 

\usepackage{fancyhdr}
\renewcommand{\chaptermark}[1]%
         {\markboth{\thechapter.\ #1}{}}
\renewcommand{\sectionmark}[1]%
         {\markright{\thesection\ #1}}
\usepackage{accents}
\newcommand\thickbar[1]{\accentset{\rule{.5em}{.5pt}}{#1}}
\newcommand{\F}{\mathcal{F}}
\newcommand{\tr}{\mathrm{Tr}} 
\newcommand{\rme}{\mathrm{e}} 
\newcommand{\rmd}{\mathrm{d}} 
\newcommand{\Tr}{\mbox{Tr}}

\newcommand{\bra}[1]{\mbox{$\langle #1 |$}}
\newcommand{\ket}[1]{\mbox{$| #1 \rangle$}}

\newcommand{\bd}[1]{\boldsymbol{#1}}
\newcommand{\wb}{\bd{w}}

\newcommand{\gh}{\hat{\gamma}}

\newcommand*\xbar[1]{\hbox{\vbox{
       \hrule height 0.6pt 
       \kern0.3ex
       \hbox{%
         \kern-0.2em
         \ensuremath{#1}%
         \kern 0.0em
         }}}}

\newcommand*\xxbar[1]{\hbox{\vbox{
       \hrule height 0.6pt 
       \kern0.3ex
       \hbox{%
         \kern-0.0em
         \ensuremath{#1}%
         \kern 0.0em
         }}}}

\newcommand{\Fw}{\mathcal{F}_{\!\wb}}
\newcommand{\Fbw}{\,\xbar{\mathcal{F}}_{\!\wb}}

\newcommand{\Ew}{\mathcal{E}^N\!(\wb)}
\newcommand{\ew}{\mathcal{E}^1_N(\wb)}

\newcommand{\Ebw}{\,\xxbar{\mathcal{E}}^N\!(\wb)}
\newcommand{\ebw}{\,\xxbar{\mathcal{E}}^1_N(\wb)}

\newcommand{\ebwS}{\,\scriptsize{\xbar{\mathcal{E}}}\normalsize^1_N\hspace{-0.3mm}(\wb)}
\newcommand{\EbwS}{\,\scriptsize{\xbar{\mathcal{E}}}\normalsize^N\hspace{-0.7mm}(\wb)}

\begin{document}

\begin{titlepage}
\begin{center}
\begin{LARGE}
\textsc{Ludwig-Maximilians-Universit{\"a}t M{\"u}nchen}\\
\end{LARGE}
\begin{Large}
\textsc{Faculty of Physics}\\
\end{Large}
\vspace{1cm}
\begin{LARGE}
\textsc{Master thesis}\\
\end{LARGE}
\vspace{1cm}
    {\parindent0cm
    \rule{\linewidth}{.5ex}}  
\begin{flushright}
    \centering
\begin{Huge}
\textbf{Reduced Density Matrix Functional Theory for Bosons: \\ Foundations and Applications}\\
\end{Huge}
\end{flushright}
\rule{\linewidth}{.5ex}
\vspace{0cm}
\end{center}
\begin{center}
\vspace{0cm}
\begin{LARGE}
Julia Liebert\\
\end{LARGE}
\vspace{1cm}
\includegraphics[width=2in]{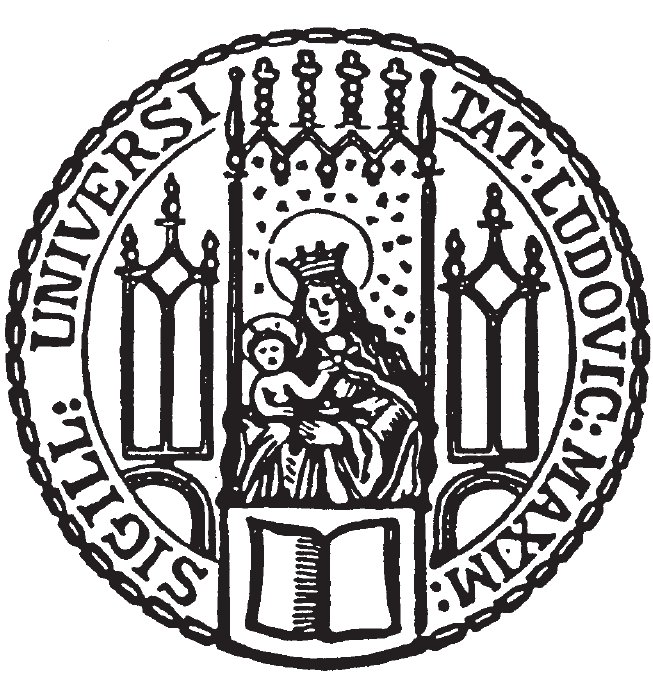}\\
\vspace{1cm}
\begin{Large}
Supervised by \\
Dr. Christian Schilling\\
\end{Large}
\vspace{1cm}
\begin{Large}
Munich, February 15, 2021
\end{Large}
\end{center}
\end{titlepage}

\clearpage{\pagestyle{empty}\cleardoublepage}
\cleardoublepage
\begin{titlepage}
\begin{center}
\begin{LARGE}
\textsc{Ludwig-Maximilians-Universit{\"a}t M{\"u}nchen}\\
\end{LARGE}
\begin{Large}
\textsc{Fakult{\"a}t f{\"u}r Physik}\\
\end{Large}
\vspace{1cm}
\begin{LARGE}
\textsc{Masterarbeit}\\
\end{LARGE}
\vspace{1cm}
    {\parindent0cm
    \rule{\linewidth}{.5ex}}  
\begin{flushright}
    \centering
\begin{Huge}
\textbf{Reduzierte Dichtematrix-Funktionaltheorie f{\"u}r Bosonen: \\ Grundlagen und Anwendungen}\\
\end{Huge}
\end{flushright}
\rule{\linewidth}{.5ex}
\vspace{0cm}
\end{center}
\begin{center}
\vspace{0cm}
\begin{LARGE}
Julia Liebert\\
\end{LARGE}
\vspace{1cm}
\includegraphics[width=2in]{lmu_siegel.pdf}\\
\vspace{1cm}
\begin{Large}
Betreut von \\
Dr. Christian Schilling\\
\end{Large}
\vspace{1cm}
\begin{Large}
M{\"u}nchen, den 15.~Februar 2021
\end{Large}
\end{center}
\end{titlepage}

\clearpage{\pagestyle{empty}\cleardoublepage}
\cleardoublepage

\pagenumbering{roman}
\setcounter{page}{1}

\section*{List of publications}

This master thesis is based on the following publications: 
\begin{enumerate}
\item[{[P1]}] J. Liebert, and C. Schilling, "Functional theory for Bose-Einstein condensates", \href{https://link.aps.org/doi/10.1103/PhysRevResearch.3.013282}{Phys. Rev. Research \textbf{3}, 013282 (2021).}
\item[{[P2]}] J. Liebert, F. Castillo, J.-F. Labb\'e, and C. Schilling, "Foundation of one-particle reduced density matrix functional theory for excited states", \href{https://doi.org/10.1021/acs.jctc.1c00561}{J. Chem. Theory Comput. \textbf{18}, 124 (2022).}
\item[{[P3]}] F. Castillo, J.-F. Labb\'e, J. Liebert, A. Padrol, E. Philippe and C. Schilling, "An effective solution to convex $1$-body $N$-representability", \href{https://arxiv.org/abs/2105.06459}{arXiv:2105.06459 (2021).}
\item[{[P4]}] J. Liebert, and C. Schilling, "Functional theory for excitations in boson systems", \\ \href{https://arxiv.org/abs/2204.12715}{arXiv:2204.12715 (2022).}
\end{enumerate}

\clearpage{\cleardoublepage}
\cleardoublepage

\tableofcontents

\clearpage{\pagestyle{empty}\cleardoublepage}
\cleardoublepage

\pagenumbering{arabic}

\setcounter{page}{1}

\chapter{Introduction}

According to quantum mechanics, all information about a quantum mechanical system of $N$ particles is contained in its many-body wave function, which is an exact solution to the Schrödinger equation. However, for wave function based methods, an analytic solution is only known for a small number of systems, and due to the exponential scaling of the underlying Hilbert space with the particle number $N$, even solving the Schrödinger equation by numerical means is only feasible for very small system sizes. To conveniently describe many-body and macroscopic systems, as they naturally appear in condensed matter physics, one resorts to different approaches like dynamical mean-field theory (DMFT) \cite{Georges96, Vollhardt12}, density matrix renormalization group studies (DMRG) \cite{Schollwoeck05, Schollwoek10}, or density functional theory (DFT) \cite{HK, Kohn65}, to only name a few. However, one is often only interested in the expectation values of observables, which do not require full knowledge of the $N$-particle wave function. The exploitation of this observation drastically simplifies the theoretical description of many-body systems, and in particular, the solution to the ground state problem. Furthermore, different fields of physics are usually characterized by a fixed pair interaction $\hat{W}$. Some prominent examples are the Coulomb interaction between electrons or (effective) hard-core interactions in ultracold atomic gases. In the context of DFT, which is widely used in quantum chemistry, it follows that every ground state observable can be expressed as a functional of the ground state density \cite{HK}. The Hohenberg-Kohn theorem \cite{HK} provides the foundation of DFT, but its success is based on the Kohn-Sham formalism \cite{Kohn65}. The idea behind Kohn-Sham DFT is to replace the interacting system with an artificial non-interacting system yielding the same ground state density. This requires the introduction of a so-called Kohn-Sham potential which is hard to predict due to the lack of its physical interpretation. Besides, DFT usually fails to describe strongly correlated systems of electrons since those systems cannot be described by a single Slater determinant and require fractional occupation numbers arising from superpositions of different Slater determinants. These examples already indicate several limitations of DFT. 

A natural extension of DFT is to include the full one-particle reduced density matrix $\gh$ (1RDM) rather than only the particle density, which is the diagonal of the 1RDM in spatial representation. Moreover, for a fixed pair interaction only the one-particle Hamiltonian $\hat{h}$ can be varied and the 1RDM is, in turn, the conjugate variable of $\hat{h}$.  The corresponding ground state theory is then called reduced density matrix functional theory (RDMFT). While both functional theories, RDMFT and DFT, abandon the complexity of the $N$-particle wave function, only RDMFT is capable of recovering quantum correlations exactly. For a $d$-dimensional one-particle Hilbert space, this results in $d^2$ degrees of freedom instead of $d$ as for the particle density, leading to a slower convergence of numerical algorithms. Nevertheless, RDMFT has many crucial advantages compared to DFT. First, since it involves the 1RDM as its natural variable, it provides direct access to occupation numbers and explicitly allows for fractional occupation numbers. As a result, RDMFT is well-suited to describe strongly correlated systems from a conceptual point of view, in contrast to DFT. In addition, the exact description of the kinetic energy through the 1RDM is known, whereas its functional dependence on the particle density has to be approximated in DFT. Combining these different aspects leads to the conclusion that RDMFT has a great potential to replace DFT in the future. However, this requires a lot of further method development to improve its viability. Furthermore, RDMFT was only developed for fermions in the past while bosonic quantum systems were rather neglected. This is surprising because bosons play an important role in quantum physics. The most prominent example is Bose-Einstein condensation (BEC), which is one of the most fascinating quantum phenomena. Einstein \cite{Einstein} predicted the existence of BEC, based on a seminal letter by Bose\cite{Bose}, already in 1925. Moreover, the realization of BEC for ultracold atoms in 1995 \cite{Anderson1995,Ketterle1995,Bradley1995} has led to a renewed interest. The development of the respective field of ultracold gases has opened new research avenues and revealed new phenomena such as the crossover from BEC-superfluidity to BCS-superconductivity \cite{Greiner2003,Bartenstein2004,Zwierlein2004,Bourdel2004}. Motivated by the significance of such bosonic quantum systems, the mathematical foundation for a bosonic RDMFT was first provided in Ref.~\cite{Benavides20} in 2020. 

In this thesis, we identify BEC as an ideal starting point to further develop a bosonic RDMFT while, at the same, time acquiring new and remarkable insights into BEC itself. According to the Penrose-Onsager criterion \cite{Penrose1956}, BEC is present whenever the largest eigenvalue  of the 1RDM is proportional to the total particle number $N$, providing the connection between functional theories and BEC.
While bosonic RDMFT would potentially be the ideal theory for describing BECs (including the regime of fractional BEC as well as  quasicondensation \cite{Popov1972}), RDMFT of course does not trivialize the ground state problem. It is a fundamental challenge in RDMFT to construct reliable approximations of the universal interaction functional $\F(\gh)$, determine its leading order behaviour in certain physically regimes or its exact form for simplified model systems. Results along any of those lines are typically quite rare, however, and their significance for the general development of RDMFT could hardly be overestimated. The latter is due to the fact that improved functional approximations often build upon previous ones (see, e.g., \cite{ML,PirisUg14,PG16} and references therein). In fermionic RDMFT, the elementary Hartree-Fock functional \cite{LiebHF} can be seen as the first level of the hierarchy of functional approximations. It has directly led to the celebrated M\"uller functional \cite{Mueller84,Buijse02} which in turn inspired more elaborated functional approximations \cite{ML,PG16}. In bosonic RDMFT even the analogue of the Hartree-Fock functional has not been established yet.
It is therefore one of the two main goals of this thesis to initiate and establish this novel bosonic RDMFT by deriving such a first-level functional in a comprehensive way. Due to the significance of BEC, we identify systems of interacting bosons in the BEC regime as the starting point for the hierarchy of functional approximations. It is worth noticing that this regime as described by the Bogoliubov theory \cite{Bogoliubov47} covers a large range of systems, including in particular the experimentally realized dilute ultracold Bose gases as well as charged bosons in the high density regime. The respective first-level functional would not only serve as a starting point for the development of further functional approximations but its concrete form will also reveal a remarkable new physical concept. Namely, the gradient of the universal functional will be found to diverge repulsively in the regime of almost complete BEC, preventing quantum systems of interacting bosons from ever reaching complete condensation. This BEC force will thus provide an alternative explanation for quantum depletion which is most fundamental because it emerges from the geometry of density matrices and the properties of the partial trace, independently from the pair-interaction between the bosons and other system-specific features. 

So far, we solely focused on the ground state problem. However, the accurate description of excited states, and in particular the energy gap between the ground state and first excited state, are of immense interest in many-body and solid-state physics. One promising approach in DFT is time-dependent DFT based on a time-dependent extension \cite{Runge84} of the Hohenberg-Kohn theorem \cite{HK}. Alternatively, Gross, Oliviera and Kohn introduced an ensemble DFT to work with excited states \cite{Gross88_1,Gross88_2,Oliveira88} in 1988, which has drawn renewed interest during the last few years \cite{F15B,YPBU17,GP17,SB18,GKP18,GP19, KF19,Fromager2020-DD,Loos2020-EDFA,GSP2020}. However, an ensemble RDMFT for excited states in fermionic quantum systems was first proposed this year \cite{Schilling21}, and it is completely missing for bosons so far. It is thus the second main goal of this thesis, besides deriving a first-level ground state functional for bosonic RDMFT, to propose a bosonic RDMFT for excited states. Clearly, the more general ensemble RDMFT for excited states has to contain the ground state RDMFT as a special case. The new $\bd\omega$-ensemble RDMFT for excites states is based on the combination of a generalization of the Rayleigh-Ritz variational principle and the constrained search formalism, similar to ground state RDMFT. As in ground state RDMFT, BEC serves as an ideal starting point to determine a first-level universal functional for excited states in a bosonic quantum system. 


This thesis contains three main chapters: In Chapter 2, we introduce all relevant theoretical concepts of RDMFT in a comprehensive way. In doing so, we provide a solid mathematical foundation and emphasize the differences between fermionic and bosonic RDMFT because both aspects are essential for the following two chapters. In the third chapter, we apply RDMFT to BEC and derive the universal functional in the regime close to complete condensation. We then consider different concrete systems to explain how RDMFT works and illustrate the universal functional. Further, we derive the new concepts of a BEC force providing an alternative and most fundamental explanation for quantum depletion. In Chapter 4, we establish a novel method, namely a bosonic ensemble RDMFT for excited states. Remarkably, we obtain a hierarchy of non-trivial linear constraints in form of inequalities on the bosonic occupation numbers interpreted as generalized exclusion principles for bosons. 

\chapter{Foundations of RDMFT\label{ch:RDMFT}}

The goal of this chapter is to introduce the theoretical framework of RDMFT and, in particular, its bosonic version. The underlying mathematical concepts presented in Sec.~\ref{sec:Math-gsRDMFT} are crucial to solve conceptual problems in the subsequent sections and develop RDMFT as a method further. Indeed, these mathematical concepts also provide the foundation for the novel bosonic RDMFT for excited states presented in Ch.~\ref{ch:w-RDMFT}. Following the name one-particle reduced density matrix functional theory, RDMFT involves the one-particle reduced density matrix (1RDM) as its natural variable. We introduce density matrices and their respective sets while focusing on their role in RDMFT, in Sec.~\ref{sec:density}. Based on the characterization of different sets of 1RDMs, two fundamental problems occur. These are the $N$-representability problem discussed in Sec.~\ref{sec:Nrepr} and the pure state $v$-representability problem. The latter arises from Gilbert's original formulation of RDMFT explained in Sec.~\ref{sec:HK} and can be circumvented by the constrained search formalism discussed in Sec.~\ref{sec:Levy}. Since this thesis is concerned with bosonic RDMFT, a particular emphasis lies on the differences for bosons compared to fermions. Moreover, we apply RDMFT to homogeneous Bose gases in Sec.~\ref{sec:RDMFT_hom} and discuss simplifications due to several symmetries in   Sec.~\ref{sec:symm}. 

\section{Mathematical Preliminaries \label{sec:Math-gsRDMFT}}

In this section, we recap the most important concepts of convex analysis that are fundamental to understand the underlying concepts of RDMFT presented in this chapter. We first recall the basic terminology, including affine set, convex sets, convex functions, and emphasize the connections between them. Further, the concepts of the duality correspondence for convex sets and biconjugation are important to gain a deeper understanding of the minimization over the set of density matrices discussed in Sec.~\ref{sec:density} and the constrained search formalism in Sec.~\ref{sec:Levy}.  

\subsection{Basic terminology\label{subsec:terminology}}

The purpose of this section is, without going into details, to review some basic terminology of convex analysis which will be used throughout this thesis. For a comprehensive discussion of convex analysis we refer the reader to the textbook Ref.~\cite{rockafellar2015}. 

An \textit{affine combination} of vectors $v_1, ..., v_k\in \mathbb{R}^d$ is a linear combination $\sum_{j=1}^k\theta_jv_j$ with $\theta_j\in \mathbb{R}$ such that $\sum_j\theta_j=1$. Note that the coefficients $\theta_j$ can be positive or negative. For example, all affine combinations of two distinct vectors define a straight line through them, whereas all linear combinations of the these two vectors would define a two-dimensional plane. 
A set $S\subseteq\mathbb{R}^d$ is called an \textit{affine set} if every affine combination of elements within the set also belongs to it. Equivalently, an affine set contains the entire line $y = \theta x_1 + (1-\theta)x_2$ with $\theta\in \mathbb{R}$ through any two distinct points $x_1$, $x_2\in S$.
The \textit{affine hull} of a set $S\subseteq \mathbb{R}^d$ is defined as the set of all affine combinations of its elements
\begin{equation}
\mathrm{aff}(S) = \left\{\sum_{j=1}^k\theta_jx_j\Big\vert \,k>0,\theta_j\in \mathbb{R}, x_j\in S, \sum_{j=1}^k\theta_j=1\right\}\,.
\end{equation} 
Equivalently, the affine hull of a set $S\subseteq\mathbb{R}^d$ can be defined as the intersection of all \textit{affine subspaces} containing the set $S$ where an affine subspace is nothing else than a translated vector space. 

A set $S\subseteq\mathbb{R}^d$ is called a \textit{convex set} if it contains the linear combination $y=qx_1 + (1-q)x_2$ with $0\leq q\leq 1$ between any two points $x_1, x_2\in S$ and is therefore related to the affine set by restricting to the line segment between the two distinct points rather than containing the full line. Clearly, every affine set is also convex. Also note that halfspaces are convex, whereas hyperplanes are both, affine and convex. A supporting hyperplane of a convex set $S$ is a hyperplane which has $S$ in one of its halfspaces and contains at least one boundary point of $S$. 

The \textit{convex hull}, denoted by $\mathrm{conv}(S)$, is defined as the set of all \textit{convex combinations} of points in $S$ where a convex combination is a linear combination $\sum_{j=1}^kq_jx_j$ of elements $x_j\in S$ with $k>0$, $0\leq q_i\leq 1$ and $\sum_jq_j=1$. Carath\'{e}odory's theorem \cite{rockafellar2015} states then that for a set $S\subseteq\mathbb{R}^d$, every element of the convex hull $\mathrm{conv}(S)$ can be written as a convex combination of $d+1$ points in $S$. Equivalent to the first definition, $\mathrm{conv}(S)$ is given by the smallest intersection of all convex subsets in $\mathbb{R}^d$ containing $S$ and thus it is the smallest convex subset which contains $S$. An extremal point of a convex subset $S\subseteq \mathbb{R}^d$ is a point $x\in S$ which is not an interior point of any line segment fully contained in $S$ and can therefore not be written as a convex combination of other elements in $S$. 
\begin{figure}[h]
\centering
\begin{subfigure}[t]{.49\textwidth}
\centering
\includegraphics[width=0.7\linewidth]{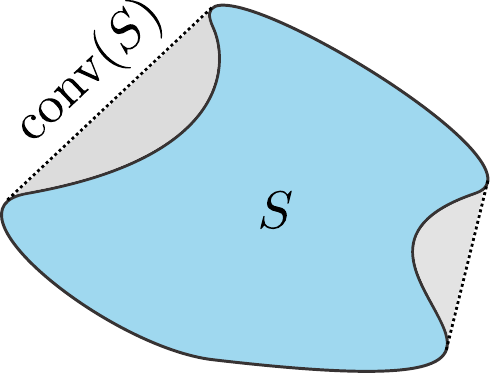}
\end{subfigure}
\begin{subfigure}[t]{.49\textwidth}
\centering
\includegraphics[width=0.7\linewidth]{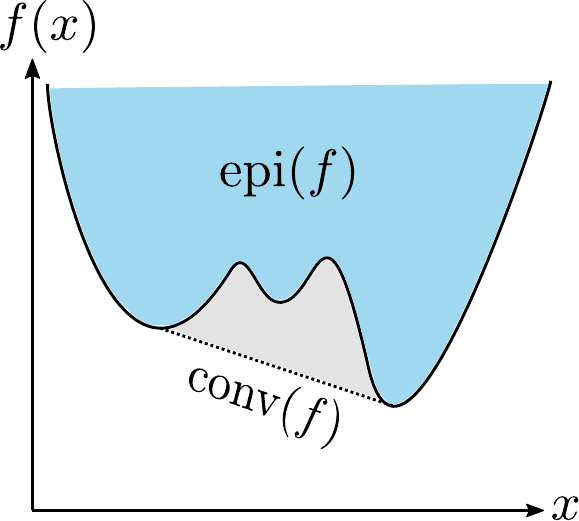}
\end{subfigure}
\medskip
\begin{minipage}[t]{\textwidth}
\caption{Left: A non-convex set $S$ (blue) and its extension to the convex hull $\mathrm{conv}(S)$ (gray and blue) illustrated by the dotted lines. Right: The epigraph $\mathrm{epi}(f)$ (blue) of the non-convex function $f(x)$ is a non-convex set. Its convex hull (gray and blue) also determines the lower convex envelope of $f(x)$.  \label{fig:conv_epi}}
\end{minipage}
\end{figure}
In the left panel of Fig.~\ref{fig:conv_epi}, we illustrate a non-convex set $S$ and its convex hull $\mathrm{conv}(S)$ which is the given by convex set containing $S$. 

The \textit{Heine-Borel theorem} states that a subset $S\subseteq \mathbb{R}^d$ is compact if and only if it is closed and bounded and according to the \textit{Kein-Milman theorem} every compact convex set $S\subseteq \mathbb{R}^d$ is given by the convex hull of its extremal elements.

Let $S\subseteq\mathbb{R}^d$ be a convex set. A function $f: S\to \mathbb{R}$ is called a \textit{convex function} if for any two points $x_1, x_2\in S$ the relation
\begin{equation}
f\left(qx_1+(1-q)x_2\right)\leq qf(x_1) + (1-q)f(x_2)\,, \quad 0\leq q\leq 1
\end{equation}
holds. 
The connection between convex sets and convex functions is provided by the \textit{epigraph} where the epigraph $\mathrm{epi}(f)$ of a function $f:\mathbb{R}^d\to\mathbb{R}$ is a subset of $\mathbb{R}^{d+1}$ defined by
\begin{equation}\label{eq:epi}
\mathrm{epi}(f) \equiv \{(x_1, x_2)\in \mathbb{R}^d \times \mathbb{R}|\,x_1\in \mathrm{dom}(f), f(x_1)\leq x_2\}\,.
\end{equation}
Note that this definition does not require the function $f$ to be convex. However, it follows that a function $f$ is convex if and only if its epigraph $\mathrm{epi}(f)$ is a convex set. In addition, the function $f$ can be reconstructed from its epigraph by determining for all $x_1 \in S$ the smallest element $x_2$ of all tuples $(x_1, x_2)\in \mathrm{epi}(f)$. 
Moreover, the \textit{lower convex envelope} $\mathrm{conv}(f)$ of a function $f$ is given by the convex function which corresponds to the convex hull of the epigraph of $f$ and thus it is defined by
\begin{equation}\label{eq:conv_fx}
\mathrm{conv}(f)(x) \equiv \inf\left\{\sum_jq_jf(x_j)\,\Big\vert\,\sum_jq_jx_j=x, \sum_jq_j=1, q_j\geq 0\right\}\,.
\end{equation}
Equivalently, the lower convex envelope is given by the largest convex function $g$ for which $g(x)\leq f(x)\,\,\forall\,x\in \mathrm{dom}(f)$ holds. The strong connection between the lower convex envelope of a function and the convex hull of its epigraph is illustrated in Fig.~\ref{fig:conv_epi}. Recall that the epigraph of the non-convex function $f(x)$ is defined by Eq.~\eqref{eq:epi}. The convex hull of the epigraph $\mathrm{epi}(f)$ and the lower convex envelope of $f(x)$ are related through $\mathrm{epi}(\mathrm{conv}(f)) = \mathrm{conv}(\mathrm{epi}(f))$ and determine each other. Moreover, the function $f(x)$ in Fig.~\ref{fig:conv_epi} emphasizes that the second derivative of a function is not sufficient to determine whether it is convex on its full domain and thus equal to its lower convex envelope or not. 

\subsection{Legendre-Fenchel transformation and biconjugation\label{subsec:LF}}

The Legendre-Fenchel transformation is an important example of a duality consideration where two mathematical objects are paired with each other leading to a strong correspondence between them. The following section summarizes the most important aspects of conjugation which are required in Sec.~\ref{sec:Levy} to establish a connection between the pure and ensemble functionals in RDMFT.

Let $f:\mathbb{R}^d\to (-\infty, \infty]$ be an extended-real-valued function which is not necessarily convex. The Legendre-Fenchel conjugate of $f$ denoted by $f^*: \mathbb{R}^d \to [-\infty, \infty]$ is defined as 
\begin{equation}\label{eq:LF}
f^*(y) = \sup_{x\in \mathbb{R}^d}\left[\langle y, x\rangle - f(x)\right]\,.
\end{equation}
Further, suppose that the domain of $f$ is non-empty, $\mathrm{dom}(f)\neq \emptyset$, and that $f$ is a proper function, which means that there exists at least one $z\in \mathbb{R}^d$ such that $f(z)<\infty$. For all $\thickbar{x}\notin \mathrm{dom}(f)$, we set $f(\bar{x})=\infty$ yielding $\langle y, \thickbar{x}\rangle - f(\thickbar{x})= -\infty$ for any $y\in \mathbb{R}^d$. As a result, we can restrict the supremum in Eq.~\eqref{eq:LF} to all $x\in \mathrm{dom}(f)$. Then, the conjugate $f^*: \mathbb{R}^d\to (-\infty, \infty]$ is a convex function \cite{rockafellar2015}. This statement can be easily proven by noticing that for any $y\in \mathbb{R}^d$
\begin{equation}
f^*(y) = \sup_{x\in \mathrm{dom}(f)}\left[\langle y, x\rangle - f(x)\right] \equiv \sup_{x\in\mathrm{dom}(f)}[F_x(y)]\,,
\end{equation}
where $F_x(y)\equiv \langle y, x\rangle - f(x)$ denotes an affine function in which $f(x)$ takes the role of the constant. Since the supremum over a family of affine functions has to be convex, the conjugate $f^*$ is always a convex function.

The biconjugation of a function $f:\mathbb{R}^d\to (-\infty, \infty]$ is defined as \cite{rockafellar2015}
\begin{equation}\label{eq:biconjugate}
f^{**}(x)\equiv (f^*)^*(x) = \mathrm{cl}\left(\mathrm{conv}(f)\right)\,.
\end{equation}
For a continous function, the closure operation can be omitted. This further implies that
\begin{equation}\label{eq:FssleqF}
f^{**}(x) \leq f(x)\quad \forall x\in \mathbb{R}^d\,.
\end{equation}
Moreover, the equality in Eq.~\eqref{eq:FssleqF} holds whenever the function $f$ is convex and lower semicontinuous. We will return to Eq.~(\ref{eq:LF}-\ref{eq:FssleqF}) in Sec.~\ref{sec:Levy}.

\subsection{Duality correspondence for convex sets\label{sec:duality}}

The support function associated with a set $S\subset \mathbb{R}^d$ is defined by
\begin{equation}\label{eq:supportfct}
\sigma_S(k) \equiv \mathrm{sup}\left(\{\langle k, x\rangle|\,x\in S\}\right)\,,\quad k\in\mathbb{R}^d
\end{equation}
and describes how the maximum of a function changes if $k$ is varied. However, minimizations can also be described by the support function through the following relation:
\begin{equation}\label{eq:supportmin}
\inf\left(\{\langle k, x\rangle|\,x\in S\}\right) = -\sup \left(\{\langle -k, x\rangle|\,x\in S\}\right) = -\sigma_S(-k)\,.
\end{equation}
It follows that even if $S$ is not convex, the support function $\sigma_S$ is always convex and $\sigma_S(k)=\sigma_{\mathrm{conv}(S)}(k)$ holds. Moreover, the support function is related to the indicator function 
\begin{equation}
\delta_S(x) \equiv \begin{cases}
  0  ,& \text{if } x\in S\\
    \infty,  & \text{otherwise}
\end{cases}
\end{equation}
through the Legendre-Fenchel transformation since for a fixed $k\in\mathbb{R}^d$ we have
\begin{equation}
\begin{split}
\delta_S^*(k) &\equiv \mathrm{sup}\left(\{\langle k, x\rangle - \delta_S(x)|\,x\in \mathbb{R}^d\}\right)\\\
&= \mathrm{sup}\left(\{\langle k, x\rangle - \delta_S(x)|\,x\in \mathrm{dom}(\delta_S)=S\}\right)\\\
&= \mathrm{sup}\left(\{\langle k, x\rangle|\,x\in S\}\right)\\\
&= \sigma_S(k)\,,
\end{split}
\end{equation}
which proves that the Legendre-Fenchel conjugate of the indicator function is the support function. 

In case $S\subset \mathbb{R}^d$ is a convex and compact set, we obtain a one-to-one correspondence between the indicator and the support function
\begin{equation}
(\sigma_S)^* = (\delta_S)^{**} = \delta_S\,.
\end{equation}
Therefore, we can use the support function as an alternative representation of a convex compact set.
In other words, this also means that a convex compact set $S\subset \mathbb{R}^d$ can be characterized equivalently through all points $x\in S$ or the intersection of all supporting halfspaces containing $S$ entirely. In the following sections, we often have to deal with the description of convex sets in the context of density matrices or reduced density matrices, where this duality consideration is applicable and strongly connected to the energy minimization.

\section{Density matrices \label{sec:density}}

According to quantum mechanics, all information about a quantum system is contained in its states. We distinguish between pure and mixed quantum states. A pure quantum state can be represented by a ray in a Hilbert space $\mathcal{H}$, which is a complete vector space of the complex numbers $\mathbb{C}$, i.e.~$\mathcal{H}\cong \mathbb{C}^d$ for $\mathrm{dim}(\mathcal{H})=d$, with scalar inner product $\langle \cdot, \cdot\rangle$. A ray is an equivalence class of vectors in $\mathcal{H}$ such that $\ket{v}\sim \ket{w}$ if and only if $\ket{v}=\lambda\ket{w}$ for some $\lambda\in \mathbb{C}$. Then, both states, $\ket{v}$ and $\ket{w}$, describe the same physics and after normalization we are left with  a non-physical global phase $\varphi$ due to $\ket{v}\sim\rme^{i\varphi}\ket{v}$. All quantum states which cannot be represented by a single ray are called mixed (or ensemble) states and are represented by a density matrix. Density matrices also naturally arise in the context of statistical ensembles at non-zero temperature. In the following, we start by recalling the definition of a density matrix and its properties in the context of an $N$-particle quantum system, which serves as a foundation for the discussion of reduced density matrices. In particular, we emphasize the relevance of the $N$-particle density matrix and the one-particle reduced density matrix in the context of RDMFT. 

\subsection{Definition and general properties}

\begin{figure}[htb]
\centering
\includegraphics[width=0.45\linewidth]{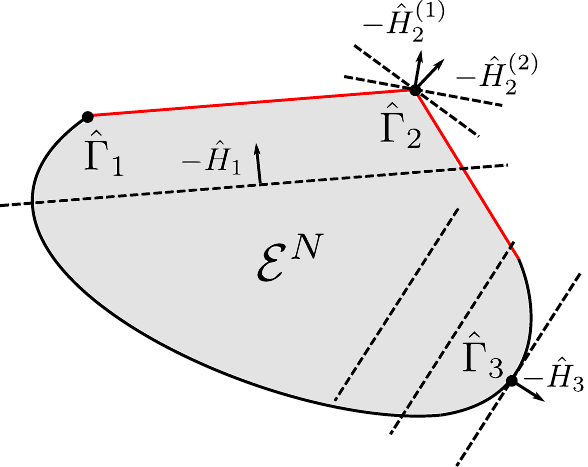}
\caption{Illustration of the set $\mathcal{E}^N$ and the energy minimization. The black segment of the boundary of $\mathcal{E}^N$ together with $\hat{\Gamma}_2$ marked by a black dot constitute the set of extremal elements of $\mathcal{E}^N$, and thus the set $\mathcal{P}^N$. All points on the red line segments, as well as all points of $\mathcal{E}^N$ not lying on the boundary, can be obtained by convex combinations of extremal elements. The energy minimization is illustrated by the dashed lines shifted along a direction determined by a Hamiltonian $\hat{H}$ until they reach the boundary of $\mathcal{E}^N$ (see text for further explanations). \label{fig:EN}}
\end{figure}

Since we are interested in the description of $N$-particle quantum systems, we first need to understand the structure of the underlying $N$-particle Hilbert space $\mathcal{H}_N$.  
Let $\mathcal{H}_1$ denote the one-particle Hilbert space with dimension $\mathrm{dim}(\mathcal{H}_1)=d$. Then, the Hilbert space for $N$ distinguishable particles is simply given by the tensor product
\begin{equation}
\mathcal{H}_N \equiv \mathcal{H}_1^{\otimes N}\,.
\end{equation}
However, indistinguishable particles in a quantum mechanical framework require a more careful treatment. For identical fermions, the states in $\mathcal{H}_N$ must be antisymmetric under the exchange of two particles and we have
\begin{equation}\label{eq:HN_fermions}
\mathcal{H}_N \equiv \wedge^N[\mathcal{H}_1]\leq \mathcal{H}_1^{\otimes N}\,.
\end{equation}
States of identical bosons have to be symmetric under the exchange of two particles and similarly to Eq.~\eqref{eq:HN_fermions} we obtain
\begin{equation}\label{eq:HN_bosons}
\mathcal{H}_N \equiv \mathcal{S}^N[\mathcal{H}_1]\leq \mathcal{H}_1^{\otimes N}\,.
\end{equation}
The definition of the density operator in the following implicitly assumes the correct choice of $\mathcal{H}_N$ depending on the type of particles under consideration.
The set of all \textit{ensemble $N$-particle density operators} is defined as
\begin{equation}\label{eq:EN}
\mathcal{E}^N \equiv \{\hat{\Gamma}:\mathcal{H}_N\to \mathcal{H}_N|\,\text{linear}, \hat{\Gamma}\geq 0, \Tr[\hat{\Gamma}]=1\}
\end{equation}
and its dimension is given by the dimension of its affine hull $\mathrm{aff}(\mathcal{E}^N)$.
Further, the set $\mathcal{E}^N$ includes the set of all rank one orthogonal projection operators which are called pure states and follow from a state $\ket{\Psi}\in\mathcal{H}_N$ as $\hat{\Gamma} = \ket{\Psi}\!\bra{\Psi}$. Thus, the set $\mathcal{P}^N$ of all \textit{pure states} is given by
\begin{equation}\label{eq:PN}
\mathcal{P}^N \equiv \{\hat{\Gamma}\in \mathcal{E}^N|\,\hat{\Gamma}^2 = \hat{\Gamma}\}\,.
\end{equation}
From the definitions in Eq.~\eqref{eq:EN} and Eq.~\eqref{eq:PN} follows that a density operator $\hat{\Gamma}$ must fulfil the following properties:
\begin{align}
&\text{1) hermiticity: } \hat{\Gamma}^\dagger = \hat{\Gamma}\,,\\\
&\text{2) positivity: } \hat{\Gamma}\geq 0\,,\\\
&\text{3) normalization: } \tr{[\hat{\Gamma}]} = 1\,.
\end{align}
In addition, pure states are characterized by $\hat{\Gamma}^2 = \hat{\Gamma}$. Since a density operator is by definition self-adjoint, it can be diagonalized leading to the spectral decomposition
\begin{equation}
\hat{\Gamma} = \sum_j p_j\ket{\Psi_j}\!\bra{\Psi_j}\,,\quad \sum_jp_j=1\,,\quad p_j\geq 0 \,,
\end{equation}
where $\{\ket{\Psi_j}\}_{j=1}^{\mathrm{dim}(\mathcal{H}_N)}$ denotes the set of orthonormal eigenstates of $\hat{\Gamma}$. Clearly, for pure states only one $p_j$ is not equal to zero. This also explains why all other density operators are called mixed (or ensemble) states because they follow from a collection of orthonormal states $\{\ket{\Psi_i}\}$ and associated probabilities $\{p_i\}$ yielding one ensemble which is equivalent to $\hat{\Gamma}$. However, the correspondence between mixed density operators and ensembles is not unique. This statement becomes obvious if we discuss the properties of the set $\mathcal{E}^N$ and its relation to $\mathcal{P}^N$ in more detail. The pure states which define the subset of extremal states within $\mathcal{E}^N$. It can be easily proven that $\mathcal{E}^N$ is convex as well as compact which means that it is bounded and closed, whereby the set $\mathcal{P}^N$ of all pure $N$-particle density operators is still compact but not convex anymore. 
The convexity of $\mathcal{E}^N$, in turn, implies that every $\hat{\Gamma}\in \mathcal{E}^N$ can be represented as a convex combination of pure states $\hat{\Gamma}_j^2=\hat{\Gamma}_j\in \mathcal{P}^N$ such that (see also Sec.~\ref{subsec:terminology})
\begin{equation}
\mathcal{E}^N\ni \hat{\Gamma} = \sum_j p_j\hat{\Gamma}_j\,,\quad \sum_jp_j=1\,,\quad p_j\geq 0\,.
\end{equation} 
Thus, there are in fact infinitely many ways to construct a mixed density operator from a convex combination of pure states. Moreover, all boundary points of $\mathcal{E}^N$ are given by those $\hat{\Gamma}$ with at least one eigenvalue equal to zero. This also means that not all $\hat{\Gamma}$ on the boundary of $\mathcal{E}^N$ are extremal points. We illustrate these properties in Fig.~\ref{fig:EN}. The set $\mathcal{P}^N$ of all pure states is given by all points on the black segment of the boundary of $\mathcal{E}^N$ and $\hat{\Gamma}_2$. The points on the two red line segments also lie on the boundary but they are not extremal points. For example, every $\hat{\Gamma}$ on the red line between $\hat{\Gamma}_1$ and $\hat{\Gamma}_2$ can be obtained by the convex combination $\lambda\hat{\Gamma}_1 + (1-\lambda)\hat{\Gamma}_2$ for a specific choice of $\lambda\in[0, 1]$. 

The knowledge of the density matrix $\hat{\Gamma}$ is sufficient to calculate the expectation value of any physical observable $\hat{O}$. Observables are hermitian linear operators on the Hilbert space, which means that they are diagonalizable with real eigenvalues. Using a density matrix $\hat{\Gamma}\in \mathcal{E}_N$, the expectation value $\langle\hat{O}\rangle$ of the observable $\hat{O}$ is defined through
\begin{equation}
\langle\hat{O}\rangle \equiv \tr [\hat{O}\hat{\Gamma}]\,.
\end{equation}
This further implies that the ground state energy $E_0$ of any Hamiltonian $\hat{H}$ is obtained from the variational principle:
\begin{equation}\label{eq:E0_varprin}
E_0 \equiv \min_{\hat{\Gamma}\in \mathcal{E}_N} \tr_N[\hat{H}\hat{\Gamma}]\,.
\end{equation}
Thus, the ground state energy for a given $\hat{H}$ follows from minimizing its expectation value $\langle\hat{H}\rangle$ over all density operators $\hat{\Gamma}\in \mathcal{E}^N$. From restricting the minimization in \eqref{eq:E0_varprin} to all $\hat{\Gamma}\in\mathcal{P}^N$ one recovers the well-known Rayleigh-Ritz variational principle, where the ground state wave function is approximated by a variational wave function to be optimized. To interpret the result in Eq.~\eqref{eq:E0_varprin} in a geometrical way, we first notice that the trace $\tr_N[\hat{H}\hat{\Gamma}]$ is simply the inner product $\langle\hat{H}, \hat{\Gamma}\rangle_{N}$ on the Hilbert space $\mathcal{H}_N$, i.e.~$\tr_N[\hat{H}\hat{\Gamma}]= \langle\hat{H}, \hat{\Gamma}\rangle_N$. Those $\hat{\Gamma}\in \mathcal{E}^N$ which lead to the same constant value of $\langle\hat{H}, \hat{\Gamma}\rangle_N$ thus determine a hyperplane whose normal vector is defined through $\hat{H}$. Since we consider a minimization process, this hyperplane in shifted in direction $-\hat{H}$ until in reaches the boundary of $\mathcal{E}^N$ determining the ground state for a specific Hamiltonian $\hat{H}$. This minimization process is also illustrated in Fig.~\ref{fig:EN}. If we shift a hyperplane, depicted by the dashed lines, along the direction $-\hat{H}_3$, it touches the boundary only at one point yielding the density operator $\hat{\Gamma}_3$ as the corresponding ground state. In contrast to $\hat{\Gamma}_3$, the point $\hat{\Gamma}_2$ is the minimizer for several Hamiltonians, but for all of them it is the unique ground state. However, the minimum along $-\hat{H}_1$ is not only attained at $\hat{\Gamma}_1$ but at all points along the corresponding red line segment. These states are then called degenerate ground states of the Hamiltonian $\hat{H}_1$.  

In the discussion above, we describe the compact and convex set $\mathcal{E}^1_N$ through all its elements. 
However, following Sec.~\ref{sec:duality} and the concept of the support function, $\mathcal{E}^1_N$ is also uniquely determined through the union of its supporting hyperplanes. This now allows us to understand the connection between the minimization described above (see also Fig.~\ref{fig:EN}) and the duality correspondence for convex sets from a different perspective. Let us consider the minimization illustrated in Fig.~\ref{fig:EN} not only for three different choices of the Hamiltonian $\hat{H}$ determining the normal vector of a hyperplane, but for all possible directions. Then, all minimizers $\hat{\Gamma}_{\hat{H}}$ fully characterize the convex set $\mathcal{E}^N$, which is obtained by taking the convex hull of all minimizers $\hat{\Gamma}_{\hat{H}}$. Of course, the same concept holds on the level of the one-particle Hamiltonian and the one-particle reduced density matrix which we discuss in the next section and appears again in Ch.~\ref{ch:w-RDMFT}.


\subsection{One-particle reduced density matrix}

All possible advantages of RDMFT compared to DFT lie in the fact that RDMFT uses the full one-particle reduced density matrix (1RDM) as its main variable rather than the spatial density as DFT does. A precise definition of the 1RDM and its properties is therefore crucial to understand the conceptual advantages of RDMFT in relation to DFT and why it has such a great potential to replace DFT at some point in the future. We start by presenting three different ways to define the 1RDM. Of course, they all lead to the same object but provide different perspectives and therefore facilitate a more comprehensive understanding of the 1RDM.

Since second quantization provides a particularly convenient way to deal with a large number of particles, it is 
widely used to work with quantum many-body systems and appears throughout this thesis. Therefore, we also introduce the 1RDM in second quantization as follows: For a one-particle Hilbert space $\mathcal{H}_1$ of dimension $d = \mathrm{dim}(\mathcal{H}_1)$, we choose an orthonormal basis set 
$\{|i\rangle\}_{i=1}^{d}$. Then, the matrix elements of the 1RDM are given by
\begin{equation}
\gamma_{ij} = \langle i|\hat{\gamma}|j\rangle \equiv \langle\Psi_N|\hat{a}_j^\dagger\hat{a}_i|\Psi_N\rangle\,,
\end{equation}
where $\ket{\Psi_N}$ denotes a properly (anti-)symmetrized N-particle wavefunction. 
The 1RDM follows directly from its matrix elements as 
\begin{equation}
\hat{\gamma} = \sum_{i,j=1}^d\gamma_{ij}\ket{i}\!\bra{j}\,.
\end{equation}

Starting from an $N$-fermion/boson quantum state $\hat{\Gamma}$, the one-particle reduced density matrix (1RDM) $\gh$ is obtained by tracing out all except one particle
\begin{equation}\label{eq:gamma_tr}
\gh \equiv N \mbox{Tr}_{N-1}[\hat{\Gamma}]\,.
\end{equation}
Due to the indistinguishability of the particles, the result for the 1RDM $\gh$ is independent of which $N-1$ particles are traced out. Note that $\hat{\gamma}$ is normalized to the total particle number $N$ rather than to one as $\hat{\Gamma}$. This normalization also has an intuitive consequence: Since $\hat{\gamma}$ is by definition self-adjoint, it can always be written in its spectral decomposition
\begin{equation}
\hat{\gamma} \equiv \sum_{\alpha=1}^d \lambda_\alpha\ket{\alpha}\!\bra{\alpha}\,,
\end{equation}
where $d=\mathrm{dim}(\mathcal{H}_1)$. Eq.~\eqref{eq:gamma_tr} now implies that the sum over all eigenvalues is equal to $N$, i.e.~$\sum_\alpha\lambda_\alpha=N$. Thus, the eigenvalues $\{\lambda_\alpha\}_{\alpha=1}^d$ of $\hat{\gamma}$ are referred to as \textit{natural occupation numbers} (NON), and the eigenstates $\{\ket{\alpha}\}_{\alpha=1}^d$ are the corresponding \textit{natural orbitals} (NO) \cite{Loewin1955}. Note that the diagonal elements of the 1RDM in spatial representation determine the particle density $\rho(\bd r)= \gamma(\bd r, \bd r)$ which is used as the natural variable in DFT.

Equivalently, using Riesz representation theorem, $\gh$ can be characterized as the mathematically most primitive object which still determines the expectation values of all one-particle observables $\hat{o}$. To explain this statement, we first denote by $\mathcal{O}_1 = \{\hat{o}\}$ the set of linear, hermitian one-particle operators $\hat{o}:\mathcal{H}_1\mapsto\mathcal{H}_1$. Lifting the one-particle observable $\hat{o}$ to the $N$-particle level yields in first quantization $\hat{O}\equiv \hat{o}\otimes\mathds{1}^{N-1} + \mathds{1}\otimes\hat{o}\otimes\mathds{1}^{N-2} + ... + \mathds{1}^{N-1}\otimes\hat{o}$, where $\hat{O}$ denotes a $N$-particle observable. In second quantization, we have $\hat{o} \equiv \sum_{i,j=1}^do_{ij}\ket{i}\!\bra{j}$ for an orthonormal basis set $\{|i\rangle\}_{i=1}^{d}$. Then, $\hat{O} = \sum_{i,j=1}^do_{ij}\hat{a}_i^\dagger\hat{a}_j$ and we eventually obtain
\begin{equation}\label{eq:1RDM_v3}
\langle\hat{O}, \hat{\Gamma}\rangle_N \equiv \mbox{Tr}_N[\hat{O}\hat{\Gamma}]=\mbox{Tr}_1[\hat{o}\hat{\gamma}]\equiv \langle\hat{o},\hat{\gamma}\rangle\,,
\end{equation}
where $\langle\cdot,\cdot\rangle$ denotes the inner product on the Euclidean space of hermitian matrices.

According to the definition of the 1RDM in Eq.~\eqref{eq:gamma_tr}, the sets $\mathcal{P}_N^1$ and $\mathcal{E}_N^1$ are obtained from $\mathcal{P}^N$ and $\mathcal{E}^N$ by tracing out $N-1$ particles 
\begin{align}
\mathcal{P}_N^1 &= N\Tr_{N-1}[\mathcal{P}^N] \label{eq:setP1}  \\\ 
\mathcal{E}_N^1 &= N\Tr_{N-1}[\mathcal{E}^N]\,.\label{eq:setE1}
\end{align}
Since $\mathcal{E}^N$ is convex and the partial trace map $\tr_{N-1}[\cdot]$ is linear, also the set $\mathcal{E}_N^1$ is convex. Recall that the extreme elements of $\mathcal{E}^N$ are the pure states $\hat{\Gamma}^2=\hat{\Gamma}\in \mathcal{P}^N$. Hence, the sets $\mathcal{P}^1_N$ and $\mathcal{E}^1_N$ are by definition related through
\begin{equation}\label{eq:P1E1subset}
\mathcal{P}^1_N\subseteq \mathcal{E}^1_N\,.
\end{equation}
In addition, Eq.~\eqref{eq:setP1} and Eq.~\eqref{eq:setE1} imply that the extremal elements of $\mathcal{E}^1_N$ are also contained in $\mathcal{P}^1_N$. 
We comment more on further relations between $\mathcal{P}^1_N$ and $\mathcal{E}^1_N$ as well as possible differences between fermions and bosons in the context of the $N$-representability problem in Sec.~\ref{sec:Nrepr}.  

Since we now understand how the 1RDM $\gh$ follows from a $N$-particle density operator $\hat{\Gamma}$, we can change our perspective and ask about the properties of the sets of all pure or ensemble $N$-particle density operators mapping to a given $\gh$. These two sets will play an important role in the constrained search formalism in Sec.~\ref{sec:Levy} and are given by
\begin{align}
\mathcal{E}^N(\hat{\gamma})&\equiv \{\hat{\Gamma}\in \mathcal{E}^N\,|\, \hat{\Gamma}\mapsto\hat{\gamma}\}\\\
\mathcal{P}^N(\hat{\gamma})&\equiv \{\hat{\Gamma}\in \mathcal{P}^N\,|\, \hat{\Gamma}\mapsto\hat{\gamma}\}\,.
\end{align}
As $\mathcal{E}^N$ and $\mathcal{P}^N$, both sets $\mathcal{E}^N(\gh)$ an $\mathcal{P}^N(\gh)$ are compact, and $\mathcal{E}^N(\gh)$ is also convex. However, as a result of the restriction of $\mathcal{E}^N$ to $\mathcal{E}^N(\gh)$, the extremal elements of $\mathcal{E}^N(\gh)$ are not necessarily pure states anymore. Similar to Eq.~\eqref{eq:P1E1subset}, but now on the $N$-particle level, we have
\begin{equation}
\mathcal{P}^N(\gh)\subseteq \mathcal{E}^N(\gh)\,.
\end{equation}


\section{N-representability problem \label{sec:Nrepr}}

In this section we consider again the two sets $\mathcal{P}^1_N$ and $\mathcal{E}^1_N$ defined in Eq.~\eqref{eq:setP1} and Eq.~\eqref{eq:setE1}, respectively, but with regard to the so-called $N$-representability problem, which amounts to answering the following question: For which 1RDM's does there exist a corresponding properly (anti-)symmetrized $N$-particle state? All 1RDMs $\gh$ for which there exists a $N$-particle density operator $\hat{\Gamma}$ such that $\mathcal{P}^N\ni\hat{\Gamma}\mapsto\hat{\gamma}$ are then called \textit{pure state $N$-representable}. This is by definition the case for all $\hat{\gamma}\in\mathcal{P}^1_N$. Similarly, all $\hat{\gamma}\in\mathcal{E}^1_N$ are called \textit{ensemble $N$-representable}. It is only if the answer to this question is known, that we are able to determine the boundaries of the two sets $\mathcal{P}^1_N$ and $\mathcal{E}^1_N$. Since fermions and bosons obey different statistics, we distinguish between them in the following discussion of $N$-representability.

\subsection{Fermions\label{subsec:N_fermions}}

The boundary of the set $\mathcal{E}^1_N$ of all ensemble $N$-representable 1RDMs is determined through the necessary and sufficient conditions
\begin{equation}\label{eq:fermions_eNcond}
0\leq \lambda_\alpha\leq 1\,,\quad \sum_{\alpha=1}^d\lambda_\alpha=N\,.
\end{equation}
Thus, the only two restrictions are that the natural occupation numbers $\lambda_\alpha$ fulfil the well-known Pauli exclusion principle $0\leq\lambda_\alpha\leq 1$ and sum up to the fixed total particle number $N$.
Moreover, the extremal points of $\mathcal{E}^1_N$ are given by those states, where $N$ natural occupation numbers are equal to one and all other $d-N$ natural orbitals are unoccupied (recall that $d=\mathrm{dim}(\mathcal{H}_1)$). The extremal elements of $\mathcal{E}^1_N$ coincide with the extremal elements of $\mathcal{P}^1_N$ and are thus also pure state $N$-representable. 

However, the boundary of $\mathcal{P}^1_N$ is in general not known because the fermionic occupation numbers are not only restricted through the well-known Pauli exclusion principle but also through further constraints, the so-called generalized Pauli constraints. These are additional constraints on the natural occupation numbers imposed by the fermionic exchange symmetry \cite{KL06, AK08}. Due to the complexity of the generalized Pauli constraints, a general solution to the pure state $N$-representability problem for fermions is unknown. 


%
\subsection{Bosons\label{subsec:N_bosons}}

For bosons, the $N$-representability problem simplifies drastically due to the bosonic statistics. However, before we start to solve the $N$-representability problem for bosons, we need to attain a deeper understanding of the set $\mathcal{E}^1_N$ and, in particular, its extremal elements. For bosons, the extremal elements of $\mathcal{E}^1_N$ are those 1RDMs, which are pure states, i.e.~$\gh = N\ket{\alpha}\!\bra{\alpha}$. It follows directly from the definition of a pure state that it cannot be written as a convex combination of other states in the corresponding convex set and is, therefore, an extremal element in this set (see also Sec.~\ref{subsec:terminology}). Further, every 1RDM $\gh$, which is extremal in $\mathcal{E}^1_N$, follows from a pure state $\hat{\Gamma}= \ket{\Phi}\!\bra{\Phi}\in\mathcal{P}^N$ with $\ket{\Phi}=\ket{\alpha, \alpha, ..., \alpha}$ by tracing out $N-1$ particles. As a result of the normalization of $\gh$ to the total particle number $N$, we have $\gh^2 = N\gh$. It follows immediately that all extremal elements of $\mathcal{E}^1_N$ are pure state $N$-representable.  

We can now discuss the more interesting case, namely those 1RDMs which are not extremal elements in $\mathcal{E}^1_N$. Recall that for those it is not possible to obtain the pure state $N$-representability constraints for fermions due to the generalized Pauli constraints, as discussed in the section above. Though similar to the fermionic case, the necessary and sufficient ensemble $N$-representability constraints are given by
\begin{equation}
\lambda_\alpha\geq 0\,,\quad \sum_{\alpha=1}^d\lambda_\alpha=N\,.
\end{equation}
Moreover, the two sets $\mathcal{E}^1_N$ and $\mathcal{P}^1_N$ are equal in the bosonic case \cite{GR19}
\begin{equation}\label{eq:P1eqE1}
\mathcal{E}^1_N = \mathcal{P}^1_N\,.
\end{equation}
To prove this statement, we need to show both directions, $\mathcal{P}^1_N\subseteq\mathcal{E}^1_N$ and $\mathcal{E}^1_N\subseteq \mathcal{P}^1_N$. The first part, $\mathcal{P}^1_N\subseteq\mathcal{E}^1_N$, which holds for fermions as well as bosons is trivial and was already explained in Eq.~\eqref{eq:P1E1subset} as a direct consequence of $\mathcal{P}^N\subset\mathcal{E}^N$. To prove $\mathcal{E}^1_N\subseteq \mathcal{P}^1_N$, we consider a $\hat{\gamma}\in \mathcal{E}_N^1$. This implies that there exists an ensemble $N$-particle density operator $\hat{\Gamma}$ such that $\mathcal{E}^N\ni\hat{\Gamma}\mapsto\gh$. Since every 1RDM is diagonalizable, it has a spectral decomposition $\gh \equiv \sum_\alpha\lambda_\alpha\ket{\alpha}\!\bra{\alpha}$, and thus the corresponding ensemble $N$-particle density operator is given by
\begin{equation}
\mathcal{E}^N\ni \hat{\Gamma} = \frac{1}{N}\sum_{\alpha}\lambda_\alpha\ket{\alpha, \alpha, ..., \alpha}\!\bra{\alpha, \alpha, ..., \alpha}\,.
\end{equation}
However, for every bosonic $\gh$ we can also write down a pure $N$-particle density operator $\mathcal{P}^N\ni\hat{\Gamma}=\ket{\Phi}\!\bra{\Phi}\mapsto\gh$ with $\ket{\Phi} = 1/\sqrt{N}\sum_{\alpha=1}^d\sqrt{\lambda_\alpha}\ket{\alpha, \alpha, ..., \alpha}$ such that
\begin{equation}
\mathcal{P}^N\ni\hat{\Gamma} = \ket{\Phi}\!\bra{\Phi} = \frac{1}{N}\sum_{\alpha,\alpha^\prime=1}^d \sqrt{\lambda_\alpha\lambda_{\alpha^\prime}}\ket{\alpha, \alpha, ..., \alpha}\!\bra{\alpha^\prime, \alpha^\prime, ...,\alpha^\prime}\,.
\end{equation}
Thus, $\gh\in \mathcal{P}^1_N$ which finishes the proof.

As a consequence of Eq.~\eqref{eq:P1eqE1}, every bosonic 1RDM is pure \textit{and} ensemble $N$-representable.
This means that the $N$-representability problem is trivial for bosons and thus will not hamper our following discussion of bosonic RDMFT. 

\section{Hohenberg-Kohn theorems and $v$-representability problem\label{sec:HK}}

In quantum mechanics, all information about a stationary quantum mechanical system can be extracted from the time-independent Schr{\"o}dinger equation 
\begin{equation}\label{eq:SG}
\hat{H}\ket{\Psi} = E\ket{\Psi}\,,
\end{equation}
which is simply an eigenvalue equation for the wave functions $\Psi$ and the eigenenergies $E$. In the following we consider quantum systems of identical fermions/bosons with Hamiltonians
\begin{equation}\label{Hfamily}
\hat{H}(\hat{h}) \equiv \hat{h} + \hat{W}\,,
\end{equation}
where $\hat{h} = \hat{v} + \hat{t}$ denotes a one-particle Hamiltonian consisting of the kinetic energy operaror $\hat{t}$ and an external potential $\hat{v}$. The last term in Eq.~\eqref{Hfamily}, $\hat{W}$, denotes the interaction between the particles. In doing so, the interaction $\hat{W}$ is usually fixed in every physical system under consideration. For example, this could be Coulomb interactions between charges particles or (effective) hard-core interactions in ultracold, atomic gases which we will discuss in further detail in Sec.~\ref{subsec:S-wave_gases}. 

A main challenge in condensed matter physics is to determine the ground state and ground state energy of a system described by a Hamiltonian $\hat{H}$. For a small total number of particles $N$, wavefunction-based methods work very well and provide an exact solution to the ground state problem. However, for large $N$, solving the Schr{\"o}dinger equation is not feasible anymore due to the exponential growth of the corresponding Hilbert space with increasing number of particles in the system. This demonstrates the need for efficient theories to circumvent this problem. The standard approach therefore in solid state physics and quantum chemistry is DFT, which is based on the observation that the calculation of ground state observables does not require the knowledge of the full wavefunction, and for local external potentials $\hat{v}$, the knowledge of the ground state density $\rho_0$ is sufficient. The theoretical foundation of DFT is provided by the Hohenberg-Kohn theorem \cite{HK} in its original formulation for local external potentials.
In 1975, Gilbert \cite{Gilbert} proved an extension of the Hohenberg-Kohn theorem also to non-local potentials, providing the foundation of RDMFT using the full 1RDM $\gh$ as its natural variable. Besides the extension of the density to the full 1RDM as the natural variable, the main difference between DFT and RDMFT lies in the fact that in DFT both, $\hat{t}$ and $\hat{W}$, are fixed and only the external potential $\hat{v}$ can be varied, whereas in RDMFT only the interaction $\hat{W}$ is fixed. We start by discussing the Hohenberg-Kohn theorem for local external potentials in Sec.~\ref{subsec:HK} before we move on to Gilbert's theorem in Sec.~\ref{subsec:Gilbert}.

\subsection{Local potentials \label{subsec:HK}}

The original formulation of the Hohenberg-Kohn theorem \cite{HK} is concerned with local potentials
\begin{equation}
v(\bd r, \bd r^\prime)= v(\bd r)\delta^{(d)}(\bd r - \bd r^\prime)\,,
\end{equation}
which are diagonal in spatial representation. The first part of the Hohenberg-Kohn theorem proves the existence of a one-to-one mapping between the local external potential $\hat{v}$ and the ground state density $\rho$ via the ground state $\ket{\Psi}$
\begin{equation}
\hat{v}\,\xleftrightarrow{\text{one-one}}\,\ket{\Psi}\,\xleftrightarrow{\text{one-one}}\, \rho\,.
\end{equation}
The direction $\hat{v}\mapsto\ket{\Psi}\mapsto\rho$ is trivial because the the external potential $\hat{v}$ determines the Hamiltonian $\hat{H}$ for fixed $\hat{t}$ and $\hat{W}$ completely which, in turn, determines the ground state wave function $\ket{\Psi}$ through the Schrödinger equation in Eq.~\eqref{eq:SG} leading to the ground state density $\rho$. Therefore, we are left with the inverse direction which splits into $\rho\mapsto\ket{\Psi}$ and $\ket{\Psi}\mapsto\hat{v}$. We start by proving the latter. Assume that the same ground state wave function $\ket{\Psi}$ corresponds to two external potential $\hat{v}$ and $\hat{v}^\prime$ which differ by more than a constant. This leads to two Hamiltonians $\hat{H} = \hat{t} + \hat{v} + \hat{W}$ and $\hat{H}^\prime= \hat{t} + \hat{v}^\prime + \hat{W}$. Subtracting the two corresponding Schr{\"o}dinger equations yields
\begin{equation}\label{eq:HK_Psiv}
(\hat{v} - \hat{v}^\prime)\ket{\Psi}=(E_0 - E_0^\prime)\ket{\Psi}\,,
\end{equation}
where $E_0$ is the ground state energy of $\hat{H}$ and $E_0^\prime$ belongs to $\hat{H}^\prime$. Eq.~\eqref{eq:HK_Psiv} can only be satisfied if either the two external potentials only differ by constant, or $\ket{\Psi}$ is zero everywhere. The latter can be excluded by the unique continuation theorem \cite{L83, reed}. Thus, our assumption that $\hat{v}$ and $\hat{v}^\prime$ differ by more than a constant has to be wrong and the ground state wave function indeed determines the local external potential uniquely.
To prove $\rho\mapsto\ket{\Psi}$, we first assume that the ground state is not degenerate and suppose that in addition to $\ket{\Psi}$ there exists a second ground state wave function $\ket{\Psi^\prime}$ yielding the same ground state density. Since the ground state of the Hamiltonian $\hat{H}$ is by assumption not degenerate, the second wave function $\ket{\Psi^\prime}$ must correspond to a different Hamiltonian $\hat{H}^\prime$. Since $\hat{t}$ and $\hat{W}$ are fixed, only the external potential can differ in the two cases. Denoting by $E_0$ the ground state energy, we obtain from the variational principle \cite{Czycholl}
\begin{equation}
E_0 = \bra{\Psi}\hat{H}\ket{\Psi} < \bra{\Psi^\prime}\hat{H}\ket{\Psi^\prime} = E_0^\prime + \int\rmd^d\bd r\,\left(v(\bd r) - v^\prime(\bd r)\right)\rho(\bd r)
\end{equation}
and similarly 
\begin{equation}
E_0^\prime = \bra{\Psi^\prime}\hat{H}^\prime\ket{\Psi^\prime} < \bra{\Psi}\hat{H}^\prime\ket{\Psi} = E_0 + \int\rmd^d\bd r\,\left(v^\prime(\bd r) - v(\bd r)\right)\rho(\bd r)\,.
\end{equation}
Combining the two inequalities above leads to the contradiction
\begin{equation}
E_0 + E_0^\prime <E_0 + E_0^\prime\,.
\end{equation}
Thus, the ground state density $\rho$ determines the ground state $\ket{\Psi}$ uniquely. This proves the map $\rho\mapsto\ket{\Psi}$, or equivalently $\rho\mapsto\hat{\Gamma}$ for $\hat{\Gamma}\in\mathcal{P}^N$. The one-to-one mapping between $\hat{v}$ and $\rho$ is the essential insight of the Hohenberg-Kohn theorem and is usually referred to as its first part in the literature. A ground state density $\rho$ is called \textit{pure state $v$-representable} if and only if there exists a local external potential $\hat{v}$ such that $\hat{H}\mapsto \ket{\Psi}\mapsto\rho$ holds. This means in particular that not every ground state density $\rho$ is $v$-representable and hence, it is not possible to find a corresponding $\hat{v}$ for every $\rho$.

The second part of the Hohenberg-Kohn theorem establishes a variational principle for the ground state energy in terms of the particle density. It follows that the ground state wave function can be written as a functional of the ground state density and thus also every ground state observable,
\begin{equation}
\langle\hat{O}\rangle_{0} = \bra{\Psi(\rho)}\hat{O}\ket{\Psi(\rho)} \equiv O(\rho)\,.
\end{equation}
This holds in particular for the ground state energy $E_0$, which is, as a direct consequence of Ritz variational principle, the unique minimum of an energy functional
\begin{equation}\label{eq:var_HK}
E_0 = \min_{\rho}E(\rho)\,.
\end{equation}
Note that the energy minimization in Eq.~\eqref{eq:var_HK} yields both, the ground state energy and the ground state density.
The energy functional follows from $\langle\hat{H}\rangle\equiv\tr[\hat{\Gamma}\hat{H}]$ with $\hat{\Gamma}\in \mathcal{P}^N$ as
\begin{equation}
E(\rho) \equiv \int \rmd^d\bd r\,v(\bd r)\rho(\bd r)  + \mathcal{F}_{\mathrm{HK}}(\rho)\,.
\end{equation}
Hence, the Hohenberg Kohn theorem proves the existence of a universal functional $\mathcal{F}_\mathrm{HK}(\rho)$, which is only a functional of the density and requires no information about the local external potential $\hat{v}$. Therefore, if the exact functional $\mathcal{F}_\mathrm{HK}(\rho)$ for a quantum system would be known, the ground state energy can be calculated for \textit{any} local external potential by the minimization in Eq.~\eqref{eq:var_HK} with almost no additional effort making DFT, as least in principle, a very efficient method. However, the exact $\mathcal{F}_\mathrm{HK}$ is usually not known and appropriate approximations must be made yielding an entire family of approximated functionals for fermions. Though the Hohenberg-Kohn theorem does not require any distinction between fermions and bosons, DFT has only been applied to bosons in a very few cases because the density itself does not provide any information about the occupation numbers of different states (for examples see Ref.~\cite{Griffin1995, Nunes1999}).

\subsection{Gilbert theorem \label{subsec:Gilbert}} 

In 1975, Gilbert \cite{Gilbert} extended the Hohenberg-Kohn theorem to non-local potentials and proved a one-to-one correspondence between the ground state wave function, i.e.~$\hat{\Gamma}\in \mathcal{P}^N$, and the ground state 1RDM $\hat{\gamma}\in \mathcal{P}^1_N$. Due to the non-locality of $\hat{v}$, the inverse direction of $\hat{v}\mapsto\hat{\Gamma}$ does not hold anymore and the correspondence between $\hat{v}$ and $\hat{\Gamma}$ is many-to-one \cite{Helbig2006}. In summary, we have
\begin{equation}\label{eq:Gilbert_HK}
\hat{v}\,\xleftrightarrow{\text{many-one}}\,\hat{\Gamma}\,\xleftrightarrow{\text{one-one}}\, \hat{\gamma}\,.
\end{equation}
As in the previous case for a local $\hat{v}$, the maps $\hat{v}\mapsto\hat{\Gamma}\in\mathcal{P}^N$ and $\hat{\Gamma}\mapsto\hat{\gamma}\in \mathcal{P}^1_N$ for a non-local potential follow directly from the time-independent Schr{\"o}dinger equation. The essential part is to show $\hat{\gamma}\mapsto\hat{\Gamma}$. The proof follows the same line of argument as for the Hohenberg-Kohn theorem. Assume that there exist two different pure $N$-particle density operator $\hat{\Gamma}$ and $\hat{\Gamma}^\prime$ which map to the same ground state 1RDM $\hat{\gamma}$. The two Hamiltonians $\hat{H}$ and $\hat{H}^\prime$ leading to $\hat{\Gamma}$ and $\hat{\Gamma}^\prime$ can again only vary in their external potentials $\hat{v}$ and $\hat{v}^\prime$. The variational principle then leads to
\begin{align}
&E_0 \equiv \tr[\hat{H}\hat{\Gamma}]< \tr[\hat{H}\hat{\Gamma}^\prime] = \tr[(\hat{H}^\prime + \hat{v} - \hat{v}^\prime)\hat{\Gamma}^\prime] = E_0^\prime + \tr[(\hat{v}-\hat{v}^\prime)\hat{\gamma}]\\\
&E_0^\prime \equiv \tr[\hat{H}^\prime\hat{\Gamma}^\prime]< \tr[\hat{H}^\prime\hat{\Gamma}] = \tr[(\hat{H} + \hat{v}^\prime - \hat{v})\hat{\Gamma}] = E_0 + \tr[(\hat{v}^\prime - \hat{v})\hat{\gamma}]
\end{align}
and we finally arrive at the contradiction $E_0  + E_0^\prime< E_0^\prime + E_0$. Thus, the initial assumption that two distinct ground state $N$-particle operators can lead to the same ground state 1RDM must be wrong and $\hat{\gamma}\mapsto\hat{\Gamma}$ holds. Consequently, every ground state observable can be written as a functional of the ground state 1RDM. 

Based on Eq.~\eqref{eq:Gilbert_HK}, Gilbert proved the existence of a universal interaction functional $\mathcal{F}_{\hat{W}}(\hat{\gamma})$ of the 1RDM: The energy and 1RDM of the ground state of $\hat{H}(\hat{h})$ for any one-particle Hamiltonian $\hat{h}$ can be determined by minimizing the total energy functional
\begin{equation}\label{energyfunc}
E(\hat{\gamma}) = \Tr_1[\hat{h}\hat{\gamma}] + \mathcal{F}_{\hat{W}}(\hat{\gamma})\,.
\end{equation}
In the equation above, we used the fact that the functional dependence of the kinetic energy on the 1RDM is known. This is contrasted with DFT, where the Hohenberg-Kohn functional $\mathcal{F}_{\mathrm{HK}}(\rho)$ also contains the kinetic energy because its dependence on the density $\rho$ is unknown. 
The significance of reduced density matrix functional theory (RDMFT) is based on the fact that the interaction functional  $\mathcal{F}_{\hat{W}}(\hat{\gamma})$ does not depend on the choice of the one-particle Hamiltonian $\hat{h}$ but only on the interaction $\hat{W}$. Since the latter is typically fixed in each scientific field (we therefore drop the index $\hat{W}$ in the following), RDMFT is a particularly economic approach for addressing the ground state problem. Indeed, any effort to approximate $\mathcal{F}(\hat{\gamma})$ contributes to the solution of the ground state problem of $\hat{H}(\hat{h})$ \emph{for all} $\hat{h}$ \emph{simultaneously}. This is in contrast to wavefunction-based methods whose application to $\hat{H}(\hat{h})$ does in general not provide any simplifying information towards solving other systems $\hat{H}(\hat{h}')$.

However, the domain of the functional $\mathcal{F}(\hat{\gamma})$ is given by all those 1RDMs which follow as ground states for a particular choice of $\hat{H}(\hat{h})$. This amounts to asking the following question: For which 1RDMs $\hat{\gamma}$ does there exist a corresponding one-particle Hamiltonian $\hat{h}$ such that the map $\hat{H}(\hat{h})\mapsto\hat{\Gamma}\mapsto\hat{\gamma}$ exists? In analogy to DFT, this is the so-called \textit{pure state $v$-representability problem} even if the name $h$-representability problem would match up to its meaning more precisely. The pure state $v$-representability problem in RDMFT is extremely hard to solve and its solution is usually unknown. The second drawback of Gilbert's theorem is that the proof of the existence of a functional $\mathcal{F}(\gh)$ does not provide any systematic approach to obtaining it. We address both problems in the following section, while keeping possible differences between fermions and bosons in mind. 

\section{Constrained search formalism\label{sec:Levy}}

Gilbert's theorem provides the conceptual foundation of RDMFT and its significance should therefore never be underestimated. However, as explained in the previous section, it shows a lack of applicability due to the unknown functional $\mathcal{F}(\gh)$ and the missing solution of the pure state $v$-representability problem for fermions and bosons. To circumvent the pure state $v$-representability problem, Levy suggested extending the domain of the universal functional from all pure state $v$-representable to all pure state $N$-representable 1RDMs $\gh \in \mathcal{P}^1_N$ \cite{LE79, Lieb2002}. For bosons, necessary and sufficient conditions for a 1RDM to be pure state $N$-representable 1RDMs $\gh \in \mathcal{P}^1_N$ are known (c.f. Sec.~\ref{subsec:N_bosons}). In contrast, for fermions, the boundary $\partial \mathcal{P}^1_N$ is usually unknown due to the too complicated generalized Pauli constraints. Therefore, Valone \cite{V80} proposed to extend the domain of $\mathcal{F}$ further to all ensemble $N$-representable 1RDMs $\gh\in\mathcal{E}^1_N$. In both cases, $\gh\in \mathcal{P}^1_N$ and $\gh\in\mathcal{E}^1_N$, the resulting constrained search formalism (also often referred to as \textit{Levy's constrained search}) is based on the following consideration 
\begin{eqnarray}\label{Levy}
E_0(\hat{h}) &=& \min_{\hat{\Gamma}}\,\Tr_N\left[(\hat{h} + \hat{W})\hat{\Gamma}\right] \\
&=& \min_{\gh} \min_{\hat{\Gamma}\mapsto \gh}\,\Tr_N\left[(\hat{h} + \hat{W})\hat{\Gamma}\right] \nonumber \\
&=& \min_{\hat{\gamma}}\,\Big[\Tr_1[\hat{h}\hat{\gamma}] + \underbracket{\min_{\hat{\Gamma}\mapsto\hat{\gamma}}\Tr_N[\hat{W}\hat{\Gamma}]}_{\equiv \mathcal{F}(\gh)}\Big]\nonumber \,.
\end{eqnarray}
The variational principle in the first line of Eq.~\eqref{Levy} can refer to either pure or ensemble $N$-particle quantum states $\hat{\Gamma}$. Depending on that choice, the constrained search formalism leads to the pure/ensemble 1RDM-functional $\mathcal{F}$ with a domain given by all pure/ensemble $N$-representable 1RDMs. In particular, we define
\begin{align}
\mathcal{F}_p(\hat{\gamma}) &\equiv \min_{\mathcal{P}^N\ni\hat{\Gamma}\mapsto\hat{\gamma}} \Tr_N[\hat{W}\hat{\Gamma}] \label{eq:Fp} \\\
\mathcal{F}_e(\hat{\gamma})&\equiv \min_{\mathcal{E}^N\ni\hat{\Gamma}\mapsto\hat{\gamma}} \Tr_N[\hat{W}\hat{\Gamma}] \label{eq:Fe}
\end{align}
As in Gilbert's formulation of RDMFT, the two functionals $\mathcal{F}_p(\gh)$ and $\mathcal{F}_e(\gh)$ are universal in the sense that they only depend on the fixed interaction $\hat{W}$ and not on the one-particle Hamiltonian $\hat{h}$.
Since the trace map $\tr_N[\cdot]$ is linear and the domain $\mathcal{E}^1_N$ of all ensemble $N$-representable 1RDMs in convex, the ensemble functional $\mathcal{F}_e(\gh)$ is also convex \cite{Zumbach1985}. 

Next, we investigate the relation between the two functionals $\mathcal{F}_p(\gh)$ and $\mathcal{F}_e(\gh)$. 
The variational principle in combination with $\mathcal{P}^1_N\subseteq \mathcal{E}^1_N$ leads for all $\gh\in\mathcal{E}^1_N$ to 
\begin{equation}\label{eq:FpleqFe}
\mathcal{F}_e(\gh)\leq \mathcal{F}_p(\gh)\,.
\end{equation}
In addition, the two universal functionals must coincide on the set of all pure state $v$-representable 1RDMs and further be equal to the universal functional defined in Gilbert's theorem on this set. To prove this statement, consider a $\hat{\Gamma}\in \mathcal{P}^1_N$ mapping to a pure state $v$-representable 1RDM $\gh_v$ which corresponds to the ground state energy $E_0(\hat{h}) = \tr[\hat{h}\gh_v] + \mathcal{F}(\gh_v)$ for a particular choice of $\hat{h}$. Then, due to Gilbert's theorem and Eq.~\eqref{eq:FpleqFe} it follows that $\mathcal{F}_e(\gh_v)=\mathcal{F}_p(\gh_v)$. 

\begin{figure}[htb]
\centering
\begin{subfigure}[t]{.49\textwidth}
\centering
\includegraphics[width=0.95\linewidth]{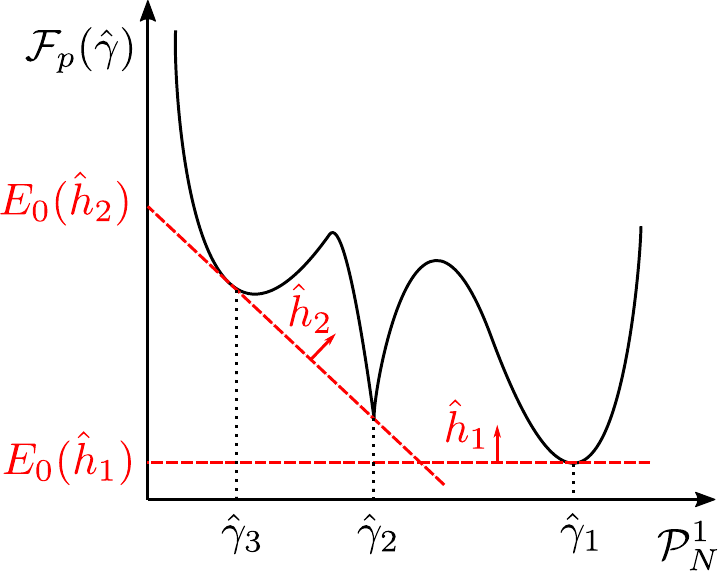}
\end{subfigure}
\begin{subfigure}[t]{.49\textwidth}
\centering
\includegraphics[width=0.95\linewidth]{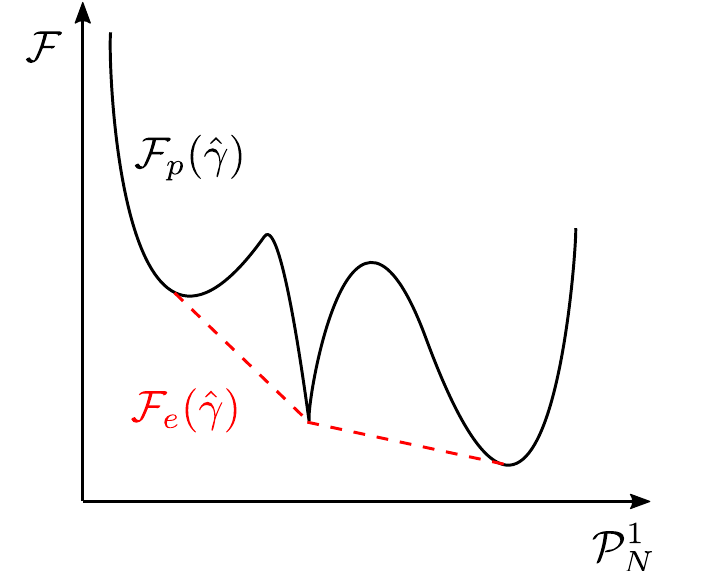}
\end{subfigure}
\medskip
\begin{minipage}[t]{\textwidth}
\caption{Illustration of the relation between the functionals $\mathcal{F}_p(\gh)$ and $\mathcal{F}_e(\gh)$.  Left: Determining for a hyperplane $\langle\hat{h}, \gh\rangle + E$ whose normal vector is determined through the one-particle Hamiltonian $\hat{h}$ the largest $E$ such that the upper halfspace contains $\mathcal{F}_p(\gh)$ entirely yields the ground state energy $E_0(\hat{h})$ as well as the ground state 1RDM $\gh$. For a non-convex functional, not all 1RDMs can be obtained as ground states for a particular choice of $\hat{h}$, and  thus not all 1RDMs are pure state $v$-representable. Right: The ensemble functional $\mathcal{F}_e(\gh)$ is obtained if the minimization procedure illustrated in the left panel is performed for all possible directions $\hat{h}$ and equal to the lower convex envelope of $\mathcal{F}_p(\gh)$.
\label{fig:LF_conv}}
\end{minipage}
\end{figure}

Using the Legendre-Fenchel transformation introduced in Sec.~\ref{subsec:LF}, we can actually find a much stronger connection between $\mathcal{F}_e(\gh)$ and $\mathcal{F}_p(\gh)$ as provided by Eq.~\eqref{eq:FpleqFe}. Since the one-particle Hamiltonian $\hat{h}$ and the 1RDM $\hat{\gamma}$ are conjugate variables (cf. Eq.~\eqref{eq:1RDM_v3}), it is natural to consider the following Legendre-Fenchel transformation of the universal functional 
\begin{equation}\label{eq:LF_F}
\begin{split}
\mathcal{F}^*(\hat{h}) &= \sup_{\gh}\left[\langle \hat{h}, \gh\rangle - \mathcal{F}(\gh)\right]\\\
&= -\inf_{\gh}\left[\mathcal{F}(\gh) - \langle\hat{h}, \gh\rangle\right]\\\
&= -E_0(-\hat{h})\,,
\end{split}
\end{equation}
where we treat both functionals, $\mathcal{F}_e(\gh)$ and $\mathcal{F}_p(\gh)$, together. In the last equality we replaced the infimum by minimum, which is valid because $\mathcal{F}$ is continuous and both sets $\mathcal{P}^1_N$ and $\mathcal{E}^1_N$ are compact. Also recall that from Eq.~\eqref{eq:LF_F} it follows that the universal functional $\mathcal{F}(\gh)$ and the ground state energy $E_0(\gh)$ are related through the Legendre-Fenchel transformation. We illustrate in the left panel of Fig.~\ref{fig:LF_conv} the geometrical interpretation of the minimization of the energy functional in Levy's constrained search for the pure state universal functional $\mathcal{F}_p(\gh)$. First, we observe that the Legendre-Fenchel transformation is motivated by the observation that a function $f(x)$ can be equivalently characterized either through the set of all tuples $(x,f(x))$ or through the set of all tangents of $f(x)$. We can now apply this idea to the energy minimization in Eq.~\eqref{Levy}. Every one-particle Hamiltonian defines a hyperplane $\langle\hat{h}, \gh\rangle$ through the inner product $\langle\cdot, \cdot\rangle$ on the Euclidean space of hermitian matrices. The hyperplane $h=\langle\hat{h}, \gh\rangle$ goes through the origin and is then shifted upwards until it touches the graph of $\mathcal{F}_p(\gh)$ such that the upper closed halfspace still contains $\mathcal{F}_p(\gh)$ entirely. According to Eq.~\eqref{eq:LF_F}, the ground state energy for the particular choice of $\hat{h}$ follows from the intersection of the hyperplane with the $\mathcal{F}$-axis and the corresponding $\gh$ is the ground state 1RDM. For $\hat{h}_1$, the ground state is unique since the hyperplane depicted by the red dashed line touches the graph of $\mathcal{F}_p(\gh)$ at a single point and the corresponding ground state energy is given by $E_0(\hat{h}_1)$. Moreover, we can even understand which 1RDMs are pure state $v$-representable using this illustration of Levy's constrained search. Consider the hyperplane with normal vector $\hat{h}_2$ which is tangent to $\mathcal{F}_p(\gh)$ at both points $\hat{\gamma}_2$ and $\gh_3$. Any 1RDM between $\gh_2$ and $\gh_3$ can never be reached by a hyperplane such that the upper halfspace contains the entire graph of $\mathcal{F}_p(\gh)$. Therefore, these 1RDMs cannot be ground states for any choice of $\hat{h}$ and thus are not pure state $v$-representable. Furthermore, the two 1RDMs $\gh_2$ and $\gh_3$ are degenerate because they correspond to the same ground state energy. 
By performing this procedure for all possible directions $\hat{h}$ we arrive at the convex hull of the pure state functional. 

We can now use Eq.~\eqref{eq:LF_F} to obtain a relation between the universal functionals $\mathcal{F}_p(\gh)$ and $\mathcal{F}_e(\gh)$. As discussed in Sec.~\ref{subsec:LF}, biconjugation leads to 
\begin{equation}\label{eq:bi_F}
\mathcal{F}^{**}(\gh) = \sup_{\hat{h}}\left[E_0(\hat{h}) - \langle\gh, \hat{h}\rangle \right]\,.
\end{equation}
Moreover, the biconjugate $\mathcal{F}^{**}(\gh)$ is equal to $\mathcal{F}(\gh)$ if and only if the universal functional is convex. Since the ensemble functional $\mathcal{F}_e(\gh)$ is convex, it follows that $\mathcal{F}_e^{**}(\gh)=\mathcal{F}_e(\gh)$. In general, the biconjugate is the closure of the lower convex envelope (see Eq.~\eqref{eq:biconjugate}). Thus, we obtain for the pure state functional $\mathcal{F}_p^{**}(\gh) = \mathrm{conv}(\mathcal{F}_p(\gh))$. Note that the closure operation can be omitted for a continuous function. 
Further, $\mathcal{F}_e(\gh)$ and $\mathcal{F}_p(\gh)$ both follow, according to Eq.~\eqref{eq:bi_F}, from the Legendre-Fenchel transformation of the ground state energy. Hence, they are related through \cite{Schilling2018}
\begin{equation}\label{eq:Fe_convFp}
\mathcal{F}_e\equiv\mathrm{conv}(\mathcal{F}_p)\,. 
\end{equation}
A detailed proof of the above equation is given in Ref.~\cite{Schilling2018} and therefore omitted at this point. Combining the result from Eq.~\eqref{eq:Fe_convFp} to the general statement from convex analysis in Eq.~\eqref{eq:FssleqF}, we arrive again at $\mathcal{F}_e(\gh)\leq \mathcal{F}_p(\gh)$ which was already obtained in Eq.~\eqref{eq:FpleqFe}.
In the right panel of Fig.~\ref{fig:LF_conv} we illustrate a non-convex functional $\mathcal{F}_p(\gh)$ and its lower convex envelope determining $\mathcal{F}_e(\gh)$. Using Eq.~\eqref{eq:conv_fx} and Eq.~\eqref{eq:Fe_convFp}, we obtain for the ensemble functional
\begin{equation}
\mathcal{F}_e(\gh) = \min\left\{\sum_j\ q_j\mathcal{F}_p(\hat{\gamma}_j)\,\right\vert\left.\,\hat{\gamma}_j\in\mathcal{P}^1_N, \hat{\gamma} = \sum_jq_j\hat{\gamma}_j, \sum_j q_j=1, q_j\geq 0   \right\}\,.
\end{equation}
Hence, the pure functional $\mathcal{F}_p(\gh)$ determines the ensemble functional $\mathcal{F}_e(\gh)$ on its entire domain, even if for fermions the domain $\mathrm{dom}(\mathcal{F}_e(\gh))=\mathcal{E}^1_N$ of $\mathcal{F}_e(\gh)$ is larger than $\mathrm{dom}(\mathcal{F}_p(\gh))=\mathcal{P}^1_N$. This remarkable consequence of Eq.~\eqref{eq:Fe_convFp} indicates that in the context of fermionic quantum systems, the complexity of the pure one-body $N$-representability conditions (generalized Pauli constraints) will hamper the calculation of either the functional's domain or the functional itself \cite{Schilling2018}. It is thus one of the major future challenges to investigate how the generalized Pauli constraints enter the universal functional and how this knowledge can be used to improve the approximated functionals.
Recall that for bosons, the domains of $\mathcal{F}_e(\gh)$ and $\mathcal{F}_p(\gh)$ coincide due to $\mathcal{E}^1_N = \mathcal{P}^1_N$, as proven in Sec.~\ref{subsec:N_bosons}.

Valone's idea to circumvent the pure state $N$-representability constrains for fermions through the relaxation of the minimization in Levy's constrained search to a minimization over a convex domain and a convex functional also has advantages for bosons, where the pure state $N$-representability constrains are known. For fermions and bosons, the relaxation of the non-convex minimization problem to a convex one has two main advantages: in case of a convex function, every local minimum is also a global one and for a strictly convex function, the minimum is even unique.

%

\section{Bosonic RDMFT for homogeneous systems\label{sec:RDMFT_hom}}

Since a huge part of this thesis (primarly Ch.~\ref{ch:RDMFT_BEC} and Sec.~\ref{sec:w-RDMFT_BEC}) is concerned with \emph{homogeneous} BECs, we discuss in this section the specific case of one-particle Hamiltonians which are diagonal in the momentum representation, i.e., there is only a kinetic energy operators $\hat{t}$ contributing to $\hat{h}$ but no external potential, $\hat{h}\equiv \hat{t}$. Implementing this within the constrained search formalism \eqref{Levy} identifies the momentum occupation numbers $\bd{n}\equiv (n_{\bd{p}})$ as the natural variables and the pure functional follows as
\begin{equation}\label{LevyP}
\F(\bd{n}) \equiv \min_{\ket{\Phi}\mapsto \bd{n}}\bra{\Phi}\hat{W}\ket{\Phi}\,.
\end{equation}
While we are focussing in the following on the pure functional, it is worth recalling that the corresponding ensemble functional would follow as the lower convex envelop of the pure functional $\F$ \cite{Schilling2018}. Also their two domains $\triangle$ coincide as shown in Sec.~\ref{subsec:N_bosons}.
To describe $\triangle$, let us first use the normalization constraint to get rid of the entry $n_{\bd{0}}= N-\sum_{\bd{p}\neq \bd{0}} n_{\bd{p}}$.
Then the functional's domain follows as
\begin{equation}\label{simplex}
\triangle = \Big\{\bd{n} \equiv (n_{\bd{p}})_{\bd{p}\neq \bd{0}} \Big| n_{\bd{p}}\geq 0, \sum_{\bd{p}\neq \bd{0}} n_{\bd{p}}\leq N \Big\}\,.
\end{equation}
In case of finite lattice models there are finitely many momenta $\bd{p}$ (forming a discrete Brillouin zone), while in case of continuous systems or infinite lattices, $\bd{n}$ will have infinitely many entries.
It will be instructive to also understand the functional's domain $\triangle$ from a geometric point of view.
Apparently, $\triangle$ is a convex set which after all takes the form of a simplex with vertices $\bd{0}$ and $\bd{v}_{\bd{p}} = N \bd{e}_{\bd{p}}$, where $\bd{e}_{\bd{p}}$ has only one non-vanishing entry $1$ at position $\bd{p}$. In Ch.~\ref{ch:RDMFT_BEC}, we are mainly interested in the regime of BEC which is characterized by an occupation number $n_{\bd{0}}$ close to $N$. This corresponds in the simplex $\triangle$  to the neighbourhood of the vertex $\bd{0}$, which can equivalently be characterized by the simultaneous saturation of the constraints $n_{\bd{p}}\geq 0$ for all $\bd{p}\neq \bd{0}$.

\section{Symmetries\label{sec:symm}}

Symmetries play an important role in physics and exploiting them can simplify the theoretical description of quantum systems tremendously. We therefore summarize in this section the most important symmetries which appear in this thesis and discuss their impact on RDMFT. It is important to notice, that whenever we exploit a symmetry of the interaction $\hat{W}$ in the derivation of a universal functional $\mathcal{F}(\hat{\gamma})$, only those one-particle Hamiltonians $\hat{h}$ are allowed to be considered in Levy's constrained search which satisfy this symmetry as well. 

\subsection{Translational invariance}

The main disadvantage of RDMFT compared to DFT is that for a $d$-dimensional one-particle Hilbert space, the 1RDM involves $d^2$ degrees of freedom in contrast to the $d$ degrees of freedom required to describe the density as in DFT. Thus, numerical calculations using RDMFT usually have a higher computational cost than in DFT. Using the spectral decomposition of the 1RDM, the constrained search formalism involves both, the natural occupation numbers and the natural orbitals, which have to be optimized. This task simplifies drastically for translational invariant systems. In Sec.~\ref{sec:RDMFT_hom} we already discussed the application of RDMFT to homogeneous Bose gases which are one example of a translational invariant system. Since translational invariance implies that the Hamiltonian $\hat{H}$ of the quantum system commutes with the total momentum operator $\hat{P}$, i.e.~$[\hat{H}, \hat{P}]=0$, the natural orbitals are given by plane waves and RDMFT reduces to a NON-functional theory omitting possible disadvantages of RDMFT in relation to DFT. 

\subsection{Parity-symmetry\label{sec:parity}}

The parity-symmetry of common physical spaces implies the additional symmetry $n_{\bd{p}}= n_{-\bd{p}}$ for all momenta $\bd{p}$. This does not really change the geometric form of the functional's domain for homogeneous Bose gases but just allows us to skip in the definition \eqref{simplex} for every pair $(\bd{p},-\bd{p})$ of momenta one of the two occupation numbers $n_{\pm \bd{p}}$.
In the context of RDMFT, respecting this common symmetry would mean to restrict the kinetic energy operators $\hat{t} \equiv \sum_{\bd{p}}\varepsilon_{\bd{p}}\hat{n}_{\bd{p}}$ to those with $\varepsilon_{\bd{p}}=\varepsilon_{-\bd{p}}$. 

\subsection{Invariance under permutations}

Let $\pi:\{\bd{p}\}\to\{\bd{p}\}$ be a permutation which leaves $\bd{p}=\bd{0}$ invariant, i.e $\pi(\bd{p}=\bd{0}) = \bd{0}$. Its unitary representation on the one-particle Hilbert space $\mathcal{H}_1$ is denoted by $\hat{u}(\pi):\mathcal{H}_1\to \mathcal{H}_1$ which acts on the momentum states $|\bd{p}\rangle\in \mathcal{H}_1$ as $|\bd{p}\rangle\mapsto \hat{u}(\pi)|\bd{p}\rangle\equiv |\pi(\bd{p})\rangle$ and on the N-particle Hilbert space we have $\hat{U}(\pi):\mathcal{H}_N\to\mathcal{H}_N$ where $\hat{U}(\pi) \equiv \hat{u}(\pi)^{\otimes^N}$. If an interaction $\hat{W}$ is invariant under permutations $\pi$, i.e.~$\left[\hat{W}, \hat{U}(\pi)\right] = 0$, then the universal 1RDM-functional $\mathcal{F}(\bd{n})$ must have the same symmetry and $\mathcal{F}(\bd{n}) = \mathcal{F}(\pi(\bd{n}))$, because
\begin{equation}
\begin{split}
\mathcal{F}(\pi(\bd{n})) &= \min_{\hat{\Gamma}\mapsto\pi(\bd{n})}\tr\left[\hat{W}\hat{\Gamma}\right] \\\
&= \min_{\hat{U}(\pi)\hat{U}^\dagger(\pi)\Gamma \hat{U}(\pi)\hat{U}^\dagger(\pi)\mapsto\pi(\bd{n})}\tr\left[\hat{W}\hat{\Gamma}\right] \\\
&=\min_{\hat{U}^\dagger(\pi)\hat{\Gamma} \hat{U}(\pi)\mapsto\bd{n}}\tr\left[\hat{W}\hat{\Gamma} \hat{U}(\pi)\hat{U}^\dagger(\pi)\right]\\\
&= \min_{\hat{U}^\dagger(\pi)\hat{\Gamma} \hat{U}(\pi)\mapsto\bd{n}}\tr\left[\hat{W}\hat{U}^\dagger(\pi)\hat{\Gamma} \hat{U}(\pi)\right]\\\
&= \min_{\hat{\Gamma}^\prime\mapsto\bd{n}}\tr\left[\hat{W}\hat{\Gamma}^\prime\right]
\end{split}
\end{equation}
where $\bd{n} \equiv (n_{\bd{p}})_{\bd{p}}$ and $\pi(\bd{n}) \equiv \left(n_{\pi(\bd{p})}\right)_{\bd{p}}$. We will encounter an interaction with the above discussed permutation invariance in Sec.~\ref{sec:Bog}.

%


\chapter{RDMFT for Bose-Einstein condensates\label{ch:RDMFT_BEC}}

In this chapter, we derive a first-level functional for bosonic ground state RDMFT, which is believed to be exact in leading order in the regime close to complete Bose-Einstein condensation (BEC). For this purpose, we summarize in Sec.~\ref{sec:prel_BEC} the most important properties of a BEC needed in the following sections.
In Sec.~\ref{sec:Bog}, we recall conventional Bogoliubov theory and explain in Sec.~\ref{sec:obstacles} why the latter is incompatible with RDMFT from a conceptual point of view. Then, in Sec.~\ref{sec:pair} we present a particle-number conserving modification of Bogoliubov's theory which eventually allows us to derive the universal functional within the BEC regime in Sec.~\ref{sec:derivfunc}. We then illustrate in Sec.~\ref{sec:appl} how bosonic RDMFT is applied and present functionals for a number of different systems. Finally, we establish and illustrate the novel concept of a BEC force in Sec.~\ref{sec:BECforce}.

\section{Preliminaries \label{sec:prel_BEC}}

\subsection{Introduction to BEC}

On a qualitative level, Bose-Einstein condensation is often explained as the transition occurring in a classical gas of bosons whose temperature is lowered until the thermal de-Broglie wavelength $\lambda_{\mathrm{dB}}$ of the particles becomes comparable to their mean inter-particle distance $l$ such that $l\approx\lambda_{\mathrm{dB}}$. At the corresponding transition temperature $T_c$, the wave packets associated with the particles start to overlap until they form a coherent matter wave at $T=0$. Remarkably, this consideration does not require any interactions between the particles in contrast to other phase transitions. Consequently, in a non-interacting Bose gas, a BEC at zero temperature is characterized by the macroscopic occupation of a single state holding even for sufficiently weak interactions. It is important to note that complete condensation occurs only for a non-interacting gas at zero temperature. As we will understand in Sec.~\ref{sec:Bog}, even weak interactions at $T=0$ cause excitations of the Bose gas due to the interactions between the particles. This phenomenon is called \textit{quantum depletion} and its degree is given by the fraction of non-condensed bosons. In principle, Bose-Einstein condensation can occur in any state. For the most prominent example of a homogeneous Bose gas, which we already discussed in the context of RDMFT in Sec.~\ref{sec:RDMFT_hom}, this would be the zero momentum state, whereas in a harmonic trap BEC occurs in the lowest energy state in both,  momentum and coordinate space. However, realizing a BEC is in general an extremely hard problem to tackle from an experimental point of view because it requires efficient cooling as well as efficient trapping methods. Following the development of laser cooling, usually used as a pre-cooling technique, magnetic trapping, and evaporate cooling, the first experimental realizations of BEC using alkali atoms were reported in 1995 \cite{Anderson1995, Ketterle1995, Bradley1995}. The detection of the BEC is usually performed by a time of flight measurement, where the atoms are initially prepared in a trap which is then suddenly switched off. Afterwards, the gas cloud expands during the time of flight period $t_\mathrm{TOF}$ before it is measured via absorption imaging. 
The signature of a BEC is then a sharp peak in the center of the velocity (or density) distribution, in contrast to a thermal gas which displays an isotropic distribution. 

We proceed in the next section by introducing criteria for the existence of a BEC providing the foundation for our functional theoretical approach to BEC. 

\subsection{Criteria for BEC}
 
In this section, we establish a connection between RDMFT and BEC through the criterion for the existence of a macroscopically occupied state. From this discussion we then conclude that bosonic reduced density matrix functional theory should be particularly well-suited to describe Bose-Einstein condensates because it involves the one-particle reduced density matrix $\hat{\gamma}$ as the natural variable. 

\subsubsection{Penrose and Onsager criterion}

The most general criterion for the existence of BEC was introduced by Penrose and Onsager already in 1956 \cite{Penrose1956}
\begin{equation}\label{eq:Penrose}
\lambda_\mathrm{max}=\max_\alpha\bra{\alpha}\hat{\gamma}\ket{\alpha}\sim \mathcal{O}(N)\,.
\end{equation}
Thus, BEC occurs whenever the largest eigenvalue $\lambda_{\mathrm{max}}$ of the 1RDM $\hat{\gamma}$ is of the order of the total particle number $N$ for a macroscopically large $N$. As a matter of fact, $\lambda_\mathrm{max}$ quantifies the number of condensed bosons, without requiring any preceding information about the maximally populated one-particle state $\ket{\alpha_\mathrm{max}}$. Moreover, the condition in Eq.~\eqref{eq:Penrose} can be further generalized because for the existence of BEC in a system it is sufficient that \textit{at least one} eigenvalue $\lambda_\alpha$ of the 1RDM is of order $N$, i.e.~$\lambda_\alpha \sim\mathcal{O}(N)$. This allows us to distinguish between two different kinds of BEC: If there is exactly one eigenvalue $\lambda_\alpha$ fulfilling $\lambda_\alpha\sim\mathcal{O}(N)$, the system is in a so-called single BEC. However, BEC can also occur in several states leading to a fragmented BEC which is then characterized by more than one $\lambda_\alpha$ satisfying $\lambda_\alpha\sim\mathcal{O}(N)$.

Since the criterion in Eq.~\eqref{eq:Penrose} defines BEC through the eigenvalues of the 1RDM, the Penrose and Onsager criterion directly indicates, that bosonic RDMFT should be well suited to describe BEC. 
Further, it applies not only to uniform systems but also to non-homogeneous and finite systems. The Penrose and Onsager criterion is therefore more general than the concept of off-diagonal long-range order of $\gamma(\bd r, \bd{r}^\prime)$ \cite{Yang1962} which we discuss in the following. 

\subsubsection{Off-diagonal long-range order}

The off-diagonal long-range order (ODLO) of a many-boson system is characterized by non-vanishing off-diagonal matrix elements of the 1RDM in coordinate space \cite{Yang1962},
\begin{equation}\label{ODLO}
\lim_{|\bd{r}-\bd{r}^\prime|\to\infty}\gamma(\bd{r},\bd{r}^\prime) = \lim_{|\bd{r}-\bd{r}^\prime|\to\infty} \bra{\bd{r}}\hat{\gamma}\ket{\bd{r}^\prime}\neq 0\,.
\end{equation}
Here we restrict to the discussion of ODLO in context of BEC, but the concept has in general a much broader scope as explained in Ref.~\cite{Yang1962}. For a translational invariant system, the N-boson density operator commutes with the momentum operator. The 1RDM is thus diagonal in momentum representation and we have $\bra{\bd p}\hat{\gamma}\ket{\bd{p}^\prime}=n_{\bd p}\delta_{\bd p,\bd{p}^\prime}$. Therefore, it is natural to consider the Fourier transform of the matrix elements $\gamma(\bd{r},\bd{r}^\prime)$ which is given by
\begin{equation}\label{eq:ODLO_FT}
\gamma(\bd{r}, \bd{r}^\prime) = \frac{1}{V} \int\rmd^3\bd{p}\,n_{\bd p}\rme^{i\bd{p}(\bd r - \bd{r}^\prime)}\,.
\end{equation}
For a many-body system with in total $N$ bosons, BEC is requires a macroscopic fraction $N\alpha$, $0<\alpha\leq 1$, of particles in a state with momentum $\bd{p}$. For non-interacting free particles, all particles would occupy the state with $\bd{p}=\bd{0}$ and in the case of sufficiently weak interactions without external potential, there the is still a macroscopic fraction of particles in the state with $\bd p=\bd 0$. Together with Eq.~\eqref{eq:ODLO_FT}, this leads to
\begin{equation}
\lim_{|\bd{r}-\bd{r}^\prime|\to\infty}\gamma(\bd{r},\bd{r}^\prime) =\frac{N\alpha}{V}\,,
\end{equation}
which is indeed a non-zero value. From the definition of ODLO in Eq.~\eqref{ODLO}, it follows immediately that this criterion for BEC can only be applied to infinite and homogeneous systems. Since the existence of ODLO implies that one eigenvalue of the 1RDM must be macroscopic, it can be understood as a special case of the Onsager and Penrose criterion in Eq.~\eqref{eq:Penrose}. 
%
%

\subsection{S-wave scattering approximation in the context of ultracold atomic gases\label{subsec:S-wave_gases}}

The s-wave scattering approximation follows from the partial wave expansion in the limit of low energies (see Appendix \ref{app:s-wave} for a more formal derivation). It is based on the assumption that for low energetic particles and short-ranged interactions, the de Broglie wavelength of the particles is large compared to the range of the interaction potential. Therefore, the particles cannot resolve the structure of the potential at small length scales, and only the potential at long length scales is important for the scattering process. In the following, we consider elastic collisions between the particles which can be described by a conservative interaction potential $V(\bd r)$ that only depends on the relative coordinate $\bd r \equiv \bd{r}_2 - \bd{r_1}$ of two particles labelled by $1$ and $2$. Since we are ultimately interested in the description of interactions in Bose-Einstein condensates, we need to understand the properties of $V(\bd r)$ in the case of ultracold, dilute Bose gases. 

In Ch.~\ref{ch:RDMFT}, we already explained that different physical systems of interest are characterized by a fixed pair interaction $\hat{W}$. In the context of ultracold, dilute atomic gases, the interaction between the particles is usually described by an attractive van der Waals interaction $V(r)=-C_6/r^6$ at large distances, and the asymptotic behaviour of the interaction is included in the van der Waals coefficient $C_6$. At small distances, the interaction potential becomes repulsive and diverges, because the orbitals of the atoms start to overlap, and Pauli's exclusion principle prohibits the electrons in these orbitals to occupy the same state. This strong repulsive interaction at small distances $r \equiv |\bd r|$ is usually modelled by a hard-core cutoff $r_c$ leading to an approximate interaction potential \cite{Bloch2008}. Since the van der Waals interaction and the strong repulsive part of the interaction are isotropic, we have $V(\bd r)=V(r)$ and the partial wave expansion can be used to solve the scattering problem.

Moreover, the van der Waals interaction and thus the total scattering potential $V(r)$ are short-ranged. This means that effects of the interaction can be neglected outside a finite scattering volume, which is required for the validity of the s-wave scattering approximation discussed below. For van der Waals interactions, the range of the interaction potential is usually given by the van der Waals length $l_\mathrm{vdW}$. In general, an interaction potential is called short-ranged if it decays faster than $1/r$ \cite{Sadeghpour00}. A prominent example of an interaction that is not short-ranged is the Coulomb interaction between charged particles, which we will encounter in Sec.~\ref{subsec:charged}. However, in the following discussion we focus only on short-ranged interactions in ultracold atomic gases.  

\begin{figure}[htb]
\centering
\includegraphics[width=0.75\linewidth]{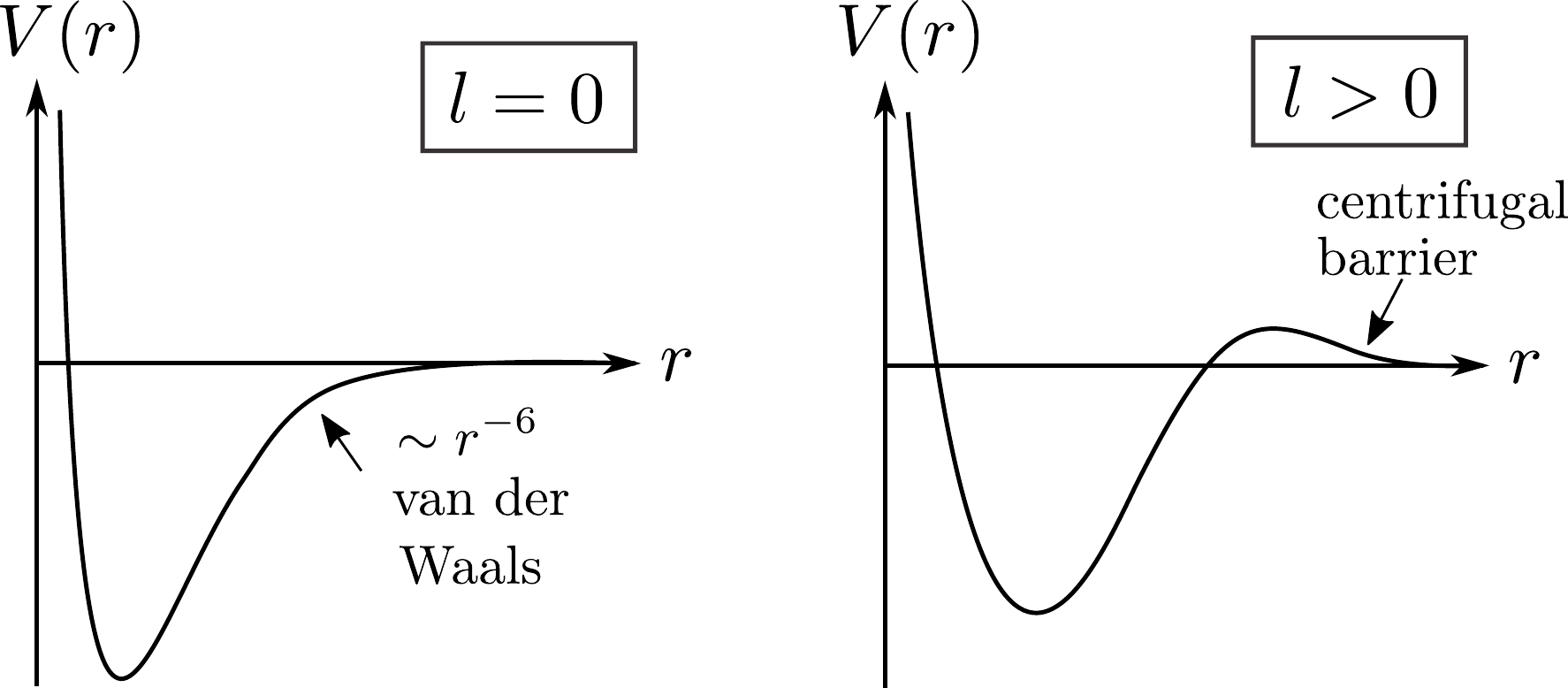}
\caption{Sketch of the effective interaction potential. Left: $l=0$ leads to s-wave scattering. Right: For $l>0$, the interaction potential is superposed with an effective $r^{-2}$-potential from angular momentum conservation leading to a centrifugal barrier at large $r$ which repels low-energetic particles. 
\label{fig:potential}}
\end{figure}

At distances much larger than the range of the interaction potential, the wave function consists of an incoming plane wave and an outgoing radial wave such that the wave function has the asymptotic form
\begin{equation}
\Psi_{\bd k}(\bd r) = \rme^{i\bd k \bd\cdot r} + f(\theta, \varphi, k)\frac{\rme^{ikr}}{kr}\,.
\end{equation}
The scattering amplitude $f(\theta, \varphi, k)$ contains all information about the scattering process. Moreover, the scattering amplitude is independent of the angle $\varphi$ due to the spherical symmetry of the potential. Since the problem is isotropic, as discussed above, one can now apply the partial wave expansion in spherical harmonics. Since the angular momentum is conserved during the interaction, partial waves with different quantum number $l$ scatter independently. Moreover, depending on the value of $l$ they experience a different effective scattering potential. For $l=0$ nothing changes and the atoms see the same interaction potential $V(r)$ which consists of a strong repulsive part at short distances and the attractive van der Waals interactions at large distances, as discussed above. For partial waves with $l=1, 2, ...$, the scattering potential changes according to $V(r)\to V(r) + l(l+1)/(2m^*r^2)$, where $m^*$ denotes the reduced mass of the two particles. The modification of the scattering potential by the effective $1/r^2$-potential follows directly from the one-dimensional radial Schrödinger equation \cite{shankar1995}. We illustrate both cases in Fig.~\ref{fig:potential}.  
It follows that for $k\to 0$, the scattering process is dominated by s-wave scattering, which means $l=0$. We can therefore neglect all partial waves, except the s-wave, in the limit of low energies. Moreover, the scattering amplitude $f$ for low momenta becomes independent of the energy and the scattering angle $\theta$ \cite{pethick08}. It is thus given by a constant value,
\begin{equation}
f(\theta, k)  \xrightarrow{k\to 0} -a\,,
\end{equation} 
defining the s-wave scattering length (or simply scattering length) $a$. Hence, the scattering process is characterized by a universal parameter $a$. The scattering length is universal in the sense that all potentials sharing the same $a$ lead to the same low-energy scattering. It thus crucially simplifies the theoretical description of the scattering process because the actual potential can then be replaced by a pseudo-potential reproducing the correct value for $a$. However, the exact value of $a$ depends on all the microscopic details of the two-body interaction. Therefore, it is hard to predict it theoretically and, usually, $a$ is obtained from experiments \cite{Bloch2008}. In three dimensions, $a>0$ corresponds to a repulsive pseudo-potential and $a<0$ describes an attractive pseudo-potential yielding the same $a$ as the actual interaction potential. 

The s-wave scattering length $a$ thus simplifies the theoretical description of the interaction between two particles tremendously and plays an important role to derive the ground state energy and low-lying energy spectrum of weakly interacting homogeneous Bose gases because it allows for a perturbative treatment of a pseudo-potential \cite{pita, pethick08}. The actual interaction potential is then usually replaced by a pseudo-potential of the form $V(\bd{r})= g\delta(\bd{r})$ with coupling constant $g$ reproducing the correct value for $a$. In Sec.~\ref{sec:appl}, we apply the s-wave scattering approximation to verify that the universal functional obtained for a dilute Bose gas in 3D leads to the well-known result for the ground state energy. 

\subsection{Weakly interacting bosons in different dimensions\label{sec:weaklydiffd}}

As explained in the sections above, the the existence of a BEC usually requires sufficiently weak interactions. The classification of the interaction strength will become even more important in the discussion of the Bogoliubov approximation in Sec.~\ref{sec:Bog} which is only valid in the limit of weak interactions. 
Heuristically, the interaction strength between particles is defined as the ratio between the interaction energy $I$ and the kinetic energy $t$ \cite{Petrov04}. It follows that a gas is called weakly interacting if $|I|\ll t$.

\subsubsection{Neutral particles}

The kinetic energy of a particle in a box with size $l$ can be approximated by $t\approx 1/m l^2$, where $m$ is the mass of the particle, and we set $\hbar=1$. For neutral bosons the interaction energy per particle is given by $I=ng$ with density $n$, coupling constant $g$ and mean inter-particle spacing $l$ \cite{Petrov04}. The coupling constant $g$ for the pair interaction is closely related to the scattering length $a$ through the Born series for the scattering length.

In three dimensions, the scattering length $a_\mathrm{3D}$ and the coupling constant $g_{\mathrm{3D}}$ for a pseudopotential $V(\bd{r}) = g_{\mathrm{3D}}\delta(\bd{r})$ are related through \cite{pita}
\begin{equation}
g_{\mathrm{3D}} = \frac{4\pi}{m}a_{\mathrm{3D}}\,.
\end{equation}
Together with $l\sim n^{-1/3}$, the condition $I\ll t$ is fulfilled if 
\begin{equation}\label{eq:weakint3D}
n|a|^3\ll 1\,.
\end{equation}
Note that this condition is equivalent to diluteness, as can easily be seen: One calls a gas of particles dilute if the mean inter-particle spacing $l$ is much larger than the characteristic range $r_0$ of the interaction potential, which means $l\gg r_0$. For low energy scattering at a short-range potential, the interaction is fully characterized by the scattering length $a$ and away from resonances $a \simeq r_0$. From these considerations, we arrive again at the condition \eqref{eq:weakint3D} for a weakly interacting gas.

For a two-dimensional pseudopotential $V(\bd{r})= g_\mathrm{2D}\delta(\bd{r})$, the coupling constant for a homogeneous Bose gas depends logarithmically in the scattering length \cite{friedrichs15, Popov84}.
\begin{equation}\label{eq:g2D}
g_{\mathrm{2D}}=\frac{4\pi}{m}\left[\ln\left(\frac{1}{na_\mathrm{2D}}\right) \right]^{-1}\,,
\end{equation}
where the occurrence of the two dimensional density $n$ in the argument of the logarithm is required to obtain a dimensionless parameter. Moreover, away from resonances, the coupling constant is positive in the dilute regime with $n\ll 1$. 
Using $l\sim n^{-1/2}$, the system is weakly interacting if $m|g_{\mathrm{2D}}|/2\pi\ll1$, or equivalently 
\begin{equation}
n|a_\mathrm{2D}|^2\ll 1\,,
\end{equation}
holds. It thus follows that weak interactions require low densities, as in the three-dimensional case.

In contrast to the two-dimensional and three-dimensional cases discussed above, a Bose gas in one dimension is weakly interacting in the limit of high densities. For a one-dimensional contact potential $V(r)=g_\mathrm{1D}\delta(r)$, we have \cite{pita}
\begin{equation}
g_\mathrm{1D} = -\frac{2\hbar^2}{ma_\mathrm{1D}}\,.
\end{equation}
Opposite to the three-dimensional case, negative scattering length $a_\mathrm{1D}$ correspond to repulsive interactions and positive $a_\mathrm{1D}$  to attractive ones. Besides, the coupling strength and scattering length are now inverse proportional to each other, whereas in three dimensions $g_\mathrm{3D}\propto a_\mathrm{3D}$. Applying the condition $t\gg|I|$ with $l\sim 1/n$ yields \cite{pita, Petrov04, pethick08}
\begin{equation}
\frac{m|g_\mathrm{1D}|}{ n}\ll 1\,.
\end{equation}
From the above equation, we conclude that weak interactions indeed require high densities in one dimension. We will return to such an example in Sec.~\ref{subsec:Hubbard5} to illustrate the universal functional obtained in Sec.~\ref{sec:derivfunc} and its domain. 

\subsubsection{Charged bosons in 3D}

Next, we discuss the condition to have weak interactions for a charged Bose gas in three dimensions because it will appear as an example in Sec.~\ref{subsec:charged}. Due to the long-range character of the Coulomb interactions, the s-wave scattering approximation is not applicable anymore. However, we can still apply the condition $|I|\ll t$ and using the average interparticle spacing $l = (3/4\pi n)^{1/3}$ we obtain that a charged Bose gas in 3D is characterized by the dimensionless coupling constant ($\hbar=1$)
\begin{equation}
r_s\equiv\left(\frac{3}{4\pi}\right)^{1/3} me^2 \frac{1}{n^{1/3}} \,.
\end{equation}
Thus, we have $r_s\ll 1$ for high densities. Moreover, it was mathematically rigorously proven in \cite{Lieb01} that the Bogoliubov theory becomes exact in the limit $n\to \infty$. 

\section{Recap of conventional Bogoliubov theory\label{sec:Bog}}
In this section be recap the most important aspects of Bogoliubov's \cite{Bogoliubov47} well-known and experimentally confirmed \cite{Lopes17} theory to describe BEC in homogeneous bosonic quantum systems and the effect of depletion of the condensate as a result of the interaction between the particles.

The Hamiltonian describing a homogeneous system of $N$ interacting spinless bosons in first quantization ($\hbar\equiv 1$) is given by
\begin{equation}\label{H1st}
\hat{H} = -\sum_{i=1}^N\frac{1}{2m}\Delta_i + \sum_{1\leq i<j\leq N}W(\bd{x}_i-\bd{x}_j)\,.
\end{equation}
Its second quantized form in momentum representation for particles in a large box of volume $V=L^3$ and size $L$ with periodic boundary conditions then reads
\begin{equation}\label{H2nd}
\hat{H} = \sum_{\bd{p}}\varepsilon_{\bd{p}}\hat{a}_{\bd{p}}^\dagger\hat{a}_{\bd{p}} + \frac{1}{2V}\sum_{\bd{p}, \bd{q}, \bd{k}} W_{\bd{p}} \hat{a}_{\bd{p}+\bd{q}}^\dagger\hat{a}_{\bd{k}-\bd{p}}^\dagger\hat{a}_{\bd{k}}\hat{a}_{\bd{q}}\,,
\end{equation}
where $W_{\bd{p}}$ is the Fourier transform of $W(\cdot)$. In case of an isotropic pair interactions, $W$ in Eq.~\eqref{H1st} would depend only on the modulus of the distance between the particles $i$ and $j$ which in turn would imply $W_{\bd{p}}\equiv W_{|\bd{p}|}$. The vector components of the momenta $\mathbf{p}$ in Eq.~\eqref{H2nd} take the discrete values $p_i = 2\pi k_i/L$, $k_i\in \mathbb{N}$ where $i=x,y,z$. 

The most crucial feature of the Hamiltonian $\hat{H}$ and the pair interaction $\hat{W}$ is that they are conserving the particle number as well as the total momentum. Since the Hamiltonian in Eq.~\eqref{H2nd} is quartic in the operators, it cannot be diagonalized directly. Assuming a BEC at $T=0$, the standard approach to determine the ground state energy (and the low lying excited states) of the Hamiltonian \eqref{H2nd} is the Bogoliubov approximation \cite{Bogoliubov47}. It is based on the assumption that for low temperatures and sufficiently weak interactions, the zero-momentum mode is macroscopically occupied and interactions between non-condensed bosons can be neglected due to the conservation of momentum: Since application of a creation/annihilation operator $a_{\mathbf{0}}^{(\dagger)}$ to the BEC ground state leads to macroscopically large prefactors of the order $\sqrt{N}$, terms in the expansion \eqref{H2nd} of $\hat{W}$ involving less then two $\bd{0}$-indices are dropped. The resulting quartic interaction is further simplified by replacing the condensate operators $\hat{a}_{\mathbf{0}}, \hat{a}_{\mathbf{0}}^\dagger \to \sqrt{n_{\mathbf{0}}} \approx\sqrt{N}$ by a c-number. This eventually leads to the quadratic Bogoliubov Hamiltonian
\begin{equation}\label{eq:H_Bogoloiubov}
\hat{H}\cong\hat{H}_\mathrm{B} = \frac{N(N-1)W_{\bd{0}}}{2V} + \sum_{\bd{p}\neq \bd{0}}\Big[\left(\varepsilon_{\bd{p}} + nW_{\bd{p}}\right)\hat{a}_{\bd{p}}^\dagger\hat{a}_{\bd{p}}+ \frac{n}{2}W_{\bd{p}}\left(\hat{a}_{\bd{p}}^\dagger\hat{a}_{-\bd{p}}^\dagger+\hat{a}_{\bd{p}}\hat{a}_{-\bd{p}}\right)\Big]\,,
\end{equation}
which involves (besides the kinetic energy $\hat{t}$ and some trivial contributions) for each pair $(\bd{p},-\bd{p})$ an anomalous term of the form $\hat{a}_{\bd{p}}^\dagger\hat{a}_{-\bd{p}}^\dagger + \hat{a}_{\bd{p}}\hat{a}_{-\bd{p}}$.
The Bogoliubov Hamiltonian can then easily be diagonalized by a Bogoliubov transformation 
\begin{equation}
\hat{U}_B = \mathrm{exp}\left\{\frac{1}{2}\sum_{\bd{p}\neq \bd 0} \theta_{\bd{p}}\left(\hat{a}_{\bd{p}}^\dagger\hat{a}_{-\bd{p}}^\dagger - \mathrm{h.c.}\right)\right\}\,.
\end{equation}
The respective ground state follows as $\ket{\Psi} = \hat{U}_B\ket{N}$, where 
\begin{equation}
\ket{N} \equiv (N!)^{-1/2}(\hat{a}_{\bd 0}^\dagger)^N\ket{0}
\end{equation}
is the ground state of the non-interacting system and $\ket{0}$ the vacuum state. The phases $\theta_{\bd{p}}$ are chosen such that the anomalous terms in the Hamiltonian, containing either two quasiparticle annihilation ($\hat{b}_{\bd{p}}\equiv \hat{U}_B^\dagger \hat{a}_{\bd{p}}\hat{U}_B$) or creation operators ($\hat{b}_{\bd{p}}^\dagger$) vanish to eventually obtain a  diagonal quadratic form in $\hat{b}_{\bd{p}}$ (see also textbook \cite{pita} for more details).
Bogoliubov's approach can also be interpreted as the variational minimization of the Bogoliubov Hamiltonian over all trial states of the form $\hat{U}_B\ket{N}$. 

\section{Incompatibility of conventional Bogoliubov theory and RDMFT\label{sec:obstacles}}

As explained in the previous section, Bogoliubov's approximation results in a Hamiltonian which is not particle-number conserving anymore.
At the same time, RDMFT defines a universal functional $\F(\bd{n})$ (or more generally $\F(\gh)$) by minimizing the interaction Hamiltonian according to \eqref{LevyP} with respect to quantum states with a \emph{fixed} total particle number $N$ and fixed momentum occupation numbers $\bd{n}$.  
To emphasize this statement even more, let us consider an interaction of the form $\hat{W} = \sum_{\thickbar{\bd{p}}}\hat{W}_{\thickbar{\bd{p}}}$, where $\hat{W}_{\thickbar{\bd{p}}}$ only contains operators acting on $(\bd 0,\bd{p}, -\bd{p})$ and $\thickbar{\bd{p}}$ denotes the pair $\thickbar{\bd{p}}=(\bd{p}, -\bd{p})$. Levy's constrained search (see Eq.~\eqref{Levy}) then leads to 
\begin{equation}
\begin{split}
\mathcal{F}(\bd{n}) &= \min_{\hat{\Gamma}\mapsto\bd{n}}\sum_{\thickbar{\bd{p}}} \tr[\hat{W}_{\thickbar{\bd{p}}}\hat{\Gamma}] \\\
&= \min_{\hat{\Gamma}\mapsto\{\hat{\Gamma}_{\bd 0, \thickbar{\bd{p}}}\}_{\thickbar{\bd{p}}}}\left.\left(\min_{\forall\thickbar{\bd{p}}:\hat{\Gamma}_{\bd 0, \thickbar{\bd{p}}}\,\mapsto(n_{\bd{p}}, n_{-\bd{p}}=n_{\bd{p}})} \sum_{\thickbar{\bd{p}}} \tr[\hat{W}_{\thickbar{\bd{p}}}\hat{\Gamma}]  \right)\right\vert_{(*)}\\\
&= \min_{\hat{\Gamma}\mapsto\{\hat{\Gamma}_{\bd 0, \thickbar{\bd{p}}}\}_{\thickbar{\bd{p}}}}\left.\left(\sum_{\thickbar{\bd{p}}} \min_{\hat{\Gamma}_{\bd 0, \thickbar{\bd{p}}}\,\mapsto(n_{\bd{p}}, n_{\bd{p}})}\tr[\hat{W}_{\thickbar{\bd{p}}}\hat{\Gamma}] \right)\right\vert_{(*)}\,.
\end{split}
\end{equation}
In the second line we used $\hat{\Gamma}_{\bd 0, \thickbar{\bd{p}}} = \tr_{\{\bd 0, \bd{p}, -\bd{p}\}^c}[\hat{\Gamma}]$ and $(*)$ denotes that the minimization over the occupation numbers of the pairs $\thickbar{\bd{p}}$ cannot be performed independently due to conservation of the total particle number and that the $\bd{p}=\bd{0}$ mode is shared by all pairs. This observation based on Levy's constrained search shows on a formal level why the different pairs of momenta $(\bd{p},-\bd{p})$ cannot be treated independently as done in Bogoliubov's theory \cite{Bogoliubov47}. Replacing in Eq.~\eqref{LevyP} $\hat{W}$ by Bogoliubov's approximated Hamiltonian would therefore erroneously ignore the important anomalous terms  $\hat{a}_{\bd{p}}^\dagger\hat{a}_{-\bd{p}}^\dagger + \hat{a}_{\bd{p}}\hat{a}_{-\bd{p}}$.
At first sight, this incompatibility of Bogoliubov's conventional approximation and RDMFT seems to be paradoxical. Yet, it is worth recalling that the merits of the unitary Bogoliubov transformation lie in the simple calculation of the (low-lying) energy spectrum while its violation of particle-number conservation can lead to conceptual difficulties beyond RDMFT as well. At the same time, since RDMFT has the distinctive goal to (partly) solve the ground state problem for $\hat{H}(\hat{h})$  for all $\hat{h}$ simultaneously, it requires apparently a mathematically more rigid and well-defined framework than the one  provided by conventional Bogoliubov theory.

Before we discuss in the following section such a well-defined mathematical framework for realizing Bogoliubov's ideas within RDMFT, we briefly comment on an alternative natural idea for circumventing the outlined difficulties. Instead of applying the constrained search formalism to a fixed particle number sector, one could also extend \eqref{LevyP} to the entire Fock space. This would result in a Fock space RDMFT and the anomalous terms would contribute to the functional. Yet, there would be a crucial drawback. The respective functional would namely allow one for any Hamiltonian \eqref{Hfamily} to only calculate the \emph{overall} ground state on the Fock space. For instance, for specific kinetic energy operators or pair interactions, this overall minimum may lie in the sector of zero or infinitely many bosons. Also adding a chemical potential term $\mu \hat{N}$ for steering the particle number to a preferred one would only work in case the Fock space functional was convex in the total particle number.

\section{Particle-number conserving Bogoliubov theory\label{sec:pair}}

As discussed in the context of Eq.~\eqref{Levy} and motivated in Sec.~\ref{sec:obstacles}, the derivation of the universal functional for BECs requires a particle-number conserving variant of conventional Bogoliubov theory. Exactly such a modification has been provided by
Girardeau \cite{Girardeau59} (see also \cite{Gardiner97,Girardeau98,Seiringer11,Seiringer14}) in the context of pair theory. In the following, we will outline and then apply this theory which in particular improves upon Bogoliubov theory by including more terms of the Hamiltonian.
The idea behind pair theory is that in the regime of BEC, excitations of pairs $(\bd{p}, -\bd{p})$ from the condensate dominate and thus the interacting ground state is well approximated by a state with a corresponding pairing structure \cite{Girardeau59, Girardeau98}. Clearly, this ansatz for the interacting ground state is only valid for sufficiently weak interactions and will break down for larger depletion of the condensate. Restricting the original Hamiltonian $\hat{H}$ to the space of such pair excitation states and assuming that the zero-momentum state is macroscopically occupied means to effectively deal with a modified interaction $\hat{W}_P$ of pair excitation type \cite{Girardeau59},
\begin{eqnarray}\label{WP}
\hat{W}_P &\equiv& \frac{N(N-1)W_{\bd{0}}}{2V} + \frac{1}{2V}\sum_{\bd{p}\neq \bd{0}}W_{\bd{p}}\left[2\hat{n}_{\bd{0}}\hat{n}_{\bd{p}} +\hat{a}_{\bd{p}}^\dagger\hat{a}_{-\bd{p}}^\dagger \hat{a}_{\bd{0}}^2 + \big(\hat{a}_{\bd{0}}^\dagger\big)^2\hat{a}_{\bd{p}}\hat{a}_{-\bd{p}} \right] \\
&&+ \frac{1}{2V}\sum_{\substack{\bd{p}, \bd{p^\prime}\neq \bd{0} \\ \bd{p}\neq \bd{p}^\prime}} W_{\bd{p}}\hat{a}_{\bd{p}^\prime}^\dagger\hat{a}_{-\bd{p}^\prime}^\dagger\hat{a}_{\bd{p}^\prime-\bd{p}}\hat{a}_{\bd{p}-\bd{p}^\prime} + \frac{1}{2V}\sum_{\substack{\bd{p^\prime}\neq \bd{0} \\ \bd{p}\neq \bd{p}^\prime, \bd{p}\neq 2\bd{p}^\prime}} W_{\bd{p}}\hat{n}_{\bd{p}^\prime-\bd{p}}\hat{n}_{\bd{p}^\prime}\nonumber\,,
\end{eqnarray}
where the restrictions in the summations in the last two lines are chosen such that no term appears twice. The terms in the first line of Eq.~\eqref{WP} give rise to the Bogoliubov Hamiltonian (after the replacement $\hat{a}_{\bd{0}}, \hat{a}_{\bd{0}}^\dagger \to\sqrt{N}$) while those in the second line improve upon Bogoliubov theory. Therefore, Eq.~\eqref{WP} can be understood as an expansion in the number of operators involving the zero-momentum mode where the terms involving two operators with zero-momentum indices are dominant and the terms involving no zero-momentum operator lead to a small correction. Besides the restriction to pair excitation states, there is no argument why these two terms in the second line of Eq.~\eqref{WP} are kept while other terms involving four operators with non-zero momentum of the full interaction $\hat{W}$ in Eq.~\eqref{H2nd} are neglected. 
For the sake of simplicity, we omit in the following derivations the constant term $\frac{N(N-1)W_{\bd{0}}}{2V}$ because it only corresponds to constant shift in energy. 
 
To determine a variational ground state energy of $\hat{H}(\hat{h})$, Girardeau's idea was then to employ a particle-number conserving analogue of Bogoliubov trial state $\hat{U}_B\ket{N}$ instead of modifying the interaction Hamiltonian in Eq.~\eqref{WP} as in the conventional Bogoliubov theory. For this, one first introduces the operators \cite{Girardeau98}
\begin{equation}\label{beta0}
\hat{\beta}_{\bd{0}} \equiv \left(\hat{n}_{\mathbf{0}}+1\right)^{-1/2}\hat{a}_{\mathbf{0}}\,,\quad
\hat{\beta}_{\mathbf{0}}^\dagger \equiv \hat{a}_{\mathbf{0}}^\dagger  \left(\hat{n}_{\mathbf{0}}+1\right)^{-1/2}
\end{equation}
which annihilate/create a boson in the condensate, yet without changing the normalization of the respective quantum state. The commutator of the operators $\hat{\beta}_{\bd{0}}$ and $\hat{\beta}_{\bd{0}}^\dagger$ is given by $[\hat{\beta}_{\bd{0}}, \hat{\beta}_{\bd{0}}^\dagger]=\hat{P}_0$, where $\hat{P}_0$ denotes the projector onto the subspace with zero particles in the zero-momentum state. 
Girardeau's $N$-boson trial states
\begin{equation}\label{Psi_int}
|\Psi\rangle \equiv \hat{U}_G|N\rangle
\end{equation}
of pair excitation form are defined by the following operators
\begin{equation}\label{U}
\hat{U}_G = \mathrm{exp}\left\{\frac{1}{2}\sum_{\bd{p}\neq \bd{0}}\theta_{\bd{p}}\left[\left(\hat{\beta}_{\mathbf{0}}^\dagger\right)^2\hat{a}_{\bd{p}}\hat{a}_{-\bd{p}} - \hat{\beta}_{\mathbf{0}}^2\hat{a}_{\bd{p}}^\dagger\hat{a}_{-\bd{p}}^\dagger\right]\right\}
\end{equation}
with $\theta_{\bd{p}}\in \mathbb{R}$ and $\theta_{\bd{p}} = \theta_{-\bd{p}}$ which reflects the invariance of the Hamiltonian under inversion $\bd{p}\to-\bd{p}$. 
The operators $\hat{U}_G$ are particle-number conserving as desired, $[\hat{U}_G, \hat{N}]=0$, which is due to the additional operators $\hat{\beta}_{\mathbf{0}}$ and $\hat{\beta}_{\mathbf{0}}^\dagger$. As $\hat{U}_\mathrm{B}$, the operators $\hat{U}_G$ commute with the total momentum operator.
Since its exponent is antihermitian, $\hat{U}_G$ is still unitary (as $\hat{U}_B$). The price one has to pay for the more complicated exponent, however, is that no compact exact expression can be found for the quasiparticle operators $\hat{U}_G^\dagger\hat{a}_{\bd{p}}\hat{U}_G$ anymore. Instead the result known from Bogoliubov theory holds only approximately,
\begin{equation}\label{qp}
\hat{U}_G^\dagger\hat{a}_{\bd{p}}\hat{U}_G \approx \frac{1}{\sqrt{1-\phi_{\bd{p}}^2}}\left(\hat{a}_{\bd{p}} - \phi_{\bd{p}}\beta_{\mathbf{0}}^2\hat{a}_{-\bd{p}}^\dagger\right) \equiv \hat{\xi}_{\bd{p}}\,,
\end{equation}
where 
\begin{equation}\label{eq:phiG}
\phi_{\bd{p}} \equiv \tanh (\theta_{\bd{p}})\,.
\end{equation}
A careful mathematical estimate of the difference between left and right side of \eqref{qp} has been provided in \cite{Seiringer11} (yet involving a slightly different but conceptually similar definition of $\hat{U}_G$).
It effectively allows us to treat \eqref{qp} and the implied Eq.~\eqref{np} as exact relations for our further derivation.
The particle number expectation value of the momentum mode $\bd{p}\neq \bd{0}$ in the interacting ground state $\ket{\Psi}$ \eqref{Psi_int} then follows as
\begin{equation}\label{np}
n_{\bd{p}} \equiv \langle\Psi|\hat{a}_{\bd{p}}^\dagger\hat{a}_{\bd{p}}|\Psi\rangle \approx \langle N|\hat{\xi}_{\bd{p}}^\dagger\hat{\xi}_{\bd{p}}|N\rangle = \frac{\phi_{\bd{p}}^2}{1-\phi_{\bd{p}}^2}\,.
\end{equation}
The occupation numbers satisfy $n_{\bd p}=n_{-\bd{p}}$ due to $\phi_{\bd{p}}=\phi_{-\bd{p}}$. It is worth stressing that this property is a result of the parity-symmetry of the system, as explained in Sec.~\ref{sec:parity}. 

\section{Calculation of the functional\label{sec:derivfunc}}

Relation \eqref{np} is the crucial ingredient for our derivation of the universal functional $\F$ in the regime of BEC. This connection between the family of variational trial states of fixed particle number and the momentum occupation numbers $\bd{n}$ will drastically simplify the constrained search \eqref{LevyP} and will allow us eventually to determine the explicit form of $\F$.
For this, we observe that relation \eqref{np} can be inverted up to binary degrees of freedom $\sigma_{\bd{p}}= \sigma_{-\bd{p}}=\pm 1$,
\begin{equation}
\phi_{\bd{p}} = \sigma_{\bd{p}} \sqrt{\frac{n_{\bd{p}}}{1+n_{\bd{p}}}}\,.
\end{equation}
This sign ambiguity is conceptually very similar to the so-called phase dilemma in fermionic RDMFT \cite{PernalPhase04}. The latter resembles the fact that general phase changes of the natural orbitals (eigenstates of the 1RDM) affect $\bra{\Psi}\hat{W}\ket{\Psi}$ in \eqref{Levy} via the $N$-fermion wavefunction $\ket{\Psi}$ while keeping the 1RDM invariant. In contrast to fermionic RDMFT, however, the minimizing signs $\{\sigma_{\bd{p}}\}$ can be found in our case of bosons in the BEC regime.

We combine now various concepts and ideas to determine the universal functional $\F(\bd{n})$ for BECs.
According to the constrained search formalism \eqref{LevyP} we need to minimize for any vector $\bd{n}$ the expectation value of the interaction $\hat{W}$ over all $N$-boson quantum states with momentum occupation numbers $\bd{n}$. Our focus on the regime of BECs then allows us to restrict this to Girardeau's $N$-boson trial states \eqref{Psi_int} with the additional effect that $\hat{W}$ simplifies to $\hat{W}_P$ in Eq.~\eqref{WP}, i.e., $\langle\Psi|\hat{W}|\Psi\rangle =\langle\Psi|\hat{W}_P|\Psi\rangle = \bra{N}\hat{U}_G^\dagger W_P\hat{U}_G\ket{N}$. The operator $\hat{U}_G^\dagger W_P\hat{U}_G$ should then be expressed in terms of the quasiparticle operators $\hat{\xi}_{\bd{p}}$ given by Eq.~\eqref{qp}, allowing us to eventually calculate its action on the state $\ket{N}$.
Since the trial states $\ket{\Psi}$ are almost uniquely determined by $\bd{n}$ according to \eqref{np}
we are only left with a minimization over all possible combinations of signs $\sigma_{\bd{p}}$.
Keeping only terms which do not vanish in the thermodynamic limit $N\to \infty$, $V\to \infty$ and $n=N/V=\mathrm{cst.}$ yields
then the final result for the Girardeau approximated functional
\begin{equation}\label{Fmin}
\mathcal{F}_G(\bd{n})
= \min_{\{\sigma_{\bd{p}}=\pm 1\}}\Bigg\{ \sum_{\bd{p}\neq \bd{0}}\Big[\frac{n_{\mathbf{0}}}{V}  W_{\bd{p}} + \frac{1}{2}I_2(\bd{p}, \bd{n})\Big]n_{\bd{p}} -\sigma_{\bd{p}}\Big[\frac{n_{\mathbf{0}}}{V} W_{\bd{p}} - \frac{1}{2}I_1(\bd{p}, \bd{n},\bd{\sigma})\Big]\sqrt{n_{\bd{p}}(n_{\bd{p}}+1)}\Bigg\} \,.
\end{equation}
where
\begin{eqnarray}\label{Iterms}
I_1(\bd{p}, \bd{n},\bd{\sigma}) &\equiv& \frac{1}{V}\sum_{\substack{\bd{p^\prime}\neq \bd{0}}}W_{\bd{p}-\bd{p}^\prime}\sigma_{\bd{p}^\prime}\sqrt{n_{\bd{p}^\prime}(n_{\bd{p}^\prime}+1)}\nonumber \\
I_2(\bd{p}, \bd{n}) &\equiv&\frac{1}{V}\sum_{\substack{\bd{p^\prime}\neq \bd{0}}}W_{\bd{p}-\bd{p}^\prime}n_{\bd{p}^\prime}\,.
\end{eqnarray}
For general $\bd{n} \in \triangle$ one cannot overcome the common phase dilemma and in particular the minimizing sign factors $\sigma_{\bd{p}}=\pm1$ in \eqref{Fmin} depend on $\bd{n}$. This in turn leads to a partitioning of the functional's domain $\triangle$ into cells characterized by different signs $\{\sigma_{\bd{p}}\}$, similarly to the Ising cells corresponding to different spin configurations (see also Fig.~\ref{fig:S} for an illustration). It is worth stressing that, according to the variational principle, the Girardeau approximated functional $\mathcal{F}_G$ provides an upper bound to the exact functional $\mathcal{F}_{\mathrm{exact}}$:
\begin{equation}
\mathcal{F}_\mathrm{exact}\leq \mathcal{F}_G
\end{equation}
and $\mathcal{F}_G$ is believed to be exact in leading order in the BEC regime.

As already explained in Sec.~\ref{sec:pair}, Girardeau's approach based on pair theory goes beyond Bogoliubov theory by including additional terms of the original Hamiltonian (see also second line of \eqref{WP}). Yet, since those involve fewer creation/annihilation operators $a_{\mathbf{0}}^{(\dagger)}$ and since the Girardeau approach uses at the end (almost) the same trial states \eqref{Psi_int} as Bogoliubov, we expect that the additional terms $I_1, I_2$ in \eqref{Fmin} have only a minor quantitative rather than a significant qualitative effect on the description of BECs. Whether this changes  beyond the regime of BEC is not clear since one still restricts to the common BEC trial states \eqref{Psi_int}.

In the regime of BEC the two terms in Eq.~\eqref{Fmin} involving $I_1$ and $I_2$, respectively, are significantly smaller than the term proportional to $n_{\mathbf{0}}$. Accordingly, in the regime of interest the minimization of various $\sigma_{\bd{p}}$ can be executed analytically, leading to
\begin{equation}\label{sgnmagic}
\sigma_{\bd{p}} = \mathrm{sgn}(W_{\bd{p}})\,,\quad\forall\,\,\bd{p}\neq \bd{0}\,.
\end{equation}
Also, the possible approximation $n_{\mathbf{0}}\approx N$ would be of the same order as neglecting the less significant Girardeau terms $I_1$ and $I_2$. Eventually, implementing those two last approximations leads to one of the key results in this thesis, the  Bogoliubov approximated functional ($n \equiv N/V$)
\begin{equation}\label{FBog}
\mathcal{F}_B(\bd{n}) = n\sum_{\bd{p}\neq \bd{0}}W_{\bd{p}}\left[n_{\bd{p}} - \mathrm{sgn}(W_{\bd{p}})\sqrt{n_{\bd{p}}(n_{\bd{p}}+1)}\right]\,.
\end{equation}
The distinctive form of the Bogoliubov functional $\mathcal{F}_B$ resembles clearly the decoupling of various momentum pairs $(\bd{p},-\bd{p})$ from each other within Bogoliubov theory. Remarkably, the Bogoliubov approximated functional $\mathcal{F}_B$ is convex, in contrast to common pure functionals in fermionic RDMFT. The pure functional $\mathcal{F}_B$ therefore coincides with the corresponding ensemble functional since the latter is given by the lower convex envelop of the former \cite{Schilling2018}.

We also would like to reiterate that due to the general significance of BECs, the functional \eqref{FBog} can be seen as the first-level approximation of the universal functional in bosonic RDMFT. In analogy to the Hartree-Fock \cite{LiebHF} and the M\"uller functional \cite{Mueller84,Buijse02} in fermionic RDMFT and the local density approximation in density functional theory \cite{PerdewJacob}, $\mathcal{F}_B$ and $\mathcal{F}_G$ will represent a promising starting point for the construction of more elaborated functional approximations. In that sense, we expect that our key results \eqref{Fmin} and \eqref{FBog} will initiate and establish eventually bosonic RDMFT. In the following, we simplify our notation by skipping the index $B,G$ of the functional, also since both functionals (almost) coincide in the relevant regime of BEC.

\section{Applications and illustrations\label{sec:appl}}
\subsection{Dilute Bose gas in 3D \label{subsec:dilute}}

We now apply the concepts of RDMFT to the homogeneous dilute Bose gas, the system for which Bogoliubov's theory \cite{Bogoliubov47} was originally developed. This will also allow us to demonstrate how the well-known expression for the ground state energy of a dilute Bose gas \cite{Brueckner57} can be obtained using RDMFT.

Let us introduce for the following considerations the degree $D$ of quantum depletion ($N_\mathrm{BEC} \equiv n_{\mathbf{0}}$)
\begin{equation}\label{Dgen}
D\equiv1-N_\mathrm{BEC}/N = \frac{1}{N}\sum_{\bd{p}\neq \bd{0}}n_{\bd{p}}\,.
\end{equation}
From a geometric point of view, $D$ is nothing else than the $l_1$-distance of $\bd{n}$ in the simplex $\triangle$ \eqref{simplex} to the vertex $\mathbf{0}$ corresponding to complete BEC. We also recall that the ground state energy of $\hat{H}(\hat{t})= \hat{t}+\hat{W}$ follows in RDMFT by minimizing the respective
energy functional over the space of occupation number vectors $\bd{n}$,
\begin{equation}\label{E0min}
E_0(\hat{t}) = \min_{\bd{n}\in \triangle}\left[\bd{\varepsilon} \cdot \bd{n} + \mathcal{F}(\bd{n})\right]\,,
\end{equation}
where $\hat{t}\equiv \sum_{\bd{p}}\varepsilon_{\bd{p}} \hat{n}_{\bd{p}}$, assuming w.l.o.g.~$\varepsilon_{\mathbf{0}} =0$, and $\bd{\varepsilon} \cdot \bd{n}\equiv \sum_{\bd{p}\neq \mathbf{0}} \varepsilon_{\bd{p}}n_{\bd{p}}$. To calculate for the \emph{realistic} dilute Bose gas the ground state energy and the degree of condensation, we would need to plug in for the kinetic energy in \eqref{E0min} the specific dispersion relation of free particles, i.e., $\varepsilon_{\bd{p}}=p^2/2m$ (ignoring boundary effects).  It is worth reiterating that in principle systems with any kinetic energy  $\hat{t}$ could be considered in RDMFT. From an experimental point of view, one could indeed imagine a modified dispersion relation due to a specific background medium and in case of lattice models both the rate and range of the hopping can be actually varied (see, e.g., Refs.~\cite{Guenter2013,Schempp2015}). Because of this, we are for the moment still considering a general $\hat{t}$ and $\bd{\varepsilon}$, respectively.

Finding the minimizer $\thickbar{\bd{n}}$ of the energy functional then means to solve
\begin{equation}\label{varE0}
\varepsilon_{\bd{p}} = -\frac{\partial \mathcal{F}}{\partial n_{\bd{p}}}(\thickbar{\bd{n}})\,,\quad \forall \bd{p}\,.
\end{equation}
Using the explicit form of the Bogoliubov functional \eqref{FBog} then leads to
\begin{equation}\label{nbar}
\thickbar{n}_{\bd{p}} = \frac{1}{2}\left(\frac{\varepsilon_{\bd{p}} + nW_{\bd{p}}}{\sqrt{\varepsilon_{\bd{p}}(\varepsilon_{\bd{p}}+2nW_{\bd{p}}})} - 1\right)\,.
\end{equation}
This is nothing else than the well-known result for the momentum occupation numbers \cite{pita}.

Considering now the specific case of a realistic dilute Bose gas then allow us to determine the ground state energy explicitly. For this, we first evaluate the universal functional at the minimum $\thickbar{\bd{n}}$, leading to (see Appendix \ref{app:Fdil})
\begin{equation}\label{Fdil}
\mathcal{F}(\thickbar{\bd{n}}) = \frac{128\sqrt{\pi}}{3m}Na_0^{5/2}n^{3/2} + \frac{4\pi Na_1 n}{m}\,.
\end{equation}
It depends only on the first two terms of the Born series of the s-wave scattering length $a$ \cite{Brueckner57},
\begin{equation}\label{a0}
a_0 = \frac{mW_{\mathbf{0}}}{4\pi} \,,\quad a_1 = -\frac{1}{4\pi V}\sum_{\bd{p}\neq \bd{0}}\frac{W_{\bd{p}}^2m^2}{p^2}\,.
\end{equation}

As it is shown in Appendix \ref{app:Fdil}, adding the kinetic energy and reintroducing the omitted constant term $W_{\mathbf{0}}N(N-1)/2V\approx W_{\mathbf{0}}Nn/2$ leads to the well-known ground state energy \cite{Brueckner57}:
\begin{eqnarray}\label{E0dilute}
E_0 &=& \frac{2\pi Nn}{m}\left(a_0+a_1 +\frac{128}{15\sqrt{\pi}}a_0(na_0^3)^{1/2}\right)\,.
\end{eqnarray}
This can be recast by using the scattering length $a$ which eventually leads (up to higher order terms) to the compact expression \cite{pita} $E_0=\frac{2\pi Nn a}{m}\left(1 +\frac{128}{15\sqrt{\pi}}(n a^3)^{1/2}\right)$.

Since the underlying domain $\triangle$ is infinite dimensional in case of a continuous system in a box it is difficult to graphically illustrate the Bogoliubov functional. Yet, to visualize at least some of its most crucial features we define two paths within $\triangle$, both starting at the physical point $\thickbar{\bd{n}}$ (corresponding to $\varepsilon_{\bd{p}} = \bd{p}^2/2m$) and terminating at the vertex $\mathbf{0}$ describing complete BEC. The first one is just the straight path $s$ between those two points, parameterized by $t\in[0,1]$,
\begin{equation}\label{nt}
\bd{n}(t) = \thickbar{\bd{n}} - t\thickbar{\bd{n}} \,.
\end{equation}
The $l_1$-distance $D(t)$ of $\bd{n}(t)$ to $\mathbf{0}$ follows directly as
\begin{equation}
D(t)=\frac{\sqrt{n}}{3\pi^2}(m W_{\mathbf{0}})^{3/2}(1-t)
\end{equation}
and the functional's concrete values $\mathcal{F}[\bd{n}(t)]$ along that path can easily be calculated by exact numerical means.
As second path $\kappa$, we consider the one experimentally realized in Ref.~\cite{Lopes17} by continuously reducing the coupling strength $\kappa$ of the pair interaction from one to zero. Since the interaction Hamiltonian $\hat{W}$ in RDMFT is fixed, this path has to be realized equivalently  by increasing the strength of the kinetic energy according to $p^2/2m\kappa$.
The respective distance $D$ of $\bd{n}(\kappa)$ to $\mathbf{0}$ follows as
\begin{equation}
D(\kappa)=\frac{\sqrt{n}}{3\pi^2}(mW_{\mathbf{0}}\kappa)^{3/2}
\end{equation}
and the functional $\mathcal{F}[\bd{n}(\kappa)]$ along that path is given by
\begin{equation}\label{Fdilkappa}
\begin{split}
\mathcal{F}[\bd{n}(\kappa)] &= 4nNW_{\mathbf{0}}D(\kappa) \\\
&\quad+ \frac{3^{2/3}4\pi^{7/3}Nn^{2/3}a_1}{m^2W_{\mathbf{0}}}D^{2/3}(\kappa)\,.
\end{split}
\end{equation}
In Eq.~\eqref{Fdilkappa} one may replace $W_{\mathbf{0}}$ by $a_0$ according to \eqref{a0}.

\begin{figure}[htb]
\includegraphics[width=0.49\linewidth]{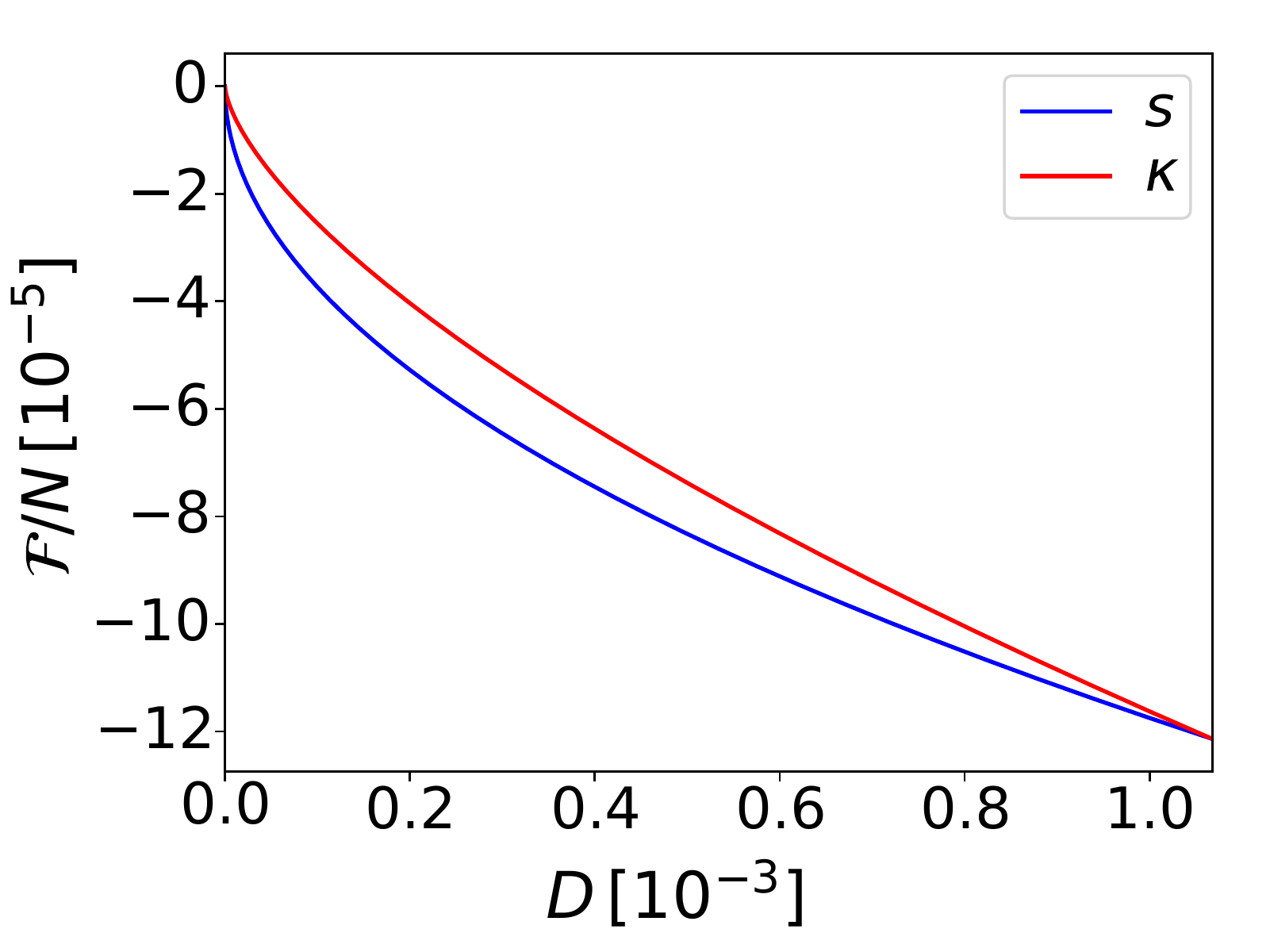}
\includegraphics[width=0.49\linewidth]{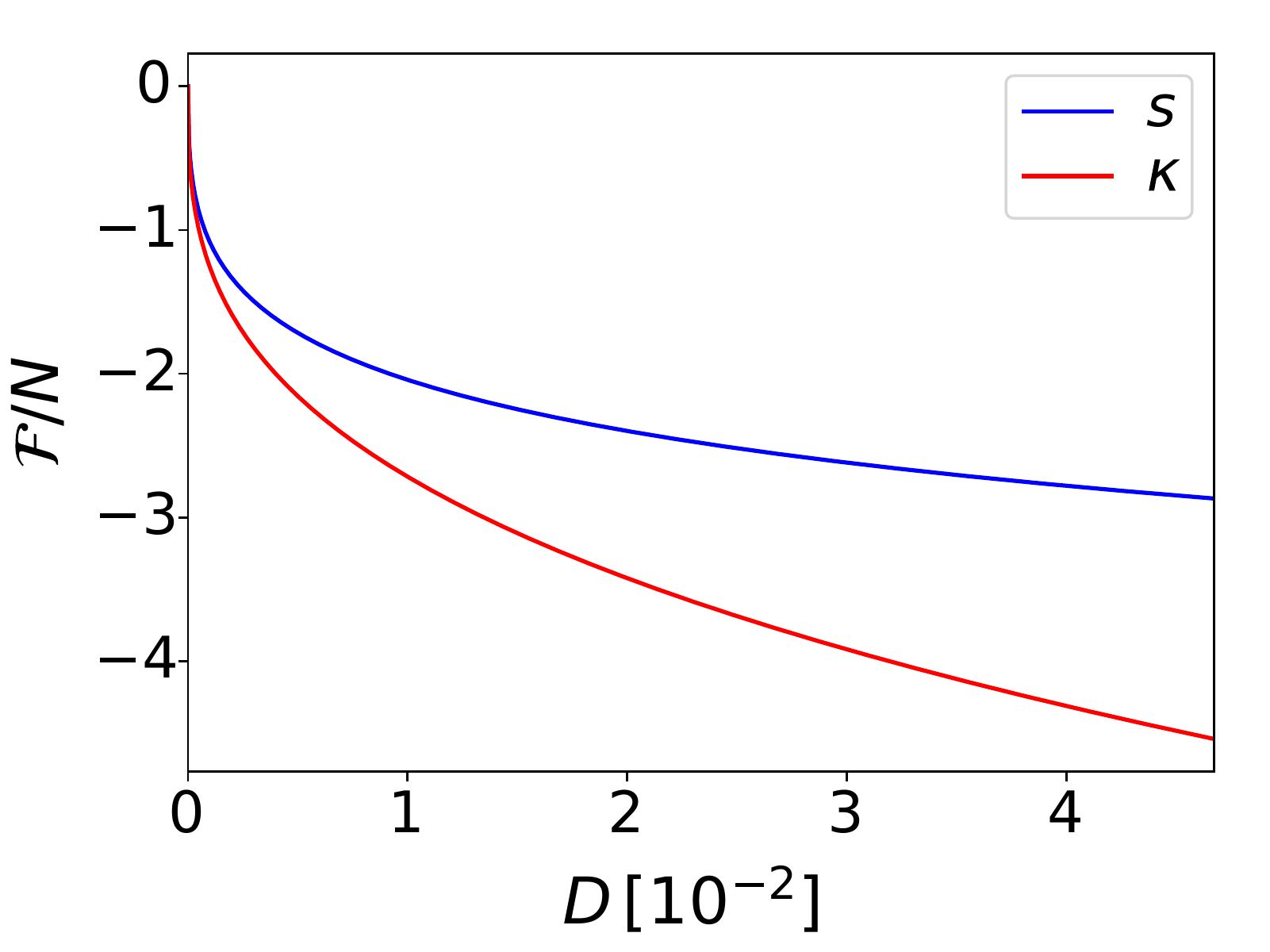}
\caption{Bogoliubov functional $\mathcal{F}$ as a function of the relative depletion $D$ along the straight path $s$ \eqref{nt} and the curved path $\kappa$. Dilute Bose gas in 3D for $n=10^{-3}$, $W_{\mathbf{0}}=m=1$ and $a_1=-0.01$ (left), charged Bose gas in 3D for $2m=e^2/2=1$ and $n=100$ (right).}
\label{fig:Fdil}
\end{figure}

The result for $\mathcal{F}$ as a function of the fraction $D$ of non-condensed bosons along the two paths $s$ and $\kappa$ is shown in Fig.~\ref{fig:Fdil} for the parameters $n=10^{-3}$, $W_{\mathbf{0}}=m=1$ and $a_1=-0.01$. This choice of parameters (recall that we set several physical constants to one) corresponds to realistic dilute Bose gases as our following results of the small degrees of depletion will confirm. We observe that the Bogoliubov functional $\mathcal{F}$ goes to zero for $D=0$ which corresponds to complete BEC. Also, it can be seen that the gradient of $\mathcal{F}$ increases for smaller distances $D$. In Sec.~\ref{sec:BECforce} we will show that the gradient of $\mathcal{F}$ actually diverges in the limit $D\to 0$ and provide a detailed discussion of this remarkable and far-reaching observation.

A generalization of $\mathcal{F}$ given by Eq.~\eqref{FBog} to dimensions $d\neq 3$ within the s-wave scattering approximation is possible if the Bogoliubov approximation for the given set of parameters, i.e.~the interaction strength and the density, is valid (c.f. Sec.~\ref{sec:weaklydiffd}). Two-dimensional dilute systems are weakly interacting if the condition $n|a|_{2\mathrm{D}}^2\ll 1$ \cite{Schick71, Lieb00} is satisfied where $a_{2\mathrm{D}}$ is now the respective two-dimensional s-wave scattering length. In contrast to higher dimensional systems, a one-dimensional Bose gas is weakly interacting in the limit of high densities and the validity of the Bogoliubov approximation in that limit was shown in Ref.~\cite{Lieb63}. Due to their distinctive role, we will study a one-dimensional model in Sec.~\ref{subsec:Hubbard5}.

\subsection{Charged Bose gas in 3D \label{subsec:charged}}

In contrast to the dilute Bose gas discussed in the previous section, the scattering of \emph{charged} bosons cannot be described within the s-wave scattering approximation anymore. This is due to the infinite range of the Coulomb interaction $W(r)\propto 1/r$. The respective Fourier coefficients $W_{\bd{p}}$ can still be determined analytically though. In case of an additional uniform background they follow as (see Appendix \ref{app:Fouriercoeff})
\begin{equation}\label{Wpch}
W_{\textbf{0}}=0\,,\quad W_{\bd{p}} = \frac{4\pi e^2}{p^2}\,,\,\forall \bd{p}\neq \mathbf{0}\,.
\end{equation}
For charged bosons the weak interaction regime corresponds to the high density limit \cite{Foldy61, Girardeau62, Lieb01}. This regime
to which Bogoliubov's approximation refers to is characterized by a small ``gas parameter'', $r_s \equiv (3/4\pi)^{1/3}me^2 n^{-1/3}\ll 1$. To illustrate again how RDMFT works, we calculate the energy and momentum occupation numbers $n_{\bd{p}}$ of the ground state for the most realistic case of a kinetic energy given by $\hat{t}=\sum_{\bd{p}}\frac{p^2}{2m} \hat{n}_{\bd{p}}$. For this, we add
the exact kinetic energy functional $\mbox{Tr}_1[\hat{t}(\cdot)]$ to the universal interaction functional $\mathcal{F}$ with Fourier coefficients $W_{\bd{p}}$ given by Eq.~\eqref{Wpch}. Then, we minimize the total energy functional with respect to all $\bd{n} \in \triangle$, leading to the minimizer $\thickbar{\bd{n}}$ which is given by Eq.~\eqref{nbar}. Evaluating then the functional at $\thickbar{\bd{n}}$ is straightforward (in contrast to the dilute neutral Bose gas) and one finds (recall $n \equiv N/V$)
\begin{equation}\label{Fc}
\mathcal{F}(\thickbar{\bd{n}}) = \frac{2 \Gamma\left(-\frac{1}{4}\right)\Gamma\left(\frac{7}{4}\right)Nn^{1/4}e^{5/2}m^{1/4}}{3\pi^{5/4}}
\end{equation}
and the respective fraction of non-condensed bosons $D=1-N_\mathrm{BEC}/N$ follows as
\begin{equation}\label{Dch}
\thickbar{D}\equiv D(\thickbar{\bd{n}}) = -\frac{\Gamma\left(-\frac{3}{4}\right)\Gamma\left(\frac{5}{4}\right)m^{3/4}e^{3/2}}{4\pi^{7/4}n^{1/4}}\,.
\end{equation}
Eq.~\eqref{Dch} verifies that the depletion of the condensate decreases with increasing density $n$.
Adding the kinetic energy to Eq.~\eqref{Fc} leads to the known result for the ground state energy ($\hbar = 4\pi\epsilon_0=1$) \cite{Foldy62}:
\begin{equation}\label{Ech}
E_0 = -\frac{4\Gamma\left[-\frac{5}{4}\right]\Gamma\left[\frac{7}{4}\right]Nn^{1/4}m^{1/4}e^{5/2}}{3\pi^{5/4}}\,.
\end{equation}
As a consistency check, this confirms the correctness of Eq.~\eqref{Fc}.

Next, in analogy to Sec.~\ref{subsec:dilute} we consider again the straight path $s$ and the curved path $\kappa$. The latter is again defined as the curve $\bd{n}(\kappa)$ obtained by reducing by  factor $\kappa \in [0,1]$ the coupling strength of the Hamiltonian above (which led to the results Eqs.~\eqref{Fc}, \eqref{Dch} and \eqref{Ech}).
Evaluating the distance $D$ along the path $\kappa$ yields
\begin{equation}
D(\kappa) = \thickbar{D}\kappa^{3/4}
\end{equation}
and the functional $\mathcal{F}$ takes the values
\begin{equation}\label{Fchkappa}
\mathcal{F}[\bd{n}(\kappa)] =
q D^{1/3}(\kappa) \,,
\end{equation}
$q \equiv 2^{5/3}\Gamma\left(-\frac{1}{4}\right)\Gamma\left(\frac{7}{4}\right)Nn^{1/3} e^2/3\left[-\Gamma\left(-\frac{3}{4}\right)\Gamma\left(\frac{5}{4}\right)\right]^{1/3}\!\pi^{2/3}$.
For the  path $s$ we have $D(t) = (1-t)\thickbar{D}$ and the concrete values of the functional $\mathcal{F}$ along that path can be evaluated by exact numerical means. The right panel of Fig.~\ref{fig:Fdil} shows $\mathcal{F}$ along the two paths $s$ and $\kappa$. The curves have qualitatively similar shapes to those for the dilute neutral Bose gas shown on the left of Fig.~\ref{fig:Fdil}. This is not surprising because both setups correspond to a weakly interacting system in which the Bogoliubov approximated functional Eq.~\eqref{FBog} is valid. However, we will see in Sec.~\ref{sec:BECforce} that the momentum dependence of $W_{\bd{p}}$ can alter the behaviour of the gradient of $\mathcal{F}$.

\subsection{Bose-Hubbard model for five lattice sites\label{subsec:Hubbard5}}

As a third example, we discuss in this section the one-dimensional Bose-Hubbard model.
For illustrative purposes, we consider the specific case of just $L=5$ lattice sites and $N=100$ bosons since this allows us to visualize
the functional and its gradient on the entire domain $\triangle$. Indeed, for $L=5$ there are only two independent momentum occupation numbers due to the general parity symmetry $n_{\bd{p}}= n_{-\bd{p}}$ and normalization $n_{\bd{0}}= N-\sum_{\bd{p} \neq \bd{0}} n_{\bd{p}}$.

We start by discussing a few conceptual aspects which are valid for any number $L$ of sites (assuming for simplicity $L$ odd). The one-dimensional Brillouin zone comprises momenta $ p= 2\pi\nu/L$ where $\nu$ takes integer values in the interval described by $|\nu| \leq  (L-1)/2$.
The bosons interact via Bose-Hubbard on-site interaction as described by the operator $\frac{U}{2}\sum_{j=1}^L\hat{n}_j\left(\hat{n}_j-1\right)$. It is worth recalling that the universal functional depends on the entire interaction Hamiltonian, i.e., it includes \emph{a priori} the coupling constant $U$ as well. Yet, due to the linear structure of the constrained search \eqref{Levy}, \eqref{LevyP} any non-negative prefactor could be separated from the interaction Hamiltonian $\hat{W}$ and added instead in front of the respective universal functional,
\begin{equation}
\F_{U \hat{W}} = |U| \, \F_{ \mathrm{sgn}(U)\hat{W}}\,.
\end{equation}
It is crucial to observe that the same does not apply to possible sign factors since otherwise this would mean to change the minimization in \eqref{Levy}, \eqref{LevyP} to a maximization. Because of this, we consider in the following the interaction Hamiltonian $\hat{W} \equiv \frac{\mathrm{sgn}(U)}{2}\sum_{j=1}^L\hat{n}_j\left(\hat{n}_j-1\right)$ and add eventually the coupling constant $|U|$
in front of the respective universal functional $\F_{\hat{W}}$. To proceed, it is then an elementary exercise to determine the corresponding Fourier coefficients $W_p=\mathrm{sgn}(U)$ which are in particular independent of the (one-dimensional) momentum $p$. The universal functional in the BEC regime is obtained by plugging the concrete result for the Fourier coefficients $W_p$ into the general formula for the Bogoliubov functional \eqref{FBog} or its extension \eqref{Fmin} based on Girardeau's approach. Just to reiterate, the respective functionals are valid in the regime of BEC, i.e., for weak interactions. In contrast to their higher dimensional counterparts, one-dimensional systems require high densities $n=N/L \gg 1$ to be weakly interacting \cite{Lieb63}. 

From a general point of view, the context of lattice models emphasizes very well the conceptual advantages of RDMFT relative to wavefunction based methods. After having determined the universal interaction functional $\F_{\hat{W}}$ (or decent approximations thereof) the ground state energy of every Hamiltonian $\hat{H}(\hat{t}) = \hat{t}+ |U|\,\hat{W}$ can be calculated with relatively little computational effort by minimizing the total energy functional with respect to all $\bd{n} \in \triangle$. In that sense, RDMFT represents a highly economic approach for solving simultaneously the ground state problem for the entire class $\{\hat{H}(\hat{t})\}$ of Hamiltonians. For continuous systems the benefits of this are less obvious since there is essentially one particularly relevant choice for the kinetic energy operator $\hat{t}$. This is quite different for lattice models since both
the rate and the range of the hopping can be varied in experiments (see, e.g., Refs.~\cite{Guenter2013,Schempp2015}). Nonetheless, we focus in the following on hopping just between neighbouring sites at a rate $t\geq 0$, i.e., we choose $\hat{t} = -2t\sum_{p}\big(\!\cos(p)-1\big)\hat{n}_p$ and w.l.o.g.~fix $|U|\equiv 1$. In analogy to the most realistic dilute and charged Bose gas as discussed in Sec.~\ref{subsec:dilute} and Sec.~\ref{subsec:charged}, respectively, we pick $t=U=1$ as a reference point for further investigations and illustrations .

\begin{figure}[htb]
\centering
\includegraphics[width=0.5\linewidth]{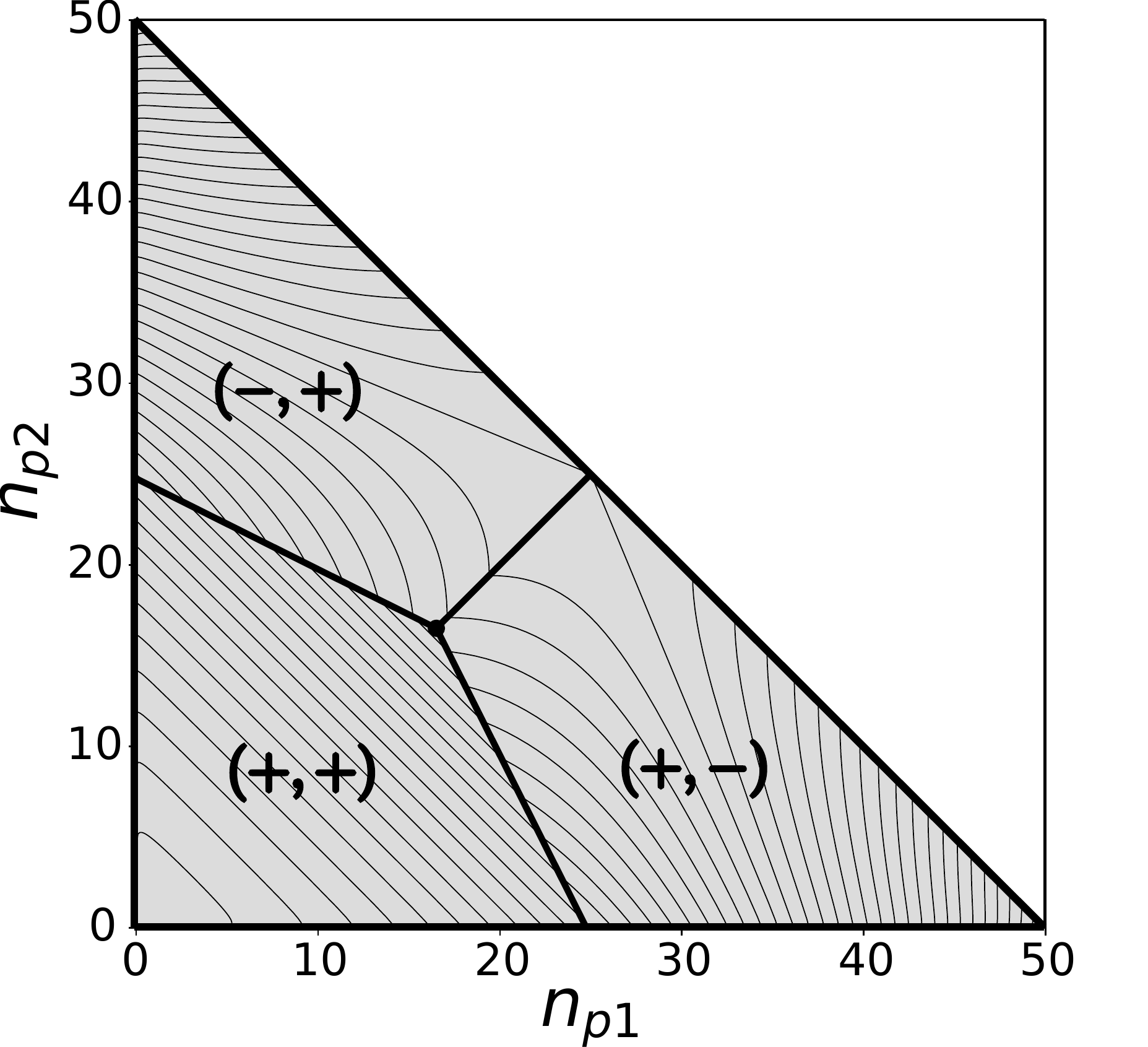}
\caption{Domain $\triangle$ of the universal functional is shown for $L=5$ sites.
Minimization of signs $(\sigma_{p_1},\sigma_{p_2})$ in \eqref{Fmin} partitions $\triangle$ into three cells (see text for details).
\label{fig:S}}
\end{figure}
To illustrate and compare the Bogoliubov- \eqref{FBog} and the Girardeau-approximated functionals \eqref{Fmin} for the specific case of $L=5$ sites we first need to execute the minimization of the sign factors in \eqref{Fmin}. As it is shown in Appendix \ref{app:signs}, this can be done analytically due to the specific Fourier coefficients and leads to a partitioning of the functional's domain $\triangle$ into three regions. Just for illustrative purposes, we present in Fig.~\ref{fig:S} the \emph{entire} domain $\triangle$ of the functional \eqref{Fmin} (recall its validity refers to the regime of BEC only) and the three cells which are characterized by different  minimizing sign configurations $(\sigma_{p_1},\sigma_{p_2})$ in \eqref{Fmin}. As the two independent occupation numbers we choose here the momenta $p_1=2 \pi/5$ and $p_2=4 \pi/5$ which can take values $n_{p_j} \in [0,50]$. The vector $(0,0)$ corresponds to complete BEC, i.e., $N_{\mathrm{BEC}}\equiv n_{0}=N=100$  and its vicinity represents the BEC-regime to which our functionals refer to.

\begin{figure}[htb]
\centering
\includegraphics[width=0.43\linewidth]{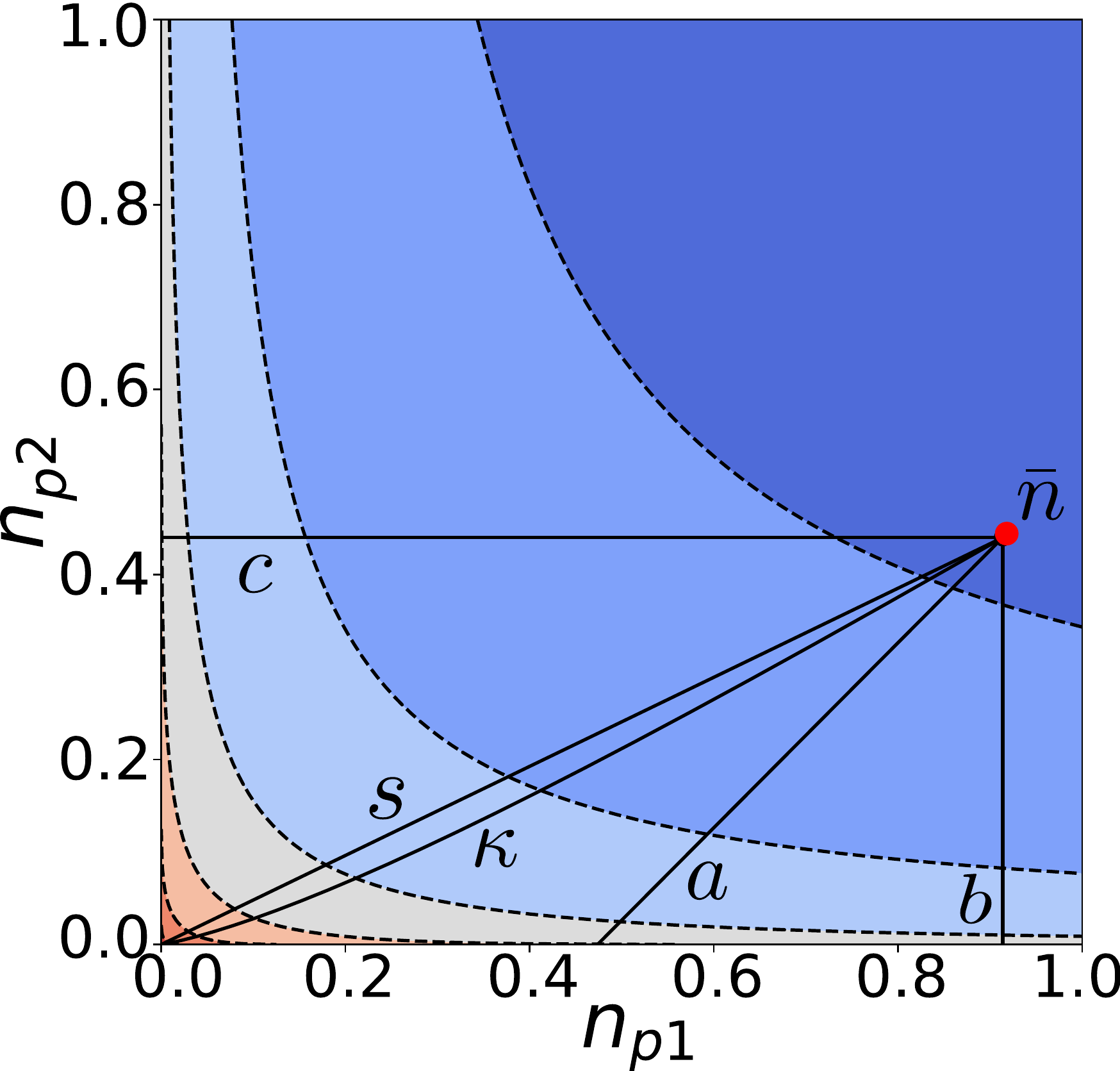}
\includegraphics[width=0.43\linewidth]{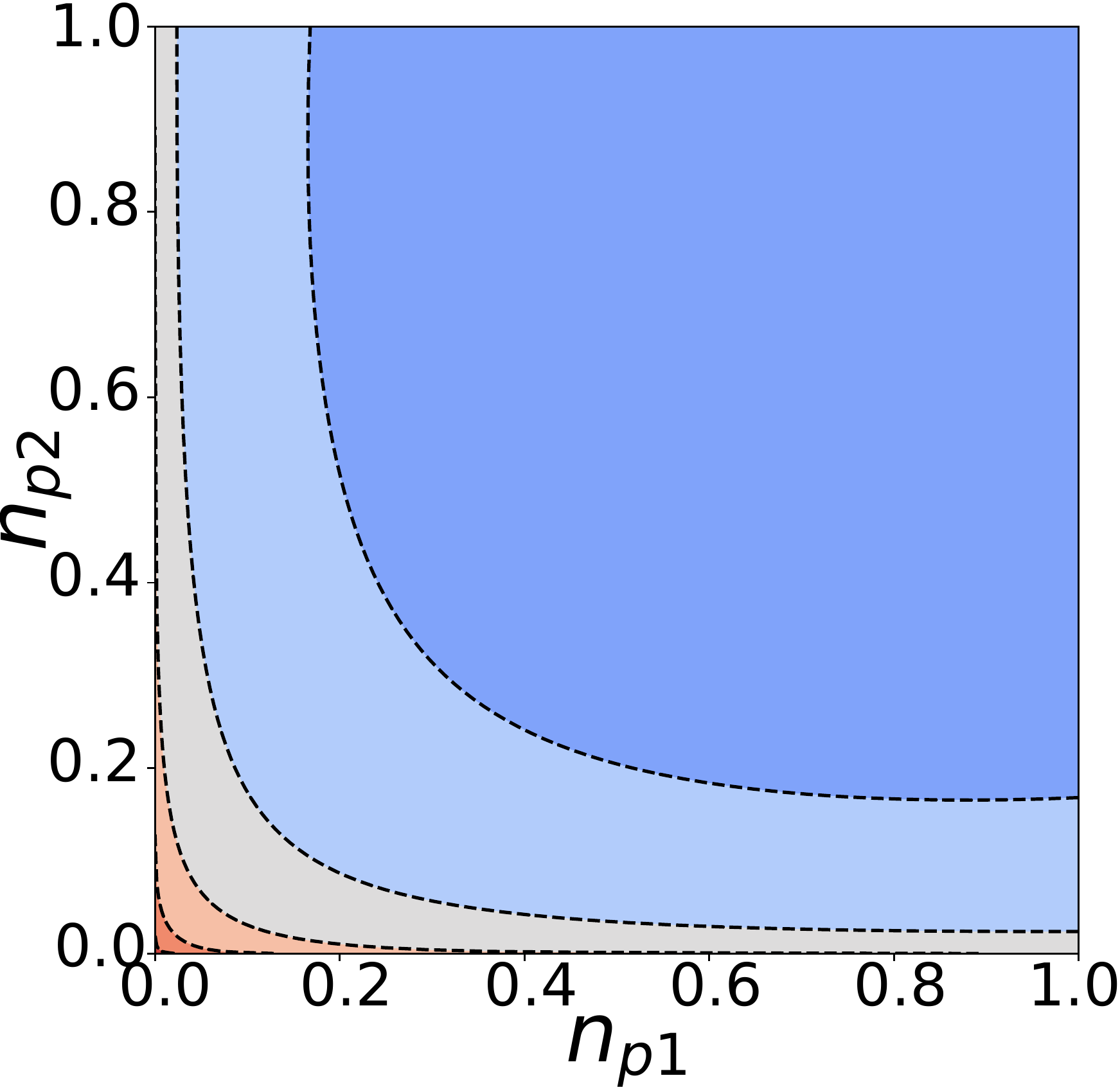}
\hspace{0.1cm}
\includegraphics[width=0.07\linewidth, trim=0cm -1.5cm 0cm +1.5cm]{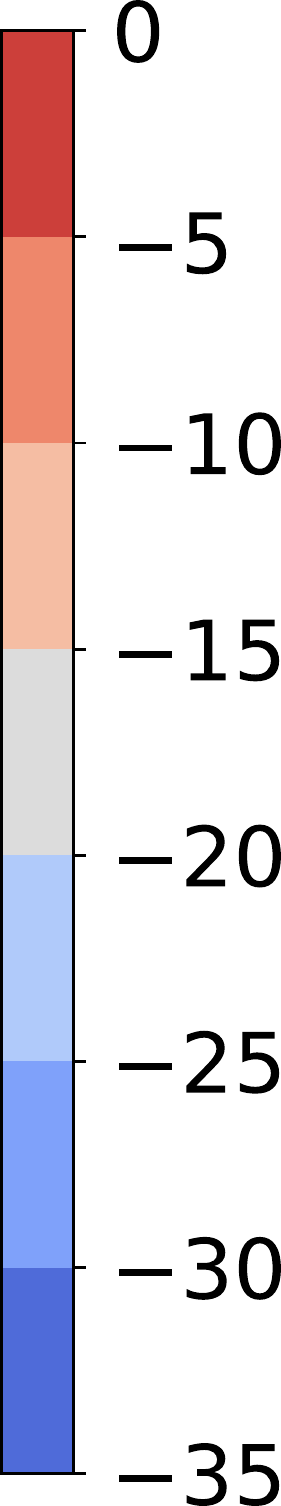}
\caption{Left: Contour plot of the Bogoliubov-approximated functional \eqref{FBog} for the Bose-Hubbard model with $N=100$ bosons on $L=5$ sites in the BEC-regime of not too large depletion. Right: Girardeau's extension \eqref{Fmin} for the same system.}
\label{fig:HubbardF}
\end{figure}
In Fig.~\ref{fig:HubbardF} we present now the Bogoliubov functional \eqref{FBog} and its extension \eqref{Fmin} based on Girardeau's approach
in the form of a contour plot in the BEC-regime of not too large quantum depletion.
The results for the two functionals are in quite good agreement for small degrees $D$ of depletion. The occupation number vector $\thickbar{n}=(0.91, 0.44)$  obtained from minimizing the total energy functional for the reference point $(t,U)=(1,1)$ is shown in Fig.~\ref{fig:HubbardF} as well. The corresponding degree of depletion, $D=2.7\%$, justifies in retrospective the treatment of the interaction $\hat{W}$ within the Bogoliubov theory and the usage of the functionals \eqref{FBog} and \eqref{Fmin}, respectively. For stronger quantum depletion the two functionals in Fig.~\ref{fig:HubbardF} begin to differ also qualitatively. Their (small) deviation already in the regime of BEC with a degree of depletion around $2\%$ emphasizes the quantitative significance of the additional terms $I_1$ and $I_2$ and the usage of the exact value $n_{\mathbf{0}}$ rather than its approximation to $N$ in Eq.~\eqref{Fmin}.

\begin{figure}[htb]
\centering
\includegraphics[width=0.7\linewidth]{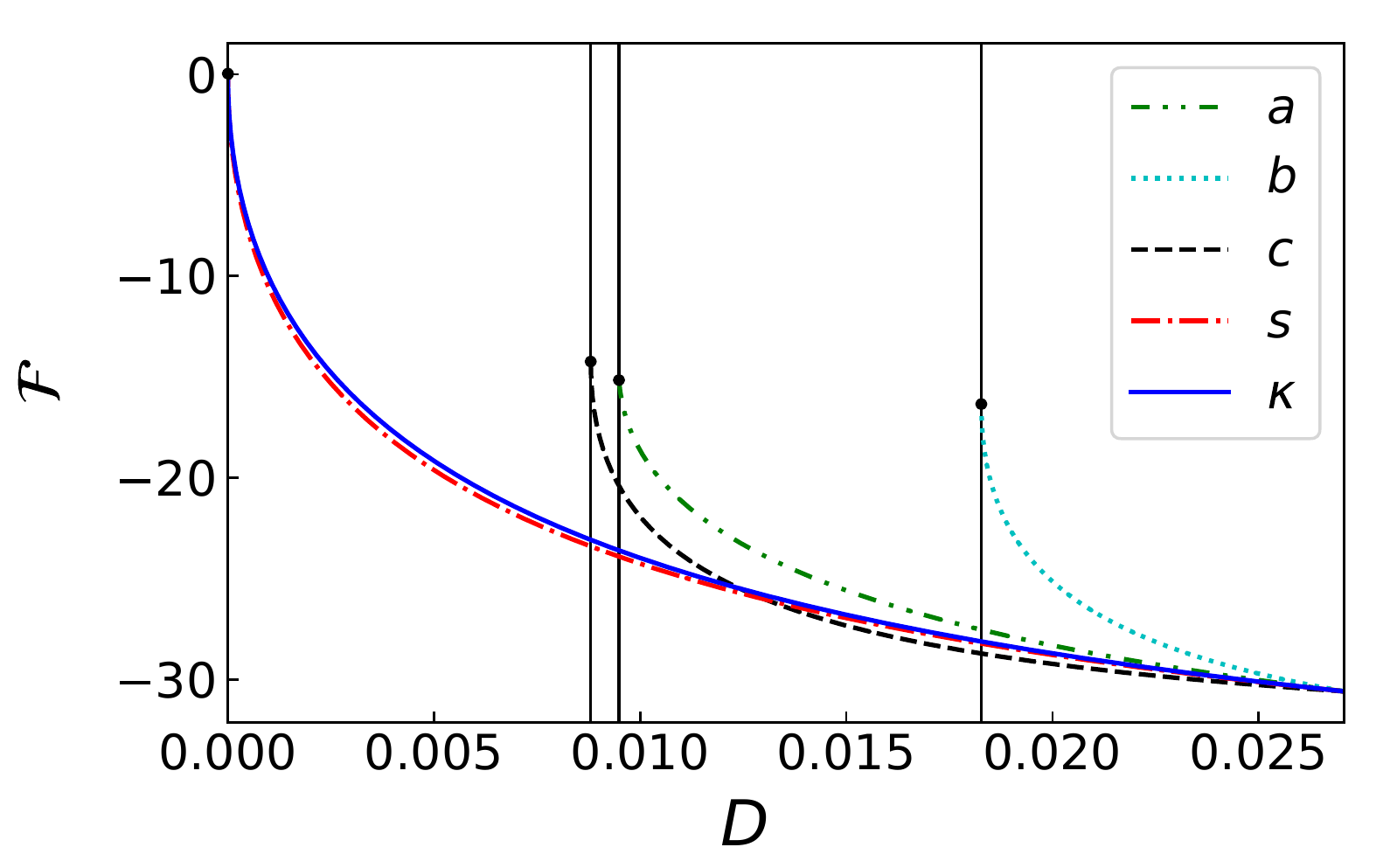}
\caption{Universal functional $\mathcal{F}$ for the Bose-Hubbard model along the five paths defined in Fig.~\ref{fig:HubbardF}}
\label{fig:FD}
\end{figure}

For the discussion in Sec.~\ref{sec:BECforce} of the new concept of a BEC force, we define in Fig.~\ref{fig:HubbardF} five qualitatively different paths towards the polytope boundary $\partial\triangle$, all starting from the point $\thickbar{n}$. The path denoted by $s$ corresponds to a straight path towards complete BEC and $\kappa$ denotes the path where the interaction strength of the model is reduced by increasing the kinetic energy by a factor $1/\kappa$ with $\kappa \in[0,1]$. The path denoted by $a$ runs perpendicular towards the hyperplane defined by $n_{p_1} + n_{p_2}=0$. Consequently, it corresponds to the path with the fastest increase of the condensate fraction (yet it will not reach complete BEC). In the cases $b$ and $c$ one occupation number is fixed while the other one is continuously decreased to zero. In Fig.~\ref{fig:FD} we present the functional $\mathcal{F}$ as a function of $D=1-N_\mathrm{BEC}/N$ along those five paths. The black dots emphasize that the value of $\mathcal{F}$ at the boundary $\partial\triangle$ remains finite (quite in contrast to its derivative as shown and discussed in the subsequent section).
The convexity of the curves in Fig.~\ref{fig:FD} corresponding to the four straight paths $a, b, c, s$ just reflects the local convexity of the exact universal functional in the regime of not too large quantum depletion.
In this context, we would like to reiterate that this convex behaviour and the repulsive character of the functional's gradient close to the boundary emerges from the minimization of the sign factors in \eqref{Fmin} leading to \eqref{sgnmagic}.

\section{Bose-Einstein force\label{sec:BECforce}}

In this section we explore in more detail the behavior of the functional and its gradient close to the boundary of their domain
$\triangle$ in the regime of BEC. This will eventually allow us to reveal and establish the novel concept of a BEC force.  Due to its potentially far-reaching consequences for our understanding of bosonic quantum systems, we will calculate and illustrate the BEC force in Sec.~\ref{subsec:forceex} for the three different systems studied in Sec.~\ref{sec:appl}.

\subsection{General results\label{subsec:forcegeneral}}

The functional \eqref{FBog} which is based on the Bogoliubov approximation is convex on its entire domain $\triangle$ \eqref{simplex}. Since this approximate functional is exact in leading order in the regime of BEC with not too large quantum depletion all conclusions drawn from it are valid for the exact universal functional of \eqref{H1st} as well. The illustrations in the previous section for dilute and charged Bose gases in 3D and the Bose-Hubbard model have also confirmed the distinctive convex behaviour of the universal functional in the BEC-regime. While the functional itself remains finite even at the point $\mathbf{0} \in \triangle$ of complete condensation, the same will not be true anymore for the functional's gradient. This can easily be deduced from the form of the Bogoliubov functional \eqref{FBog}. To be more specific, approaching the vertex $\mathbf{0}$ of the simplex  $\triangle$ means to simultaneously send all momentum occupation numbers $n_{\bd{p}}$ with $\bd{p}\neq \mathbf{0}$ to zero. Taking then the derivative of \eqref{FBog} (or of its extension \eqref{Fmin}) with respect to $n_{\bd{p}}$ sufficiently close to $\mathbf{0}$ yields in leading order
\begin{equation}\label{npforce}
\frac{\partial \F}{\partial n_{\bd{p}}}(\bd{n}) \sim - \frac{n |W_{\bd{p}}|}{2} \frac{1}{\sqrt{n_{\bd{p}}}}\,.
\end{equation}
It is worth noticing that the divergence of this derivative for $ n_{\bd{p}} \rightarrow 0$ is always repulsive for \emph{any} interaction $\hat{W}$. This remarkable feature follows directly from the minimization of the sign factors in \eqref{Fmin}, leading to \eqref{sgnmagic}. The repulsive nature of the diverging gradient also proves universally that occupation numbers in interacting bosonic quantum systems can never attain the exact mathematical value $0$. Although this chapter refers to homogeneous systems in their BEC regime only, we have little doubt that this conclusion is also valid for any generic nonhomogeneous interacting bosonic quantum system, also beyond the BEC regime.

The general result \eqref{npforce} implies that the point $\mathbf{0}$ of complete condensation can never be reached, independent of the path towards $\mathbf{0}$ that is envisaged. Since the functional $\F$ is finite this seems to be paradoxical as far as the energy is concerned. Yet, the reader shall note that it is the kinetic energy which will need to diverge according \eqref{varE0} to enforce such a path towards $\mathbf{0}$.

We also would like to emphasize that the divergence of the gradient of $\F$ along a straight path is always proportional to $1/\sqrt{D}$ and its prefactor depends on the direction of the path, i.e., the angle at which $\mathbf{0}$ is approached. To confirm this in a quantitative way, let us consider a general straight path from a starting point $\thickbar{n}$ in the regime of BEC to $\mathbf{0}$, linearly parameterized by $t \in [0,1]$,
\begin{equation}\label{nstraight}
\bd{n}(t) = (1-t)\, \thickbar{\bd{n}}\,.
\end{equation}
The degree $D$ \eqref{Dgen} of quantum depletion along that path reduces according to
\begin{equation}\label{Dstraight}
D(t) = (1-t)\, \thickbar{D} \equiv (1-t)\,\frac{1}{N}\sum_{\bd{p}\neq \mathbf{0}} \thickbar{n}_{\bd{p}}\,.
\end{equation}
The gradient of $\F$ projected onto that path is then nothing else than the weighted sum of individual contributions \eqref{npforce} from every $\bd{p}$,
\begin{eqnarray}\label{BECforce}
\frac{\partial \F}{\partial D}\Big\vert_{\mathrm{path}}  &=& \nabla_{\!\bd{n}}\F \cdot \frac{\partial \bd{n}}{\partial D}\Big\vert_{\mathrm{path}} \\
&=& \nabla_{\!\bd{n}}\F \cdot \frac{\thickbar{\bd{n}}}{\thickbar{D}}
\sim -\frac{n}{2} \sum_{\bd{p}\neq \mathbf{0}} |W_{\bd{p}}|\sqrt{\frac{\thickbar{n}_p}{\thickbar{D}}}\,  \frac{1}{\sqrt{D}}\,. \nonumber
\end{eqnarray}
This second key result in this chapter establishes the new concept of a BEC force which prevents interacting bosonic quantum systems from ever exhibiting complete BEC. This novel concept is conceptually very similar to the fermionic exchange force that we have recently revealed and established in fermionic lattice models \cite{Schilling2019}.

\subsection{Examples\label{subsec:forceex}}
In this section we illustrate the novel concept of a BEC force \eqref{BECforce} for various systems introduced in Sec.~\ref{sec:appl}.

\subsubsection{Bose gases in 3D}
We revisit the 3D Bose gas for neutral atoms in the low density and for charged atoms in the high density regime. The aim is to calculate for those concrete systems the explicit values of the BEC force \eqref{BECforce}. For both systems, the derivative of $\mathcal{F}$ with respect to the degree $D$ of quantum depletion along the path $s$ defined by \eqref{nstraight} is given by Eq.~\eqref{BECforce}.
The summation over $\bd{p}\neq \bd{0}$ can be converted into an integral in the thermodynamic limit where $N\to \infty$, $V\to \infty$ and $n=N/V=\mathrm{cst.}$. This eventually allows us (see Appendix \ref{app:forcedil}) to obtain a compact analytic expression  for the BEC force,
\begin{equation}\label{DFsdilute}
\left.\frac{\rmd \mathcal{F}(\thickbar{\bd{n}})}{\rmd D}\right\vert_s\sim
N \left[\eta(a_0, n, m) + \frac{2\pi n a_1}{m\sqrt{\thickbar{D}}}\right]\frac{1}{\sqrt{D}}
\end{equation}
where $\eta(a_0, n, m)$ is a positive constant and $a_0$ and $a_1$ are the first two terms in the Born series for the scattering length $a$.
It is worth reiterating that according to the general result \eqref{BECforce} the BEC force is always repulsive. Since
only the second term in Eq.~\eqref{DFsdilute} is negative (recall $a_1<0$) this imposes in turn a bound on the maximal valid distance $\thickbar{D}$ of the starting point $\thickbar{n} \in \triangle$ to the regime of complete BEC.

For the charged Bose gas the last expression in the second line in Eq.~\eqref{BECforce} can only be calculated by exact numerical means. Nonetheless, this also allows us to confirm the square root dependence of the divergence. In general, the functional's gradient diverges as $1/\sqrt{\mbox{dist}(\bd{n},\partial \triangle )}$ along straight paths reaching any \emph{arbitrary} point on the boundary $\partial \triangle$ in the regime of BEC.

\begin{figure}[htb]
\centering
\includegraphics[width=0.48\linewidth]{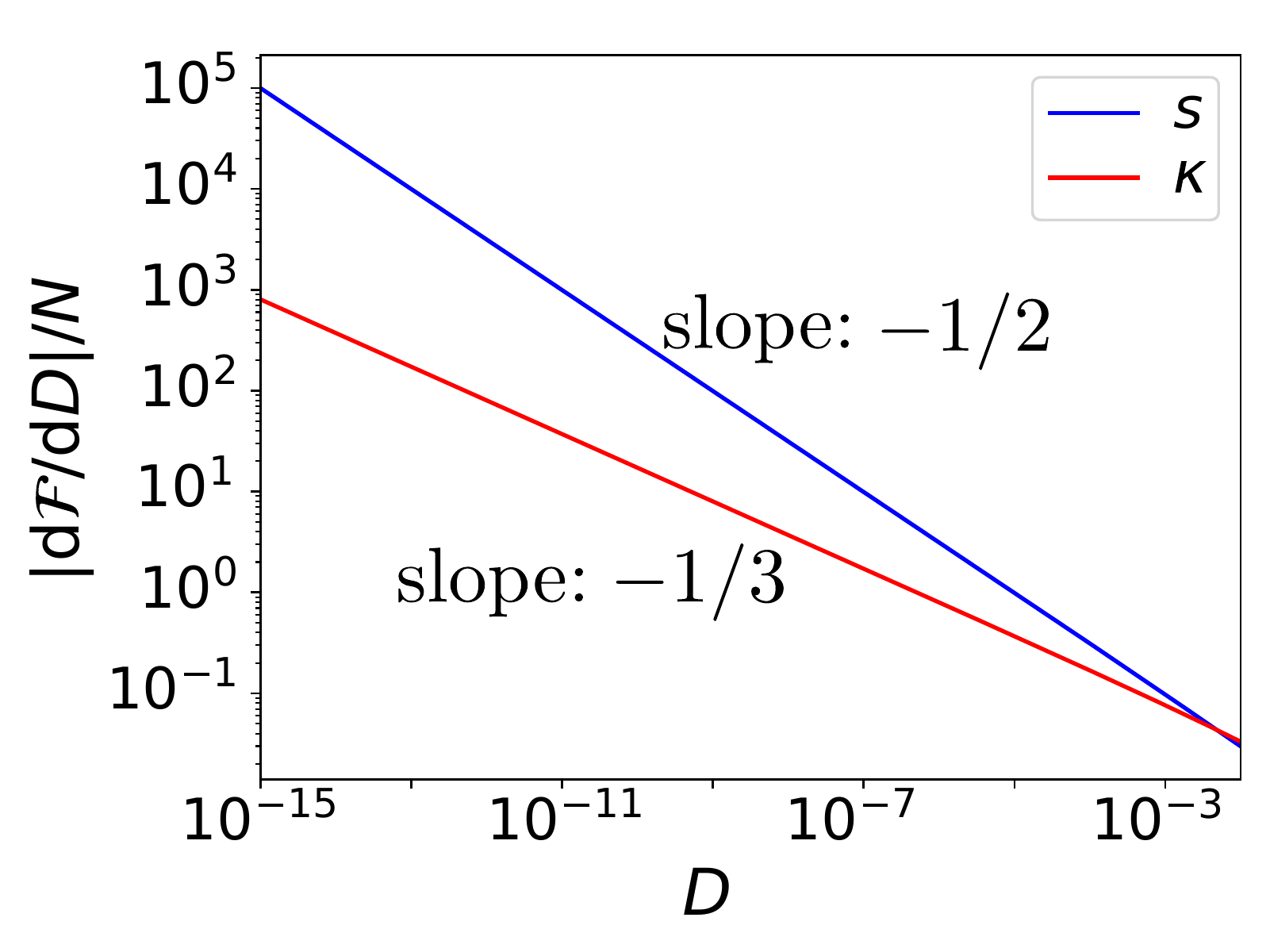}
\hspace{0.1cm}
\includegraphics[width=0.48\linewidth]{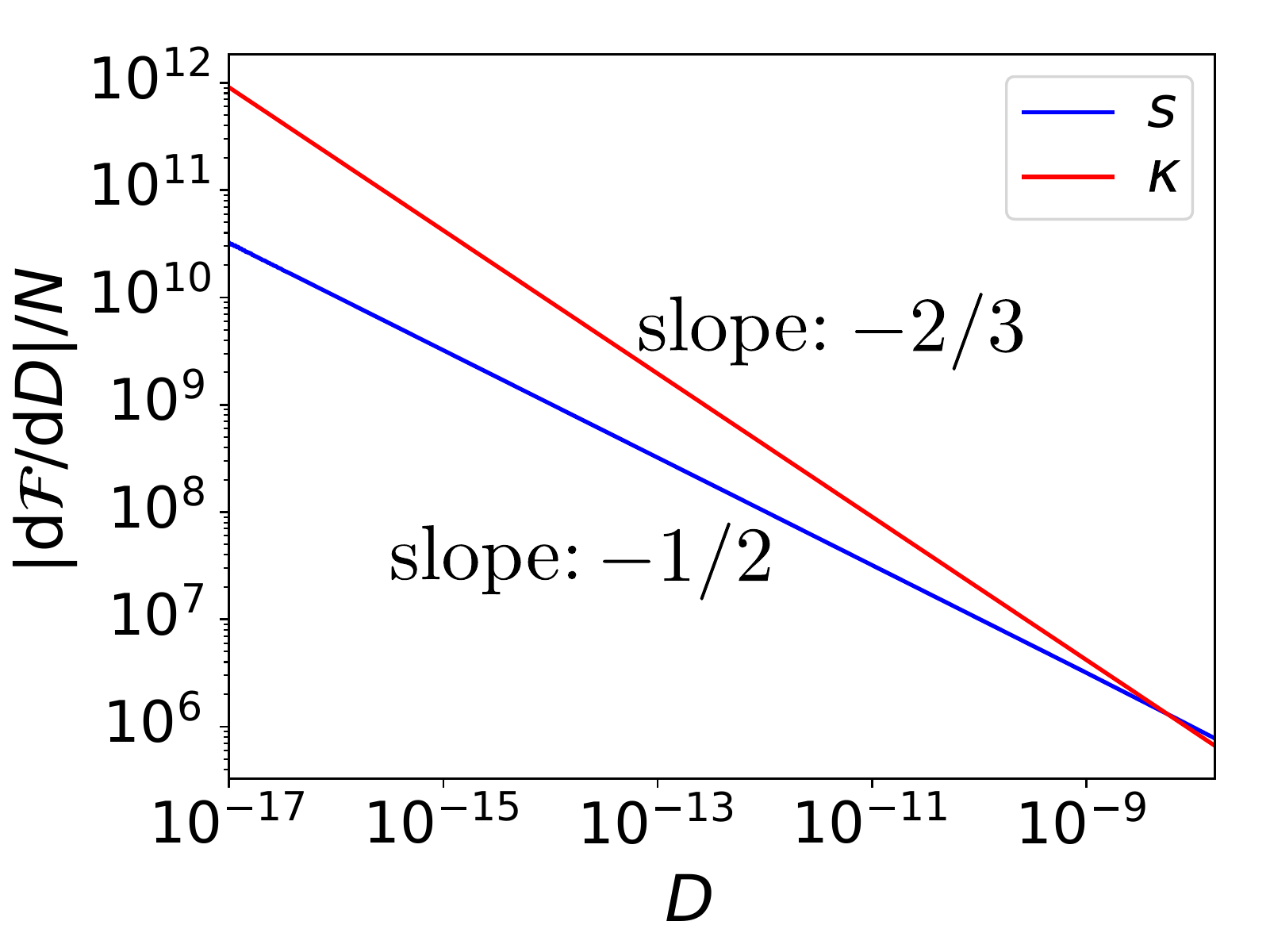}
\caption{BEC force $|\rmd \mathcal{F}/\rmd D|$ along the straight path $s$ (blue) and the curved path $\kappa$ (red) is shown for the dilute Bose gas in 3D with $n=10^{-3}$, $W_{\mathbf{0}}=m=1$ and $a_1=-0.01$ (left) and for the charged Bose gas in 3D with $2m=e^2/2=1$ and $n=100$ (right).
\label{fig:dF}}
\end{figure}

Moreover, we determine for both systems the BEC force along the curved path $\bd{n}(\kappa)$ which is defined by reducing an additional coupling constant $\kappa$ in front of $\hat{W}$ from one to zero. Since exactly this path has been implemented in a very recent experiment \cite{Lopes17}
this may suggest a first experimental setup for realizing and visualizing our novel concept of a BEC force.
The explicit calculation of the BEC force along the $\kappa$-path follows directly from differentiation of the expressions in \eqref{Fdilkappa} and \eqref{Fchkappa}, respectively, leading to
\begin{equation}\label{dFdilkappa}
\left.\frac{\rmd \mathcal{F}_\mathrm{dilute}}{\rmd D}\right\vert_\kappa\propto -\frac{1}{D^{1/3}}
\end{equation}
and
\begin{equation}\label{dFchkappa}
\left.\frac{\rmd \mathcal{F}_\mathrm{charged}}{\rmd D}\right\vert_\kappa\propto -\frac{1}{D^{2/3}}\,.
\end{equation}
Fig.~\ref{fig:dF} displays the BEC force along the straight $s$-path and the curved $\kappa$-path for both ultracold gas systems. The linear behaviour shown in this log-log plot confirms the algebraic dependence of the BEC force on the degree $D$ of quantum depletion along both paths.
For the dilute Bose gas, the gradient of $\mathcal{F}$ according to Eq.~\eqref{DFsdilute} and Eq.~\eqref{dFdilkappa} diverges faster along the path $s$ than along the path $\kappa$. For the charged Bose gas we observe the opposite behaviour.

\subsubsection{Bose-Hubbard model}

\begin{figure}[htb]
\centering
\includegraphics[width=0.7\linewidth]{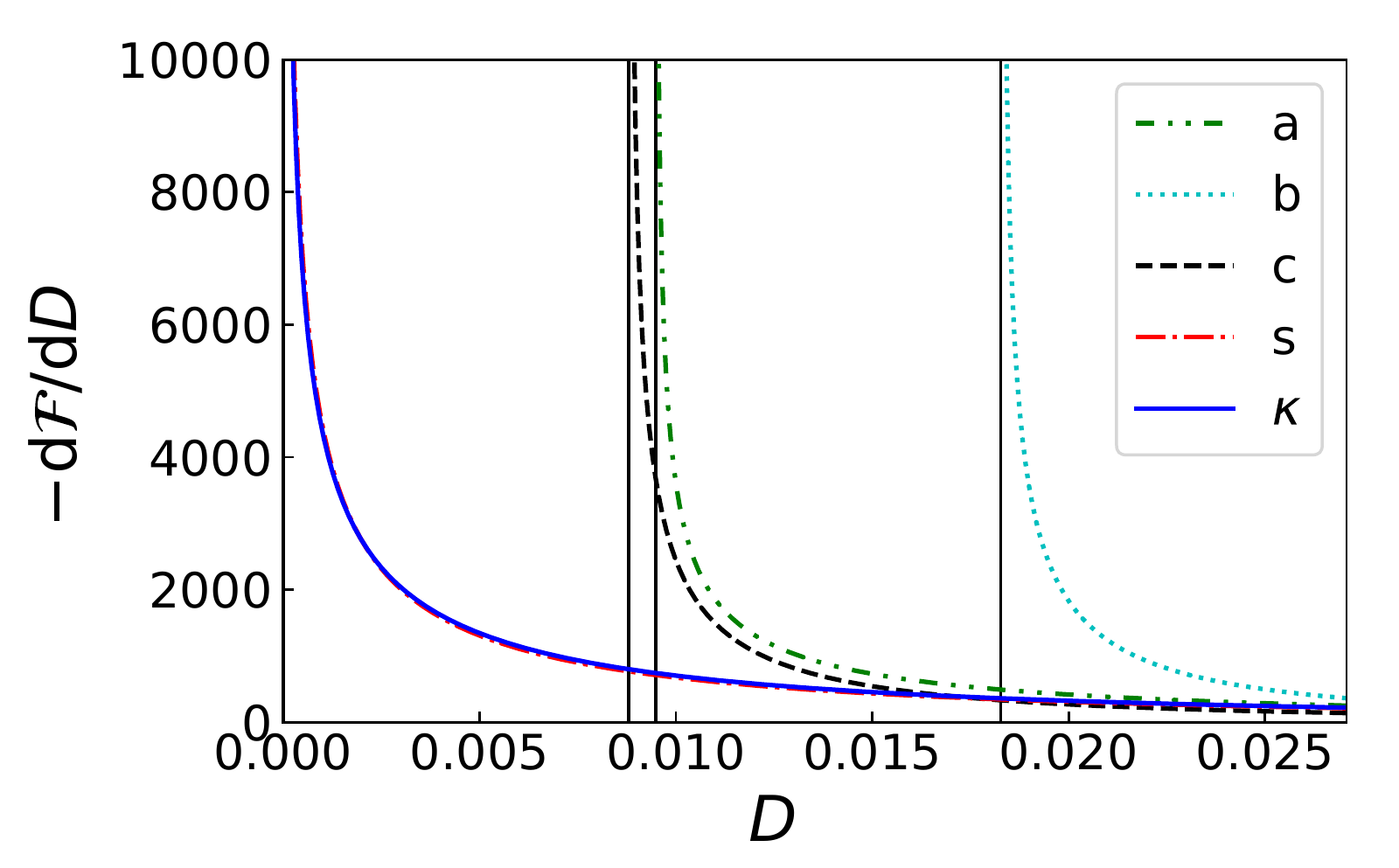}
\caption{Gradient of the universal functional $\mathcal{F}$ for the Bose-Hubbard model along the five paths defined in Fig.~\ref{fig:HubbardF}. The results for $\kappa$ and $s$ almost coincide.
\label{fig:dFD}}
\end{figure}

We illustrate the BEC force and the diverging behaviour of the functional's gradient close to the boundary $\partial \triangle$ of its domain in general for the Bose-Hubbard model. For this we consider again as in Sec.~\ref{subsec:Hubbard5} the case of $N=100$ bosons on $L=5$ sites. We then determine the directional derivative of the functional along the five paths which were defined in Fig.~\ref{fig:HubbardF}. Since for all five paths the distance $D$ of the occupation number vector $\bd{n}$ to $\mathbf{0}$ is monotonously decreasing we can parametrize the functional's derivative along each path by $D$. The respective results are depicted in Fig.~\ref{fig:dFD}. There, the vertical solid lines correspond to the values of $D$ at which the respective paths reach the boundary of $\triangle$ (see also Figs.~\ref{fig:HubbardF}, \ref{fig:FD}).
We first observe that for all five paths $-\partial \F/\partial D$ is diverging at the end point of each path on the boundary $\partial \triangle$. As a rather elementary analysis reveals (not shown here) this divergence is always proportional to $1/\sqrt{\mbox{dist}(\bd{n},\partial \triangle)}$. As far as the four straight paths $a, b, c, s$ are concerned, this was expected given the general results of Sec.~\ref{subsec:forcegeneral}. In contrast to the two continuous Bose gases, however, the same applies in the Bose-Hubbard model also for the curved $\kappa$-path which is obtained by just reducing the coupling strength.

\chapter{Excited State RDMFT\label{ch:w-RDMFT}}

In the previous two chapters, we focused on the foundations of ground state RDMFT for bosons and successfully applied it to BECs. In this chapter, we introduce a novel method, namely a bosonic ensemble RDMFT for excited states. Just like ground state RDMFT, the development of excited state RDMFT requires a solid mathematical foundation. Therefore, we first extend in Sec.~\ref{sec:Math_wRDMFT} the collection of mathematical tools, mainly from convex analysis, started in Sec.~\ref{sec:Math-gsRDMFT}. Based on these concepts, we introduce the new $\bd\omega$-ensemble RDMFT in Sec.~\ref{sec:intro_wRDMFT}. Furthermore, we apply a relaxation scheme to turn $\bd\omega$-ensemble RDMFT into a viable method in Sec.~\ref{sec:relaxation}. The main task is to characterize the domain of the relaxed $\bd\omega$-ensemble universal functional in Sec.~\ref{sec:setE1N}. Since Sec.~\ref{sec:setE1N} is quite technical, we proceed in Sec.~\ref{sec:Examples_wRDMFT} by discussing several examples and illustrations. Applying the excited state RDMFT to homogeneous Bose gases allows us to derive an excited state universal functional for BECs in Sec.~\ref{sec:w-RDMFT_BEC}. Moreover, we illustrate the $\bd\omega$-ensemble functional and its domain for the Bose-Hubbard Dimer in Sec.~\ref{sec:HubbardDimer_omega}. 

\section{Mathematical Preliminaries\label{sec:Math_wRDMFT}}

\subsection{Vector majorization\label{sec:majorization}}

For any $\bd{x}=(x_1, x_2, ..., x_d)\in\mathbb{R}^d$, we denote the vector of decreasingly ordered entries by $\bd{x}^\downarrow$ such that $x_1^\downarrow\geq x_2^\downarrow\geq ...\geq x_d^\downarrow$. The superscript $\downarrow$ might be omitted in the following sections if it is clear from the context which of the two vectors $\bd{x}$, $\bd{x}^\downarrow$ is meant.  

A vector $\bd{x}\in \mathbb{R}^d$ is said to be majorized by $\bd{y}\in \mathbb{R}^d$, denoted by $\bd{x}\prec\bd{y}$, if and only if the two conditions
\begin{align}
&\sum_{i=1}^kx_i^\downarrow \leq \sum_{i=1}^ky_i^\downarrow\,,\quad 1\leq k\leq d-1 \label{eq:majcond1}\\\
& \sum_{i=1}^dx_i^\downarrow = \sum_{i=1}^dy_i^\downarrow \label{eq:majcond2}
\end{align}
hold. Replacing the equality in \eqref{eq:majcond2} by an inequality with $\leq$ leads to the concept of weak majorization, $\bd{x}\prec_\mathrm{w}\bd{y}$. Clearly, the original order of the entries in $\bd{x}$, $\bd{y}$ does not enter the majorization conditions and therefore it does not follow from $\bd{x}\prec\bd{y}$ and $\bd{y}\prec\bd{x}$ that $\bd{x}=\bd{y}$ because the two vectors can have the same entries but in a different order. 

Hardy, Littlewood, and P\'{o}lya \cite{HLP53} provided an alternative definition of majorization, than through the partial sums in Eq.~\eqref{eq:majcond1} and Eq.~\eqref{eq:majcond2}, which might be more practical for some purposes (see also Sec.~\ref{sec:genNd}). We first recall that a matrix $P\in\mathbb{R}^{d\times d}$ is called doubly stochastic if all its entries are nonnegative and the entries in each row and column sum up to one. Then, there is a strong connection between doubly stochastic matrices and the majorization of vectors provided by the following theorem \cite{HLP53}: For any $\bd{x}, \bd{y}\in \mathbb{R}^d$ we have $\bd{x}\prec\bd{y}$ if and only if $\bd{x}=P\bd{y}$ for some doubly stochastic matrix $P\in\mathbb{R}^{d\times d}$.

Since we intent to use the concept of majorization in the context of RDMFT and thus apply it to quantum systems, we require a connection to quantum states and in particular density matrices. Alberti and Uhlmann \cite{AU82} proved that for two density matrices $\hat{\Gamma},\hat{\Gamma}^\prime:\mathcal{H}\to\mathcal{H}$ acting on a $D$-dimensional Hilbert space $\mathcal{H}$, the density operator $\hat{\Gamma}^\prime$ can be decomposed into a convex combination
\begin{equation}
\hat{\Gamma}^\prime= \sum_{i=1}^D q_i\hat{U}_i^{\phantom{\dagger}}\hat{\Gamma}\hat{U}_i^\dagger\,,
\end{equation}
where $\hat{U}_i$ are unitary operators, $0\leq q_i\leq 1$ and $\sum_{i=1}^Dq_i=1$, if and only if the spectrum of $\hat{\Gamma}^\prime$ is majorized by the spectrum of $\hat{\Gamma}$, $\mathrm{spec}(\hat{\Gamma}^\prime)\prec \mathrm{spec}(\hat{\Gamma})$. 

A special subclass of doubly stochastic matrices are the permutation matrices which are square matrices with exactly one entry equal to one in each row and column, and zeros in all other entries. Birkhoff's theorem \cite{B46, vN53} then states that the doubly stochastic matrices represent the convex hull of the permutation matrices. Further, the permutation matrices constitute the extremal points of the convex set of doubly stochastic matrices. For $P\in\mathbb{R}^{d\times d}$, there exist $d!$ many permutation matrices $\pi_i, i=1, ..., d!$.

\subsection{Permutohedra and Rado's theorem \label{sec:Rado}}

A polyhedron is a subset $P\subset \mathbb{R}^d$ which is given by the intersection of finitely many halfspaces and thus the set of solutions $\bd x$ to finitely many inequalities $A\bd{x} \leq \bd{y}$ where $A$ is a $m\times d$ matrix and $\bd{x}, \bd{y}\in \mathbb{R}^d$. Note that following this definition any polyhedron is convex. A bounded polyhedron is called a polytope. This leads to the so-called \textit{hyperplane representation} ($H$-representation) of a polytope. In general, one finds arbitrary many different $H$-representations for one polytope but there always exists a unique minimal $H$-representation up to scaling. The inequalities belonging to the minimal $H$-representation are called facet-defining. Moreover, every $H$-representation can be turned into a \textit{vertex representation} ($V$-representation) in which the polytope is given by the convex hull of a finite set of extremal points called vertices. A special case of polytopes are simplices and another important subclass of polytopes is the permutohedron which is defined as 
\begin{equation}\label{eq:permutohedron}
P_{\bd{x}} \equiv \mathrm{conv}(\{\pi(\bd{x})\,|\,\pi\in \mathcal{S}^d\})
\end{equation}
for a vector $\bd{x}\in \mathbb{R}^d$ and $\mathcal{S}^d$ denotes the group of permutations $\pi$ of $d$ elements. Thus, $P_{\bd{x}}$ is given by the convex hull of all possible permutations of the entries of $\bd{x}$. Using Eq.~\eqref{eq:permutohedron}, we obtain the following theorem by Rado \cite{R52} relating permutohedra to the majorization of vectors: Let $\bd{x}, \bd{y}\in \mathbb{R}^d$ be two vectors in $\mathbb{R}^d$. Then, $\bd{x}\prec \bd{y}$  if and only if $\bd{x}\in P_{\bd{y}}$.
To prove Rado's theorem we first show that for $\bd{x}, \bd{y}\in \mathbb{R}^d$ the majorization $\bd{x}\prec\bd{y}$ implies $\bd{x} \in P_{\bd{y}}$. It follows directly from combining the theorem by Hardy, Littlewood,  P\'{o}lya with Birkhoff's theorem, that 
\begin{equation}
\begin{split}
\bd{x} &= P\bd{y}= \sum_{i=1}^{d!}p_i(\pi_i\bd{y})\,,
\end{split}
\end{equation}
where $\pi_1, ..., \pi_{d!}$ are the permutation matrices, $p_i\geq 0$ and $\sum_{i=1}^{d!} p_i=1$. Therefore, $\bd{y}$ lies in the convex hull of all permutations of the entries of $\bd{y}$ generating the set $P_{\bd{y}}$. To prove the opposite direction $\bd{x}\in P_{\bd{y}}\Rightarrow \bd{x}\prec\bd{y}$ one simply has to reverse the above argument. 

\subsection{Generalization of Rayleigh-Ritz variational principle}

One of the most well-known methods to derive the ground state energy of a quantum system is the Rayleigh-Ritz variational principle, which we already explained in context of the energy minimization in Eq.~\eqref{eq:E0_varprin}. Moreover, this variational principle is the underlying concept of Levy's constrained search discussed in Sec.~\ref{sec:Levy} providing a more viable formulation of RDMFT for ground states than Gilbert's theorem. Also, the RDMFT for excited states, which we will develop in this chapter, is based on a variational principle and a constrained search formalism. In 1988, Gross, Oliviera, and Kohn \cite{Gross88_1, Gross88_2, Oliveira88} provided such a generalization of the Rayleigh-Ritz variational method for DFT. In the following, we refer to this generalization as the GOK variational principle. The GOK variational principle holds for observables $\hat{H}$ acting on a $D$-dimensional Hilbert space $\mathcal{H}$. We denote the increasingly ordered eigenvalues by $E_1\leq E_2\leq ...\leq E_D$. Let $\boldsymbol{\omega}\in \mathbb{R}^D$ be a vector with decreasingly arranged entries $\omega_1\geq\omega_2\geq ...\geq \omega_D\geq 0$ and $\sum_j \omega_j=1$. Then, the weighted sum of the eigenvalues follows from \cite{Gross88_1, Gross88_2, Oliveira88}:
\begin{equation}\label{Evar_GOK}
E_{\boldsymbol{\omega}} \equiv \sum_{j=1}^D\omega_j E_j = \min_{\hat{\Gamma}\in \mathcal{E}^N(\boldsymbol{\omega)}}\Tr\left[\hat{\Gamma}\hat{H}\right]\,.
\end{equation}
The minimizer state in Eq.~\eqref{Evar_GOK} is given by 
\begin{equation}\label{eq:Gamma_minimizer}
\hat{\Gamma}_{\boldsymbol{\omega}} = \sum_{j=1}^D\omega_j \ket{\Psi_j}\!\bra{\Psi_j}
\end{equation}
with $\hat{H}\ket{\Psi_j}= E_j\ket{\Psi_j}$, and $\mathcal{E}^N(\boldsymbol{\omega)}$ denotes the set of all $N$-particle density operators with spectrum $\boldsymbol{\omega}$. 
In the case of degenerate eigenvalues, one has to assign an arbitrary ordering to the orthonormal states in the corresponding degenerate subspace of states and keep the labels fixed afterwards \cite{Gross88_1}. 

In the case that $\hat{H}$ is a Hamiltonian, the $E_j$'s are the corresponding eigenenergies, as already anticipated by the notation.
In physics, one is often interested in the first few low-lying excited state energies and thus chooses only finitely many weights to be non-zero. We denote the number of non-vanishing weights by $r$.
Then, knowing the expression for $E_{\bd \omega}$ in Eq.~\eqref{Evar_GOK} allows extracting the eigenvalues $E_j$ for $j\leq r$ by taking appropriate derivatives. Alternatively, one can evaluate $E_{\bd \omega}$ for different $\bd \omega$ and apply appropriate gradient triangles to determine the eigenenergies. For example, $r=2$ is already sufficient to calculate the energy gap $\Delta E$ between the ground state and the first excited state. In that case, we have $\bd \omega= (\omega, 1-\omega,0, ...)$ and evaluating $E_{\bd \omega}$ for $\omega =1$ and a second $1/2\leq\omega^\prime<1$ yields
\begin{equation}
\Delta E\equiv E_2-E_1 = \frac{E_{(\omega^\prime, 1-\omega^\prime, 0, ...)} - E_{(1, 0, ...)}}{1-\omega^\prime}\,.
\end{equation}

\section{Introduction of $\bd{\omega}$-ensemble RDMFT\label{sec:intro_wRDMFT}}

In the following derivation of an RDMFT for excited states, we adopt the GOK variational principle established for DFT and combine it with a constrained search formalism in the spirit of Levy's \cite{LE79} and Valone's \cite{V80} ideas. However, this procedure is far from being trivial and several obstacles occur. For fermions, such a theory was just recently proposed and explained in Ref.~\cite{Schilling21}. In the following, we focus mainly on the bosonic version which uses the same underlying concepts and ideas up to modifications due to the bosonic statistics. 

In Eq.~\eqref{Evar_GOK}, we already introduced the set $\mathcal{E}^N(\boldsymbol{\omega)}$ of all $\boldsymbol{\omega}$-ensemble $N$-particle density operators $\hat{\Gamma}$ which is defined by
\begin{equation}\label{def:ENomega}
\mathcal{E}^N(\boldsymbol{\omega)}\equiv \{\hat{\Gamma}\in \mathcal{E}^N\,|\,\mathrm{spec}^\downarrow (\hat{\Gamma})=\boldsymbol{\omega}\}\,.
\end{equation}
In analogy to Eq.~\eqref{eq:setP1} and Eq.~\eqref{eq:setE1}, we obtain by tracing out $N-1$ particles
\begin{equation}\label{eq:EN1omega}
\mathcal{E}_N^1(\boldsymbol{\omega}) = N\Tr_{N-1}[\mathcal{E}^N(\boldsymbol{\omega})]\,.
\end{equation} 
The 1RMDs $\hat{\gamma} \in \mathcal{E}^1_N(\boldsymbol{\omega})$ are then called \textit{$\boldsymbol{\omega}$-ensemble $N$-representable} in contrast to the ensemble $N$-representable 1RDMs $\hat{\gamma}\in \mathcal{E}_N^1$ introduced in Sec.~\ref{sec:density}. Thus, a practical description of the set $\mathcal{E}^1_N(\bd \omega)$ requires the knowledge of these additional constraints. For one non-vanishing weight, i.e.~$\boldsymbol{\omega}_0 = (1, 0, ...)$, there exists only one minimizer state in Eq.~\eqref{Evar_GOK} which is pure. Then, Eq.~\eqref{Evar_GOK} reduces to the minimization performed to obtain the ground state energy and ground state 1RDM. From the definition of $\mathcal{E}^N(\boldsymbol{\omega)}$ in Eq.~\eqref{def:ENomega} follows that for $\boldsymbol{\omega}_0$, the set of all $\bd\omega$-ensemble $N$-particle density operators reduces to
\begin{equation}\label{eq:Ew0}
\mathcal{E}^N(\boldsymbol{\omega}_0) = \mathcal{P}^N\,.
\end{equation} 
Moreover, by tracing out all except one particle we obtain
\begin{equation}\label{eq:E1w0}
\mathcal{E}_N^1(\boldsymbol{\omega}_0)=\mathcal{P}_N^1\,,
\end{equation}
and thus the set of all pure state $N$-representable 1RDMs $\hat{\gamma}\in\mathcal{P}_N^1$. Recall that for fermions, the set $\mathcal{P}_N^1$ is usually unknown due to the generalized Pauli constraints. This hints that for general $r$, the additional $\bd \omega$-ensemble $N$-representability constraints increase the complexity of the problem drastically. From a conceptual point of view, this has an even more remarkable effect in case of bosons: Solving the pure state $N$-representability problem for bosons is trivial because the ground state occupation numbers are only restricted through positivity and normalization of the 1RDM. However, for a bosonic $\gh\in \mathcal{E}^1_N(\bd \omega)$, the description of $\mathcal{E}^1_N(\bd \omega)$ leads to additional constraints on the natural occupation numbers interpreted as generalized exclusion constraints for bosons. Altogether, we conclude that finding the solution to the $\bd \omega$ ensemble $N$-representability problem is an extremely difficult task and tackle this problem in Sec.~\ref{sec:relaxation}.

Next, we focus on some important properties of the two sets in Eq.~\eqref{def:ENomega} and Eq.~\eqref{eq:EN1omega}.  
First, note that the restriction of the set $\mathcal{E}^N$ of all ensemble $N$-particle density operators to the subset $\mathcal{E}^N(\boldsymbol{\omega})$ of all $N$-particle density operators with fixed spectrum $\bd \omega$ is non-linear. As a result, the two sets $\mathcal{E}^N(\boldsymbol{\omega})$ and $\mathcal{E}_N^1(\boldsymbol{\omega})$ are in general not convex, contrarily to $\mathcal{E}^N$ and $\mathcal{E}_N^1$. 
Furthermore, we can exploit the invariance of the spectrum of a density operator under unitary transformations to simplify the description of the set $\mathcal{E}^1_N(\bd \omega)$. We observe that if $\hat{\gamma}\in \mathcal{E}^1_N(\boldsymbol{\omega})$ it follows that also the unitary transformation  $\hat{u}: \mathcal{H}_1\to\mathcal{H}_1$ of $\gh$ satisfies $\hat{\gamma}^\prime \equiv\hat{u}\hat{\gamma}\hat{u}^\dagger \in \mathcal{E}^1_N(\boldsymbol{\omega})$. To proof this statement we first lift the unitary transformation of $\hat{\gamma}$ to the $N$-particle level through $\hat{\gamma}^\prime \equiv N\Tr_{N-1}[\hat{u}^{\otimes^N}\hat{\Gamma}(\hat{u}^\dagger)^{\otimes^N}]$. If $\hat{\Gamma} \in \mathcal{E}^N(\boldsymbol{\omega})$ is fulfilled, it follows that $\hat{\Gamma}^\prime \equiv \hat{u}^{\otimes^N}\hat{\Gamma}(\hat{u}^\dagger)^{\otimes^N} \in \mathcal{E}^N(\boldsymbol{\omega})$ because the set $\mathcal{E}^N(\boldsymbol{\omega})$ is defined only through the spectral constraint $\mathrm{spec}^\downarrow (\hat{\Gamma})=\boldsymbol{\omega}$. Therefore, it is invariant under unitary transformations of its elements. The conclusion that $\hat{\gamma}^\prime\in \mathcal{E}^1_N(\bd \omega)$ follows directly from tracing out $N-1$ particles of $\hat{\Gamma}^\prime$.
Thus, to determine whether a 1RDM $\hat{\gamma}$ belongs to the set $\mathcal{E}^1_N(\boldsymbol{\omega})$ or not, it is sufficient to know its spectrum. We will return to the consequences of this unitary invariance in Sec.~\ref{sec:setE1N} and explain how it facilitates the description of $\mathcal{E}^1_N(\bd\omega)$.

Inspired by Levy's constrained search and the GOK variational principle, we now introduce a constrained search on the set $\mathcal{E}^N(\bd \omega)$ such that
\begin{equation}\label{eq:Levyomega}
\begin{split}
E_{\boldsymbol{\omega}}(\hat{h}) &= \min_{\hat{\Gamma}\in \mathcal{E}^N(\boldsymbol{\omega})}\Tr_N[(\hat{h} + \hat{W})\hat{\Gamma}]\\\
&= \min_{\hat{\gamma}\in \mathcal{E}_N^1(\boldsymbol{\omega})}\Big[  \min_{\mathcal{E}^N(\boldsymbol{\omega})\ni\hat{\Gamma}\mapsto\hat{\gamma}}\Tr_N[(\hat{h}+\hat{W})\hat{\Gamma}]\Big] \\\
&= \min_{\hat{\gamma}\in \mathcal{E}_N^1(\boldsymbol{\omega})}\Big[\Tr_1[\hat{h}\hat{\gamma}] + \underbracket{\min_{\mathcal{E}^N(\boldsymbol{\omega})\ni\hat{\Gamma}\mapsto\hat{\gamma}}\Tr_N[\hat{W}\hat{\Gamma}]}_{\equiv\mathcal{F}_{\boldsymbol{\omega}}(\hat{\gamma})}\Big]\,.
\end{split}
\end{equation}
The above equation provides a definition of the new universal functional $\mathcal{F}_{\bd \omega}(\gh)$ in the same spirit as in Eq.~\eqref{Levy}. Note that Eq.~\eqref{eq:Levyomega} does not require a generalization of the Gilbert theorem to excited states to prove the existence of a $\boldsymbol{\omega}$-ensemble functional. In that case, the domain would be defined by all 1RDMs which follow as $\boldsymbol{\omega}$-minimizers for a particular Hamiltonian $\hat{H}(\hat{h})$, leading again to a sort of modified $v$-representability problem. However, the constrained search formalism in Eq.~\eqref{eq:Levyomega} allows us to circumvent this $v$-representability problem from the very beginning and avoid difficulties due to possible degenerate states by extending the domain of $\mathcal{F}_{\boldsymbol{\omega}}(\gh)$ to the set $\mathcal{E}_N^1(\boldsymbol{\omega})$ of all $\boldsymbol{\omega}$-ensemble $N$-representable 1RDMs.  

Due to the nonlinear restriction of $\mathcal{E}^N$ to $\mathcal{E}^N(\boldsymbol{\omega})$, the $\boldsymbol{\omega}$-ensemble functional $\mathcal{F}_{\boldsymbol{\omega}}$ is in general not convex on the entire domain $\mathcal{E}_N^1(\boldsymbol{\omega})$. This shall be contrasted with the ensemble ground state functional $\mathcal{F}_e(\gh)$ in Eq.~\eqref{eq:Fe}. As a consistency check, we observe that for $r=1$, the $\bd \omega$-ensemble functional $\mathcal{F}_{\bd \omega}(\gh)$ reduces to the pure state universal functional
\begin{equation}
\mathcal{F}_p(\hat{\gamma}) = \mathcal{F}_{\boldsymbol{\omega}_0}(\hat{\gamma})\,.
\end{equation}
In the more interesting case $\boldsymbol{\omega}\neq \boldsymbol{\omega}_0$, the minimization in the second line of Eq.~\eqref{eq:Levyomega} is performed over all ensemble $N$-particle density operators $\hat{\Gamma}\in \mathcal{E}^N(\boldsymbol{\omega})$ and thus the ensemble functional $\mathcal{F}_e(\gh)$ is related to $\mathcal{F}_{\boldsymbol{\omega}}(\gh)$ through
\begin{equation}\label{eq:FeFw}
\begin{split}
\mathcal{F}_e(\hat{\gamma}) &= \min_{\mathcal{E}^N\ni\hat{\Gamma}\mapsto\hat{\gamma}}\Tr_N[\hat{W}\hat{\Gamma}]\\\
&= \min_{\boldsymbol{\omega}}\min_{\mathcal{E}^N(\boldsymbol{\omega})\ni\hat{\Gamma}\mapsto\hat{\gamma}}\Tr_N[\hat{W}\hat{\Gamma}]\\\
&\equiv \min_{\boldsymbol{\omega}}\mathcal{F}_{\boldsymbol{\omega}}(\hat{\gamma})\,.
\end{split}
\end{equation}
In Eq.~\eqref{eq:FeFw}, we extend the domain of $\mathcal{F}_{\boldsymbol{\omega}}$ to the full set $\mathcal{E}_N^1$ by defining 
\begin{equation}
\mathcal{F}_{\boldsymbol{\omega}}(\hat{\gamma}) \equiv
\begin{cases}
  \min_{\mathcal{E}^N(\boldsymbol{\omega})\ni\hat{\Gamma}\mapsto\hat{\gamma}}\Tr_N[\hat{W}\hat{\Gamma}]  ,& \forall \hat{\gamma}\in \mathcal{E}_N^1(\boldsymbol{\omega})\,,\\
    \infty,  & \forall \hat{\gamma}\notin \mathcal{E}_N^1(\boldsymbol{\omega})\,.
\end{cases}
\end{equation}
Thus, it follows that the extension of the domain of $\mathcal{F}_{\bd \omega}(\gh)$ to the entire set of all ensemble $N$-representable 1RDMs does not affect any minimization process to calculate $E_{\bd \omega}(\hat{h})$ because the minimum will always be attained in $\mathcal{E}^1_N(\bd \omega)$. 
It is worth noticing that whenever a 1RDM $\hat{\gamma}$ is ground state $v$-representable, the minimizer $\boldsymbol{\omega}_\mathrm{min}$ in Eq.~\eqref{eq:FeFw} is given by $\boldsymbol{\omega}_\mathrm{min}=\boldsymbol{\omega}_0$.

\section{Relaxation of $\bd{\omega}$-ensemble RDMFT\label{sec:relaxation}}

The $\bd \omega$-ensemble RDMFT introduced in the above section is not practically feasible due to the involved $\bd\omega$-ensemble $N$-representability constraints. For this purpose, we resort in this section to the well-known convex relaxation method to circumvent this problem. It is based on the fact, that it is always possible to replace a non-convex minimization problem by the corresponding convex minimization problem (c.f. Sec.~\ref{subsec:terminology} and Sec.~\ref{sec:Levy}). Applied to the constrained search formalism in Eq.~\eqref{eq:Levyomega}, this means replacing the $\bd\omega$-ensemble functional $\Fw(\gh)$ by its lower convex envelope
\begin{equation}\label{eq:defFbar1}
\Fbw(\hat{\gamma})\equiv\mathrm{conv}\left(\mathcal{F}_{\boldsymbol{\omega}}(\hat{\gamma})\right)\,.
\end{equation}
In addition, the domain of the relaxed $\bd\omega$-ensemble functional is given by the convex hull of $\mathcal{E}^1_N(\bd\omega)$, and thus
\begin{equation}
\ebw \equiv \mathrm{conv} \left(\mathcal{E}_N^1(\boldsymbol{\omega})\right)\,.
\end{equation}
We will refer to a 1RDM $\hat{\gamma}\in \ebw$ in the following as being \textit{relaxed $\boldsymbol{\omega}$-ensemble $N$-representable}. Then, the weighted sum of lowest eigenenergies, $E_{\bd \omega}$, follows from minimizing the new energy functional $\Tr_1[\hat{h}\hat{\gamma}]+\Fbw(\hat{\gamma})$ as
\begin{equation}
E_{\boldsymbol{\omega}} = \min_{\hat{\gamma}\in \thickbar{\mathcal{E}}_N^1(\boldsymbol{\omega})}\left[\Tr_1[\hat{h}\hat{\gamma}]+\Fbw(\hat{\gamma})\right]\,.
\end{equation}
The above expression yields not only $E_{\boldsymbol{\omega}}$ but also the minimizer 1RDM $\hat{\gamma}= N\Tr_{N-1}[\hat{\Gamma}_{\boldsymbol{\omega}}]$ corresponding to the $\boldsymbol{\omega}$-minimizer $\hat{\Gamma}_{\boldsymbol{\omega}}$ in Eq.~\eqref{eq:Gamma_minimizer}. The convex relaxation has two main advantages: First, every local minimum of $\Fbw(\gh)$ is also a global one facilitating the minimization of the energy functional. However, even more important is that we can find a concrete description of the compact convex set $\ebw$ which is not hampered by the $\bd \omega$-ensemble $N$-representability problem. We present a concrete strategy to obtain $\ebw$ in Sec.~\ref{sec:setE1N}. 

Alternatively, we can introduce the relaxed $\bd\omega$-ensemble RDMFT following Valone's ideas \cite{V80} of ensemble RDMFT by extending on the $N$-particle level the set $\mathcal{E}^N(\bd \omega)$ to its convex hull,
\begin{equation}\label{eq:ENbar_conv}
\Ebw \equiv \mathrm{conv}\left(\mathcal{E}^N(\boldsymbol{\omega})\right)\,.
\end{equation}
Before applying the constrained search formalism to $\Ebw$, we further investigate the properties of $\Ebw$ as well as their important consequences.
Since according to Eq.~\eqref{def:ENomega}, the non-convex set $\Ew$ is fully characterized through the spectral constraint $\mathrm{spec}(\hat{\Gamma}^\downarrow)=\bd\omega$, and the spectrum is invariant under unitary transformations, we can apply Uhlmann's theorem in Sec.~\ref{sec:majorization} to show that
\begin{equation}\label{eq:ENunion}
\Ebw = \bigcup\limits_{\bd\omega^\prime\prec\,\bd\omega} \mathcal{E}^N(\bd\omega^\prime) \equiv \{\hat{\Gamma}\in \mathcal{E}^N\,|\,\mathrm{spec}(\hat{\Gamma})\prec\bd\omega\}\,. 
\end{equation}
Thus, the set $\Ebw$ constitutes of all $N$-particle density operators $\hat{\Gamma}$ whose spectrum is majorized by $\bd\omega$. Since the partial trace map $\tr_{N-1}[\cdot]$ is linear, we are allowed to change its order with the convex hull operation $\mathrm{conv}(\cdot)$. From this consideration it follows immediately that
\begin{equation}\label{eq:EN1union}
\ebw = N\tr_{N-1}[\Ebw]=\bigcup\limits_{\bd\omega^\prime\prec\,\bd\omega}\mathcal{E}^1_N(\bd \omega^\prime)\,.
\end{equation}
Moreover, the last equality in Eq.~\eqref{eq:EN1union} has another striking consequence
\begin{equation}\label{eq:inclusion}
\bd\omega^\prime\prec\bd\omega\,\,\Leftrightarrow\,\,\xbar{\mathcal{E}}^1_N(\bd\omega^\prime)\subset\ebw\,.
\end{equation}
Thus, the set of all relaxed $\bd\omega$-ensemble $N$-representable 1RDMs becomes smaller if a vector $\bd\omega^\prime$ is majorized by another vector $\bd\omega$. We illustrate this inclusion relation of the set $\ebw$ for the Bose-Hubbard dimer in Sec.~\ref{sec:HubbardDimer_omega}.
Next, recall from our discussion in context of Eq.~\eqref{eq:Ew0} and \eqref{eq:E1w0} that we recover the ground state RDMFT for $\bd\omega=\boldsymbol{\omega}_0\equiv(1, 0, ...)$. Then, it follows that
\begin{equation}\label{eq:setsw0}
\mathcal{E}_N^1 = \xbar{\mathcal{E}}_N^1(\boldsymbol{\omega}_0)\equiv \mathrm{conv}(\mathcal{E}_N^1(\boldsymbol{\omega}_0)) = \mathrm{conv}(\mathcal{P}_N^1)\,.
\end{equation} 
We prove the statement in Eq.~\eqref{eq:setsw0} in a general way without distinguishing between bosons and fermions. First, we show that $\mathcal{E}_N^1 \subset  \xbar{\mathcal{E}}_N^1(\boldsymbol{\omega}_0)$. For every $\hat{\gamma}\in \mathcal{E}_N^1$ we consider a $N$-particle operator $\hat{\Gamma}$ with $\mathcal{E}^N\ni\hat{\Gamma} \mapsto \hat{\gamma}$. Due to the Krein-Milman theorem every element in the convex and compact set $\mathcal{E}^N$ can be expressed as a convex combination of its extremal points which are given by pure states $\hat{\Gamma}_i\in \mathcal{P}^N$ such that $\hat{\Gamma} = \sum_j p_j\hat{\Gamma}_j$. Since $\xbar{\mathcal{E}}^N(\boldsymbol{\omega}_0)= \mathrm{conv}(\mathcal{E}^N(\boldsymbol{\omega}_0)) = \mathrm{conv}(\mathcal{P}^N)$, it follows that $\hat{\Gamma}\in \xbar{\mathcal{E}}^N(\boldsymbol{\omega}_0)$ and thus $\hat{\gamma}\in \xbar{\mathcal{E}}^1_N(\bd\omega_0)$ proving that $\mathcal{E}_N^1 \subset  \xbar{\mathcal{E}}_N^1(\boldsymbol{\omega}_0)$. Next, we have to show that $\xbar{\mathcal{E}}_N^1(\boldsymbol{\omega}_0)\subset \mathcal{E}_N^1$. Every $\hat{\Gamma}\in \xbar{\mathcal{E}}^N(\boldsymbol{\omega}_0)$ with $\hat{\Gamma}\mapsto \hat{\gamma}\in \xbar{\mathcal{E}}_N^1(\boldsymbol{\omega}_0)$ can be written as a convex combination $\hat{\Gamma}= \sum_j p_j\hat{\Gamma}_j$ with $\hat{\Gamma}_j\in \mathcal{E}^N(\boldsymbol{\omega}_0) = \mathcal{P}^N$. This implies that $\hat{\Gamma}\in \mathcal{E}^N$ and therefore $\hat{\gamma}\in \mathcal{E}_N^1$ which finishes the proof of Eq.~\eqref{eq:setsw0}. Note that this elaborated proof is redundant for bosons since Eq.~\eqref{eq:setsw0} follows directly from $\mathcal{P}_N^1=\mathcal{E}_N^1$ derived in Sec.~\ref{subsec:N_bosons}.

In analogy to the definition of the ensemble functional $\mathcal{F}_e(\gh)$ in Eq.~\eqref{eq:Fe}, we obtain the relaxed functional $\Fbw(\gh)$ through the constrained search formalism 
\begin{equation}
\begin{split}
\min_{\EbwS\ni\hat{\Gamma}\mapsto\hat{\gamma}}\Tr_N[\hat{W}\hat{\Gamma}]&=  \min_{\substack{\sum_jp_j\hat{\Gamma}_j\mapsto\hat{\gamma}, \\ \hat{\Gamma}_j\in \mathcal{E}^N(\boldsymbol{\omega})}} \sum_jp_j\Tr_N[\hat{W}\hat{\Gamma_j}]\\\
&=  \min_{\substack{\sum_jp_j\hat{\gamma}_j=\hat{\gamma}, \\ \hat{\gamma}_j\in \mathcal{E}_N^1(\boldsymbol{\omega})}}\,\,\min_{\mathcal{E}^N(\boldsymbol{\omega})\ni\hat{\Gamma}_j\mapsto\hat{\gamma}_j}\sum_jp_j\Tr_N[\hat{W}\hat{\Gamma_j}]\\\
&\equiv \min_{\substack{\sum_jp_j\hat{\gamma}_j=\hat{\gamma}, \\ \hat{\gamma}_j\in \mathcal{E}_N^1(\boldsymbol{\omega})}} \sum_jp_j\mathcal{F}_{\boldsymbol{\omega}}(\hat{\gamma}_j)\\\
&=\mathrm{conv}\left(\mathcal{F}_{\boldsymbol{\omega}}(\hat{\gamma})\right)\\
&\equiv \Fbw(\gh)\,.
\end{split}
\end{equation}
In the first line, we use that every $N$-particle density operator $\hat{\Gamma}\in \Ebw$ can be written as a convex combination $\hat{\Gamma}=\sum_jp_j\hat{\Gamma}_j$ of $\hat{\Gamma}_j\in \mathcal{E}_N(\boldsymbol{\omega})$. In the third line, we insert the definition of $\mathcal{F}_{\bd\omega}(\gh)$ leading directly to the expression for the lower convex envelope of $\mathcal{F}_{\bd\omega}(\gh)$, defined as $\Fbw(\gh)$ in agreement with Eq.~\eqref{eq:defFbar1}. To summarize, the constrained search formalism provides a concrete approach to derive the relaxed functional resulting into a convex minimization problem with
\begin{equation}\label{eq:defFbar2}
\Fbw(\hat{\gamma})\equiv \min_{\EbwS\ni\hat{\Gamma}\mapsto\hat{\gamma}}\Tr_N[\hat{W}\hat{\Gamma}]\,.
\end{equation}
Moreover, the domain of $\Fbw(\gh)$ is the compact, convex set $\ebw$ of all relaxed $\bd\omega$-ensemble $N$-representable 1RDMs. It is thus our next task to determine a conclusive description of $\ebw$, which provides a convenient procedure to check whether a given 1RDM belongs to the set or not. 

\section{Characterization of $\ebw$ \label{sec:setE1N}}

In this section, we derive a systematic approach to characterize the set $\ebw$. To be more specific, we will show that it is sufficient to describe a spectral polytope which is only determined through the spectra of all 1RDMs $\gh\in\ebw$. We will then obtain a concrete description of this spectral polytope in its vertex representation in Sec.~\ref{sec:ebw_vertex}. Unfortunately, it turns out that the vertex representation is in many cases still not sufficient for practical purposes. Thus, we need to translate the vertex representation of the spectral polytope into its halfspace representation in Sec.~\ref{sec:h-repr_ebw}. 
Although both representations are equivalent, as discussed in Sec.~\ref{sec:Rado}, it can be quite complicated to transform the two representations into each other. 

\subsection{General procedure\label{sec:ebw_vertex}}

Let us start with the general procedure to determine the set $\ebw$ which is based on various mathematical concepts explained in Sec.~\ref{sec:Math-gsRDMFT} and Sec.~\ref{sec:Math_wRDMFT}. 

\subsubsection{Duality principle}

\begin{figure}[htb]
\centering
\begin{subfigure}[t]{.49\textwidth}
\centering
\includegraphics[width=0.8\linewidth]{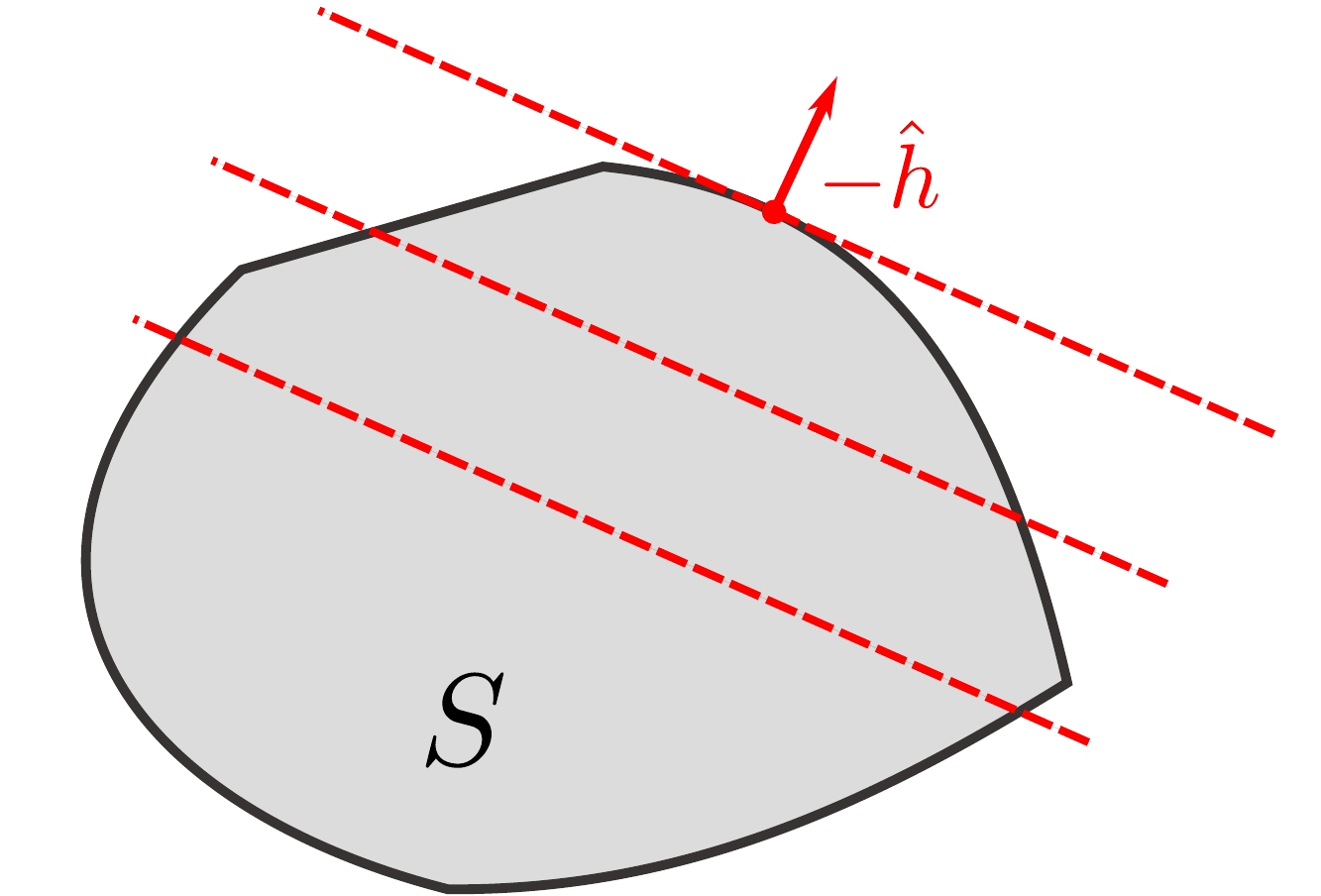}
\end{subfigure}
\begin{subfigure}[t]{.49\textwidth}
\centering
\includegraphics[width=0.8\linewidth]{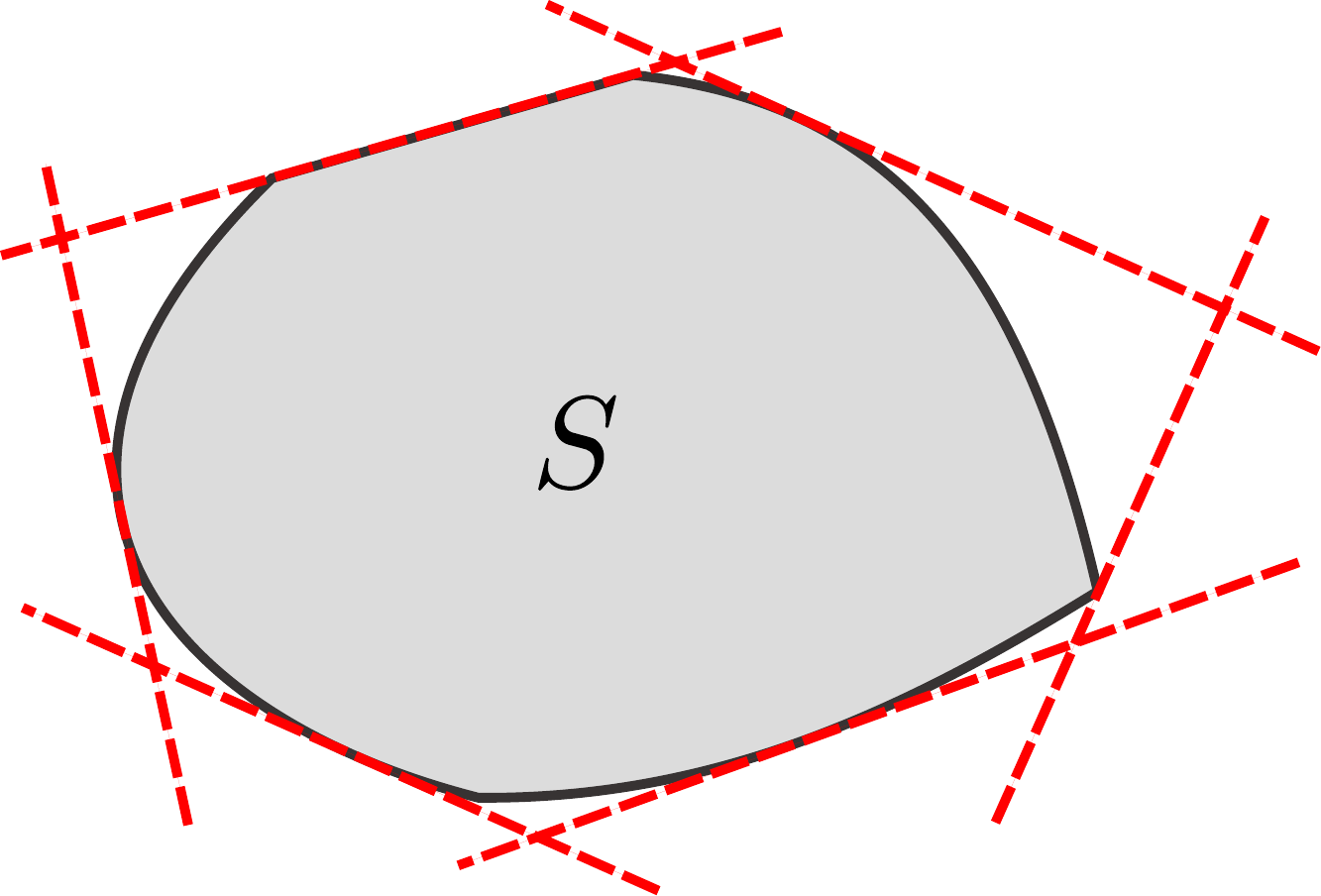}
\end{subfigure}
\medskip
\begin{minipage}[t]{\textwidth}
\caption{Left: Schematic illustration of the minimization over a convex set $S$ as described in the text. Right: Performing the minimization for all possible directions determines the boundary of $S$.\label{fig:minconvS}}
\end{minipage}
\end{figure}

The duality correspondence for convex sets, discussed in Sec.~\ref{sec:duality}, states that $\ebw$ can be characterized through its support function $\sigma_{\ebwS}(h)$. We illustrate the duality correspondence for the compact convex set $\ebw$ in Fig.~\ref{fig:minconvS}. Therefore, Eq.~\eqref{eq:supportmin} shows that the convex set $\ebw$ follows from minimizing $\Tr_1[\hat{h}\hat{\gamma}]$ for all possible hermitian one-particle Hamiltonians $\hat{h}$ and determining the corresponding minimizers $\hat{\gamma}_{\hat{h}}$. Since $\Tr_1[\hat{h}\hat{\gamma}]\equiv \langle\hat{h}, \hat{\gamma}\rangle_1$, the inner product $\langle\hat{h}, \hat{\gamma}\rangle_1$ on the Euclidean space of hermitian matrices defines a hyperplane in the set of 1RDMs $\hat{\gamma}\in \ebw$. Moreover, the one-particle Hamiltonian $\hat{h}$ determines the normal vector of the hyperplane. Therefore, minimizing $\Tr_1[\hat{h}\hat{\gamma}]$ over all $\hat{\gamma}\in \ebw$ corresponds to shifting the hyperplane in direction $-\hat{h}$ until it touches the boundary of the compact convex set $\ebw$. Taking the convex hull of all minimizers $\hat{\gamma}_{\hat{h}}$ eventually leads to
\begin{equation}\label{eq:E1Nconv}
\ebw = \mathrm{conv}\left(\left\{\underset{\hat{\gamma}\in \ebwS}{\mathrm{arg\,min}} \,\Tr_1[\hat{h}\hat{\gamma}]\Big\vert\,\hat{h} \text{ herm.}\right\}\right)\,.
\end{equation}
The extremal points of the set $\ebw$ are given by those 1RDMs $\hat{\gamma}\in \ebw$, which follow as unique minimizers for a particular choice of $\hat{h}$. If there exists more than one $\hat{\gamma}_{\hat{h}}$ for a given $\hat{h}$, the boundary points are not extremal and can be expressed as convex combinations of extremal elements in $\ebw$.

\subsubsection{Lifting minimization to $N$-particle level}

Since the set $\Ebw$ contains all information about the spectral vector $\boldsymbol{\omega}$ on the $N$-particle level, we lift the minimization to the $N$-boson level. This is implemented through
\begin{equation}
\min_{\hat{\gamma}\in \ebwS}\Tr_1[\hat{h}\hat{\gamma}] = \min_{\hat{\Gamma}\in \EbwS}\Tr_N[\hat{h}\hat{\Gamma}]\,.
\end{equation}
For simplicity, we drop the identities attached to $\hat{h}$ due to the lifting process to the $N$-particle level on the right hand side. 
Further, the extremal points of $\mathcal{E}^N(\boldsymbol{\omega})$ and $\Ebw$ coincide, leading to
\begin{equation}
\min_{\hat{\Gamma}\in \EbwS}\Tr_N[\hat{h}\hat{\Gamma}] = \min_{\hat{\Gamma}\in \mathcal{E}^N(\boldsymbol{\omega})}\Tr_N[\hat{h}\hat{\Gamma}]\,. 
\end{equation}
The above equality is based on the observation that relaxing the minimization from $\hat{\Gamma}\in\mathcal{E}^N(\boldsymbol{\omega})$ to $\hat{\Gamma}\in \Ebw$ does not change the outcome of the minimization. We then apply the GOK variational principle in Eq.~\eqref{Evar_GOK} to obtain 
\begin{equation}
\begin{split}
\min_{\hat{\gamma}\in \ebwS}\Tr_1[\hat{h}\hat{\gamma}] &= \min_{\hat{\Gamma}\in \mathcal{E}^N(\boldsymbol{\omega})}\Tr_N[\hat{h}\hat{\Gamma}] \\\
&= \sum_j\omega_jE_j\,.
\end{split}
\end{equation}
Thus, the minimization on the one-particle level on the left hand-side of the above equation, is determined through the weighted sum of the eigenenergies of the one-particle Hamiltonian $\hat{h}$ on the $N$-particle level. It is thus the next step to determine the eigenstates of $\hat{h}$ on the $N$-boson Hilbert space $\mathcal{H}_N$ defined in Eq.~\eqref{eq:HN_bosons}. This also requires that we restrict solely to bosonic quantum systems in the following. 

\subsubsection{Configuration states}

Using an orthonormal basis $\{\ket{i}\}_{i=1}^d$ with $d=\mathrm{dim}(\mathcal{H}_1)$, the eigenstates of $\hat{h}$ on the $N$-boson Hilbert space are given by the configuration states 
\begin{equation}
\ket{\bd{i}}\equiv\ket{i_1, ..., i_N}\equiv \hat{a}_{i_1}^\dagger...\hat{a}_{i_N}^\dagger\ket{0}\,,
\end{equation}
where $\ket{0}$ denotes the vacuum state and $\hat{a}_i^\dagger$ is a bosonic creation operator creating a boson in state $\ket{i}$. The one-particle Hamiltonian $\hat{h}$ on $\mathcal{H}_1$ in its spectral decomposition reads
\begin{equation}
\hat{h}\equiv \sum_{i=1}^dh_i\ket{i}\!\bra{i}\,.
\end{equation}
Since bosons are allowed to occupy the same quantum state, the set of all possible configurations in the $N$-boson Hilbert space follows as
\begin{equation}
\mathcal{I}_{N, d} \equiv \{ \bd{i}\equiv(i_1, ..., i_N)|\,1\leq i_1\leq i_2\leq ...\leq i_N\leq d\}\,.
\end{equation}
Note that referring to a $N$-boson configuration state $\ket{\bd i}$ is always associated with a respective reference basis on the one-particle Hilbert space $\mathcal{H}_1$. Further, in the case of unique minimizers in Eq.~\eqref{eq:E1Nconv}, the eigenstates of $\gh_h$ and $\hat{h}$ have to be equal. 

\subsubsection{Spectral polytope}

Next, we choose a fixed set of natural orbitals (NO) by resorting to the unitary invariance
\begin{equation}
\hat{u}\ebw\hat{u}^\dagger = \ebw\,,
\end{equation}
which holds for any unitary operator $\hat{u}:\mathcal{H}_1\to\mathcal{H}_1$. 
Let $\{\mathcal{B}_1\}$ denote the set of all possible ordered orthonormal bases for $\mathcal{H}_1$. Then, we split $\ebw$ into subsets for each basis $\mathcal{B}_1\in \{\mathcal{B}_1\}$. As a result, $\ebw$ is given by 
\begin{equation}
\begin{split}
\ebw&\equiv \bigcup\limits_{\mathcal{B}_1}\ebw\Big\vert_{\mathrm{NOs}=\mathcal{B}_1}\\\
&= \mathrm{spec}^\downarrow\left(\ebw\right)\times \{\mathcal{B}_1\}\,,
\end{split}
\end{equation}
where $\mathrm{spec}^\downarrow(\ebw)$ denotes the set of all possible spectra of 1RDMs $\hat{\gamma}\in \ebw$ with decreasingly ordered entries. The larger set of all possible (unordered) spectra is denoted by $\mathrm{spec}(\ebw)$ and we define the spectral polytopes
\begin{equation}
\begin{split}
\Sigma(\boldsymbol{\omega}) &\equiv \mathrm{spec}\left(\ebw\right)\,,\\\
\Sigma^\downarrow(\boldsymbol{\omega}) &\equiv \mathrm{spec}^\downarrow\left(\ebw\right)\,.
\end{split}
\end{equation}
The sets $\Sigma(\boldsymbol{\omega})$ and $\Sigma^\downarrow(\boldsymbol{\omega})$ are called spectral polytopes because they only depend on spectral constraints on the 1RDMs $\gh \in \ebw$. We will show below that they are indeed polytopes. 

To characterize the set $\ebw$ it is sufficient to determine the spectral polytope $\Sigma(\bd\omega)$. It is important to note that this finding simplifies our task to determine the domain $\ebw$ of the relaxed $\bd\omega$-ensemble functional $\Fbw(\gh)$ tremendously because we can restrict to a fixed choice of NOs.  
Next, we introduce the set
\begin{equation}\label{eq:Delta}
\Delta = \{\boldsymbol{\lambda}\in \mathbb{R}^d\,|\,N\geq\lambda_1\geq \lambda_2\geq ...\geq \lambda_d\geq 0\}
\end{equation}
of all possible 1RDMs with decreasingly ordered natural occupation numbers to relate the two sets $\Sigma(\boldsymbol{\omega})$ and $\Sigma^\downarrow(\boldsymbol{\omega})$. Then, it follows that
\begin{equation}
\Sigma^\downarrow(\boldsymbol{\omega}) = \Sigma(\boldsymbol{\omega}) \cap \Delta\,.
\end{equation}
Therefore, it is sufficient to determine the natural occupation number (NON) vector $\boldsymbol{\lambda}$ for all $\boldsymbol{\omega}$-minimizers $\hat{\Gamma}_{\boldsymbol{\omega}}$ for all possible choices of $\hat{h}$ in order to obtain the set $\Sigma^\downarrow(\boldsymbol{\omega})$.
In the next step we show that it is indeed sufficient to determine $\Sigma^\downarrow(\boldsymbol{\omega})$ instead of the full spectral polytope $\Sigma(\boldsymbol{\omega})$.

\subsubsection{Generating vertices\label{sec:genvert}}
 
The class of all one-particle Hamiltonians $\hat{h}$ considered in the minimization \eqref{eq:E1Nconv} can be restricted to those with arbitrary, but in the following fixed, eigenbasis $\mathcal{B}_1\equiv \{\ket{i}\}_{i=1}^d$ and $h_1<h_2<...<h_d$. It follows that for a finite number $r$ of non-vanishing weights $\omega_j$ there will be only a finite number $R$ of NON vectors $\bd{v}^{(j)}$ with $j=1, 2, ..., R<\infty$. The entries of the NON vectors $\bd{v}^{(j)}$ are by definition ordered decreasingly due to $h_i<h_{i+1}$ and $\omega_i>\omega_{i+1}$. Thus, the spectral sets $\Sigma(\boldsymbol{\omega})$ and $\Sigma^\downarrow(\boldsymbol{\omega})$ take indeed the form of convex polytopes and $\Sigma(\boldsymbol{\omega})$ follows as the convex hull of all possible permutations of the vector entries of all $\bd{v}^{(j)}$, $j=1, ..., R$:
\begin{equation}\label{eq:Sigmaomega}
\Sigma(\boldsymbol{\omega}) = \mathrm{conv}\Big(\big\{\pi(\bd{v}^{(j)})\,\big\vert\,j=1, ..., R, \pi\in \mathcal{S}^d\big\}\Big)\,.
\end{equation}
Since the spectral polytope $\Sigma(\boldsymbol{\omega})\subset\mathbb{R}^d$ is invariant under permutations, we can from now on restrict to the set $\Sigma^\downarrow(\boldsymbol{\omega})$, which contains all generating vertices $\bd{v}^{(j)}$.

\subsubsection{Partial ordering of configurations}

The next step is to determine in a systematic way the excitation spectrum for different choices of $N, d$ and $r$. 
For a system with $N$ bosons, all of them can occupy the lowest orbital of a fixed eigenbasis $\mathcal{B}_1\equiv \{|i\rangle\}_{i=1}^d$ and $h_1<h_2<...<h_d$. Clearly, the lowest configuration for an arbitrary number of bosons is given by $(1, ..., 1)$ followed by $(1, ..., 1, 2)$. For more than two non-vanishing weights we have to consider all possible excitations while taking the indistinguishability of the bosons into account. To achieve this we first introducing a total ordering on $\mathcal{I}_{N, d}$ through
\begin{equation} \label{eq:orderh}
\bd{i}\leq_{\bd{h}}\bd{j}:\,\, \sum_{k=1}^Nh_{i_k}\leq  \sum_{k=1}^Nh_{j_k}\,,
\end{equation}
where $\bd{h}\equiv\bd h^\uparrow\equiv (h_1, ..., h_d)$, $h_1<h_2<...<h_d$.
Performing this ordering for all configurations in a sequence of length $r$ leads for a fixed $\bd h$ to 
\begin{equation}
\bd{i}_1\leq_{\bd{h}} \bd{i}_2\leq_{\bd{h}} ...\leq_{\bd{h}} \bd{i}_r\,.
\end{equation}
This ordering has to be performed for all possible, and thus in general infinitely many $\bd h$.  However, depending on the number $r$ of non-vanishing weights, many one-particle Hamiltonians lead to the same sequence and thus are considered as equivalent. As a result, we only obtain a finite number $R$ of distinct sequences and to each sequence we assign a vector $\bd{v}$ according to
\begin{equation}\label{eq:NONvector}
\bd{v} = \sum_{j=1}^r\omega_j\bd{n}_{\bd{i}_j}\,.
\end{equation}
This NON vector $v$ contains all the information about the occupancies in the consecutive configurations of the sequence. Further, the $k$-th entry of the occupation number vector $\bd{n}_{\bd{i}}$ of the configuration $\bd{i}$ can take values $0\leq n_{\bd{i}}^{(k)}\leq N$. Thus, the lowest configuration is given by $\bd{n}_{(1, ..., 1)} = (N, 0, ...)$. 

To determine all possible sequences for a given number $r$ of non-vanishing weights, we introduce the following partial ordering on the set $\mathcal{I}_{N, d}$
\begin{equation}\label{eq:partialordering_sequence}
\bd{i}\leq \bd{j}:\,\,\Leftrightarrow\bd{i}\leq_{\bd{h}}\bd{j}\,,\quad \forall\bd{h}\,.
\end{equation}
From the ordering $h_1<h_2<...<h_d$ and Eq.~\eqref{eq:orderh} it follows immediately that $\bd{i}\leq \bd{j}$ is fulfilled if and only if $i_k\leq j_k$ for all $k=1, 2, ..., N$. Moreover, the condition $i_k\leq j_k$ for all $k=1, 2, ..., N$ holds if and only if $\sum_{k=1}^Nh_{i_k}\leq  \sum_{k=1}^Nh_{j_k}$ for all $\bd h$. Thus, the order described by Eq.~\eqref{eq:partialordering_sequence} is determined by the eigenenergies of the one-particle Hamiltonian. The so-called Gale poset \cite{bookGale} in Eq.~\eqref{eq:partialordering_sequence} results into lineups of length $r$. The number of lineups is equal to the number $R$ of generating vertices defined in Eq.~\eqref{eq:NONvector} which, in turn, determine the vertices of the spectral polytope $\Sigma(\bd\omega)$ according to Eq.~\eqref{eq:Sigmaomega}.

Let us now summarize in detail the procedure to construct the sequence
\begin{equation}\label{sequence}
\bd h\mapsto \hat{\Gamma} \mapsto \hat{\gamma}\mapsto \bd v
\end{equation}
needed to obtain the spectral polytope $\Sigma(\bd\omega)$ from the partial ordering in Eq.~\eqref{eq:partialordering_sequence} in a systematic way. Every vector $\bd h$ determines a lineup 
\begin{equation}
\bd i_1\to \bd i_2 \to \ldots\to \bd i_r
\end{equation}
of configurations. It is important to notice that, even if there exist uncountably many $\bd h$, we only obtain a finite number $R$ of lineups since many $\bd h$ are equivalent concerning the linear order introduced in Eq.~\eqref{eq:orderh}. For every lineup, the $N$-boson density operator $\hat{\Gamma}$ follows as
\begin{equation}
\hat{\Gamma} = \sum_{j=1}^r w_j\ket{\bd i_j}\!\bra{\bd i_j}\,,
\end{equation}
and tracing out $N-1$ particles yields the corresponding $\hat{\gamma}$. Since the occupation number vectors $\bd n_{\bd i}$ introduced in Eq.~\eqref{eq:NONvector} are nothing else than the spectrum of $\hat{\gamma}$, we eventually obtain
\begin{equation}
\bd  v = \sum_{j=1}^r w_j\mathrm{spec}\left(N\Tr_{N-1}\left[\ket{\bd i_j}\!\bra{\bd i_j}\right]\right)\,.
\end{equation} 

\begin{figure}[htb]
\centering
\includegraphics[width=0.5\linewidth]{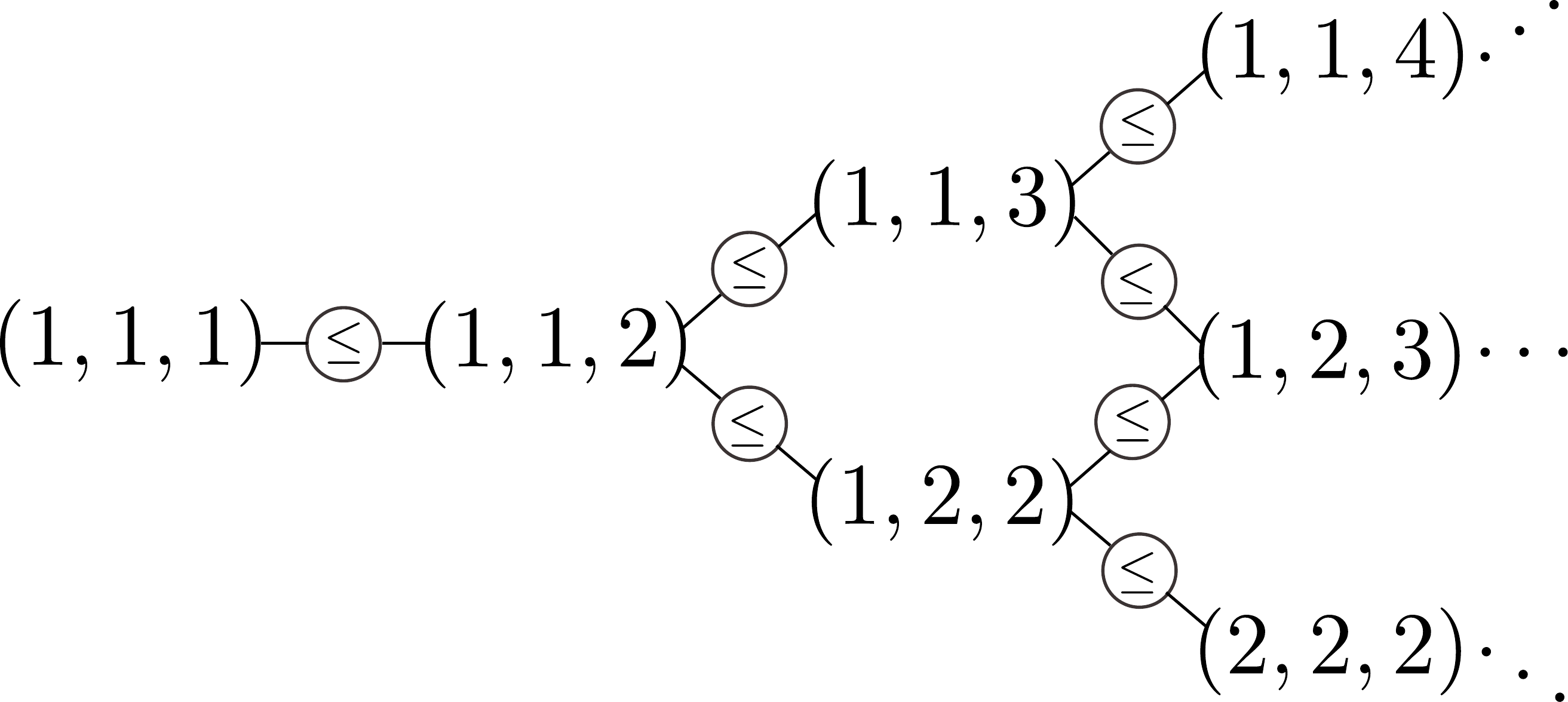}
\caption{Illustration of the excitation spectrum and linear ordering for $N=3$ bosons.\label{fig:specN3}}
\end{figure}

To illustrate the partial ordering in Eq.~\eqref{eq:partialordering_sequence}, we show in Fig.~\ref{fig:specN3} the excitation spectrum for $N=3$ bosons. 
The first configuration corresponds to the ground state where all three bosons occupy the same state. The corresponding lineup consists of one configuration, namely $(1,1,1)$. Also the excitation pattern for only one excitation in the system is unique and is represented by the lineup $(1,1,1)\to(1,1,2)$. Thus, for $r=1$ and $r=2$ we obtain in each case one NON vector according to Eq.~\eqref{eq:NONvector}. For a larger number of excitations, it depends on the values of the eigenenergies $h_i$ which configuration follows. For $r=3$, this consideration results into two possible lineups
\begin{align}
&(1,1,1)\to(1,1,2)\to (1,1,3)\,,\\\
&(1,1,1)\to(1,1,2)\to (1,2,2)\,.
\end{align}
This also implies that in every excitation pattern with $r\geq 4$, both configurations $(1,1,3)$ and $(1,2,2)$ must appear before $(1,2,3)$, but the exact sequence depends on the particular choice of $\bd{h}$. We discuss all those examples in more detail in Sec.~\ref{sec:Examples_wRDMFT}.

It is worth noticing, that the corresponding excitation spectrum for fermions shown in Ref.~\cite{LCLS21} can be obtained from the bosonic excitation spectrum in Fig.~\ref{fig:specN3} by adding the tuple $(0,1,2)$ to each bosonic configuration. For example, the lowest configuration for fermions is given by $(1,2,3)$ followed by $(1,2,4)$. 

\subsection{Hyperplane representation of spectral polytopes\label{sec:h-repr_ebw}}

In the above section, we succeeded in determining all generating vertices $\bd{v}^{(j)}$, $j=1, ..., R$, and thus the spectral set $\Sigma(\boldsymbol{\omega})$ through Eq.~\eqref{eq:Sigmaomega} leading to $\ebw$. However, in practice, this vertex representation of the spectral polytope is not very convenient since it is extremely complicated to check whether a NON vector $\boldsymbol{\lambda}$ lies inside the polytope or not. Therefore, the vertex description is of little use in practice. In contrast, the hyperplane representation of the permutation-invariant polytope $\Sigma(\bd\omega)$ provides a more convenient way to test a given NON vector because one has to check only finitely many linear hyperplane conditions $D_k(\boldsymbol{\lambda})\geq 0$. Thus, in the next step, we need to turn the vertex representation of the spectral polytope $\Sigma(\bd\omega)$ into a hyperplane representation.

The procedure for $r=1, 2$ is trivial because in each case there is only one lineup, as discussed in the section above. Consequently, we also have only one generating vertex $\bd{v}$. The spectral polytope $\Sigma(\bd\omega)$ is then obtained by considering all possible permutations of the entries of $\bd v$. Note that even for $R=1$, the total number of vertices in $\Sigma(\bd\omega)$ can become arbitrarily large because it depends on the dimension $d$ of the one-boson Hilbert space. As a result, already $r=1,2$ demonstrates clearly that the vertex representation is not efficient to test whether a given occupation number vector $\bd\lambda$ lies inside the permutohedron $\Sigma(\bd\omega)$ or not. However, for maximally two non-vanishing weights, we can directly apply Rado's theorem \cite{R52} explained in Sec.~\ref{sec:Rado} to obtain
\begin{equation}
\Sigma(\boldsymbol{\omega})= \{\boldsymbol{\lambda}\in\mathbb{R}^d\,|\,\boldsymbol{\lambda}\prec \bd{v}\}\,.
\end{equation}
The spectral polytope $\Sigma(\boldsymbol{\omega})$ is then determined by the $d$ inequalities following from the majorization condition $\boldsymbol{\lambda}\prec\bd{v}$. 
For more than one lineup, the spectral polytope $\Sigma(\bd\omega)$ defined in Eq.~\eqref{eq:Sigmaomega} is not a permutohedron anymore, and thus Rado's theorem does not apply. However, $\Sigma(\bd\omega)$ is still permutation-invariant. To translate its vertex into a hyperplane representation, we thus require a modification of Rado's theorem adjusted to the new situation. Such a crucial generalization of Rado's theorem, introduced and proven in Ref.~\cite{LCLS21}, states that for finitely many vectors $\bd v^{(1)}, ..., \bd v^{(R)}\in \mathbb{R}^d$, the polytope 
\begin{equation}
\mathcal{P} = \mathrm{conv}\left(\left\{\pi(\bd v^{(j)})\,\Big\vert\,j=1, ..., R, \pi\in \mathcal{S}^d\right\}\right)
\end{equation}
is equivalent to
\begin{equation} \label{eq:genRado}
\mathcal{P} = \Big\{\boldsymbol{\lambda}\,\Big\vert\, \exists\,\,\text{conv. comb. }\sum_{j=1}^Rp_j\bd{v}^{(j)}\equiv \bd{v}: \boldsymbol{\lambda}\prec \bd{v}\Big\}\,.
\end{equation}
In Sec.~\ref{subsec:N3r3}, we show for the concrete example of $N=3$ and $r=3$ how the generalized Rado theorem in Eq.~\eqref{eq:genRado} can be used to determine a minimal hyperplane description up to scaling. The minimal hyperplane representations for $r>3$ require a mathematically more complex derivation presented in Ref.~\cite{CLLS21}, introducing a  mathematically rigorous derivation of the unique minimal hyperplane representation for arbitrary values of $N$ and $d$, based on the concept of the normal fan of a polytope and its extremal rays. However, $r\leq 2$ already incorporates the ground state energy and its gap, and large values of $r$ are often less interesting from the physical point of view. 

Furthermore, the system stabilizes in the bosonic case for $N\geq r-1$ and $d\geq r$. This means, that for sufficiently large $N$ and $d$, increasing the particle number $N$ or dimension $d$ does not result into new inequalities \cite{CLLS21}. We comment on the relation between the spectral potytopes for different settings $(N,d)$ in more detail in Sec.~\ref{sec:genNd}, and several examples are also provided in Sec.~\ref{sec:Examples_wRDMFT}.

Remarkably, increasing $r$ leads to a hierarchy of inequalities \cite{CLLS21} which means that for increasing $r$, all inequalities obtained for a smaller $r$ are still facet-defining and only finitely many additional inequalities occur. It is worth noticing that these additional constraints are entirely independent of the interaction itself and arise directly from the geometry of the underlying set of density matrices. This shall be contrasted with the quantum depletion due to interactions in Ch.~\ref{ch:RDMFT_BEC}. Thus, these additional constraints are interpreted as a generalized Pauli exclusion principle for bosons. The number of generating vertices $\bd{v}^{(j)}$ and the corresponding number of inequalities for $r=1, ..., 12$ \cite{CLLS21} are presented in Tab.~\ref{tab:Noineq}. 

\begin{table}[htb]
\centering
\begin{tabular}{ | c | c | c | c | c |c |c |c |c |c |c |c |c |c |}
\hline
$r$ & $1$  & $2$ &$3$ & $4$ & $5$&$6$ &$7$ &$8$ &$9$ &$10$&$11$ &$12$ \\
\hline
$\# \bd{v}^{(j)}$ & $1$  & $1$ &$2$ & $4$ & $8$&$17$ &$37$ &$82$ &$184$ &$418$&$967$ &$2278$ \\
\hline
$\# \text{ineq}$ & $1$  & $2$ &$3$ & $5$ & $8$&$13$ &$22$ &$36$ &$59$ &$99$&$171$ &$299$ \\
\hline
\end{tabular}
\caption{Number of generating vertices $\bd{v}^{(j)}$ and number of facet-defining inequalities of $\Sigma^\downarrow(\bd\omega)$ for $r\leq 12$.} 
\label{tab:Noineq}
\end{table}

\subsection{Generalization to larger particle numbers \label{sec:genNd}}

In this section, we derive a concrete, general relation between the spectral polytopes for different particle numbers. This will further explain why no new inequalities appear in the minimal hyperplane representation for increasing particle number $N$ (and $d$). We keep $d$ fixed from now on and only vary $N$. Moreover, we restrict our discussion to large enough $N$ and $d$ such that the number of configurations, and thus the number of lineups, are independent of $N$ and $d$. Since in the lowest configuration all bosons occupy the same orbital, the highest occupied orbital can be $i_N=r$ for a fixed number of non-vanishing weights $r$ (c.f. Fig.~\ref{fig:specN3}). Thus, for the lineups to be independent of $N$ and $d$, they must fulfil $N\geq r-1$ and $d\geq r$.

For $N\geq r-1$, increasing the particle number to $N^\prime > N$ amounts to placing more bosons in the lowest orbital. Thus, it does not change the number of lineups, and for each of them it only changes the first entry of the corresponding NON vector $\bd{v}^{(i)}$ which depends explicitly on $N$. This also implies that for fixed $d\geq r$ the generating vertices $\bd{v}^{(i)}$ for $N$ and $N^\prime > N$ are related through
\begin{equation} \label{eq:vNNprime}
\bd{v}^{(i)}_N = \bd{v}^{(i)}_{N^\prime}-\delta\bd{e}_1\,,
\end{equation}
where $\delta=N^\prime-N$. Thus, generalizing the vertex representation of the spectral polytope $\Sigma(\bd{\omega})$ from $N$ to $N^\prime$ is trivial. A higher dimensional setting with $d^\prime > d$ can be obtained from $\Sigma_N(\bd{\omega})$ for N particles by first extending the $d$-dimensional vector $\bd{v}^{(i)}_N$ to a $d^\prime$-dimensional vector by adding zero entries and only afterwards using the relation in Eq.~\eqref{eq:vNNprime}.

Recall that according to Eq.~\eqref{eq:Sigmaomega}, we obtain the spectral polytopes $\Sigma(\bd\omega)$ by considering all possible permutations of the entries of all generating vertices. The key result, which we will prove in this section, is that the spectral polytope $\Sigma_{N^\prime}(\bd\omega)$ with $\delta= N^\prime-N>0$ is the Minkowski sum of $\Sigma_N(\bd\omega)$ and a permutation-invariant rescaled simplex
\begin{equation}\label{eq:SigmaNNprimeC}
\Sigma_{N^\prime}(\bd\omega) \equiv \Sigma_{N}(\bd\omega) + \mathcal{C}\,,
\end{equation}
where $\mathcal{C}$ is a rescaled simplex given by
\begin{equation}
\mathcal{C} \equiv \mathrm{conv}(\{\pi(\delta\bd{e}_1)\,|\,\pi\in \mathcal{S}^d\})\,.
\end{equation}
To prove Eq.~\eqref{eq:SigmaNNprimeC}, we start by deriving the relation between the polytope $\Sigma_{N}(\bd\omega)$ and $\Sigma_{N^\prime}(\bd\omega)$ for only one generating vertex $\bd{v}$ which corresponds to $N$ bosons. Due to Eq.~\eqref{eq:vNNprime}, we also have only one generating vertex $\bd{v} + \delta\bd{e}_1$ for $N^\prime$ bosons. According to a theorem by Hardy, Littlewood and P\'olya \cite{HLP53} presented in Sec.~\ref{sec:majorization}, we have $\Sigma_{N^\prime}(\bd\omega)\ni\bd{\mu}\prec \bd{v} + \delta\bd{e}_1$ if and only if $\bd{\mu} = P(\bd{v} + \delta\bd{e}_1)$ for some double stochastic matrix $P\in \mathbb{R}^{d\times d}$. Combining this statement with Birkhoff's theorem (see Sec.~\ref{sec:majorization}) leads to
\begin{equation}
\Sigma_{N^\prime}(\bd\omega)\ni \bd{\mu} = \sum_{i=1}^{d!}p_i\left(\pi_i(\bd{v} + \delta\bd{e}_1)\right) =  \sum_{i=1}^{d!}p_i(\pi_i\bd{v}) +   \delta \sum_{i=1}^{d!}p_i(\pi_i\bd{e_1})\,,
\end{equation}
where $\pi_1, ..., \pi_{d!}$ are permutation matrices, $p_i\geq 0$ and $ \sum_{i=1}^{d!}p_i=1$. It follows that $\bd{\mu} \in \Sigma_N(\bd\omega) + \mathrm{conv}(\{\pi(\delta\bd{e}_1)|\pi\in \mathcal{S}^d\}$ where the set 
\begin{equation}\label{eq:Csimplex}
\mathcal{C} \equiv \mathrm{conv}(\{\pi(\delta\bd{e}_1)\,|\,\pi\in \mathcal{S}^d\})
\end{equation}
is a rescaled simplex with edge length $\delta$. Then, $\Sigma_{N^\prime}(\bd\omega) = \Sigma_{N}(\bd\omega) + \mathcal{C} = \{\bd{\lambda} + \bd{c}\,|\,\bd{\lambda}\in\Sigma_N(\bd\omega), \bd{c}\in \mathcal{C}\}$ is the Minkowski sum of two permutation invariant polytopes. Clearly, the sum of two convex sets is also convex. Also, every $\bd{\mu}\in \Sigma_{N^\prime}(\bd\omega)$ is correctly normalized since all $\bd{\lambda}\in \Sigma_N(\bd\omega)$ are normalized to $N$ and all $\bd{c}\in \mathcal{C}$ are normalized to $\delta = N^\prime-N$. 

Next, we consider the more general case of more than one generating vertex, i.e.~$R>1$. Using the generalized Rado theorem in Eq.~\eqref{eq:genRado}, we show that vector $\bd{\mu}\in \Sigma_{N^\prime}(\bd\omega)$ can be related to a vector $\bd{\lambda}\in \Sigma_{N}(\bd\omega)$, where  
\begin{equation}\label{eq:SigmagenRado}
\begin{split}
\Sigma_{N}(\bd\omega) &= \Big\{\boldsymbol{\lambda}\,\Big\vert\, \exists\,\,\text{conv. comb. }\sum_{j=1}^Rp_j\bd{v}^{(j)}\equiv \bd{v}: \boldsymbol{\lambda}\prec \bd{v}\Big\}\,,\\\
\Sigma_{N^\prime}(\bd\omega) &= \Big\{\boldsymbol{\mu}\,\Big\vert\, \exists\,\,\text{conv. comb. }\sum_{j=1}^Rp_j(\bd{v}^{(j)} + \delta \bd{e}_1)\equiv \bd{u} : \boldsymbol{\mu}\prec \bd{u} \Big\}\,.
\end{split}
\end{equation}
From Eq.~\eqref{eq:genRado} follows that every $\bd{\mu}\in \Sigma_{N^\prime}(\bd\omega)$ can be written as a convex combination
\begin{equation}
\begin{split}
\Sigma_{N^\prime}(\bd\omega)\ni\bd{\mu} &= \sum_{i=1}^R\sum_{\pi\in\mathcal{S}^d}q_{i, \pi}\pi(\bd{v}^{(i)} + \delta\bd{e}_1) =\sum_{i=1}^R\sum_{\pi\in\mathcal{S}^d}q_{i, \pi}\left(\pi(\bd{v}^{(i)}) +\pi(\delta\bd{e}_1)\right)\\\
&\equiv \sum_{i=1}^Rp_i\bd{\lambda}_i + \delta\sum_{i=1}^R\sum_{\pi\in\mathcal{S}^d}q_{i, \pi}\pi(\bd{e}_1)\\\
&= \bd{\lambda} + \delta\sum_{i=1}^R\sum_{\pi\in\mathcal{S}^d}q_{i, \pi}\pi(\bd{e}_1)\,,
\end{split}
\end{equation}
where $\bd{\lambda}\in \Sigma_{N}(\bd\omega)$, and we defined $p_i\equiv \sum_{\pi\in\mathcal{S}^d}q_{i, \pi}$ and $\bd{\lambda}_i \equiv \sum_{\pi\in\mathcal{S}^d}q_{i, \pi}\pi(\bd{v}^{(i)})/p_i$. Then, we eventually obtain $\Sigma_{N^\prime}(\bd\omega) \equiv \Sigma_{N}(\bd\omega) + \mathcal{C}$, where $\mathcal{C}$ is the rescaled simplex given by Eq.~\eqref{eq:Csimplex} with edge length $\delta$. Hence, the relation between the spectral polytopes belonging to different total particle numbers stated in \eqref{eq:SigmaNNprimeC} holds for any $R>1$. Note that this is in agreement with the observation that every extremal point in $\Sigma_{N^\prime}(\bd\omega)$ can be obtained from adding a vertex in $\Sigma_N(\bd\omega)$ and a vertex in $\mathcal{C}$. 

Eq.~\eqref{eq:SigmaNNprimeC} can now be used to explain how the hyperplane representation for $N^\prime>N$ follows from the inequalities for the smaller setting $N$. Combining $\bd{\mu}\prec \bd{\mu}^\downarrow$ with
\begin{equation}\label{eq:mumajorizationcond}
\begin{split}
&\sum_{j=1}^k\mu_j^\downarrow = \sum_{j=1}^k(\lambda_j + c_j)^\downarrow \leq \sum_{j=1}^k\lambda_j^\downarrow + c_j^\downarrow\,,\quad 1\leq k\leq d-1\,,\\\
&\sum_{j=1}^d\mu_j^\downarrow = \sum_{j=1}^d\lambda_j^\downarrow + c_j^\downarrow
\end{split}
\end{equation}
leads to the majorization condition $\bd{\mu} \prec \bd{\lambda}^\downarrow + \bd{c}^\downarrow$. Since in addition $0\leq c_j\leq \delta$ for all $j=1, ..., d$, we have 
\begin{equation}\label{eq:mumajorization}
\bd{\mu}\prec \bd{\lambda}^\downarrow + \delta\bd{e}_1
\end{equation} 
and it follows immediately that 
\begin{equation}\label{eq:lamda1geq}
\lambda_1^\downarrow\geq \mu_1^\downarrow  -\delta\,.
\end{equation}
The above equation can then be used together with Eq.~\eqref{eq:mumajorization} for the other vector entries to replace $\bd{\lambda}^\downarrow$ in the linear constraints $D_k(\bd{\lambda}^\downarrow)\geq 0$ by $\bd{\mu}^\downarrow-\delta\bd{e}_1$ to obtain the set of linear constraints for the larger setting with $N^\prime$ bosons. 
Since increasing the particle number does not lead to new generating vertices and those are related for different particles numbers through Eq.~\eqref{eq:vNNprime}, also no additional inequalities appear as already explained in Sec.~\ref{sec:h-repr_ebw}. For a mathematically rigorous approch including the extension of the one-particle Hilbert space to larger dimensions, we refer the reader to Ref.~\cite{CLLS21}.
 
Next, we present a few examples illustrating how the minimal hyperplane representation for a larger particle number $N^\prime$ can be obtained from the inequalities for a smaller particle number $N$ ($d$ is still fixed). First, we use the generalized Rado theorem \eqref{eq:genRado} together with Eq.~\eqref{eq:vNNprime} to show that a vector $\bd\mu\in\Sigma_{N^\prime}(\bd\omega)$ (see Eq.~\eqref{eq:SigmagenRado}) satisfies $\bd\mu\prec \bd u + \delta\bd e_1$ for a convex combination $\bd u=\sum_{i=1}^Rp_i\bd v_N^{(i)}$. Hence, it follows for the ordered NON vector $\bd{\mu}^\downarrow \in \Sigma^\downarrow_N(\bd\omega)$ that
\begin{equation}
\tilde{\lambda}_1 \equiv \mu_1^\downarrow-\delta \leq  u_1^\downarrow\,.
\end{equation}
Thus, the new vector $\tilde{\bd{\lambda}}$ with the entries $\tilde{\lambda}_1 \equiv \mu_1-\delta$ and $\tilde{\lambda}_i = \mu_i$, $1<i\leq d$ is normalized to $N$, but its entries are not necessarily ordered decreasingly anymore. To illustrate that this statement is in agreement with Eq.~\eqref{eq:lamda1geq}, we start by discussing the two simplest examples $R=1,2$.
Recall that it is sufficient to restrict to the ordered case $\lambda_1\geq \lambda_2\geq...\geq\lambda_d$ since the polytopes are permutation invariant.  
For $r=1$ and the example of $N=3$ and $N^\prime = 5$ we get the following two conditions for the ordered NON vectors $\bd{\lambda}^\downarrow$ ($N=3$) and $\bd{\mu}^\downarrow$ ($N=5$)
\begin{equation}
\begin{split}
\lambda_1^\downarrow&\leq 3\,, \quad  \mu_1^\downarrow\leq 5
\end{split}
\end{equation}
which are equivalent to the linear hyperplane conditions
\begin{equation}
D(\bd{\lambda}^\downarrow)\equiv 3 - \lambda_1^\downarrow\geq 0 \,,\quad D(\bd{\mu}^\downarrow)\equiv 5 - \mu_1^\downarrow\geq 0\,.
\end{equation}
Thus, $D(\bd{\mu}^\downarrow)$ can be obtained from $D(\bd{\lambda}^\downarrow)$ by replacing $\bd{\lambda}^\downarrow\to \bd{\mu}^\downarrow - \delta\bd{e}_1$. This statement holds obviously also for $r=2$ where there is also only one possible sequence of configurations. Next, we consider the case of multiple vectors $\bd{v}^{(i)}, \,\, i=1, ..., R$, and denote again by $\bd{\lambda}$ the NON vector in the polytope of $N$ particles. The NON vector $\bd{\mu}$ corresponds to a larger particle number $N^\prime>N$. Tab.~\ref{tab:r2NNprime} compares the linear inequalities obtained from the majorization condition in the generalized Rado theorem for $r=3$, where we have two generating vertices, $R=2$, by using the relation of the vectors $\bd{v}^{(i)}$ for different total particle numbers in Eq.~\eqref{eq:vNNprime}.
\begin{table}[t]
\resizebox{\textwidth}{!}{
\centering
\begin{tabular}{ c | c }
\makecell{$N$}  & \makecell{$N^\prime > N$} \\
\hline
\makecell{$\lambda_1^\downarrow \leq qv_1^{(1)} + (1-q)v_1^{(2)}$} & \makecell{$\mu_1^\downarrow\leq qv_1^{(1)} + (1-q)v_1^{(2)} + \delta$   }\\
\makecell{$\lambda_1^\downarrow  + \lambda_2^\downarrow \leq q(v_1^{(1)} + v_2^{(1)}) + (1-q)(v_1^{(2)} + v_2^{(2)})$}  & \makecell{$\mu_1^\downarrow  + \mu_2 ^\downarrow\leq q(v_1^{(1)} + v_2^{(1)}) + (1-q)(v_1^{(2)} + v_2^{(2)}) + \delta$}   \\
\makecell{$\sum_{i=1}^{d-1}\lambda_i^\downarrow\leq N$} & \makecell{$\sum_{i=1}^{d-1}\mu_i^\downarrow\leq N^\prime$} \\
\end{tabular}
}
\caption{Comparison of the linear inequalities obtained from the majorization condition in the generalized Rado theorem for $r=3$ and two different boson numbers $N$ and $N^\prime$.\label{tab:r2NNprime}} 
\end{table}
In accordance with Tab.~\ref{tab:Noineq}, we obtain three inequalities for the minimal hyperplane representation of $\Sigma^\downarrow(\bd\omega)$. We then compare the linear conditions, 
\begin{equation}
\begin{split}\label{eq:DlambdaR2}
D_1(\bd{\lambda}^\downarrow) &\equiv qv_1^{(1)} + (1-q)v_1^{(2)} - \lambda_1^\downarrow \geq 0\,,\\\
D_2(\bd{\lambda}^\downarrow) &\equiv q\left(v_1^{(1)} + v_2^{(1)}\right) + (1-q)\left(v_1^{(2)}  + v_2^{(2)}\right)- \lambda_1^\downarrow - \lambda_2^\downarrow\geq 0\,,\\\
D_r(\bd{\lambda}^\downarrow) &\equiv N - \sum_{i=1}^{d-1}\lambda_i^\downarrow\geq 0\,,
\end{split}
\end{equation}
corresponding to the smaller particle number $N$ in the left column of Tab.~\ref{tab:r2NNprime}, to the linear hyperplane conditions 
\begin{equation}\label{eq:DmuR2}
\begin{split}
D_1(\bd{\mu}^\downarrow) &\equiv qv_1^{(1)} + (1-q)v_1^{(2)} + \delta - \mu_1^\downarrow\geq 0\,,\\\
D_2(\bd{\mu}^\downarrow) &\equiv q\left(v_1^{(1)} + v_2^{(1)}\right) + (1-q)\left(v_1^{(2)}  + v_2^{(2)}\right) + \delta - \mu_1^\downarrow - \mu_2^\downarrow\geq 0\,,\\\
D_r(\bd{\mu}^\downarrow) &\equiv N^\prime - \sum_{i=1}^{d-1}\mu_i^\downarrow\geq 0\
\end{split}
\end{equation}
for $N^\prime$ bosons in the right column. Then, \eqref{eq:DmuR2} can be obtained from Eq.~\eqref{eq:DlambdaR2} by replacing $\lambda_1^\downarrow$ by $\mu_1^\downarrow - \delta$. It can be easily shown that the same observation holds for arbitrary $R$ using $\sum_{i=1}^Rq_i=1$ and the generalized Rado theorem in Eq.~\eqref{eq:genRado}. To obtain the minimal hyperplane representation one needs to eliminate the $q_j$'s arising from the convex combinations of generating vertices which is only feasible for $r\leq 3$ as shown in Sec.~\ref{subsec:N3r3}. However, the purpose of these examples for small $R$ is to illustrate Eq.~\eqref{eq:lamda1geq} and to show how the inequalities for a larger particle number can be obtained from the inequalities for a smaller particle number. A mathematically more elaborate treatment is provided in Ref.~\cite{CLLS21}, where the facet-defining inequalities are derived for arbitrary $N$ and $d$ right from the beginning.  

\section{Examples and illustrations\label{sec:Examples_wRDMFT}}

In this section, we discuss the description of the spectral set $\Sigma(\bd \omega)$ proposed in Sec.~\ref{sec:setE1N} for several small settings. Moreover, for $r\leq 3$, we illustrate how the redundant inequalities in the hyperplane representation are identified and how the facet-defining inequalities can be generalized to arbitrary particle number $N$ and dimension $d$ of the one-particle Hilbert space. 

\subsection{Illustration of the spectral polytopes for two particles and three non-vanishing weights}

The main purpose of our first example is to illustrate the concepts of lineups and partial ordering introduced in Sec.~\ref{sec:setE1N} for a small system size which does not satisfy the stabilization conditions $N\geq r-1$ and $d\geq r$. Therefore, the number of inequalities does not necessarily match Tab.~\ref{tab:Noineq}.

\begin{figure}[h]
\centering
\includegraphics[width=0.5\linewidth]{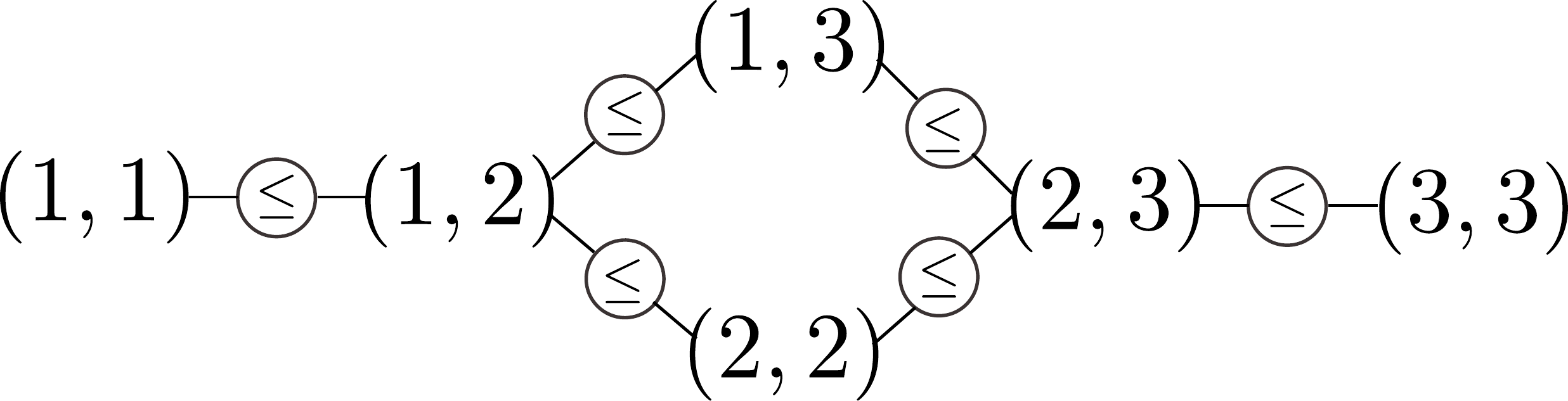}
\caption{Excitation spectrum with partial ordering for $d=3$ and $N=2$ bosons.}\label{fig:specN2d3}
\end{figure}

\begin{figure}[h]
\centering
\begin{subfigure}[t]{.49\textwidth}
\centering
\includegraphics[width=\linewidth]{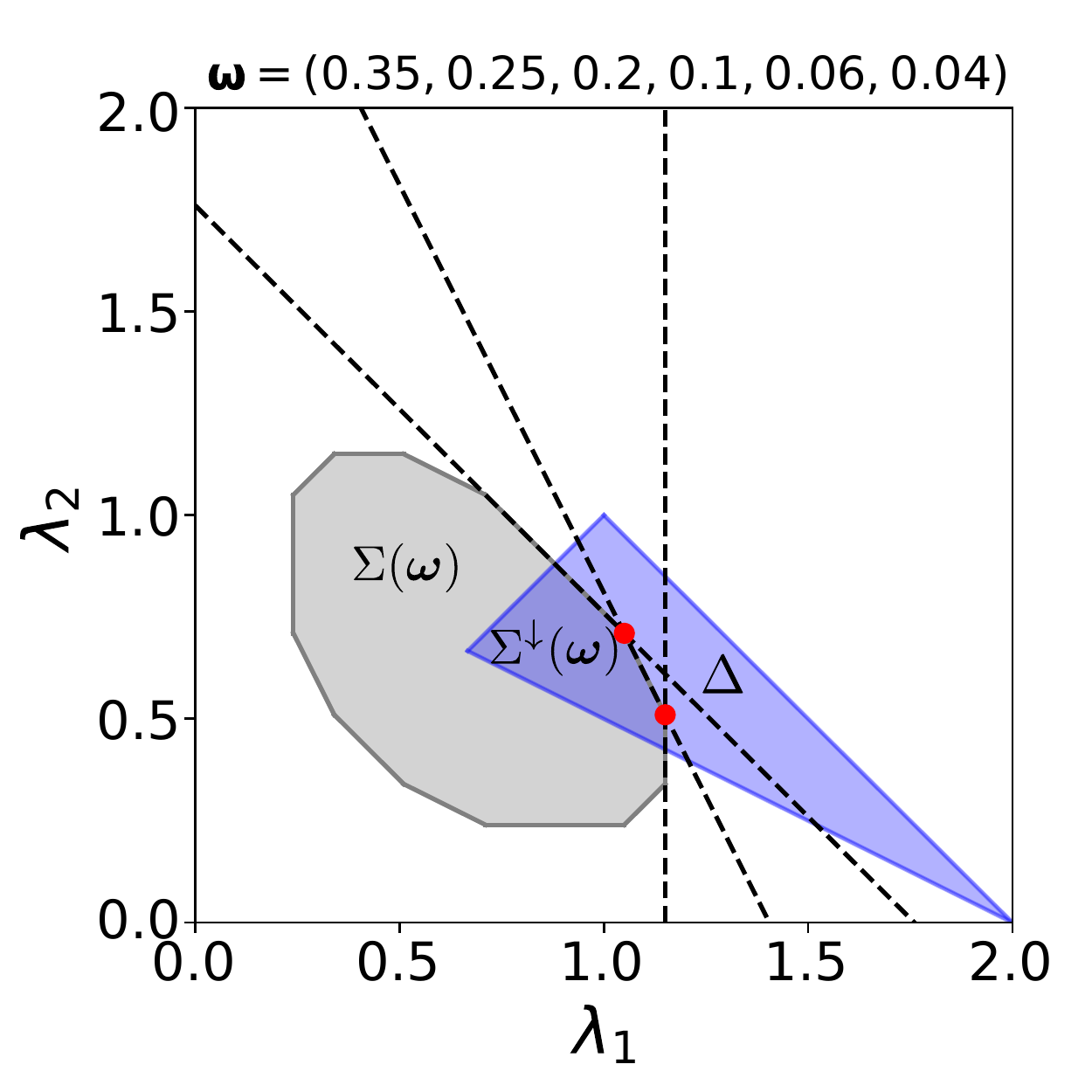}
\end{subfigure}
\begin{subfigure}[t]{.49\textwidth}
\centering
\includegraphics[width=\linewidth]{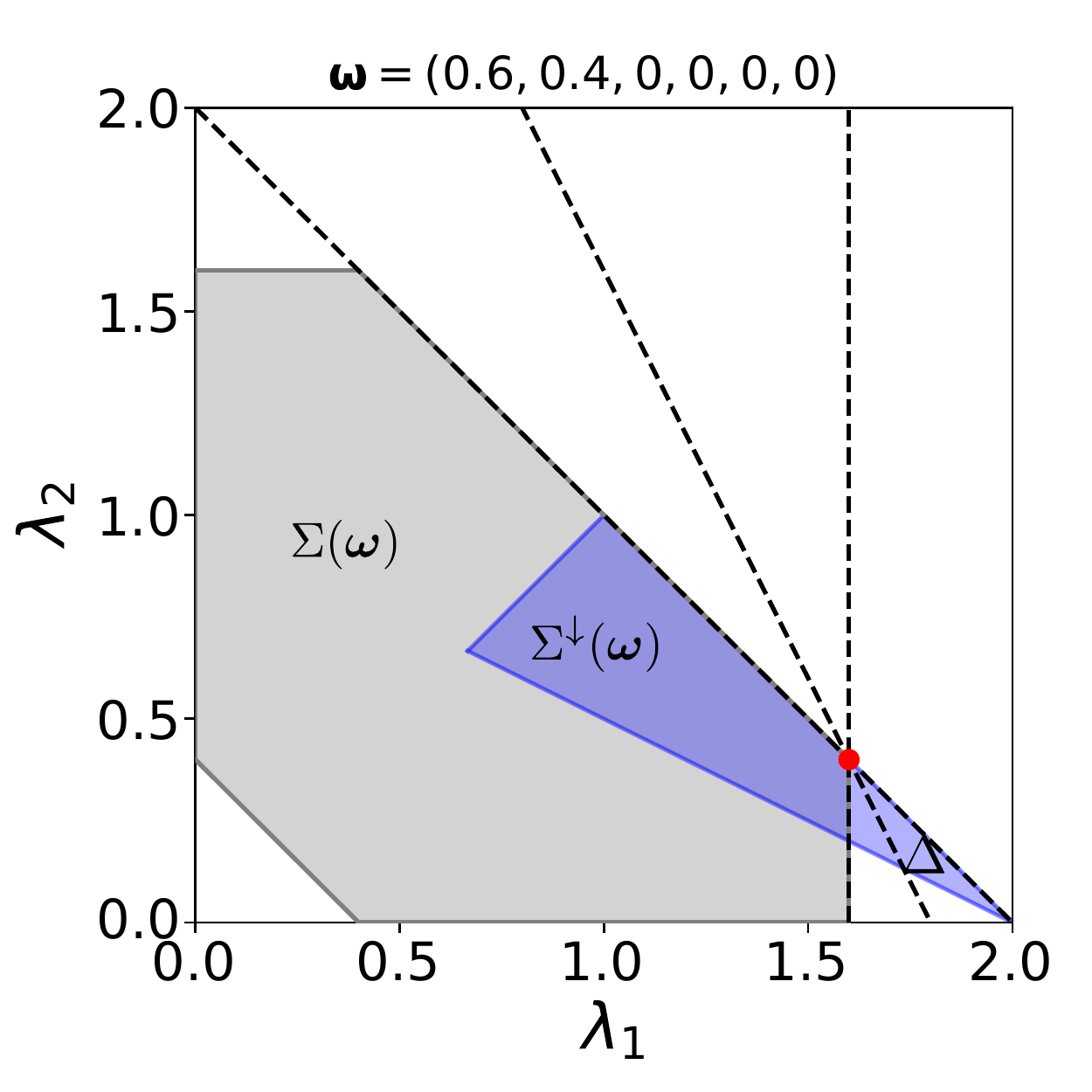}
\end{subfigure}
\medskip
\begin{minipage}[t]{\textwidth}
\caption{Illustration of the spectral polytopes $\Sigma(\boldsymbol{\omega})$ and $\Sigma^\downarrow(\boldsymbol{\omega}) = \Sigma(\boldsymbol{\omega})\cap \Delta$ for  $d=3$ and $N=2$ bosons for two different choices of $\boldsymbol{\omega}$, and thus $r$.\label{fig:specpol}}
\end{minipage}
\end{figure}

The excitation spectrum with partial ordering for $N=3$ bosons and $d=\mathrm{dim}(\mathcal{H}_1)=2$ is shown in Fig.~\ref{fig:specN2d3}. Since both configurations, $(1,3)$ and $(2,2)$, are lower than $(2,3)$ according to Eq.~\eqref{eq:partialordering_sequence}, we obtain the following two possible sequences of length $6$:
\begin{equation}
\begin{split}
&(1, 1)\to (1,2)\to(1,3)\to(2,2)\to(2,3)\to(3,3)\,,\\
&(1,1)\to(1,2)\to(2,2)\to(1,3)\to(2,3)\to (3,3)\,.
\end{split}
\end{equation}
For $r\geq 3$ non-vanishing weights $\omega_j$, the two sequences are distinct, leading to two NON vectors $\bd{v}^{(1)}$ and $\bd{v}^{(2)}$, whereas for $r\leq 2$ there is only one such NON vector. For the maximal number of non-vanishing weights, $r=6$, they are given by
\begin{equation}
\begin{split}
\bd{v}^{(1)} &= (2\omega_1 + \omega_2 + \omega_3, \omega_2 + 2\omega_4 + \omega_5, \omega_3 + \omega_5 + 2\omega_6)\,,\\\
\bd{v}^{(2)} &= (2\omega_1 + \omega_2 + \omega_4, \omega_2 + 2\omega_3 + \omega_5, \omega_4 + \omega_5 + 2\omega_6)\,.
\end{split}
\end{equation}
Fig.~\ref{fig:specpol} shows the spectral polytopes $\Sigma(\boldsymbol{\omega})$ and $\Sigma^\downarrow(\boldsymbol{\omega}) = \Sigma(\boldsymbol{\omega})\cap \Delta$ for $r=6$ non-vanishing weights on the left and $r=2$ on the right hand side. Using the normalization of $\lambda_1 + \lambda_2 + \lambda_3$ we can omit the third dimension and substitute $\lambda_3 = 2-\lambda_1-\lambda_2$ to obtain the simplex $\Delta$ defined in Eq.~\eqref{eq:Delta}. The two NON vectors $\bd{v}^{(1)}$ and $\bd{v}^{(2)}$ for $r=6$ are marked by two red dots in the left panel of Fig.~\ref{fig:specpol}. According to Eq.~\eqref{eq:Sigmaomega}, the extremal points of the spectral polytope $\Sigma(\omega)$ follow from all possible permutations of the vector entries of $\bd{v}^{(1)}$ and $\bd{v}^{(2)}$. In contrast, for $r=2$ there is only one NON vector $\bd{v}$ whose permutations lead to the full spectral polytope $\Sigma(\omega)$ since we have just a single corresponding sequence in Fig.~\ref{fig:specN2d3}. 

The facet-defining inequalities for $r=6$ are depicted by the three dashed lines in the left subfigure of Fig.~\ref{fig:specpol}. To derive them, we first observe that, according to the generalization of Rado's theorem in Eq.~\eqref{eq:genRado}, we need to determine the set of all NON vectors $\bd{\lambda}$ for which there exists at least one convex combination $\bd{u} = q\bd{v}^{(1)} + (1-q)\bd{v}^{(2)}$ such that $\bd{\lambda}\prec \bd{u}$. This leads to the following set of constraints on the NONs $\lambda_j^\downarrow$,
\begin{equation}
\begin{split}
\lambda_1^\downarrow &\leq 2\omega_1 + \omega_2 + \omega_4 + q(\omega_3-\omega_4)\,,\\\
\lambda_1^\downarrow + \lambda_2^\downarrow &\leq 2(\omega_1 + \omega_2 + \omega_3)+\omega_4 + \omega_5 - q(\omega_3-\omega_4)\,,\\\
\lambda_1^\downarrow + \lambda_2^\downarrow + \lambda_3^\downarrow &\leq 2\,,
\end{split}
\end{equation}
where the last inequality is automatically fulfilled by the normalization. Since $q\in [0, 1]$, the upper bound on the largest NON $\lambda_1^\downarrow$ can vary between $2\omega_1 + \omega_2 + \omega_4$  and $2\omega_1 + \omega_2 + \omega_3$ where the latter bound must always be satisfied and the value of $\lambda_1^\downarrow + \lambda_2^\downarrow$ can never exceed $2(\omega_1 + \omega_2 + \omega_3)+\omega_4 + \omega_5$. Combining these bounds with the inequality arising from the elimination of the parameter $q$ leads to
\begin{equation}
\begin{split}\label{eq:minhN2d3}
\lambda_1^\downarrow &\leq 2\omega_1 + \omega_2 + \omega_3\,,\\\
\lambda_1^\downarrow + \lambda_2^\downarrow&\leq 2\omega_1 + 2\omega_2 + 2\omega_3 + \omega_4 + \omega_5\,,\\\
2\lambda_1^\downarrow + \lambda_2^\downarrow&\leq 4\omega_1 + 3\omega_2 + 2\omega_3 + 2\omega_4 + \omega_5\,,\\\
\lambda_1^\downarrow + \lambda_2^\downarrow + \lambda_3^\downarrow &\leq 2\,,
\end{split}
\end{equation}
where the last inequality is automatically fulfilled by the normalization of $\bd\lambda$.

For $r=2$ shown in the right panel of Fig.~\ref{fig:specpol}, we only have one NON vector $\bd{v}$ and therefore the first two inequalities (with $\omega_3=\omega_4=\omega_5=0$) in Eq.~\eqref{eq:minhN2d3} are sufficient to describe $\Sigma^\downarrow(\bd{\omega})$. The two dashed lines then correspond to the saturation of the bounds on $\lambda_1^\downarrow$ and $\lambda_1^\downarrow + \lambda_2^\downarrow$.

\subsection{Two non-vanishing weights}

For the case of $r=1$ and $N$ bosons, we recover ground state RDMFT with only one configuration given by $(1, 1, ..., 1)$, i.e.~all bosons in the ground state, and $\omega$-minimizer $\hat{\Gamma}_{\boldsymbol{\omega}_0}=\ket{1, 1, ..., 1}\!\bra{1, 1, ...,1}$. The corresponding NON vector reads $\bd{v}=(N, 0, ...)$.

The only sequence for $N=3$ particles and two non-vanishing weights is given by $(1, 1, 1)\to (1, 1, 2)$, yielding the corresponding minimizer $\hat{\Gamma} = \omega_1\ket{1, 1, 1}\!\bra{1, 1, 1} + \omega_2\ket{1, 1, 2}\!\bra{1, 1, 2}$ and 
\begin{equation}
\bd{v}= (3\omega_1 + 2\omega_2, \omega_2, 0, ...)\,,
\end{equation} 
with $\omega_1+\omega_2=1$. Then, applying Rado's theorem, introduced in Sec.~\ref{sec:Rado}, leads to
\begin{equation}
\ebw = \{\hat{\gamma}\in \mathcal{E}^1_N\,|\,\mathrm{spec}(\hat{\gamma})\prec \bd{v}\}\,.
\end{equation}
The constraints on the natural occupation numbers of  $\hat{\gamma} \in \ebw$ for arbitrary $(N, d)$ with $d=\mathrm{dim}(\mathcal{H}_1)$ follow as
\begin{equation}\label{eq:consr=2}
\begin{split}
&\sum_{i=1}^{d-1} \lambda_i^\downarrow\leq N\,,\\\
&\lambda_1^\downarrow \leq N-1+\omega_1\,.
\end{split}
\end{equation}
All further possible constraints are automatically fulfilled by the normalization of $\hat{\gamma}$ to the total particle number. Note that we obtain only one additional constraint in the last line of Eq.~\eqref{eq:consr=2} compared to $r=1$ in agreement with Tab.~\ref{tab:Noineq}. Thus, the facet-defining inequalities for $r=1$ and $r=2$ represent the first two levels of the hierarchy of facet-defining inequalities introduced in Sec.~\ref{sec:h-repr_ebw}. 

\subsection{Three non-vanishing weights\label{subsec:N3r3}}

For $N=3$, we have two sequences 
\begin{equation}
\begin{split}
&(1):\,\, (1, 1, 1)\to (1, 1, 2)\to (1, 1, 3)\,,\\\
&(2):\,\, (1, 1, 1)\to (1, 1, 2)\to (1, 2, 2)\,,
\end{split}
\end{equation}
and therefore the two NON vectors
\begin{equation}
\begin{split}
&\bd{v}^{(1)} = (2 + \omega_1, \omega_2, \omega_3, 0, ...)\,,\\\
&\bd{v}^{(2)} = (1 + 2\omega_1 + \omega_2, 2 - 2\omega_1-\omega_2, 0, ...)\,.
\end{split}
\end{equation}
The majorization condition $\mathrm{spec}(\hat{\gamma})\prec \bd{u}$ from the generalized Rado theorem (see Eq.~\eqref{eq:genRado}) for a convex combination $\bd{u}= q\bd{v}^{(1)} + (1-q)\bd{v}^{(2)}$ requires that 
\begin{equation}\label{eq:ineq_q_N3}
\begin{split}
\lambda_1^\downarrow &\leq 1 + 2\omega_1 + \omega_2 + q(1-\omega_1-\omega_2)\,,\\\
\lambda_1^\downarrow + \lambda_2^\downarrow &\leq 3 - q(1-\omega_1-\omega_2)\,,\\\
\lambda_1^\downarrow + \lambda_2^\downarrow + \lambda_3^\downarrow &\leq 3\,.
\end{split}
\end{equation}
To derive the facet-defining inequalities of $\Sigma^\downarrow(\bd\omega)$, we need to eliminate the parameter $q$ in the next step. Therefore, we first notice that the upper bound on $\lambda_1^\downarrow$ can vary between $1 + 2\omega_1 + \omega_2$ and $2+\omega_1$ since $q$ is restricted to $q\in[0,1]$. Moreover, the upper bound on $\lambda_1^\downarrow + \lambda_2^\downarrow$ can vary between $3$ and $2 + \omega_1 + \omega_2$. Thus, it must always hold that $\lambda_1^\downarrow\leq 2+\omega_1$ which requires to adjust the value of $q$ according to
\begin{equation}
q\geq \frac{\lambda_1^\downarrow-1-2\omega_1-\omega_2}{1-\omega_1-\omega_2}\,.
\end{equation}
The above inequality is then used to obtain the facet-defining upper bound on $\lambda_1^\downarrow + \lambda_2^\downarrow$. Thus, we obtain for the minimal hyperplane representation of $\Sigma^\downarrow(\bd\omega)$,
\begin{equation}
\begin{split}
\lambda_1^\downarrow &\leq 2 + \omega_1\,,\\\
2\lambda_1^\downarrow + \lambda_2^\downarrow &\leq 4 + 2\omega_1 + \omega_2\,,\\\
\lambda_1^\downarrow + \lambda_2^\downarrow + \lambda_3^\downarrow &\leq 3\,,
\end{split}
\end{equation}
where the last inequality is automatically fulfilled by the normalization of the 1RDM. Generalizing the inequalities to arbitrary $N$ and $d$ leads to
\begin{equation}\label{eq:r3}
\begin{split}
\lambda_1^\downarrow &\leq N-1 + \omega_1\\\
\lambda_1^\downarrow - \sum_{j=3}^d\lambda_j^\downarrow &\leq N-2 + 2\omega_1 + \omega_2\\\
\sum_{j=1}^{d-1}\lambda_j^\downarrow &\leq  N\,.
\end{split}
\end{equation}
Thus, the second inequality is the only additional one compared to $r=2$ illustrating again the hierarchy of a generalized exclusion principle for bosons explained in Sec.~\ref{sec:h-repr_ebw}. In the next example, we thus only show the additional inequalities for $r=4$ and for the remaining inequalities refer to the minimal hyperplane representation for smaller values of $r$. 

\subsection{Four non-vanishing weights}

For four non-vanishing weights and $N=3$ particles we obtain four sequences,
\begin{equation}
\begin{split}
&(1, 1, 1) \to (1, 1, 2)\to (1, 1, 3)\to(1, 1, 4)\,,\\\
&(1, 1, 1)\to (1, 1, 2)\to (1, 1, 3) \to (1, 2, 2)\,,\\\
&(1, 1, 1)\to (1, 1, 2)\to (1, 2, 2)\to (1, 1, 3)\,,\\\
&(1, 1, 1)\to (1, 1, 2)\to (1, 2, 2)\to (2, 2, 2)\,.
\end{split}
\end{equation}
The four corresponding generating vertices for arbitrary particle number $N\geq 3$ are summarized in Tab.~\ref{tab:r4}. Thus, different particle numbers $N$ only affect the first entry $v_1$ by adding different constants depending on $N$ to the weights $\omega_j$. This is generally true for $r$ non-vanishing weights, $N\geq r-1$ particles and $d\geq r$.
\begin{table}
\resizebox{\textwidth}{!}{
\centering
\begin{tabular}{ | c | c | c | c | c |}
\hline
& $\bd{v}^{(1)}$  & $\bd{v}^{(2)}$ &$\bd{v}^{(3)}$ & $\bd{v}^{(4)}$\\
\hline
$v_1$ & $\omega_1 + N-1$ & $2\omega_1 + \omega_2 + \omega_3+ N-2$ & $2\omega_1 + \omega_2 + \omega_4 + N-2$ &  $3\omega_1 + 2\omega_2 + \omega_3 + N-3$  \\
\hline
$v_2$&  $\omega_2$ & $\omega_2 + 2\omega_4$ & $ \omega_2 + 2\omega_3$ &$ \omega_2 + 2\omega_3 + 3\omega_4$   \\
\hline
$v_3$& $\omega_3$& $\omega_3$ & $\omega_4$ &$0$     \\
\hline
$v_4$ & $\omega_4$ & $0$& $0$ &  $0$ \\
\hline
\end{tabular}
}
\caption{Entries of the NON vectors $\bd{v}^{(i)}$ for $r=4$ and arbitrary $N\geq r-1$ and $d\geq r$.} 
\label{tab:r4}
\end{table}
Due to the larger number of generating vertices $\bd v^{(i)}$ than for $r\leq 3$, turning the vertex representation into a halfspace representation requires a mathematically more complex framework than in the previous examples. We present the formalism in Ref.~\cite{CLLS21} and provide only the inequalities in the following. For the setting $(N,d)=(3,4)$, the minimal hyperplane representation consists of the three facet-defining inequalities for $r=3$, as well as
\begin{equation}\label{eq:N3r4}
\begin{split}
2\lambda_1^\downarrow + \lambda_2^\downarrow + \lambda_3^\downarrow &\leq 4+2\omega_1 + \omega_2 + \omega_3 \,,\\\
3\lambda_1^\downarrow + 2\lambda_2^\downarrow &\leq 6 + 3\omega_1 + 2\omega_2 + \omega_3\,.
\end{split}
\end{equation}
Thus, we obtain in total five constraints on the NON vector $\bd\lambda^\downarrow$ and two new inequalities in Eq.~\eqref{eq:N3r4} in agreement with Tab. \ref{tab:Noineq}. For arbitrary $N$ and $d$, one obtains the following inequalities \cite{CLLS21},
\begin{equation}
\begin{split}
2\lambda_1^\downarrow + \lambda_2^\downarrow + \lambda_3^\downarrow &\leq 2N-2+2\omega_1 + \omega_2 + \omega_3 \,,\\\
3\lambda_1^\downarrow + 2\lambda_2^\downarrow &\leq 3N-3 + 3\omega_1 + 2\omega_2 + \omega_3\,.
\end{split}
\end{equation}
Recall that the minimal hyperplane representation for $\Sigma(\bd\omega)$ follows from the facet-defining inequalities for $\bd\lambda^\downarrow$ by permuting the coefficients in the inequalities in all possible ways and replacing $\lambda_i^\downarrow$ by $\lambda_i$. 

\section{Application to Bose-Einstein condensation \label{sec:w-RDMFT_BEC}}

In this section, we apply the excited state RDMFT to BEC's using the Bogoliubov approximated interaction (see Eq.~\eqref{eq:H_Bogoloiubov})
\begin{equation}
\hat{W}_\mathrm{B} =  \frac{N(N-1)W_{\bd{0}}}{2V} + \frac{1}{2V}\sum_{\bd{p}\neq 0}W_{\bd{p}}\left[2\hat{n}_{\bd{0}}\hat{n}_{\bd{p}} +\hat{a}_{\bd{p}}^\dagger\hat{a}_{-\bd{p}}^\dagger \hat{a}_{\bd{0}}^2 + \big(\hat{a}_{\bd{0}}^\dagger\big)^2\hat{a}_{\bd{p}}\hat{a}_{-\bd{p}} \right] \,.
\end{equation}
The goal is then, similar to Ch.~\ref{ch:RDMFT_BEC}, to derive a first-level functional for excited states valid in the regime close to complete condensation.
The interacting ground state of a BEC at $T=0$ has the form $\ket{\Psi_0} = \hat{U}\ket{N}$ where $\ket{N} = (N!)^{-1/2}(\hat{a}_{\bd{0}}^\dagger)^N\ket{0}$ and $\hat{U}$ is a unitary operator. The GOK variational principle requires that $E_0\leq E_1\leq...$, where $\hat{H}\ket{\Psi_0}=E_0\ket{\Psi_0}$ is the eigenvalue equation for the ground state and ground state energy. Thus, the $\bd{\omega}$-minimizer for $r=1$ would be $\hat{\Gamma}_{\bd{\omega}}= \ket{\Psi_0}\!\bra{\Psi_0}$. Following this consideration, the first excited state is the excitation of one quasiparticle. Just as Levy's constrained search for ground state RDMFT, the constrained search formalism for $\bd\omega$-ensemble RDMFT requires fixed total particle numbers $N$. Thus, we have to use particle number conserving quasiparticle operators introduced by Girardeau \cite{Girardeau59} (see also Sec.~\ref{sec:pair}). 

\subsection{Derivation of the $\bd{\omega}$-ensemble functional}

The excitation spectrum of a homogeneous BEC in 3D in the thermodynamic limit is gapless. However, for finite systems there will be a gap between the ground state and the first excited state because the momenta $\bd{p}$ are discrete in that case. We will therefore restrict to finite systems to derive the excited state functional for Bogoliubov approximated systems in the following.

The particle number conserving quasiparticles operators, appearing in the diagonal Bogoliubov Hamiltonian, read
\begin{equation}\label{eq:qp2}
\hat{c}_{\bd{q}}^\dagger = \hat{U}\hat{a}_{\bd{q}}^\dagger\hat{U}^\dagger\hat{\beta}_0 \,.
\end{equation}
As discussed in Sec.~\ref{sec:obstacles}, RDMFT requires that the unitary operator $\hat{U}$ lives on the $N$-boson Hilbert space and conserves the total particle number. 
The Bogoliubov approximated Hamiltonian $\hat{H}_\mathrm{B}$ consists of a constant term yielding the ground state energy plus a second term which is diagonal in the Bogoliubov quasiparticle operators with dispersion $\omega_{\bd{q}}$ \cite{Seiringer11}. The quasiparticle dispersion is given by
\begin{equation}
\omega_{\bd{q}} = \sqrt{\varepsilon_{\bd q}(\varepsilon_{\bd q} + 2 n W_{\bd q})}\,,\quad \bd q\neq \bd 0\,.
\end{equation}
Thus, the momentum $\bd q$ leading to the lowest $\omega_{\bd p}$ defines the first excited state $\ket{\Psi_1}$ such that $\hat{H}_\mathrm{B}\ket{\Psi_1} = E_1\ket{\Psi_1}$ with $E_0< E_1$. Moreover, this reveals a fundamental obstacle in RDMFT because a possible crossing of energy levels splits the domain of the $\bd \omega$-ensemble functional into different cells, each corresponding to a different first excited state and 
thus a different functional $\mathcal{F}_{\bd \omega}^{\bd q}(\gh)$. This immediately implies that we cannot obtain a closed expression for the universal functional but rather have
\begin{equation}
\mathcal{F}_{\bd \omega}(\hat{\gamma}) = \min_{\bd q} \mathcal{F}_{\bd \omega}^{\bd q}(\hat{\gamma})\,.
\end{equation}
Thus, while deriving the expression for ${F}_{\bd \omega}^{\bd q}(\gh)$ in the following, we implicitly assume that we already made the correct choice of the momentum $\bd q$ corresponding to the first exited state eigenenergy $E_1$.
Since the Hamiltonian is invariant under inversion $\bd{p}\to -\bd{p}$, also the excited states $\hat{c}_{\bd{q}}^\dagger \ket{\Psi_0}$ and $\hat{c}_{-\bd{q}}^\dagger \ket{\Psi_0}$ correspond to the same energy. Thus, this holds in particular for any superposition 
\begin{equation}
\ket{\Psi_1}_x = \left(x\hat{c}_{\bd{q}}^\dagger + \sqrt{1-|x|^2}\hat{c}_{-\bd{q}}^\dagger\right)\ket{\Psi_0}= \hat{U}\left(x\hat{a}_{\bd{q}}^\dagger + \sqrt{1-|x|^2}\hat{a}_{-\bd{q}}^\dagger\right) \ket{N-1}
\end{equation}
with $x\in \mathbb{C}$ and the quasiparticle operators given by Eq.~\eqref{eq:qp2}.
The eigenvalues of the inversion operator can take the values $\pm 1$ and in the following we thus use the excited state 
\begin{equation}
\ket{\Psi_1} = \frac{1}{\sqrt{2}}\left(\hat{c}_{\bd{q}}^\dagger + \hat{c}_{-\bd{q}}^\dagger\right)\ket{\Psi_0}= \frac{1}{\sqrt{2}}\hat{U}\left(\hat{a}_{\bd{q}}^\dagger + \hat{a}_{-\bd{q}}^\dagger\right) \ket{N-1}
\end{equation}
which corresponds to the symmetric case. The expectation value of $\hat{H}_\mathrm{B}$ in the state $\ket{\Psi_1}$ reads
\begin{equation}\label{eq:Psi1HBPsi1}
\begin{split}
E_1 &= \bra{\Psi_1}\hat{H}_B\ket{\Psi_1} = E_0 + \frac{1}{2}\left(\sqrt{\varepsilon(\bd{q})(\varepsilon(\bd{q}) + 2 n W_{\bd{q}})} + \sqrt{\varepsilon(-\bd{q})(\varepsilon(-\bd{q}) + 2 n W_{-\bd{q}})}\right)\\\
&=  E_0 + \sqrt{\varepsilon(\bd{q})(\varepsilon(\bd{q}) + 2 n W_{\bd{q}})}\,,
\end{split}
\end{equation}
where we used the degeneracy in the last line. To derive the excited state functional, $\bd{q}$ and $-\bd{q}$ can be either treated together or independently. Both approaches work, but it is important to remember which convention has been used to be consistent. This will be particularly important when we derive $\mathcal{F}_{\bd{\omega}}$ using the Legendre-Fenchel transform because, if the last line in Eq.~\eqref{eq:Psi1HBPsi1} is used, also the kinetic energy must be treated as $\langle\hat{t}\rangle = 2\sum_{\bd{p}>\bd{0}}\varepsilon(\bd{p})n_{\bd{p}}$.
For the weighted sum of the two lowest energies $E_0$ and $E_1$ we eventually obtain from Eq.~\eqref{eq:Psi1HBPsi1}
\begin{equation}\label{eq:EomegaBog}
\begin{split}
E_{\bd{\omega}} &= \omega E_0 + (1-\omega)\left(E_0 +  \sqrt{\varepsilon(\bd{q})(\varepsilon(\bd{q}) + 2 n W_{\bd{q}})}\right)\\\
&= E_0 + (1-\omega)\sqrt{\varepsilon(\bd{q})(\varepsilon(\bd{q}) + 2 n W_{\bd{q}})}\,,
\end{split}
\end{equation}
where 
\begin{equation}
E_0 = -\sum_{\bd{p}>\bd{0}}\left(\varepsilon(\bd{p}) + n W_{\bd{p}} - \sqrt{\varepsilon(\bd{p})(\varepsilon(\bd{p}) + 2 n W_{\bd{p}})}\right)
\end{equation}
is the well-known result for the ground state energy \cite{pita,Seiringer11}.

In the following, we present three different, but equivalent, approaches to derive the excited state functional $\mathcal{F}_{\bd{\omega}}^{\bd{q}}(\gh)$ which lead to the same functional, as required. 
We start by performing the Levy's constrained search. Then, we derive $\mathcal{F}_{\bd{\omega}}^{\bd{q}}(\gh)$ by calculating the Legendre-Fenchel transform of \eqref{eq:EomegaBog}. As a third option to calculate the functional we use the solution for the variational parameters $\phi_{\bd{p}}$ from the ground state problem. We believe that this extended discussion of several approaches to obtain the excited state universal functional illustrate different perspectives on how RDMFT works and thus promotes a more comprehensive understanding of the method.

\subsubsection{Constrained search\label{sec:LevyBog}}

For two non-vanishing weights, $r=2$, the minimization in Eq.~\eqref{eq:Levyomega} is performed over all $N$-boson density operators
\begin{equation}
\hat{\Gamma}_{\bd{\omega}} = \omega\ket{\Psi_0}\!\bra{\Psi_0} + (1-\omega)\ket{\Psi_1}\!\bra{\Psi_1}\,.
\end{equation}
Then, the average numbers of particles with different momenta $\bd{p}$ representing the diagonal elements of the 1RDM $\hat{\gamma}$, follow as
\begin{equation}\label{eq:npGammaw}
n_{\bd{p}} = \Tr[\hat{\Gamma}_{\bd{\omega}}\hat{n}_{\bd{p}}] = \begin{cases}
\frac{\phi_{\bd{p}}^2}{1-\phi_{\bd{p}}^2}\,,&\quad\text{if } \bd{p}\neq\pm\bd{q}\,,\\
\omega\frac{\phi_{\bd{p}}^2}{1-\phi_{\bd{p}}^2} + \frac{1}{2}(1-\omega)\frac{1 + 3\phi_{\bd{p}}^2}{1-\phi_{\bd{p}}^2}\,,&\quad\text{if } \bd{p}=\pm\bd{q}\,,
\end{cases}
\end{equation}
where we used the expression for the unitary operator in Eq.~\eqref{U} as well as Eq.~\eqref{eq:phiG}, determining the connection between the phases $\phi_{\bd p}$ and the variational parameters in the unitary operator $\hat{U}$.
Inverting the above expression yields
\begin{equation}
\phi_{\bd{p}} = \begin{cases}
\sigma_{\bd{p}} \sqrt{\frac{n_{\bd{p}}}{1+n_{\bd{p}}}}\,,&\quad\text{if } \bd{p}\neq\pm\bd{q}\,,\\
\sigma_{\bd{p}}\sqrt{\frac{2n_{\bd{p}}+\omega-1}{2n_{\bd{p}} + 3 - \omega}}\,,&\quad\text{if } \bd{p}=\pm\bd{q}\,.
\end{cases}
\end{equation}
Using these results, the excited state universal functional follows from Levy's constrained search formalism as
\begin{equation}\label{eq:FwLevyBEC}
\begin{split}
\mathcal{F}_{\bd{\omega}}(\bd{n}) &= \min_{\hat{\Gamma}_{\bd{\omega}}\mapsto\hat{\gamma}} \Tr\left[\hat{\Gamma}_{\bd{\omega}}\hat{W}_\mathrm{B} \right] \\\
&= \mathcal{F}_0(\bd{n}) + \sum_{\pm\bd{q}}n|W_{\bd{q}}|\left(\sqrt{n_{\bd{q}}(1+n_{\bd{q}})} - \frac{1}{2}\sqrt{4n_{\bd{q}}(n_{\bd{q}}+1) + \omega(4-\omega)-3}\right)\\\
&= \mathcal{F}_0(\bd{n})  +  n|W_{\bd{q}}|\left(2\sqrt{n_{\bd{q}}(1+n_{\bd{q}})} - \sqrt{4n_{\bd{q}}(n_{\bd{q}}+1) + \omega(4-\omega)-3}\right)\,.
\end{split}
\end{equation}
where $\mathcal{F}_0$ denotes the ground state functional
\begin{equation}
\mathcal{F}_0(\bd{n})  = n \sum_{\bd{p}\neq \bd{0}} W_{\bd{p}}\left(n_{\bd{p}} - \mathrm{sgn}(W_{\bd{p}})\sqrt{n_{\bd{p}}(n_{\bd{p}} + 1)}\right)\,.
\end{equation}
Consequently, Eq.~\eqref{eq:FwLevyBEC} reduces to $\mathcal{F}_0$ for $\omega=1$, as required. 

As a consistency check with the weighted sum of energies given in Eq.~\eqref{eq:EomegaBog}, we minimize the energy functional 
\begin{equation}\label{eq:EfunctionalBog}
\mathcal{E}_{\bd{\omega}}(\bd{n}) = \sum_{\bd{p}\neq \bd{0}}\varepsilon(\bd{p}) n_{\bd{p}} + \mathcal{F}_{\bd{\omega}}(\bd{n}) 
\end{equation}
with respect to the occupation number vector $\bd n$. The solution of $\varepsilon(\bd{p}) = - \partial\mathcal{F}_{\bd{\omega}}/\partial n _{\bd{p}}$ for every $\bd{p}$ is denoted by $\thickbar{n}_{\bd{p}}$, leading to
\begin{equation}\label{eq:npBogmin}
\thickbar{n}_{\bd{p}} = \frac{1}{2}\left(\frac{\varepsilon(\bd{p}) + n W_{\bd{p}}}{\sqrt{\varepsilon(\bd{p})(\varepsilon(\bd{p})+2nW_{\bd{p}})}}-1 \right)
\end{equation}
and
\begin{equation}\label{eq:nqBogmin}
\thickbar{n}_{\bd{q}} = \frac{1}{2}\left(\frac{(\varepsilon(\bd{q}) + n W_{\bd{q}})(2-\omega)}{\sqrt{\varepsilon(\bd{q})(\varepsilon(\bd{q})+2nW_{\bd{q}})}}-1 \right)\,.
\end{equation}
Note that we treat $\bd{q}$ and $-\bd{q}$ separately at this point which, as already explained above, does not make any difference, if done consistently. After collecting the contributions from $\bd{q}$ and $-\bd{q}$ we eventually obtain
\begin{equation}
\begin{split}
E_{\bd{\omega}} &= -\sum_{\bd{p}>\bd{0} } \left(nW_{\bd{p}} + \varepsilon(\bd{p}) - \sqrt{\varepsilon(\bd{p})(\varepsilon(\bd{p})+2nW_{\bd{p}})}\right)+ (1-\omega)\sqrt{\varepsilon(\bd{q})(\varepsilon(\bd{q})+2nW_{\bd{q}})}\,,
\end{split}
\end{equation}
which is in agreement with Eq.~\eqref{eq:EomegaBog} and thus verifies that $\mathcal{F}_{\bd{\omega}}$ obtained in Eq.~\eqref{eq:FwLevyBEC} is indeed correct.

\subsubsection{Legendre-Fenchel transform\label{sec:LFBog}}

Since the energy $E_{\bd{\omega}}$ and and the universal functional $\mathcal{F}_{\bd{\omega}}$ are related through the Legendre-Fenchel transform by (see Sec.~\ref{sec:Levy})
\begin{equation}\label{eq:Leg}
\mathcal{F}_{\bd{\omega}}^*(\hat{h}) = \sup_{\hat{\gamma}\in \ew} \left[\langle \hat{h}, \hat{\gamma}\rangle - \mathcal{F}_{\bd{\omega}}(\hat{\gamma})\right] = -\inf_{\hat{\gamma}\in \ew}\left[\mathcal{F}_{\bd{\omega}}(\hat{\gamma}) + \langle -\hat{h}, \hat{\gamma}\rangle\right] = -E_{\bd{\omega}}(-\hat{h})\,,
\end{equation}
we obtain the excited state functional $\mathcal{F}_{\bd{\omega}}$ for $\hat{h}=\hat{t}$ from
\begin{equation}\label{eq:LFFw}
\mathcal{F}_{\bd{\omega}}(\hat{\gamma}) = \sup_{\hat{t}} \left[E_{\bd{\omega}}(\hat{t}) - \langle\hat{\gamma}, \hat{t}\rangle  \right]\,,
\end{equation}
where $E_{\bd{\omega}}(\hat{t})$ is given by Eq.~\eqref{eq:EomegaBog}. In the following, we assume that $W_{\bd{p}}\geq 0$ for all $\bd{p}$. Solving $n_{\bd{p}}=\partial E_{\bd{\omega}}/\partial \varepsilon(\bd{p})$ for $\varepsilon(\bd{p})$ yields
\begin{equation}\label{eq:epLFBog}
\begin{split}
\thickbar{\varepsilon}(\bd{p})&= \frac{nW_{\bd{p}}}{2}\left(\frac{1 + 2 n_{\bd{p}}}{\sqrt{n_{\bd{p}}(n_{\bd{p}} + 1)}} - 2\right)\quad\text{if } \bd{p}\neq \pm \bd{q}\,,\\\
\thickbar{\varepsilon}(\bd{p}) &= nW_{\bd{p}}\left(\frac{1 + 2n_{\bd{p}}}{\sqrt{(3 + 2n_{\bd{p}} - \omega)(2n_{\bd{p}}+ \omega - 1)}}-1\right) \quad \text{if } \bd{p}=\pm \bd{q}\,.
\end{split}
\end{equation}
By calculating the second derivative, it can be easily checked that the solution above corresponds to a maximum. Inserting Eq.~\eqref{eq:epLFBog} into \eqref{eq:LFFw} yields
\begin{equation}\label{eq:FLFBEC}
\mathcal{F}_{\bd{\omega}}(\bd{n}) = \mathcal{F}_0(\bd{n})  +  nW_{\bd{q}}\left(2\sqrt{n_{\bd{q}}(1+n_{\bd{q}})} - \sqrt{4n_{\bd{q}}(n_{\bd{q}}+1) + \omega(4-\omega)-3}\right)\,,
\end{equation}
which is equal to the excited state functional in Eq.~\eqref{eq:FwLevyBEC} under the assumption that $W_{\bd{p}}\geq 0 \text{ } \forall \bd{p}$. It was already shown in Sec.~\ref{sec:LevyBog} that this functional leads to the correct result for the energy. However, it is important to notice that the Legendre-Fenchel transform only coincide with the universal functional obtained from the constrained search formalism because $\Fw$ is convex. In case of a non-convex functional we would have obtained its lower convex envelope as discussed in Sec.~\ref{subsec:LF}.

\subsubsection{Solution from ground state problem\label{sec:GSBog}}

The variational parameters $\phi_{\bd{p}}$ from the solution of the ground state problem are given by
\begin{equation}\label{eq:phiGZ}
\phi_{\bd{p}} = \frac{1}{nW_{\bd{p}}}\left(\varepsilon(\bd{p}) + nW_{\bd{p}} - \sqrt{\varepsilon(\bd{p})(\varepsilon(\bd{p}) + 2nW_{\bd{p}})}\right)\,,
\end{equation}
determining both states, $\ket{\Psi_0}$ and $\ket{\Psi_1}$. Inserting \eqref{eq:phiGZ} into \eqref{eq:npGammaw} and solving for $\varepsilon(\bd{p})$ leads to the same result for the dispersion $\varepsilon(\bd{p})$ as in Eq.~\eqref{eq:epLFBog}. Note that the following discussion is equivalent to the Legendre-Fenchel transform discussed in the subsection above. 
Using Eq.~\eqref{eq:epLFBog} and the energy $E_{\bd{\omega}}$ in Eq.~\eqref{eq:EomegaBog} we obtain the functional $\mathcal{F}_{\bd{\omega}}$ from
\begin{equation}
\begin{split}
\mathcal{F}_{\bd{\omega}}(\bd{n})  &= E_{\bd{\omega}} - \sum_{\bd{p}> \bd{0}}\varepsilon(\bd{p})n_{\bd{p}}\\\
&= \mathcal{F}_0(\bd{n}) +  nW_{\bd{q}}\left(2\sqrt{n_{\bd{q}}(1+n_{\bd{q}})} - \sqrt{4n_{\bd{q}}(n_{\bd{q}}+1) + \omega(4-\omega)-3}\right)\,.
\end{split}
\end{equation}
Note that we obtain the same result as already derived in Eq.~\eqref{eq:FwLevyBEC} and Eq.~\eqref{eq:FLFBEC}.

\subsection{Bose-Einstein force}

In this section, we calculate the gradient of the excited state functional $\mathcal{F}_{\bd{\omega}}^{\bd q}(\gh)$ and show that it diverges repulsively at the boundary of the domain of $\mathcal{F}_{\bd{\omega}}^{\bd q}(\gh)$ in most cases. For $r=2$, we have only one NON vector 
\begin{equation}
\bd{v} = (N-1 + \omega, 1-\omega, ...)
\end{equation}
representing the single generating vertex of the spectral polytope $\Sigma(\bd\omega)$ (see also Sec.~\ref{subsec:N3r3}), all others are obtained from permutations of its entries. The set $\Sigma^\downarrow(\bd \omega)$ is not a simplex (as $\Sigma(\bd \omega)$), but we already used the thermodynamic limit in the derivation of $\mathcal{F}_{\bd \omega}^{\bd q}(\gh)$ and for large enough $N$, the additional constraints on the domain of $\mathcal{F}_{\bd\omega}^{\bd q}(\gh)$ become negligible.

In the following we show that the divergence of the gradient of $\mathcal{F}_{\bd{\omega}}^{\bd q}(\gh)$ along a straight path towards the vertex $\bd v$ is proportional to $1/\sqrt{D}$. 
For the occupation numbers $n_{\bd{p}}$ with $n_{\bd p}=n_{-\bd p}$ we obtain close to the generating vertex $\bd v$
\begin{align}
\frac{\partial \mathcal{F}_{\omega}^{\bd q}(\gh)}{\partial n_{\bd{p}}}(\bd{n})&\sim - \frac{n |W_{\bd{p}}|}{2} \frac{1}{\sqrt{n_{\bd{p}}}} \quad\text{if } \bd{p}\neq \pm \bd{q}\label{eq:BEC_N0}\,,\\\
\frac{\partial \mathcal{F}_{\omega}^{\bd q}(\gh)}{\partial n_{\bd{p}}}(\bd{n})&\sim - \frac{n |W_{\bd{p}}|\sqrt{2-\omega}}{2} \frac{1}{\sqrt{n_{\bd{p}} - \frac{1-\omega}{2}}} \quad\text{if } \bd{p}=\pm \bd{q}\,.\label{eq:BEC_Nq}
\end{align}
The distance of an occupation number vector $\bd n$ to $\bd v$ is given by
\begin{equation}
D = \frac{1}{N}\sum_{\bd{p}\neq \bd{0}} n_{\bd{p}} - D_0\,,
\end{equation} 
where
\begin{equation}
D_0 = \frac{1-\omega}{N}
\end{equation} 
is the minimal fraction of non-condensed bosons. We parametrize the path, starting at the point $\thickbar{\bd{n}}$ by $t\in [0, 1]$, such that the occupation number vector changes according to 
\begin{equation}
\bd{n}(t) = \thickbar{\bd{n}} + t(\bd{n}^{(0)} - \thickbar{\bd{n}})\,.
\end{equation}
For $\bd{p}\neq \bd 0, \pm \bd{q}$, the entries $n^{(0)}_{\bd p}$ of the final NON vector $\bd{n}^{(0)}$ are equal to zero, whereas for momenta $\pm \bd{q}$ they take the values $n^{(0)}_{\pm\bd q} = \frac{1-\omega}{2}$. Then, the distance $D$ along the path can be written as
\begin{equation}
D(t) = (1-t)D(0)\equiv (1-t)\thickbar{D}
\end{equation}
and we eventually obtain 
\begin{equation}
\begin{split}
\frac{\partial \mathcal{F}_{\bd{\omega}}^{\bd q}(\bd n)}{\partial D}\Big\vert_{\mathrm{path}} 
&\approx - \frac{n}{\sqrt{\thickbar{D}}}\left(\sum_{\bd{p}\neq \bd{0}, \pm\bd{q}} \frac{|W_{\bd p}|\sqrt{\thickbar{n}_{\bd p}}}{2}  + \sum_{\pm \bd{q}} \frac{|W_{\bd q}|\sqrt{2 - \omega}}{2}\sqrt{\thickbar{n}_{\bd q} - \frac{1 - \omega}{2}} \right)\frac{1}{\sqrt{D}}\,.
\end{split}
\end{equation}
From this result it follows that the gradient of $\mathcal{F}_{\bd\omega}^{\bd q}(\bd n)$ diverges repulsively proportional to $1/\sqrt{D}$ in the vicinity of the NON vector $\bd v$, whereby the information about the interaction between the particles and the weights $\omega$ is contained in its prefactor. 

Next, we explain the behaviour of the gradient of $\mathcal{F}^{\bd q}_{\bd \omega}(\gh)$ close to the facet of the functional's domain. Since the functional distinguishes the momentum corresponding to the first excited state from all other $\bd p\neq \bd 0$, we restrict the following discussion to $\Sigma^\downarrow(\bd \omega)$. Compared to the domain of the universal functional in ground state RDMFT, which takes the form of a simplex $\Delta$, we now have one additional constraint on the largest occupation number, given by (see also Eq.~\eqref{eq:consr=2})
\begin{equation}
\lambda_1^\downarrow \leq N-1 + \omega\,.
\end{equation}
This defines a hyperplane which cuts the simplex $\Delta$ in Eq.~\eqref{simplex}, leading to an additional facet. Therefore, the resulting spectral polytope is not a simplex anymore and thus we cannot express the functional $\mathcal{F}_{\bd \omega}^{\bd q}$ by uniquely defined distances to the facets as in ground state RDMFT explained in Sec.~\ref{sec:RDMFT_hom}. In case of a BEC with $\varepsilon_{\bd p}=p^2/2m$, the zero momentum state has the largest occupation number, and we have $\lambda^\downarrow_1\equiv n_{\bd 0}$. Thus, if the upper bound on $\lambda_1^\downarrow$ is saturated, we can still occupy all remaining natural orbitals such that no NO is completely empty. For example, we could choose $n_{\bd p}=\frac{1-\omega}{d-1}$ for all $\bd p\neq 0$, where $d= \mathrm{dim}(\mathcal{H}_1)$. As a result, we see from Eq.~\eqref{eq:BEC_N0} and Eq.~\eqref{eq:BEC_Nq} that the gradient of the functional $\mathcal{F}^{\bd q}_{\bd \omega}$ does not diverge on this facet of the polytope, at least for a small system size. However, if $d$ is increased, the occupation numbers of the respective orbitals will decrease to avoid a zero occupation of a state and thus we obtain a collective force from all facets.
%

\section{Bose-Hubbard dimer\label{sec:HubbardDimer_omega}}

Even in ground state RDMFT it is almost impossible to determine an exact ground state universal functional for most quantum systems and sufficiently good approximations are needed. One well-known exception for bosons as well as fermions is the Hubbard dimer \cite{Benavides20, Cohen1}. In the following, we discuss the bosonic Hubbard dimer constituting the building block of the Bose-Hubbard model. 
The Hamiltonian for two spinless bosons on two lattices sites reads 
\begin{equation}\label{eq:H_Dimer}
\hat{H} = -t\left(\hat{a}_L^\dagger\hat{a}_R^{\phantom{\dagger}} + \hat{a}_R^\dagger\hat{a}_L^{\phantom{\dagger}}\right) + \sum_{j=L,R}\epsilon_j\hat{n}_j + \frac{U}{2}\sum_{j=L, R}\hat{n}_j\left(\hat{n}_j-1\right)\,,
\end{equation}
where the first term describes hopping with strength $t$ between the left ($L$) and right side ($R$), $U>0$ is the on-site interaction and $\hat{n}_j = \hat{a}_j^\dagger\hat{a}_j^{\phantom{\dagger}}$ the occupation number operator. In the following, we consider the asymmetric Hubbard dimer where $\epsilon_L\neq \epsilon_R$.

In Sec.~\ref{subsec:N3r3}, we derived the spectral polytope $\Sigma(\boldsymbol{\omega})$, and thus $\ebw$, for $r=3$ non-vanishing weights and arbitrary $N$. However, for the Hubbard dimer we have $d = \mathrm{dim}(\mathcal{H}_1)=2$ and thus only one sequence of configurations, $(1, 1)\to (1, 2)\to (2, 2)$, as well as one NON vector 
\begin{equation}
\bd{v} = (2 \omega_1 + \omega_2, 2-2\omega_1-\omega_2)\,.
\end{equation}
As in the previous sections, we denote by $\bd\lambda\equiv \mathrm{spec}(\gh)\in\ebw$ the spectrum of a 1RDM $\gh$ and eventually obtain
\begin{equation}\label{eq:r3HubbardDimer}
\lambda_1^\downarrow\leq 2\omega_1 + \omega_2\,,\quad \lambda_1^\downarrow + \lambda_2^\downarrow=2\,,
\end{equation}
in agreement with Eq.~\eqref{eq:r3}. This is, up to a factor $1/2$, the same result as for the fermionic Hubbard dimer restricted to the singlet sector. In the following we normalize the 1RDM $\hat{\gamma}$ to one. Thus, its largest eigenvalue is restricted by 
\begin{equation}\label{eq:lambda1_Dimer}
1/2\leq \lambda_1^\downarrow\leq (2\omega_1 + \omega_2)/2\,.
\end{equation}

\begin{figure}[h]
\centering
\begin{subfigure}[t]{.49\textwidth}
\centering
\includegraphics[width=\linewidth]{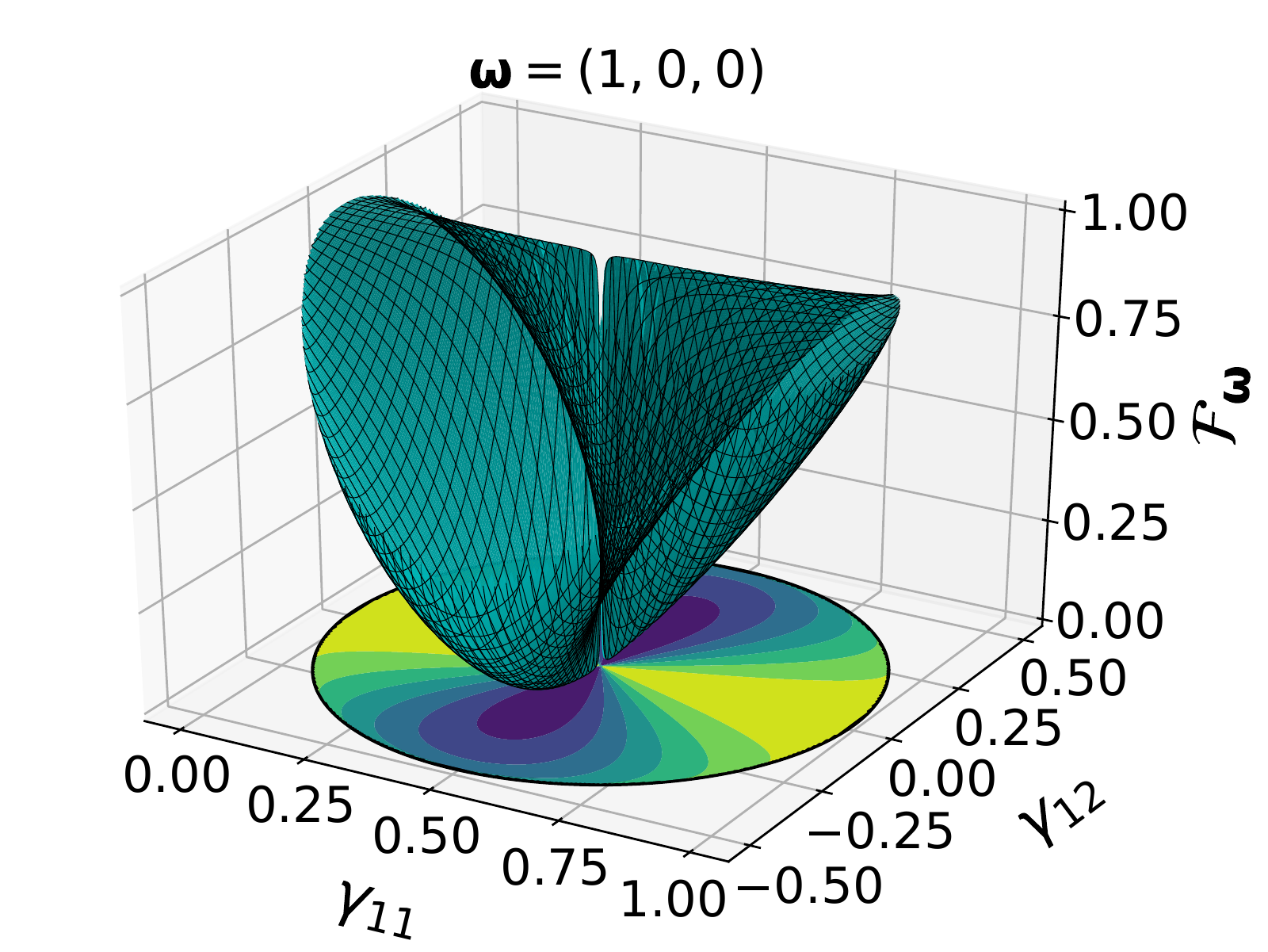}
\caption{\label{Fw_100000}}
\end{subfigure}
\begin{subfigure}[t]{.49\textwidth}
\centering
\includegraphics[width=\linewidth]{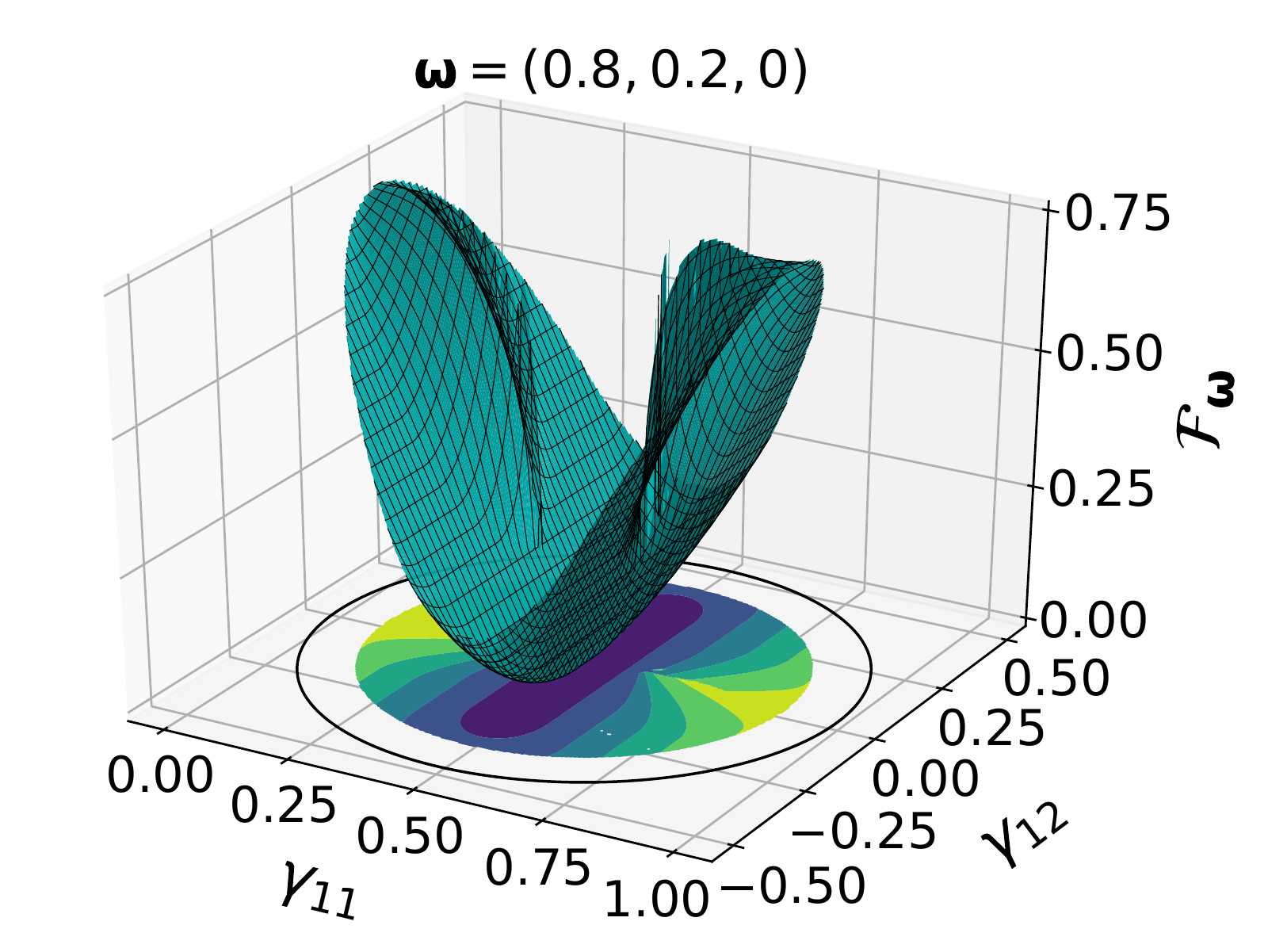}
\caption{\label{Fw_080200}}
\end{subfigure}
\medskip
\begin{subfigure}[t]{.49\textwidth}
\centering
\includegraphics[width=\linewidth]{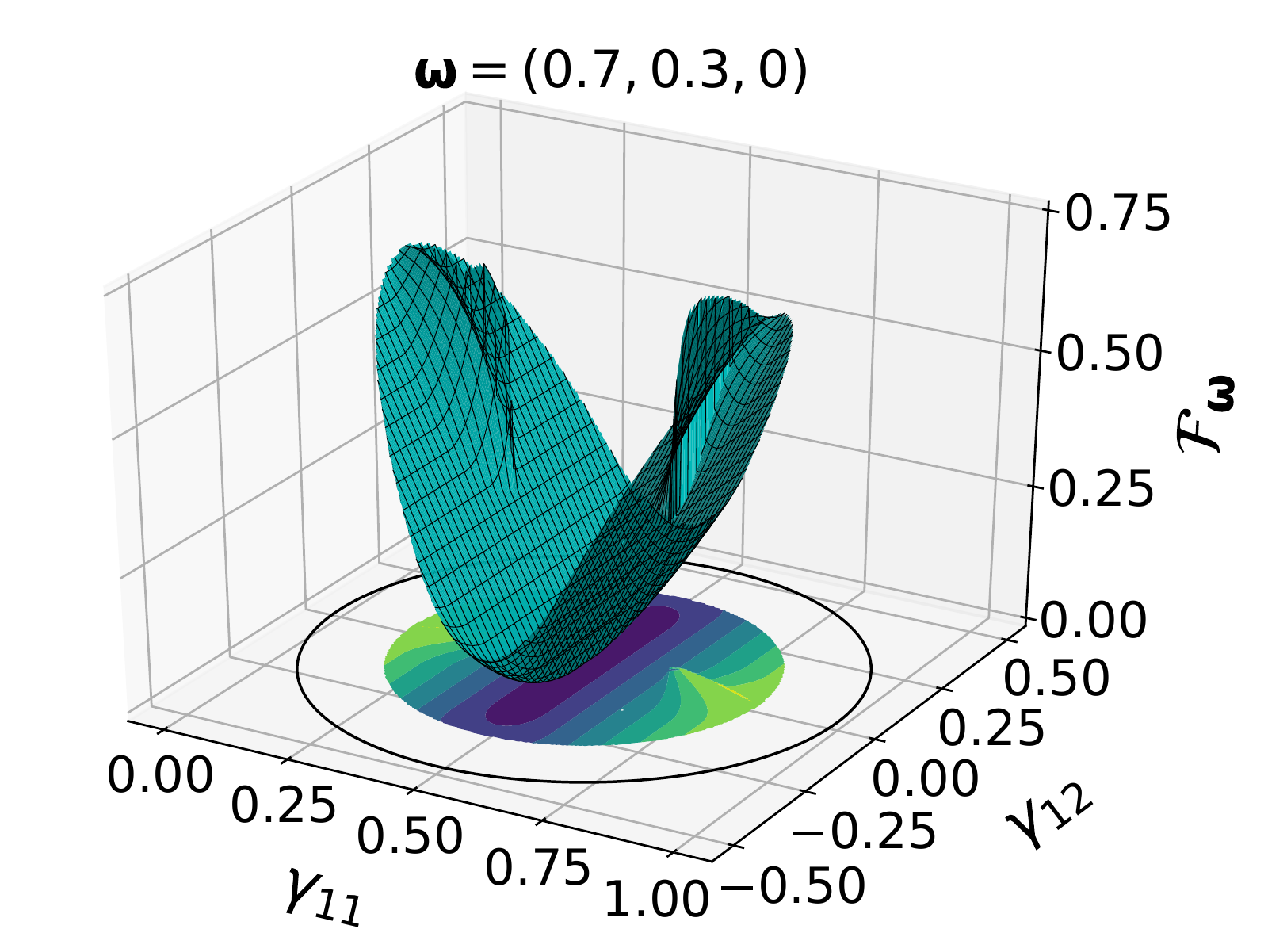}
\caption{\label{Fw_070300}}
\end{subfigure}
\begin{subfigure}[t]{.49\textwidth}
\centering
\includegraphics[width=\linewidth]{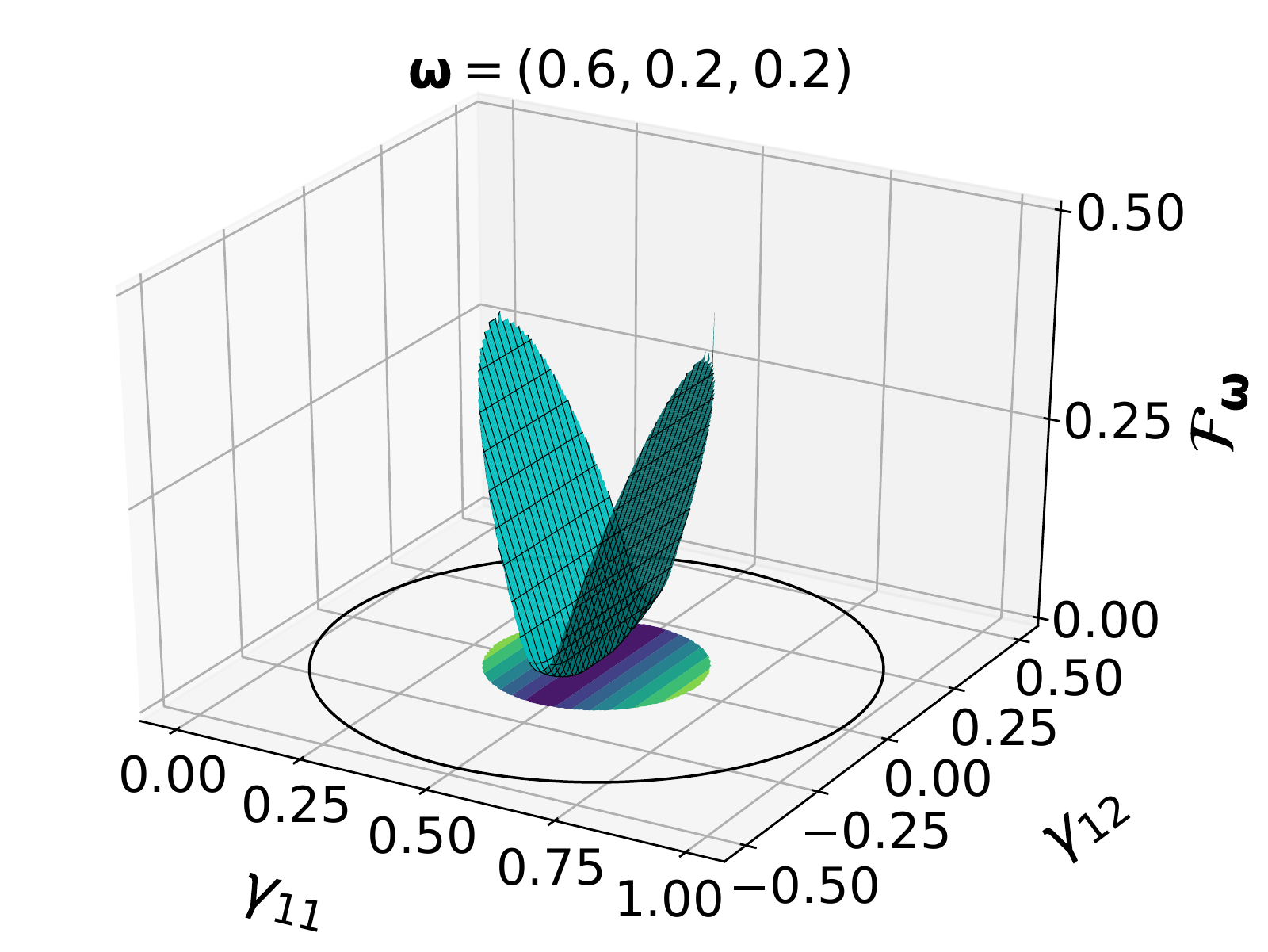}
\caption{\label{Fw_060202}}
\end{subfigure}
\medskip
\begin{minipage}[t]{\textwidth}
\caption{$\bd\omega$-ensemble functional for different $\bd\omega$ obtained from a numerical minimization. $(a) $ confirms that we recover the correct result for $\omega_1=1$. For $\omega_1<1$, the domain of $\mathcal{F}_{\bd \omega}$ becomes smaller according to Eq.~\eqref{eq:lambda1_Dimer}. The functionals in $(b)$ and $(c)$ still have non-$v$-representable regions, whereas for an appropriate choice of the weights $\omega_i$ we obtain a functional which is convex on the entire domain in $(d)$.
\label{fig:Fw_Dimer}}
\end{minipage}
\end{figure}

%
The Hilbert space $\mathcal{H}_2 = \mathrm{span}(\{\ket{i}\}_{i=1}^3)$ is spanned by the basis states $\ket{1} = \ket{2,0}$, $\ket{2}=\ket{0, 2} $ and $\ket{3} = \ket{1, 1}$. We plot the $\bd\omega$-ensemble functional obtained from a numerical minimization in Fig.~\ref{fig:Fw_Dimer}. Depending on $\bd \omega$, not only the domain of $\mathcal{F}_{\bd{\omega}}(\gh)$ but also the shapes of the $v$-representable regions change. 
Since we choose the interaction strength in Eq.~\eqref{eq:H_Dimer} to be $U/2$, we would obtain the same result in the fermionic case. Fig.~\ref{fig:Fw_Dimer} also illustrates the inclusion relation in Eq.~\eqref{eq:inclusion}, namely that the domain of the $\bd \omega$-ensemble functional becomes smaller for a new vector $\bd \omega^\prime\prec\bd\omega$ because the allowed interval for the largest eigenvalue $\lambda_1$ changes according to Eq.~\eqref{eq:lambda1_Dimer}. 

\chapter{Summary and Conclusion}

In this thesis, we have initiated and established a bosonic RDMFT for ground states as well as excited states. A particular emphasis of the former lied on the application of RDMFT to Bose-Einstein condensates (BECs) and the derivation of a first-level functional. 

First, we introduced in Ch.~\ref{ch:RDMFT} the mathematical and conceptual foundations of RDMFT. While Levy's constrained search formalism \cite{LE79} allows us to circumvent the pure state $v$-representability problem in Gilbert's RDMFT, it leads to the pure-state $N$-representability problem. Although its solution is unknown for fermions, the situation simplifies drastically for bosons because every bosonic 1RDM is pure state $N$-representable. Nevertheless, the geometrical interpretation of Levy's constrained search and the Legendre-Fenchel transformation reveals that the convex relaxation of the minimization problem, introduced originally for fermions by Valone \cite{V80}, also facilitates bosonic RDMFT since it turns a non-convex optimization problem into a convex one. Further, for homogeneous bosonic quantum systems, the domain of the universal functional $\mathcal{F}(\gh)$ takes the form of a simplex and we discussed how symmetries simplify the implementation of RDMFT. This is in particular relevant for the application of RDMFT to BECs presented in the subsequent chapter.

Since Levy's constrained search requires a fixed total particle number, conventional Bogoliubov theory and RDMFT are conceptually incompatible. Thus, to apply RDMFT to homogeneous Bose gases, we need a mathematically more rigorous framework. By using a particle-number conserving modification of the conventional Bogoliubov theory, we eventually succeeded in deriving the respective universal interaction functional $\F(\bd{n})$ for homogeneous BECs. Similar to the Hartree-Fock functional in fermionic RDMFT \cite{LiebHF} and the local density approximation in density functional theory \cite{GD95}, the Bogoliubov functional could then serve as a starting point for such approximations. That is particularly promising, since, in contrast to the Hartree-Fock functional, our functional already involves some quantum correlations arising from fractional occupation numbers while the former always leads to occupation numbers identical to zero or one \cite{LiebHF}. As the most striking feature of the universal functional $\F(\bd{n})$, its gradient has been found to diverge repulsively as $1/\sqrt{1-N_{\mathrm{BEC}}/N}$ in the regime close to complete condensation. The associated universal BEC force provides an alternative and most fundamental explanation for the absence of complete condensation in bosonic quantum systems. The BEC force is universal in the sense that it is merely based on the geometry of density matrices and the properties of the partial trace.

Since a bosonic RDMFT for excited states is completely missing in the literature so far, we introduced a new method that allows for calculating excited state energies and energy gaps in Ch.~\ref{ch:w-RDMFT}. Just like ground state RDMFT, $\bd\omega$-ensemble RDMFT for excited states is based on the combination of an appropriate variational principle and a constrained search formalism. By resorting to convex relaxation, we avoid the involved $\bd\omega$-ensemble $N$-representability constraints and turn $\bd\omega$-ensemble RDMFT into a practical method. In addition to the development of a new method, we obtain, as a key result, non-trivial constraints on the excited state bosonic occupation numbers. 
Dependent on the number of excitations in the system, there exists a full hierarchy of such inequalities. 
Thus, we interpret this key finding as a generalized exclusion principle for bosons overlooked in the past.

\begin{appendices}

\chapter{S-wave scattering approximation\label{app:s-wave}}

The s-wave scattering amplitude, usually denoted by $a$ in the literature, plays a crucial to derive the ground state energy and low-lying energy spectrum of weakly interacting homogeneous Bose gases. Moreover, in Sec.~\ref{sec:appl} it is used to verify that the universal functional obtained for a dilute Bose gas in 3D leads to the well-known result for the ground state energy. For this purpose, we show in this section how the s-wave scattering approximation follows from the partial wave expansion in the limit of slow particles. 

\subsubsection{Partial wave expansion}
In the following, we consider a spherical symmetric potential $V(\bd r)=V(r)$, as explained in Sec.~\ref{subsec:S-wave_gases}. Thus, we are interested in solutions of the Schrödinger equation in spherical coordinates
\begin{equation}\label{eq:SG_sp}
\left[-\frac{\hbar^2}{2m}\left(\frac{\partial^2}{\partial r^2} + \frac{2}{r}\frac{\partial}{\partial r}\right) + \frac{\textbf{L}^2}{2mr^2} + V(\bd{r}) - E \right]\Psi(r, \theta, \phi) = 0,
\end{equation}
where 
\begin{equation}
\textbf{L}^2 = \hbar^2\left[ \frac{1}{\mathrm{sin}(\theta)}\frac{\partial}{\partial\theta}\left(\mathrm{sin}(\theta)\frac{\partial}{\partial\theta}\right) + \frac{1}{\mathrm{sin}^2(\theta)}\frac{\partial^2}{\partial\phi^2}\right]\,.
\end{equation}
The eigenfunctions of $\textbf{L}^2$ are the spherical harmonics $Y_l^m(\theta, \phi)$ with eigenvalues $l(l+1)$, where $l=0, 1, ...$ and $m=-l, ..., l$. 
The expansion of an incoming plane wave with wave vector $\bd{k}||\bd e_z$ in spherical harmonics (Rayleigh expansion) is given by \cite{Varshalovich:1988}
\begin{equation}\label{eq:rayleigh_exp}
\rme^{ikz} = \rme^{ikr\mathrm{cos}(\theta)} = \sum_{l=0}^\infty i^l(2l+1)j_l(kr)P_l(\mathrm{cos}(\theta))\,,
\end{equation}
where $\bd e_z$ is the unit vector in $z$-direction and $P_l(\mathrm{cos}(\theta))$ are the Legendre polynomials. 
Since scattering at the potential $V(r)$ conserves angular momentum, different channels of angular momenta $l$ are decoupled. 
In the following, we assume that $V(r)$ is a spherical symmetric potential with a finite scattering volume described by the radius $r_0$. Then, the wave function is independent of $\phi$ and can be expanded as
\begin{equation}
\Psi(r, \theta) = \sum_{l=0}^\infty i^l(2l+1)R_l(r)P_l(\mathrm{cos}(\theta))\,,
\label{Psi_Rl}
\end{equation}
in analogy to the plane wave in Eq.~\eqref{eq:rayleigh_exp}.
Next, we separate the wave function into its radial and angle dependent part. According to the Schrödinger equation in Eq.~\eqref{eq:SG_sp}, the radial functions $R_l(r)$ are determined through
\begin{equation}\label{eq:SG_radial}
\left[\frac{\rmd^2}{\rmd r^2} + \frac{2}{r}\frac{\rmd }{\rmd r}- \frac{l(l+1)}{r^2} - v(r) + k^2\right] R_l(r) = 0, \quad v(r) = \frac{2m}{\hbar^2}V(r), \quad k^2 = \frac{2mE}{\hbar^2}\,.
\end{equation} 
The two linearly independent solutions for each energy $E$ outside of the scattering volume are given by the spherical Bessel functions $j_l(kr)$ and the spherical Neumann function $n_l(kr)$, whose asymptotic behavior is given by
\begin{align}
&j_l(kr) \xrightarrow{kr\ll 1} \frac{(kr)^l}{(2l+1)!!}, \quad j_l(kr) \xrightarrow{kr\gg 1} \frac{1}{kr}\,\mathrm{sin}\left(kr - l\pi/2\right)\\\
&n_l(kr) \xrightarrow{kr\ll 1} \frac{(2l-1)!!}{(kr)^{l+1}}, \quad n_l(kr) \xrightarrow{kr\gg 1} -\frac{1}{kr}\,\mathrm{cos}\left(kr - l\pi/2\right)\,.
\end{align}
The radial part of solution $R_l(r)$ of Eq.~\eqref{eq:SG_radial} is a superposition of $j_l(kr)$ and $n_l(kr)$ leading to the spherical Hankel functions
\begin{equation}
h_l^{\pm}(kr) = n_l(kr)\pm ij_l(kr), \quad h_l^{\pm}(kr) \xrightarrow{kr\gg 1} \frac{1}{kr}\rme^{(\pm kr-l\pi/2)}\,.
\end{equation}
Thus, the incident wave can be written as a superpositon of incoming ($h_l^{(-)}$) and outgoing ($h_l^{(+)}$) spherical Hankel functions,
\begin{equation}
\begin{split}
\Psi(r, \theta) &= \sum_{l=0}^\infty i^l(2l+1)j_l(kr)P_l(\mathrm{cos}(\theta))\\\
&= \frac{i}{2}\sum_{l=0}^\infty i^l(2l+1)\left(h_l^{(-)}(kr) - h_l^{(+)}(kr)\right)P_l(\mathrm{cos}(\theta))\,.
\end{split}
\end{equation}
Next, we need to explain the effect of the scattering potential on the wave function. It follows that the radial wave functions must reduce to the free-particle wave functions in the limit $r\to\infty$, although the potential can attach a phase factor to the outgoing wave. A real phase shift $\delta_l(k)$ will only change the phase of the scattered wave not its amplitude, but it alters the angular distribution. Moreover, the phase shifts $\delta_l$ contain all information about the scattering process. Outside of the scattering volume, we obtain for the wave function after scattering
\begin{equation}\label{Psi_sca}
\begin{split}
\Psi(r, \theta) &= \frac{i}{2}\sum_{l=0}^\infty i^l(2l+1)\left(h_l^{(-)}(kr) - \rme^{i2\delta_l(k)}h_l^{(+)}(kr)\right)P_l(\mathrm{cos}(\theta))\\\ 
&= \sum_{l=0}^\infty i^l(2l+1)\left[j_l(kr) + \frac{1}{2i}\left(\rme^{i2\delta_l(k)} - 1\right)h_l^{(+)}(kr)\right]P_l(\mathrm{cos}(\theta))\\\
&= \rme^{i\bd{kr}} + \sum_{l=0}^\infty i^l(2l+1)T_l(k)\frac{1}{kr}\rme^{i(kr - l\pi/2)}P_l(\mathrm{cos}(\theta))\\\
&\equiv \rme^{i\bd{kr}} + f(k, \theta)\frac{\rme^{ikr}}{r}\,.
\end{split}
\end{equation} 
In Eq.~\eqref{Psi_sca}, we split the wave function into the incident plane wave and an outgoing spherical wave and obtained an asymptotic form of the solution to the Schrödinger equation with the scattering amplitude
\begin{equation}\label{scatt_ampl}
f(k, \theta) = \sum_{l=0}^\infty (2l+1)\frac{T_l(k)}{k}P_l(\mathrm{cos}(\theta))\,.
\end{equation}
Furthermore, $f_l(k) = T_l(k)/k$ is the partial wave scattering amplitude and $S_l(k) = \rme^{i2\delta_l(k)}$ and $T_l(k) = \frac{1}{2i}\left(S_l(k) - 1\right) = \rme^{i\delta_l(k)}\mathrm{sin}(\delta_l(k))$ are the so-called partial wave scattering matrix element and partial wave transition matrix element, respectively. The total cross section for distinguishable particles is given by \cite{shankar1995}
\begin{equation}
\begin{split}
\sigma(k) &= \int \rmd\Omega\, \vert f(k, \theta)\vert^2 = \frac{4\pi}{k^2}\sum_{l=0}^\infty (2l+1)\mathrm{sin}^2(\delta_l(k)) = \sum_{l=0}^\infty \sigma_l\,,
\end{split}
\end{equation}
where we used the orthogonality relations for the Legendre polynomials ensuring that the cross terms involving different components of $l$ vanish. For identical particles we need a totally symmetrized asymptotic wave function for bosons and an antisymmetrized one for fermions. Thus, we have
\begin{equation}
\Psi(r, \theta) = \frac{\rme^{i\bd{kr}} + \zeta\rme^{-i\bd{kr}}}{\sqrt{2}} + \frac{f(k, \theta) + \zeta f(k, \pi-\theta)}{\sqrt{2}}\frac{\rme^{ikr}}{r}\,,
\end{equation}
where $\zeta = -1$ for spinless fermions and $\zeta = +1$ for spinless bosons. Thus, the differential cross section for indistinguishable particles reads
\begin{equation}
\frac{\rmd\sigma}{\rmd\Omega} = \vert f(k, \theta) + \zeta f(k, \pi-\theta)\vert^2, \quad 0\leq\theta \leq \pi/2\,.
\end{equation}
Since $P_l(-x) = (-1)^{l}P_l(x)$, the sum over all $l$ in the partial wave expansion reduces to a sum over even $l$ for bosons and a sum over odd $l$ for fermions. We eventually obtain for the total cross sections \cite{friedrichs15}
\begin{align}
\mathrm{bosons:}&\quad\sigma(k) = \frac{8\pi}{k^2}\sum_{l\,\mathrm{even}}(2l+1)\mathrm{sin}^2(\delta_l(k))\label{sigma_bosons}\\\
\mathrm{fermions:}&\quad\sigma(k) = \frac{8\pi}{k^2}\sum_{l\,\mathrm{odd}}(2l+1)\mathrm{sin}^2(\delta_l(k))\,.\label{sigma_fermions}
\end{align}
From Eq.~\eqref{sigma_fermions} follows directly that there is no s-wave scattering for fermions. 

By comparing Eq.~(\ref{Psi_Rl}) and Eq.~(\ref{Psi_sca}) we can read off the radial part of the wave function, which is given by 
\begin{equation}
\begin{split}
R_l(r) &= j_l(kr) + h_l^{(+)}(kr)\rme^{i\delta_l(k)}\mathrm{sin}(\delta_l(k))\\\
&= \rme^{i\delta_l(k)}\left[\mathrm{cos}(\delta_l(k))j_l(kr) + \mathrm{sin}(\delta_l(k))n_l(kr) \right]\,,
\end{split}
\end{equation}
and in the asymptotic limit $kr\gg 1$ it reduces to
\begin{equation}\label{eq:R_L}
\begin{split}
R_l(r) &\approx \rme^{i\delta_l(k)}\frac{1}{kr}\left[\mathrm{cos}(\delta_l(k))\mathrm{sin}(kr-l\pi/2) + \mathrm{sin}(\delta_l(k))\mathrm{cos}(kr-l\pi/2)\right]\\\
&= \rme^{i\delta_l(k)}\frac{1}{kr}\mathrm{sin}\left(kr-l\pi/2 +\delta_l(k)\right) \,.
\end{split}
\end{equation}
Thus, in the far-field, $kr\gg 1$, the scattered wave differs from the incoming plane wave $\propto j_l(kr)=\mathrm{sin}(kr-l\pi/2)/(kr)$ by a phase $\delta_l(k)$.

\subsubsection{Low-energy limit}

In the following, we discuss the scattering of particles in the low-energy limit, which is the regime applying to the dilute Bose gases, where it is assumed that the distance between the particles is much larger than the range of the potential denoted by $r_0$, as previously (c.f. Sec.~\ref{subsec:S-wave_gases}) .
For low enough velocities, the de-Broglie wavelength of the particles is large compared to the scattering volume, described by the radius $r_0$, outside which the effect of the potential $V(r)$ is negligible. Thus, this limit is described by $kr_0 \ll1$. We can then neglect the energy $E$ in the Schrödinger equation and are left with \cite{landau1974}
\begin{equation}
\left[\frac{\rmd^2}{\rmd r^2} + \frac{2}{r}\frac{\rmd}{\rmd r} - \frac{l(l+1)}{r^2} - \frac{2m}{\hbar^2}V(r)\right]R_l(r) = 0\,.
\end{equation}
Moreover, in the intermediate region $r_0\ll r\ll 1/k$, we can also neglect the term $\propto V(r)$ and obtain
\begin{equation}
\left[\frac{\rmd^2}{\rmd r^2} + \frac{2}{r}\frac{\rmd}{\rmd r} - \frac{l(l+1)}{r^2}\right]R_l(r) = 0\,.
\label{SG_2}
\end{equation}
However, for radii $r \sim 1/k$, the energy term needs to be included, but $V(r)$ can still be neglected, leading to the well-known Schrödinger equation for a free particle
\begin{equation}
\left[\frac{\rmd^2}{\rmd r^2} + \frac{2}{r}\frac{\rmd}{\rmd r} - \frac{l(l+1)}{r^2} + k^2 \right]R_l(r) = 0\,.
\label{SG_3}
\end{equation}
A solution of Eq.~(\ref{SG_3}) taking into account that it should match the solution of Eq.~(\ref{SG_2}) at the boundary is derived in §33 and §130 of Ref.~\cite{landau1974}. For small momenta, it follows that $\delta_l(k) \propto k^{2l+1}$ \cite{landau1974}. Thus, we obtain for the partial cross section for bosons in Eq.~(\ref{sigma_bosons})
\begin{equation}
\sigma_l(k) \propto k^{4l}\,,
\end{equation}
which goes to zero for small $k$ except for $l=0$. Moreover, the partial scattering amplitudes for small $k$ are given by
\begin{equation}
f_l(k) \approx \delta_l(k)/k \propto k^{2l}
\end{equation}
which means that the partial scattering amplitude with $l=0$ is large compared to those with $l\neq 0$. Neglecting all $f_{l\neq 0}$ is called s-wave scattering approximation. Then, the scattering amplitude $f(\theta, k)$ (see Eq.~(\ref{scatt_ampl})) is given by
\begin{equation}
f(\theta, k) \approx f_0(k) = \frac{\delta_0}{k} = -a\,,
\label{defaf}
\end{equation}
where the scattering length $a$ is defined through
\begin{equation}
a = - \lim \limits_{k\to 0}\frac{\mathrm{tan}(\delta_0)}{k}\,.
\label{defadelta}
\end{equation}
Thus, the cross section for bosons reduces to $\sigma = 8\pi a^2$ in the limit $k\to 0$.
Since $P_{l=0}(\mathrm{cos}(\theta)) = 1$, the low-energy scattering is isotropic and the cross section is independent of the energy. However, it is important to recall that the entire derivation of this low-energy limit is based on the assumption that the potential decreases sufficiently fast at large distances. A more detailed discussion of this issue is presented in §130 of Ref.~\cite{landau1974}. 

\chapter{Functional and ground state energy for the dilute Bose gas}\label{app:Fdil}
In this section we derive the functional $\mathcal{F}(\thickbar{\bd{n}})$ for the dilute Bose gas in 3D and use this result to obtain the well-known expression for the ground state energy.

Let us first emphasize that replacing already in Eq.~\eqref{FBog} all Fourier coefficients $W_{\bd{p}}$ by $W_{\mathbf{0}}$
would make the respective sum divergent. Instead, we rewrite \eqref{FBog} as
\begin{eqnarray}
\mathcal{F}(\thickbar{\bd{n}}) &=& n\sum_{\bd{p}\neq \bd{0}}W_{\bd{p}}\left(\thickbar{n}_{\bd{p}}-\sqrt{\thickbar{n}_{\bd{p}}(\thickbar{n}_{\bd{p}}+1)} + \frac{nW_{\bd{p}}m}{p^2} \right) \nonumber \\
&& \quad - \sum_{\bd{p}\neq \bd{0}}\frac{n^2W_{\bd{p}}^2m}{p^2}
\end{eqnarray}
since then one is allowed to replace $W_{\bd{p}}$ by $W_{\mathbf{0}}$ in the first term. This yields (also replacing the sum by an integral)
\begin{eqnarray}\label{Fphysnp}
\mathcal{F}(\thickbar{\bd{n}}) &=& \frac{V}{2\pi^2}n\int_0^\infty\rmd p\,p^2 W_{\mathbf{0}} \left(\thickbar{n}_{\bd{p}}-\sqrt{\thickbar{n}_{\bd{p}}(\thickbar{n}_{\bd{p}}+1)}+ \frac{nW_{\mathbf{0}}m}{p^2} \right)\nonumber \\
&& \quad - \sum_{\bd{p}\neq \bd{0}}\frac{n^2W_{\bd{p}}^2m}{p^2}\\
&=& \frac{V}{2\pi^2}(2m)^{3/2}\frac{2}{3}\sqrt{2}(nW_{\mathbf{0}})^{5/2} - \sum_{\bd{p}\neq \bd{0}}\frac{n^2W_{\bd{p}}^2m}{p^2}\,. \nonumber
\end{eqnarray}
The second term can be rewritten in terms of $a_1$ given by Eq.~\eqref{a0}. Including also the constant term which we neglected so far and replacing $W_{\mathbf{0}}$ by $a_0$ through Eq.~\eqref{a0} yields
\begin{eqnarray}
\mathcal{F}(\thickbar{\bd{n}}) &=& \frac{nN2\pi}{m}a_0 + \frac{nN\sqrt{\pi}}{m}\frac{128}{3}a_0(na_0^3)^{1/2} + \frac{4\pi nN}{m}a_1\nonumber \\
&=& \frac{2\pi nN}{m}\left(a_0 + \frac{64}{3\sqrt{\pi}}a_0(na_0^3)^{1/2} +2a_1\right)\,.
\end{eqnarray}

As a consistency test we start now from Eq.~\eqref{Fphysnp} and add the kinetic energy.
The second term in Eq.~\eqref{Fphysnp} can be split into two parts such that it cancels the divergence in the integral for the kinetic energy as follows:
\begin{eqnarray}
E_0 &=& \sum_{\bd{p}\neq 0}\frac{p^2}{2m}n_{\bd{p}} + \mathcal{F}(\thickbar{\bd{n}})\\
&=& \frac{nNW_{\mathbf{0}}}{2} + \frac{V}{2\pi^2}(2m)^{3/2}\frac{2}{3}\sqrt{2}(nW_{\mathbf{0}})^{5/2} \nonumber \\
&&\quad+ \frac{V}{2\pi^2}\int_0^\infty\rmd p\,p^2\left(\frac{p^2}{2m}n_{\bd{p}} - \frac{n^2W_{\mathbf{0}}^2m}{2p^2}\right) - \sum_{\bd{p}\neq 0}\frac{n^2W_{\bd{p}}^2m}{2p^2}\nonumber \\
&=& \frac{nNW_{\mathbf{0}}}{2} +\frac{V}{2\pi^2}(2m)^{3/2}\frac{4}{15}\sqrt{2}(nW_{\mathbf{0}})^{5/2} - \sum_{\bd{p}\neq 0}\frac{n^2W_{\bd{p}}^2m}{2p^2}\,. \nonumber
\end{eqnarray}
Inserting $a_0$ and $a_1$ leads to the ground state energy
\begin{equation}
E_0 = \frac{4\pi Nn}{2m}(a_0+a_1) + \frac{4\pi Nn}{2m}a_0\frac{128}{15\sqrt{\pi}}(na_0^3)^{1/2}
\end{equation}
which is in agreement with Ref.~\cite{Brueckner57}.

\chapter{BEC force for the dilute Bose gas}\label{app:forcedil}
We calculate the derivative of $\mathcal{F}$ with respect to the distance along a straight path denoted by $s$ towards complete BEC starting at the occupation number vector $\thickbar{\bd{n}}$. Then, $\thickbar{n}_{\bd{p}}(t) = \thickbar{n}_{\bd{p}}(1-t)$ and for $t\approx 1$ or equivalently $D(t)\ll 1$ we can approximate
\begin{eqnarray}\label{dFsum}
\left.\frac{\rmd \mathcal{F}(\thickbar{\bd{n}})}{\rmd D}\right\vert_s &=& \frac{1}{D(t)} \sum_{\bd{p}\neq \bd{0}}nW_{\bd{p}}\thickbar{n}_{\bd{p}}(t)\left(1 - \frac{2\thickbar{n}_{\bd{p}}(t)+1}{2\sqrt{\thickbar{n}_{\bd{p}}(t)(\thickbar{n}_{\bd{p}}(t)+1)}}\right)\nonumber \\
&\approx& -\left(\frac{n}{2\sqrt{D(0)}}\sum_{\bd{p}\neq \bd{0}}W_{\bd{p}}\sqrt{\thickbar{n}_{\bd{p}}}\right)\frac{1}{\sqrt{D(t)}}\,.
\end{eqnarray}
The summation in Eq.~\eqref{dFsum} can be replaced by an integral ($\sum_{\bd{p}}\to \frac{V}{(2\pi)^3}\int \rmd^3\bd{p}$) in the thermodynamic limit where $N\to \infty$, $V\to\infty$ and $n=N/V=\mathrm{cst.}$. To evaluate the integral over the momentum $\bd{p}$ we rewrite Eq.~\eqref{dFsum} as follows:
\begin{equation}\label{F_nbar_a1}
\begin{split}
\left.\frac{\rmd \mathcal{F}(\thickbar{\bd{n}})}{\rmd D}\right\vert_s &\approx -\frac{n}{4\pi^2\sqrt{D(0)}}\int_0^\infty\rmd p\,p^2 \left(W_{\bd{p}}\sqrt{\thickbar{n}_{\bd{p}}} \right.\left.- \frac{(nW_{\bd{p}})^2m}{p^2}\right)\frac{1}{\sqrt{D(t)}}\\\
&\quad- \frac{1}{2\sqrt{D(0)}}\sum_{\bd{p}\neq \bd{0}}\frac{(nW_{\bd{p}})^2m}{p^2}\frac{1}{\sqrt{D(t)}}
\end{split}
\end{equation}
such that the integral over $p$ is converging after replacing $W_{\bd{p}}$ by the constant value $W_{\mathbf{0}}$. The first two terms in the Born series for the scattering length $a$ for identical particles are given by Eq.~\eqref{a0}. Thus, the summation in the second line of Eq.~\eqref{F_nbar_a1} can be identified with $a_1$ and the result of the integration in the first line will depend on $W_{\mathbf{0}}$ which can be replaced by $a_0$ through Eq.~\eqref{a0}. Since the integral can only be evaluated numerically we define a positive constant $\eta(a_0, n, m)$ for its value and obtain for the derivative of $\mathcal{F}$ along the path $s$:
\begin{equation}\label{F_nbar_result}
\left.\frac{\rmd \mathcal{F}(\thickbar{\bd{n}})}{\rmd D}\right\vert_s\approx \frac{N\eta(a_0, n, m)}{\sqrt{D(t)}} + \frac{2\pi nNa_1}{m\sqrt{D(0)}}\frac{1}{\sqrt{D(t)}}\,.
\end{equation}

\chapter{Fourier coefficients of charged Bose gas in 3D \label{app:Fouriercoeff}}

In this section, we derive the Fourier coefficients of the charged Bose gas in an oppositely charged uniform background ensuring total charge neutrality. We consider a system of $N$ positively charged bosons in a volume $V$. The charge density of the background is homogeneous and given by $-en=-eN/V=\mathrm{cst.}$. Then, the interaction $\hat{W}$ splits into three terms,
\begin{equation}
\hat{W} = \hat{W}_{\mathrm{pp}} + \hat{W}_\mathrm{pb} + \hat{W}_\mathrm{bb}\,,
\end{equation}
where $\hat{W}_\mathrm{pp}$ describes the interaction between the charged bosons, $\hat{W}_\mathrm{pb}$ the interaction between the bosons and the background charge and $\hat{W}_\mathrm{bb}$ the interaction between background particles. To avoid divergences in the Fourier transform of the long-range Coulomb potential, we introduce a cutoff $\mu>0$ such that $V(r)\propto \rme^{-\mu r}/r$. Later, $\mu$ will be sent to zero. Denoting by $\bd{x}_j$ the position of boson j, for the interaction between the bosons, we have 
\begin{equation}
\hat{W}_\mathrm{pp}= \frac{e^2}{2}\sum_{\substack{i, j=1\\i\neq j}}^N \frac{\rme^{-\mu|\bd{x}_i - \bd{x}_j|}}{|\bd{x}_i - \bd{x}_j|}\,.
\end{equation}
The background-background interaction is given by 
\begin{equation}
\begin{split}
\hat{W}_{\mathrm{bb}} &= \frac{e^2}{2}\int\rmd^3\bd{x}\rmd^3\bd{y}\, \frac{n(\bd{x})n(\bd{y})\rme^{-\mu|\bd{x} - \bd{y}|}}{|\bd{x}-\bd{y}|} \\\
&= \frac{4 \pi e^2 N^2}{2V\mu^2}\,,
\end{split}
\end{equation}
where we used that the densities $n(\bd{x})$ are independent of position and constant, i.e.~$n(\bd{x})=n$. Similarly, we find for the boson-background interaction
\begin{equation}
\begin{split}
\hat{W}_\mathrm{pb} &= -e^2\sum_{i=1}^N\int\rmd^3\bd{y} \frac{n(\bd{y})\rme^{-\mu|\bd{x}_i - \bd{y}|}}{|\bd{x}_i-\bd{y}|}\\\
&= -\frac{4 \pi e^2 N^2}{V\mu^2}\,.
\end{split}
\end{equation}
Hence, the sum of $\hat{W}_\mathrm{pb}$ and $\hat{W}_\mathrm{bb}$ reduces to $\hat{W}_\mathrm{pb} + \hat{W}_\mathrm{bb} = -4\pi e^2N^2/2V\mu^2$. In the next step, we express $\hat{W}_\mathrm{pp}$ in its second quantized form in momentum representation
\begin{equation}
\hat{W}_\mathrm{pp} = \frac{1}{2V}\sum_{\bd p, \bd k, \bd q}\frac{4\pi e^2}{q^2 + \mu^2}\hat{a}_{\bd p + \bd q}^\dagger\hat{a}_{\bd k - \bd q}^\dagger\hat{a}_{\bd p}\hat{a}_{\bd{k}}
\end{equation}
and split the summation over $\bd q$ into $\bd q\neq\bd 0$ and $\bd q =\bd 0$. Using the bosonic commutation relation $[\hat{a}_{\bd p}, \hat{a}_{\bd k}^\dagger] = \delta_{\bd p, \bd k}$, we finally obtain
\begin{equation}\label{Wpp}
\begin{split}
\hat{W}_\mathrm{pp} &= \frac{1}{2V}\sum_{\substack{\bd p, \bd k,\\ \bd q\neq \bd 0}}\frac{4\pi e^2}{q^2 + \mu^2}\hat{a}_{\bd p + \bd q}^\dagger\hat{a}_{\bd k - \bd q}^\dagger\hat{a}_{\bd p}\hat{a}_{\bd{k}} + \frac{4\pi e^2}{2V\mu^2}\left(\hat{N}^2 - \hat{N}\right)\,.
\end{split}
\end{equation}
Since we are working with a fixed total number of bosons, we replace the operators $\hat{N}$ by the c-number $N$. 
Further, we are interested in the thermodynamic limit, where $N\to \infty$, $V\to\infty$ and $n=N/V=\mathrm{cst.}$, such that $N(N-1)/V\approx N^2/V$. Thus, the second term in Eq.~\eqref{Wpp} cancels exactly with $\hat{W}_\mathrm{pb} + \hat{W}_\mathrm{bb}$ and we are left with the first term of Eq.~\ref{Wpp}, where $\bd{q}\neq \bd 0$. After taking the limit $\mu\to 0$, the resulting interaction $\hat{W}$ is equivalent to 
\begin{equation}
\hat{W} = \frac{1}{2V}\sum_{\bd p, \bd k, \bd q}W_{\bd q}\hat{a}_{\bd p + \bd q}^\dagger\hat{a}_{\bd k - \bd q}^\dagger\hat{a}_{\bd p}\hat{a}_{\bd{k}}
\end{equation}
with the Fourier coefficients
\begin{equation}
W_{\bd 0}=0\,,\quad W_{\bd p}=\frac{4\pi e^2}{p^2}\text{ }\forall\bd p\neq \bd 0\,.
\end{equation}

\chapter{Universal functional for the Bose-Hubbard model}\label{app:signs}
In this section we solve the minimization in Eq.~\eqref{Fmin} for any pair of occupation numbers $(n_1, n_2)$ for the Bose-Hubbard model with $N$ bosons on $L=5$ lattice sites and $U>0$. Since in that case the Fourier coefficients $W_p= \mathrm{sgn}(U)$ are independent of the momentum $p$, they can be pulled out of the summation over $p$. The four different combinations of the signs are $(\sigma_1, \sigma_2)=(+,+), (+, -), (-,+), (-, -)$ and the four corresponding functionals are denoted by $\mathcal{F}_{(\sigma_1, \sigma_2)}$. The functional $\mathcal{F}_{(-,-)}$  can be neglected in the following discussion since it comprises only positive terms and thus $\mathcal{F}_{(\sigma_1,\sigma_2)}  \leq \mathcal{F}_{(-,-)}$ for all $(\sigma_1,\sigma_2)$. The remaining three functionals $\mathcal{F}_{\sigma_1, \sigma_2}$ are then split into $\mathcal{F}_{(\sigma_1, \sigma_2)} = 2(\mathcal{F}^{(1)} + \mathcal{F}_{(\sigma_1, \sigma_2)}^{(2)})/L$ where $\mathcal{F}^{(1)}$ is independent of the choice of signs $(\sigma_1, \sigma_2)$. Therefore, to find the minimizing configuration for any $\bd{n} \in \triangle$ we only have to compare
\begin{equation}
\begin{split}
\mathcal{F}_{(+,+)}^{(2)} &= -\sum_{\nu=1}^2\left(n_{\mathbf{0}} - \sum_{\nu=1}^2\sqrt{n_\nu(n_\nu+1)}\right)\sqrt{n_\nu(n_\nu+1)}\\\
\mathcal{F}_{(+,-)}^{(2)} &= -\left(n_{\mathbf{0}}- \sqrt{n_1(n_1+1)}+\sqrt{n_2(n_2+1)}\right)\\\
&\quad\times\left(\sqrt{n_1(n_1+1)} - \sqrt{n_2(n_2+1)}\right)
\end{split}
\end{equation}
and the third functional $\mathcal{F}^{(2)}_{(-,+)}$ follows from $\mathcal{F}^{(2)}_{(-,+)}$ by replacing everywhere $1\leftrightarrow 2$.
The minimizing configuration $(\sigma_1, \sigma_2)$  can then easily be determined analytically leading to the cells shown in Fig.~\ref{fig:S}.
There, the black point in the middle marks the distinctive occupation number vector for which all three functionals take the same value  $\mathcal{F}_{(+,+)}=\mathcal{F}_{(-,+)}=\mathcal{F}_{(+,-)}$. It is given by
\begin{equation}
\tilde{n}\equiv n_1=n_2=\frac{1}{6}\left(1+2N-\sqrt{1+N(4+N)}\right) \,.
\end{equation}
The border between regions $(-,+)$ and $(+,-)$ is determined by
\begin{equation}
n_2=n_1 \geq \tilde{n}\,.
\end{equation}
The border separating region $(+,+)$ and $(+,-)$ is obtained from $\mathcal{F}_{(+,+)}(\bd{n})=\mathcal{F}_{(+,-)}(\bd{n})$, leading to
\begin{equation}
n_2 = \frac{1}{2}\left(N - 2\left(n_1+\sqrt{n_1(n_1+1)}\right)\right)\,,\quad n_1\geq \tilde{n}\,.
\end{equation}
The solution for $\mathcal{F}_{(+,+)}=\mathcal{F}_{(-,+)}$ is obtained by exchanging the two occupation numbers $n_1$ and $n_2$ in the result for $\mathcal{F}_{(+,+)}=\mathcal{F}_{(+,-)}$.

\end{appendices}

\printbibliography

\cleardoublepage
\pagestyle{plain}

\chapter*{Acknowledgments}

First of all, I wish to express my deepest gratitude to my supervisor Dr.~Christian Schilling for proposing this fascinating project to me and for his continuous support throughout this thesis. 
I am very thankful to Dr.~Federico Castillo and Dr.~Jean-Philippe Labb\'e for the fruitful collaboration on RDMFT for excited states.
Moreover, I am very grateful to all group members for many stimulating discussions during the last year and the nice and productive working atmosphere. 
In addition, I would like to thank my family and friends for their constant support during my studies. 

\cleardoublepage
\pagestyle{plain}
\noindent Erklärung:
\vspace{1cm}

\noindent Hiermit erkläre ich, die vorliegende Arbeit selbständig verfasst zu haben und keine anderen als die in der Arbeit angegebenen Quellen und Hilfsmittel benutzt zu haben.
\vspace{3cm}

\noindent München, 15.02.2021
\vspace{2cm}

\noindent Julia Liebert

\end{document}